\documentclass[a4paper,fleqn,usenatbib]{mnras}

\usepackage{graphicx}
\usepackage{rotating}
\usepackage{float}
\usepackage{aecompl}
\usepackage{color}
\usepackage{pdflscape}
\title[Infall and Outflow Motions in Massive Star Forming Regions]{Infall and Outflow Motions towards a Sample of Massive Star Forming Regions from the RMS Survey}

\author[N. Cunningham et al.]{
N. Cunningham,$^{1,2,3}${\thanks{E-mail: cunningham@iram.fr}}
S L. Lumsden,$^{3}$
T. J. T. Moore,$^{4}$
L T. Maud,$^{5}$
I. Mendigut\'\i{}a$^{6,3}$\\
$^{1}$Institut de Radioastronomie Millimetrique (IRAM), 300 rue de la Piscine, 38406 Saint Martin d'H$\grave{e}$res, France\\
$^{2}$Green Bank Observatory, 155 Observatory Rd, P.O. Box 2, Green Bank, WV, 24944, USA\\
$^{3}$School of Physics and Astronomy, University of Leeds, LS2 9JT, UK \\
$^{4}$Astrophysics Research Institute, Liverpool John Moores University, IC2, Liverpool Science Park, 146 Brownlow Hill, Liverpool L3 5RF, UK\\
$^{5}$Leiden Observatory, Leiden University, PO Box 9513, 2300 RA Leiden, The Netherlands\\
$^{6}$Centro de Astrobiolog\'{\i}a, Departamento de Astrof\'{\i}sica (CSIC-INTA), ESA-ESAC Campus, P.O. Box 78, 28691\\ Villanueva de la
Ca$\tilde{n}$ada, Madrid, Spain.}

\date{Accepted XXX. Received YYY; in original form ZZZ}

\pubyear{2018}

\begin{document}
\label{firstpage}
\pagerange{\pageref{firstpage}--\pageref{lastpage}}
\maketitle

\begin{abstract}
We present the results of an outflow and infall survey towards a distance limited sample of 31 massive star forming regions drawn from the RMS survey. The presence of young, active outflows is identified from SiO\,(8-7) emission and the infall dynamics are explored using HCO$^+$/H$^{13}$CO$^+$\,(4-3) emission. We investigate if the infall and outflow parameters vary with source properties, exploring whether regions hosting potentially young active outflows show similarities or differences with regions harbouring more evolved, possibly momentum driven, ``fossil" outflows. SiO emission is detected towards approximately 46\% of the sources. When considering sources with and without an SiO detection (i.e. potentially active and fossil outflows respectively), only the $^{12}$CO outflow velocity shows a significant difference between samples, indicating SiO is more prevalent towards sources with higher outflow velocities. Furthermore, we find the SiO luminosity increases as a function of the {\em Herschel} 70\,$\mu$m to {\em WISE} 22$\mu$m flux ratio, suggesting the production of SiO is prevalent in younger, more embedded regions. Similarly, we find tentative evidence that sources with an SiO detection have a smaller bolometric luminosity-to-mass ratio, indicating SiO (8-7) emission is associated with potentially younger regions.  We do not find a prevalence towards sources displaying signatures of infall in our sample. However, the higher energy HCO$^+$ transitions may not be the best suited tracer of infall at this spatial resolution in these regions.

\end{abstract}


%%%%%%%%%%%%%%%%%%%%%%%%%%%%%%%%%%%%%%%%%%%%%%%%%%

%%%%%%%%%%%%%%%%% BODY OF PAPER %%%%%%%%%%%%%%%%%%
\newcommand{\htco}{H$^{13}$CO$^{+}$\,}
\newcommand{\hco}{HCO$^{+}$\,}
\def\kms{\hbox{${\rm\thinspace km\,s^{-1}}$\,}}
\section{Introduction}

Infall and outflow motions are an important part of the star-formation process. However, a comprehensive understanding of both processes, particularly towards massive star forming regions, is still lacking. This is due, in part, to the larger distances and typically more clustered and complex nature of such regions, making it difficult to disentangle the infall and outflow properties of individual objects in a given cluster. 

Observationally, young stellar objects (YSOs) of all masses are known to drive bipolar molecular outflows and SiO emission has been effectively used to detect outflows driven by low (M$_{sun}<$2\,M$_{*}$), intermediate (2\,M$_{*}<$M$_{sun}<$8\,M$_{*}$) and high-mass (M$_{sun}>$8\,M$_{*}$) stars (e.g., \citealt{Gibb2004}; \citealt{Gibb2007}; \citealt{DuarteCabral2014}; \citealt{Klaassen2012}; \citealt{Cunningham2016}). The passage of fast shocks are required to disrupt and release SiO from the solid grains into the gas phase (e.g., \citealt{Gusdorf2008b}; \citealt{Guillet2009}; \citealt{Schilke1997}; \citealt{Flower2012}). Thus, SiO emission, particularly the higher energy transitions, is likely to be an excellent tracer of an active outflow located close to the stellar driving source. \citet{Gibb2004} found SiO emission was preferentially detected towards Class 0 sources in their sample of low-mass stars. Furthermore, those sources with an SiO detection were associated with higher outflow velocities and higher densities, suggesting shock velocity and ambient density are likely to play an important role in the production of SiO in the early stages of low-mass star-formation. \citet{Bontemps1996} observed more powerful outflows to be associated with Class 0 sources in their sample of 45 embedded YSOs. Similarly, a decrease of the outflow force with source evolution was observed by \citet{Mottram2017} towards a sample of Class 0 and Class I sources. In the high-mass regime, \citet{Gibb2007} found SiO emission was preferentially detected towards sources with higher outflow velocities, but were unable to establish the evolutionary nature of individual sources. Further work by \citet{Klaassen2012} found an increase in the integrated intensity of the SiO emission with evolutionary stage, contrary to the observations in the low-mass regime, detecting both infall and outflow signatures towards ultra-compact H{\scriptsize II} (UCH{\scriptsize II}) regions. 
As CO is more readily excited in the ambient medium, it has been suggested (e.g., \citealt{Klaassen2012}; \citealt{Bally1999}) that emission from CO may potentially trace a remnant, momentum driven, outflow cavity that is no longer being actively driven by the central star. In comparison, SiO, which requires a fast shock and higher critical density to be excited, may be tracing an active outflow close to the central star. A major aim of this work is to explore systematic differences in the environment, age and evolutionary nature between massive star forming regions hosting outflows traced by both CO and SiO emission (i.e. potentially active outflows) compared with regions that have an outflow traced by CO and show no associated SiO emission (i.e. potentially momentum driven fossil outflows).

In addition, we purposely observed the dense-gas tracer \hco as a means of probing the infall dynamics in these regions. Infall is believed to form an important role in the high-mass star formation process. However, exactly how mass is accumulated on the clump/cloud scales and finally accreted onto the central cores in massive star forming regions is still unclear (e.g. see \citealt{Motte2017} for a recent review). There are two dominant theoretical scenarios for the formation of massive stars; turbulent core accretion \citet{McKeeandTan2003} and competitive accretion \citet{Bonnell2001}. In the former, the infall dynamics would likely be localized on individual core/binary type scales, whereas in the latter the cloud and high mass prototstars form simultaneously (e.g. \citealt{Tige2017}) and global collapse on clump/cloud scales is expected. In this formation scheme, the gas is likely to be channeled along converging flows onto central clouds that are undergoing global collapse on parsec scales. Several recent observations (e.g. \citealt{Peretto2014}; \citealt{Williams2018}) have observed velocity gradients along filamentary structures converging onto a central hub. In the observations presented here, we expect to probe signatures of global infall on scales of 1-2 pc, if present. 

We present the results of an \hco, \htco\,J=4-3, and SiO\,J=8-7 molecular line survey performed using the James Clerk Maxwell Telescope (JCMT) towards a sample of 33 high-mass star forming regions selected from the {\sc RMS} MSX survey \citep{Lumsden2013}. In Section 2 we summarize the observations presented in this paper. The results are presented in Section 3, the discussion in Section 4 and the main conclusions of the work are outlined in Section 5.
%%%%%%%%%%%%%%%%%%%%%%%%%%%%%%%%%%%%%%%%%%%%%%%%%%%%%%%%%%%%%%%%%%%%%%%%%%%%%%%%%%%%%%%%%%%%%%%%%%%%%%%%%%%%%%%%%%%%%%%%%%%%%%%%%%%%
%%%%%%%%%%%%%%%%%%%%%%%%%%%%%%%%%%%%%%%%%%%%%%%%%%%%%%%%%   Observations and Data Reduction%%%%%%%%%%%%%%%%%%%%%%%%%%%%%%%%%%%%%%%%
%%%%%%%%%%%%%%%%%%%%%%%%%%%%%%%%%%%%%%%%%%%%%%%%%%%%%%%%%%%%%%%%%%%%%%%%%%%%%%%%%%%%%%%%%%%%%%%%%%%%%%%%%%%%%%%%%%%%%%%%%%%%%%%%%5
\section{Sample and observations}

\subsection{Sample selection}
The sample includes 33 massive star forming regions, selected from a previous outflow survey by \citet{Maud2015outflows}, where 27 of the sources observed have an outflow detection traced by $^{12}$CO\,(3-2). For completeness, we also include 6 regions that have no confirmed $^{12}$CO\,(3-2) outflow detection in \citet{Maud2015outflows}, but have associated C$^{18}$O\,(3-2) emission (see \citealt{Maud2015cores}) and therefore retain a dense massive core. All sources are part of the RMS survey and were selected to probe both evolutionary nature and cover a range in luminosity. The sample includes; 20 YSOs, 11 compact H{\scriptsize II} and 2 H{\scriptsize II}/YSO RMS classified regions \citep{Lumsden2013}. Objects labeled as H{\scriptsize II}/YSO regions were found to display characteristics of both YSOs and compact H{\scriptsize II} regions (see \citealt{Lumsden2013} for a full discussion of the classification of RMS sources). Furthermore, the source selection was chosen to be distance limited ($<$4.5\,kpc) to minimise distance-related bias. However, since the observations were undertaken, the distances of two sources, G020.7617 and G045.0711, have been corrected. The distance to G020.7617 has been updated to the far kinematic distance of 11.8kpc, and the distance to G045.0711 has been corrected to 7.75$\pm$0.4kpc (\citealt{Wu2014}, obtained from parallax and proper motion measurements).
To keep the sample distance limited we omit these sources from the remaining analysis. Table \ref{table:sourcelist} presents the source properties taken from the RMS survey. The sources are labelled by their Galactic name (Column 1), and properties such as the RMS survey classification (e.g., YSO and H{\scriptsize II}), source V$_{\rm LSR}$, distance and bolometric luminosity are given. Where possible the IRAS name and/or more commonly used name(s) for each source are provided. 
\begin{table*}
\begin{minipage}{180mm}
\begin{small}
\begin{center}
\caption{\label{table:sourcelist} Source parameters for all objects in the sample taken from the RMS survey online archive (\url{http://rms.leeds.ac.uk/cgi-bin/public/RMS_DATABASE.cgi}).}
\begin{tabular}{l c c c c c c c}

\hline\hline
Source          &  RMS  &  RA      &   Dec          & V$_{\rm LSR}$ & Distance    & Luminosity  & IRAS/Common\\
Name            & Classification	  &  (J2000)     & (J2000)    & (km\,s$^{-1}$)    & (kpc)& (L$_{\odot}$)&    Name        \\

\hline 
& &  \multicolumn{4}{c}{CO outflow detection$^{a}$} & &  \\
%CO Outflow  $^a$       &		  &              &                 &       &           &              &                \\
\hline
G010.8411$-$02.5919  & YSO  &  18:19:12.09 &  -\,20:47:30.9    &    12.3         & 1.9  &   2.4e+04    & 18162-2048      \\
G012.9090$-$00.2607  & YSO  &  18:14:39.56 &  -\,17:52:02.3    &    36.7         & 2.4  &   3.2e+04    & 18117-1753/ W33A\\
G013.6562$-$00.5997  & YSO  &  18:17:24.38 &  -\,17:22:14.8    &    47.4         & 4.1  &   1.4e+04    & 18144-1723      \\
G017.6380$+$00.1566  & YSO  &  18:22:26.37 &  -\,13:30:12.0    &    22.1         & 2.2  &   1.0e+05    & 18196-1331      \\
G018.3412$+$01.7681  & YSO  &  18:17:58.11 &   -\,12:07:24.8   &    33.1         & 2.8  &   2.2e+04    & 18151-1208      \\
G020.7617$-$00.0638  & H{\scriptsize II}/YSO  & 18:29:12.36  &  -\,10:50:38.4 &  56.9           & 11.8$^b$ &  1.3/3.6e+04 &                  \\
G043.3061$-$00.2106$^c$  & H{\scriptsize II}  & 19:11:16.97  &  +\,09:07:28.9     &   59.6          & 4.4 & 1.1e+04     &19088+0902         \\
G045.0711$+$00.1325  & H{\scriptsize II}  & 19:13:22.10  &  +\,10:50:53.4     &   59.2          & 7.8$^b$ & 6.2e+05     &19110+1045         \\ 
G050.2213$-$00.6063  & YSO  &  19:25:57.77 &   +\,15:02:59.6    &    40.6         & 3.3  &   1.3e+04    & 19236+1456      \\
G078.1224$+$03.6320  & YSO  &  20:14:25.86 &   +\,41:13:36.3    &   -3.9          & 1.4  &   4.0e+03    & 20126+4104      \\
G079.1272$+$02.2782  & YSO  &  20:23:23.83 &   +\,41:17:39.3    &   -2.0          & 1.4  &   1.6e+03    &20216+4107       \\
G079.8749$+$01.1821  & H{\scriptsize II}  & 20:30:27.45  &  +\,41:15:58.5     &  -4.3           & 1.4  & 1.1e+03     & 20286$+$4105  \\
G081.7133$+$00.5589  & H{\scriptsize II}  & 20:39:02.36  &  +\,42:21:58.7     & -3.8   &   1.4    &  1.9e+03 &	\\
G081.7220$+$00.5699  & H{\scriptsize II}  & 20:39:01.01     & +\,42:22:50.2   &   -4.7            & 1.4  &   1.2e+04    &  DR21 OH \\
G081.7522$+$00.5906  & YSO  &  20:39:01.98 &   +\,42:24:59.1    &    -4.0         & 1.4  &   9.0e+03    &              \\
G081.7624$+$00.5916  & YSO  & 20:39:03.72  &  +\,42:25:29.6     &    -4.4         & 1.4  &   2.6e+03      &  \\
G081.8652$+$00.7800  & YSO  & 20:38:35.36	 & +\,42:37:13.7      &    9.4          & 1.4  &  3.6e+03       &  \\
G081.8789$+$00.7822  & H{\scriptsize II}  & 20:38:37.71  &  +\,42:37:58.6     &   8.1           & 1.4  &  1.1e+04    &                   \\
G083.0936$+$03.2724  & H{\scriptsize II}  & 20:31:35.44  &  +\,45:05:45.8     &  -3.1           & 1.4  &  1.2e+04    &                   \\
G083.7071$+$03.2817  & YSO  &  20:33:36.51 &   +\,45:35:44.0    &    -3.6         & 1.4  &   3.9e+03    &                 \\
G083.7962$+$03.3058  & H{\scriptsize II}  & 20:33:48.02  &  +\,45:40:54.5     &  -4.3           & 1.4  &  4.8e+03    &                   \\
G103.8744$+$01.8558  & YSO  &  22:15:09.08 &   +\,58:49:07.8        &   -18.3         & 1.6  &   6.8e+03    &22134+5834       \\
G109.8715$+$02.1156  & YSO  &  22:56:17.98 &   +\,62:01:49.7        &   -11.1         & 0.7  &   1.5e+04    &22543+6145/Cep A \\
G192.6005$-$00.0479  & YSO  &  06:12:54.01&   +\,17:59:23.1         &     7.4          & 2.0  &   4.5e+04    & 06099+1800/ S255 IR\\
G194.9349$-$01.2224  & YSO  &  06:13:16.14 &   +\,15:22:43.3        &    15.9         & 2.0  &   3.0e+03    & 06103+1523      \\
G203.3166$+$02.0564  & YSO  &  06:41:10.15 &   +\,09:29:33.6        &    7.4          & 0.7  &   1.8e+03    & 06384+0932/NGC2264-C\\
G207.2654$-$01.8080  & H{\scriptsize II}/YSO  & 06:34:37.74  &  +\,04:12:44.2     &  12.6           & 1.0  &  1.3/9.1e+03 &06319+0415        \\
\hline
 & &  \multicolumn{4}{c}{No CO outflow detection$^{a}$} &  &\\
% No CO Outflow $^a$    &     	  &              &             &           &      &              &                  \\
\hline
G080.8645$+$00.4197 & H{\scriptsize II}  & 20:36:52.16  &  +\,41:36:24.0   &  -3.1           & 1.4  &  9.1e+03    &                   \\
G080.9383$-$00.1268 & H{\scriptsize II}  & 20:39:25.91  &  +\,41:20:01.6   &  -2.0           & 1.4  &  3.2e+04    &                    \\
G081.7131$+$00.5792 & YSO  &  20:38:57.19 &   +\,42:22:40.9  &   -3.6          & 1.4  &   4.9e+03    &                 \\
G196.4542$-$01.6777 & YSO  &  06:14:37.06 &   +\,13:49:36.4 &    18.0         & 4.1$^b$  &   5.4e+04    & 06117+1350      \\
G217.3771$-$00.0828 & H{\scriptsize II}  & 06:59:15.73  &  -\,03:59:37.1   &   25.1          & 1.3  &  8.0e+03    &06567-0355         \\
G233.8306$-$00.1803 & YSO  &  07:30:16.72 &  -\,18:35:49.1   &    44.6         & 3.3  &   1.3e+04    & 07280-1829       \\
\hline
\end{tabular}
\end{center}
\end{small}

{\bf Notes}\\
{\it (a)} The CO outflow sources have either a confirmed $^{12}$CO(3-2) outflow or in the case of two sources, G017.6380 and G083.7962, show evidence of an outflow, whereas the No CO outflow sources have no observed emission consistent with an outflow in \citet{Maud2015outflows}.\\
{\it (b)} The distance to G020.7617 has been updated to the far distance since the observations were undertaken. A distance of 7.75$\pm$0.4\,kpc to G045.0711 has recently been identified through measurements of parallax and proper motions by \citet{Wu2014}. The distance to G196.4542 has been since updated to  4.05$^{+0.65}_{-0.49}$\,kpc \citep{Asaki2014}. The corrected distances for these sources are used in the remainder of the analysis. \\ 
{\it (c)} G043.3061-00.2106 was observed as part of the $^{12}$CO outflow survey \citep{Maud2015outflows}. However, as G043.3061-00.2106 was not observed in the C$^{18}$O core properties survey by \citet{Maud2015cores}, this source was subsequently excluded from the $^{12}$CO\,(3-2) outflow survey \citep{Maud2015outflows}. Inspection of the $^{12}$CO\,(3-2) data shows emission indicative of outflow motions, thus we include this source as a CO outflow candidate in this work.\\

\end{minipage}
\end{table*}
\subsection{JCMT observations}

SiO J=8-7, \htco J=4-3 and \hco J=4-3 were observed using the Heterodyne Array Receiver program (HARP) \citep{Buckle2009} at the 15\,m James Clerk Maxwell Telescope\footnote{The James Clerk Maxwell Telescope has historically been operated by the Joint Astronomy Centre on behalf of the Science and Technology Facilities Council of the United Kingdom, the National Research Council of Canada and the Netherlands Organisation for Scientific
Research.} (JCMT) as part of the projects M09AU18 (SiO J=8-7, and \htco J=4-3)
and M10AU04 (\hco\,J=4-3). Due to time limitations, only 25 sources were
observed as part of project M10AU04 (\hco J=4-3). Project M09AU18 was
observed between 12/04/2009 - 05/04/2010, and project M10AU04 between
16/04/2010 and 01/09/2010. The HARP array consists of 16 receiver elements but
during both projects receiver H14 was not operational and is subsequently
missing from the data. The observations were taken in position switched jiggle
chop mode \citep{Buckle2009}, creating $\sim$\,2\,arcminute by 2\,arcminute maps. We observed each
source for between 30-60 minutes, and the pointing was checked every hour on a
known bright molecular source and is accurate to within $\sim$5$\arcsec$. \htco
and SiO were observed simultaneously in the same frequency set-up, where the
Auto-Correlation Spectral Imaging System (ACSIS) was configured with an
operational bandwidth of 1000\,MHz$\times$2048 channels, providing a velocity
resolution of 0.42\,\kms. For \hco the bandwidth was set-up at
250\,MHz$\times$4096 channels, providing a velocity resolution of
0.05\,\kms. At the observed frequency range of $\sim$345\,GHz the JCMT has a beam size of $\sim$15$\arcsec$. The average atmospheric opacity ($\tau_{(225GHz)}$) obtained from the Caltech Submillimeter Observatory (CSO) during both sets of observations was 0.07.

The HARP/ACSIS data reduction was undertaken using the Starlink software packages \textsc{SMURF}, \textsc{KAPPA}, and \textsc{GAIA} \citep{Jenness2015}. The data were initially converted to spectral (RA-DEC-velocity) cubes using the \textsc{SMURF} command \textsc{MAKECUBE}. The data were gridded on to cubes with a pixel size of 7.5\,$\arcsec$ by 7.5\,$\arcsec$ using the function ``SincSinc", which is a weighting function using a sinc($\pi\,x$sinc$k\pi\,x$) kernel. The noisy channels at the edges of the band were removed, and a linear baseline was subtracted. The data were converted from the antenna temperature scale T$^*_A$ \citep{Kutner1981} to main-beam brightness temperature T$_{mb}$ using $T_{mb}=T^*_A/\eta_{mb}$, where the main beam efficiency $\eta_{mb}$ has a value of 0.61 \citep{Buckle2009}. To increase the signal to noise ratio of the SiO\,(8-7) line, we re-sampled the velocity resolution to 1.68\kms using the \textsc{KAPPA} command \textsc{SQORST}. The 1$\sigma$ rms T$_{\rm mb(rms)}$ per channel was determined from line free channels excluding any noisy pixels towards the edges of the map; the typical values are 0.08\,K, 0.04\,K, and 0.6\,K for \htco\,(0.42\kms), SiO\,(1.68\kms) and \hco\,(0.05\kms), respectively. 
As mentioned in the previous section, the \hco observations were not completed towards all sources in this survey; sources that were not observed are noted in Table \ref{table:detections_work}.

\subsection{Archival data}

To complement the JCMT HARP observations, we utilise archival far-infrared (IR) data. The far-IR 70$\mu$m observations, performed with the ESA \emph{Herschel Space Observatory}\footnote{{\it Herschel} is an ESA space observatory with science instruments provided by European-led Principal Investigator consortia and with important participation from NASA.} \citep{Pilbratt2010} using the PACS instrument \citep{Poglitsch2010}, were obtained from the {\em Herschel} archive in standard product generation form\footnote{http://www.cosmos.esa.int/web/herschel/science-archive}. The majority of the data were taken from the HOBYS \citep{Motte2010} or HiGal \citep{Molinari2010} surveys. Only two regions, G018.3412 and G078.1224, were not observed as part of these two surveys, and were observed under the PIs; Krauss (observation ID:1342191813) and Cesaroni (observation ID:1342211514) respectively (see Table \ref{table:detections_work} for a summary of the sources covered).

%%%%%%%%%%%%%%%%%%%%%%%%%%%%%%%%%%%%%%%%%%%%%%%%%%%%%%%%%%%%%%%%%%%%%%%%%%%%%%%%%%%%%%%%%%%%%%%%%%%%%%%%%%%%%%%%%%%%%%%%%%%%%%%%%%%%
%%%%%%%%%%%%%%%%%%%%%%%%%%%%%%%%%%%%%%%%%%%%%%%%%%%%%%%%%   RESULTS %%%%%%%%%%%%%%%%%%%%%%%%%%%%%%%%%%%%%%%%%%%%%%%%%%%%%%%%%%%%%%%
%%%%%%%%%%%%%%%%%%%%%%%%%%%%%%%%%%%%%%%%%%%%%%%%%%%%%%%%%%%%%%%%%%%%%%%%%%%%%%%%%%%%%%%%%%%%%%%%%%%%%%%%%%%%%%%%%%%%%%%%%%%%%%%%%5
\section{Results}

\subsection{Determining the source extents and properties from the \hco and \htco emission}

\begin{figure*}
\includegraphics[width=0.44\textwidth]{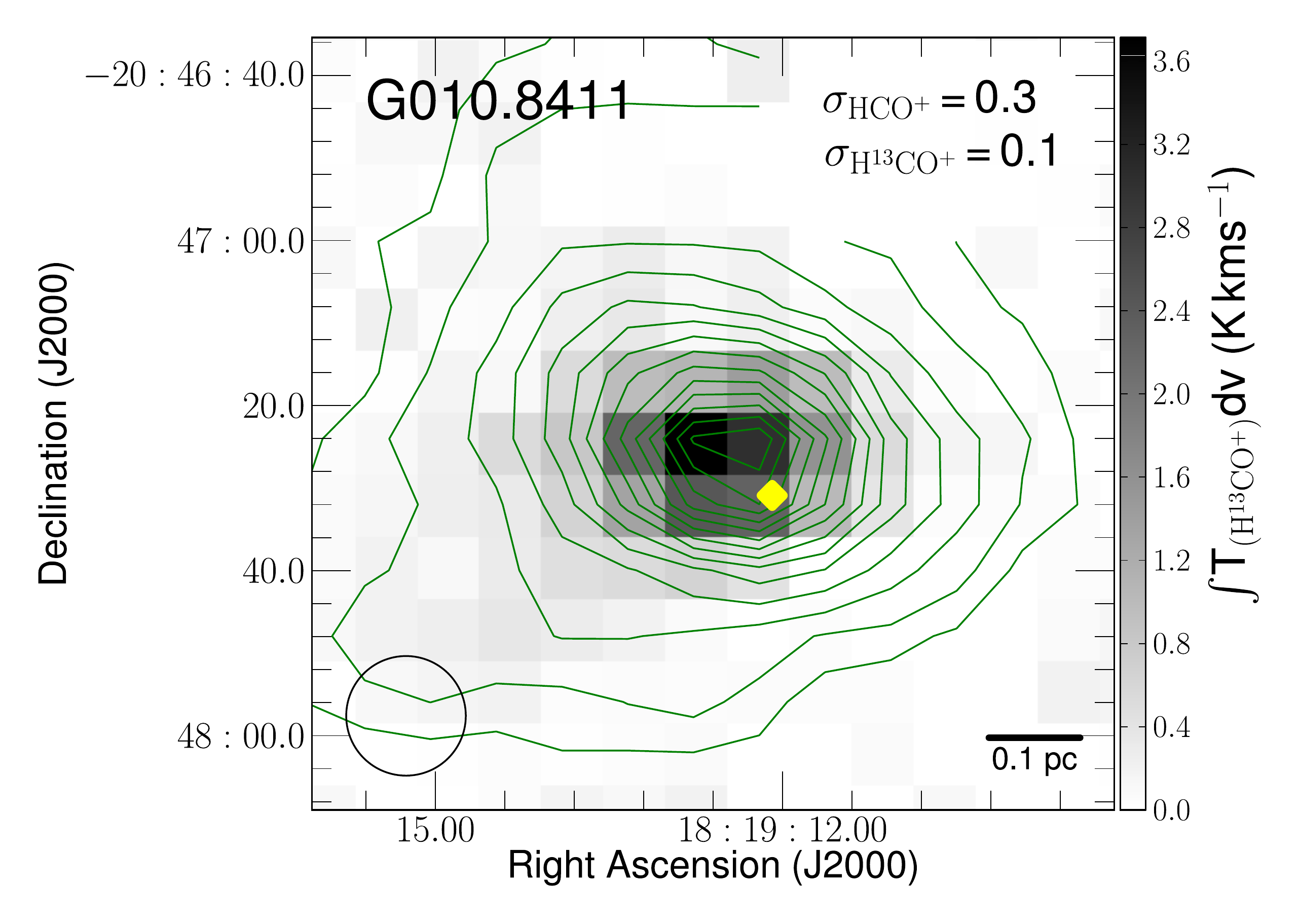}
\includegraphics[width=0.44\textwidth]{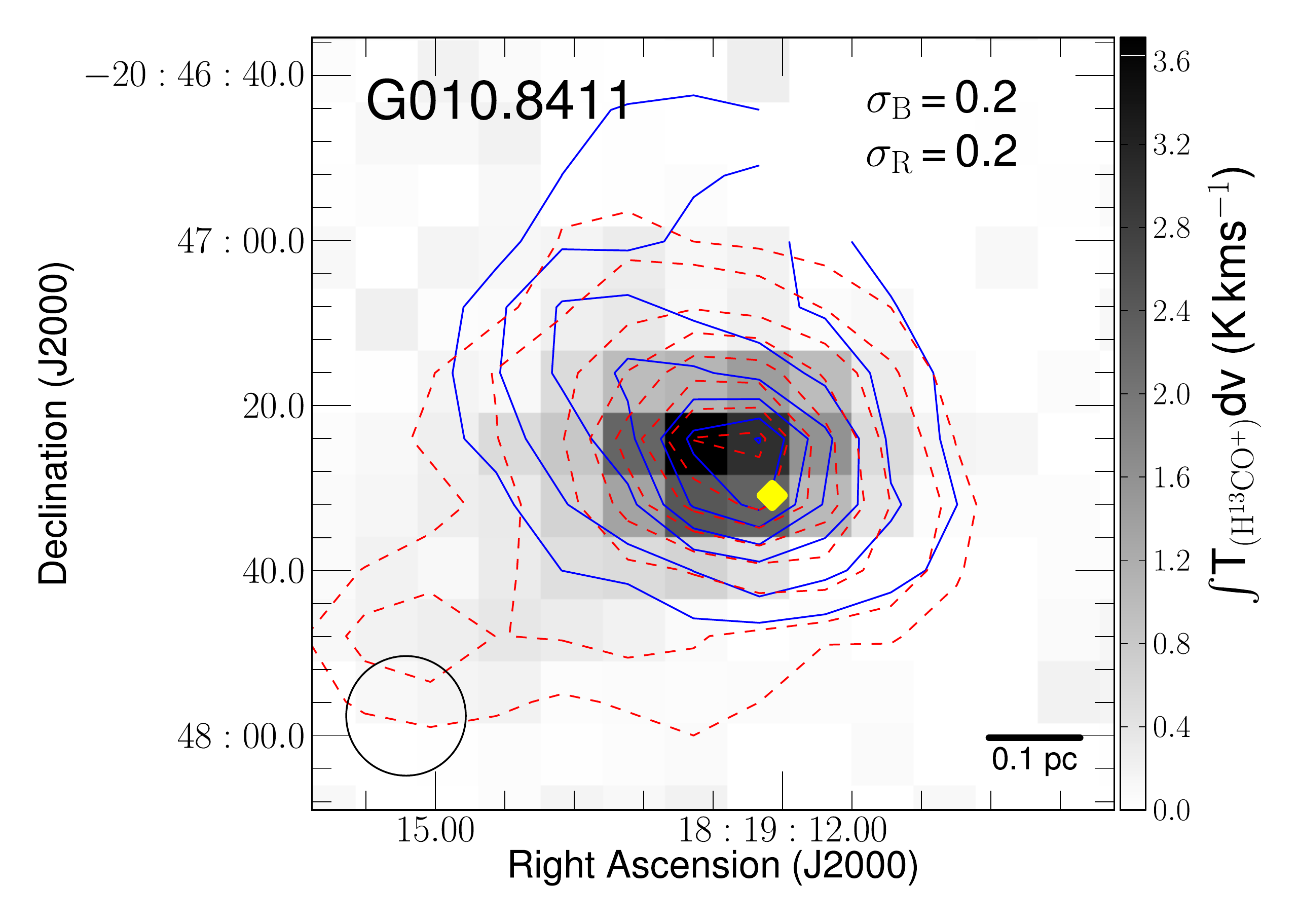}
\includegraphics[width=0.44\textwidth]{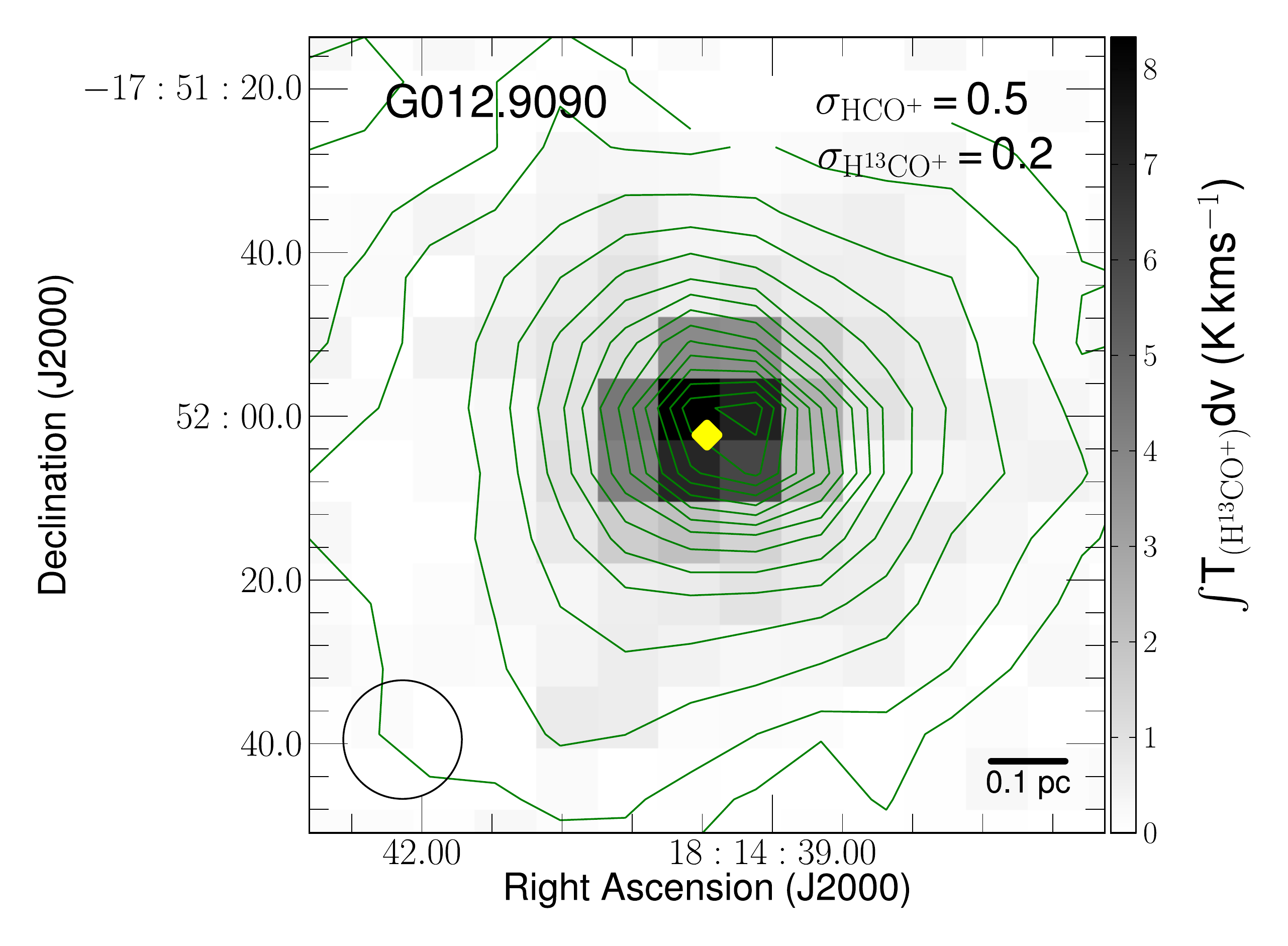}
\includegraphics[width=0.44\textwidth]{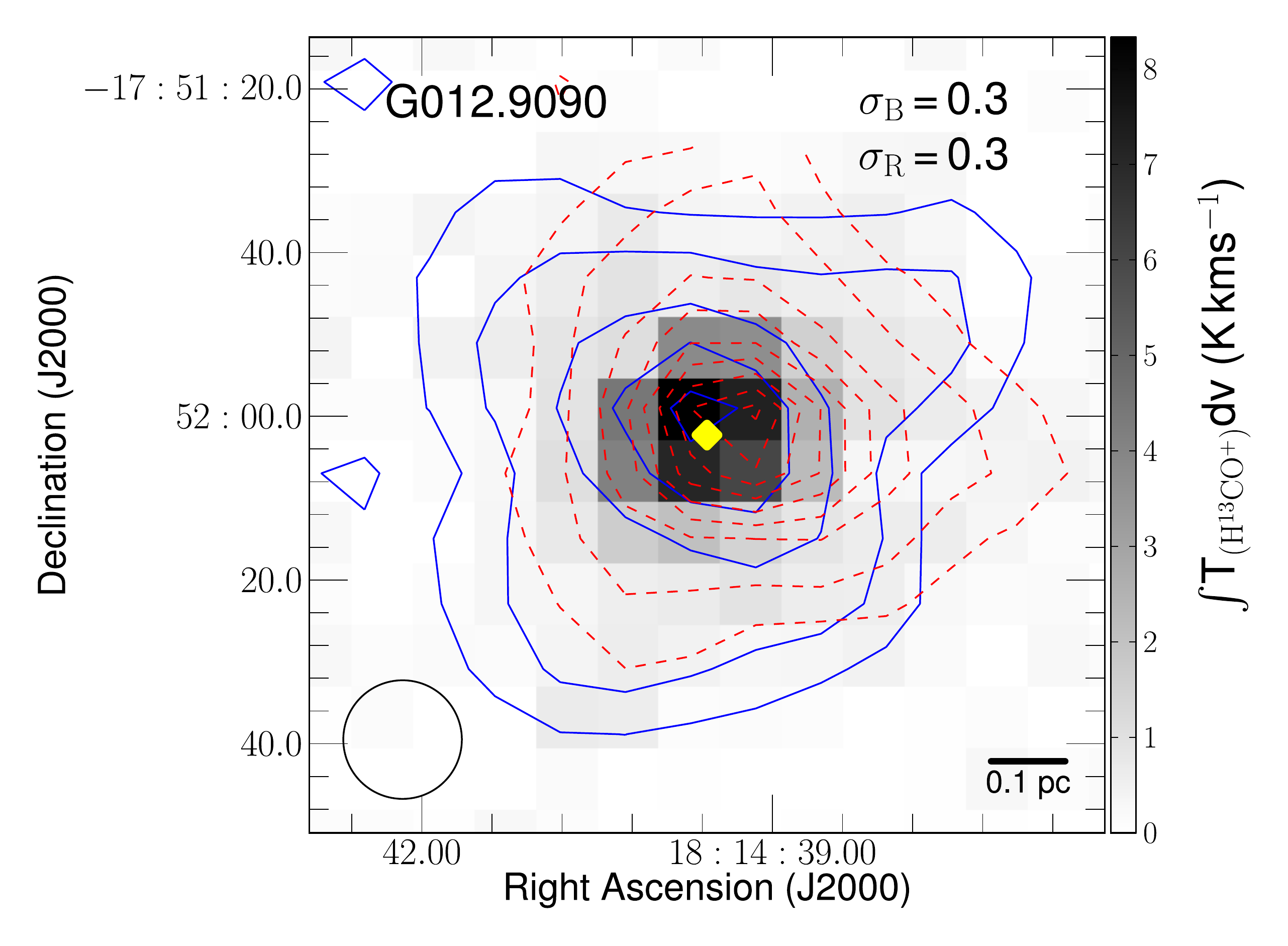}

\caption{\htco and \hco zeroth order moment maps. The \htco maps are shown in greyscale and are the total integrated emission ($\int{T_{mb}dv}$ in units of K.\kms), integrated from the minimum to maximum channels with 3$\sigma$ emission. The yellow diamonds mark the RMS source positions. The JCMT beam is shown in the bottom left corner, and the source name is shown in the top left corner. {\bf Left:} The \hco emission is overlaid in green solid contours for the total moment maps (again integrated from the minimum and maximum channels with 3$\sigma$ emission in the \hco maps) where the 1$\sigma$ rms (in units of K.\kms) for the \hco ($\sigma$$_{\rm HCO^+}$) and the \htco ($\sigma$$_{\rm H^{13}CO^+}$) integrated intensity maps are given in the top right corner. The \hco contour levels are from 1$\sigma\times$(5,10,20,... to peak in-steps of 10$\sigma$). {\bf Right}: The red- and blue-shifted \hco emission  is shown by the red (dashed) and blue (solid) contours, respectively. The blue- and red-shifted contours are taken from the minimum and maximum channels with 3$\sigma$ emission respectively, excluding the central emission which is defined by the \htco FWHM (see Table \ref{aa:tab:htcogaus} for the \htco FWHM values). The 1$\sigma$ levels for the red- ($\sigma$$_{\rm R}$) and blue-shifted ($\sigma$$_{\rm B}$) emission are given in the top right corner, where the contour levels are from 1$\sigma\times$(5,10,20,... to peak in-steps of 10$\sigma$). The velocity ranges used to integrate the HCO$^+$ emission are 9.5$-$15.6\,\kms for G010.8411, and 30.3$-$44.0\,\kms for G012.9090. The remainder of the sources are presented in the online data.} \label{figure:emission_maps_body}
\end{figure*}

\begin{figure*}
\includegraphics[width=0.21\textwidth]{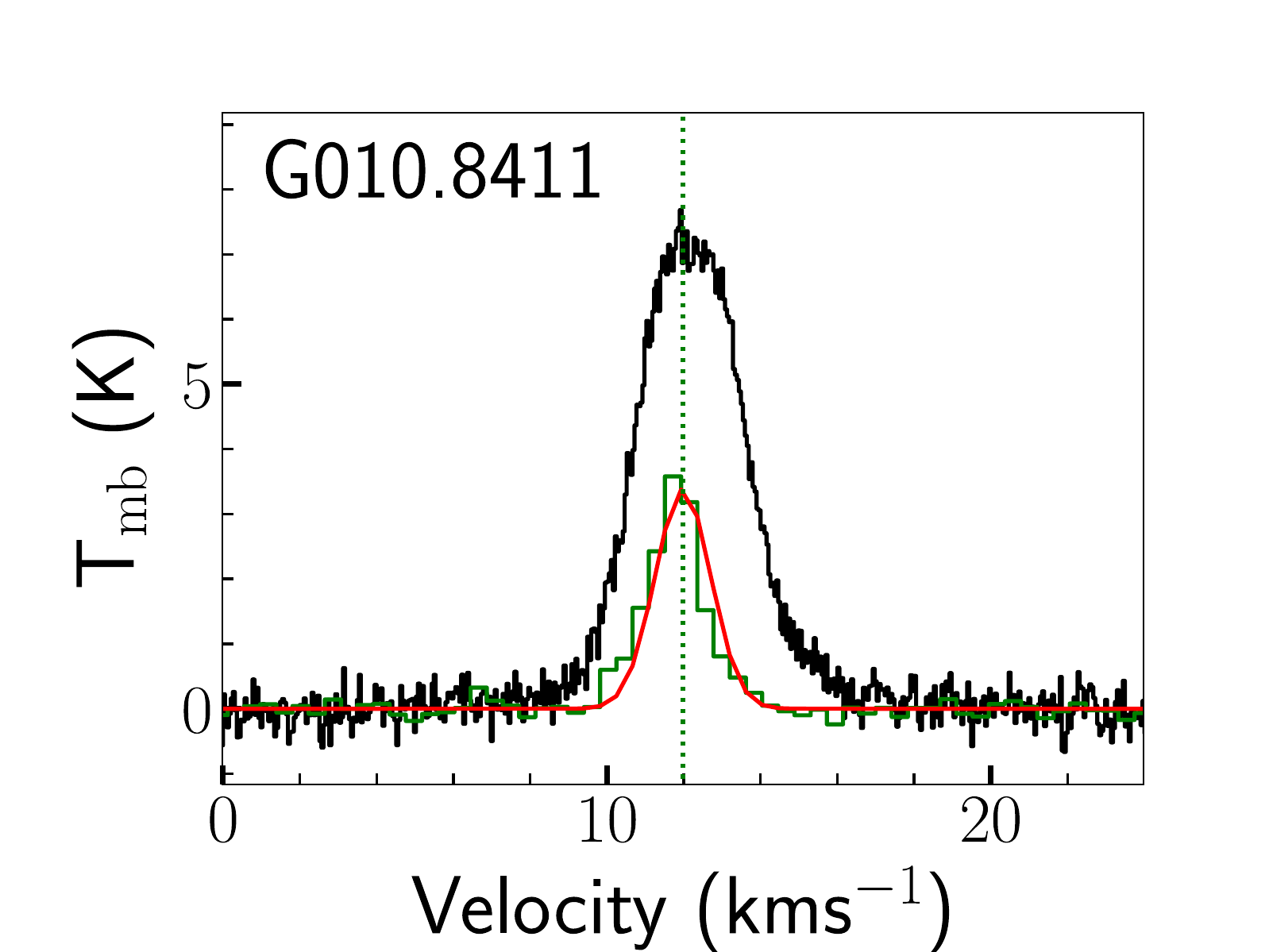}
\includegraphics[width=0.21\textwidth]{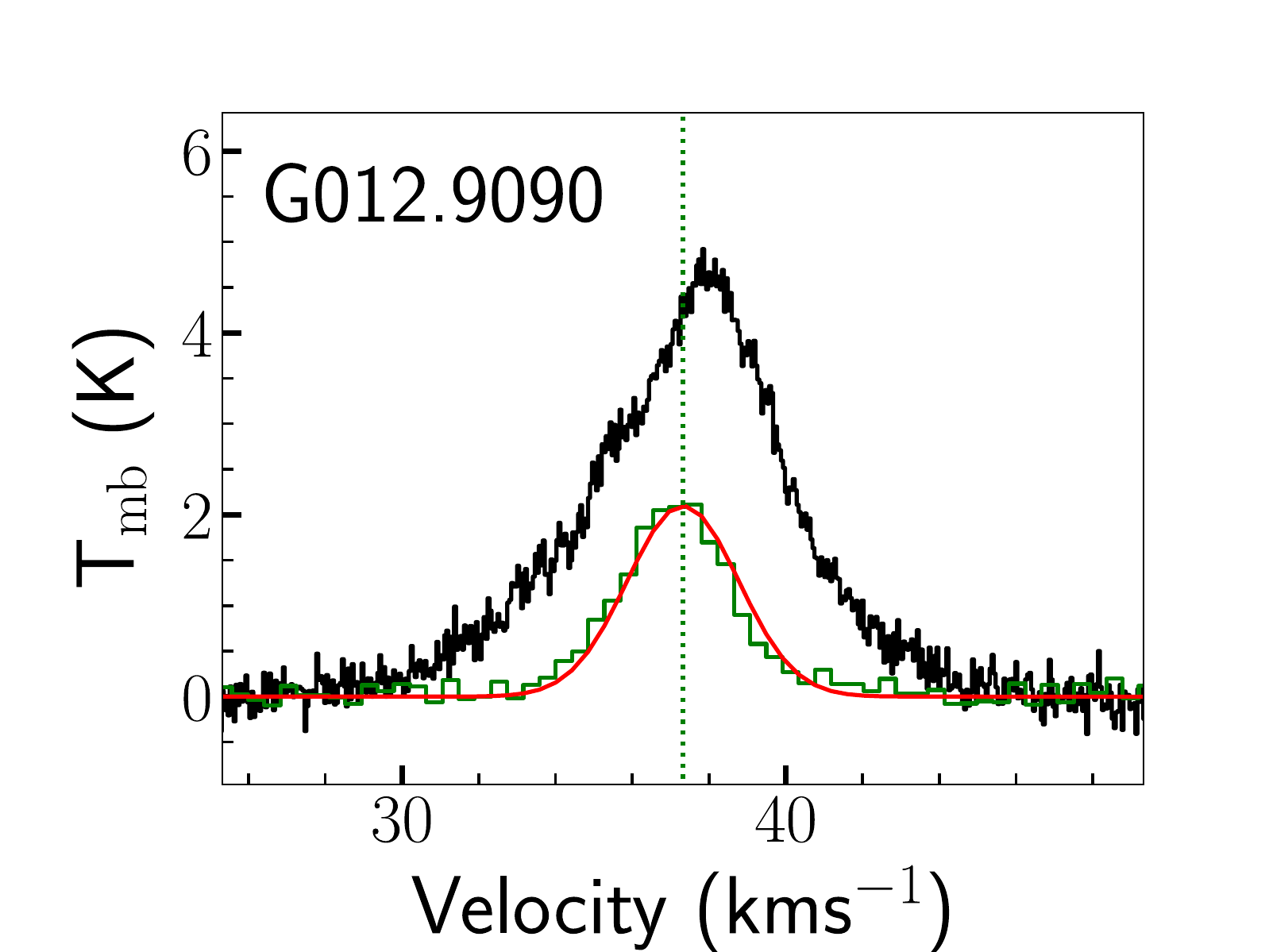}
\includegraphics[width=0.21\textwidth]{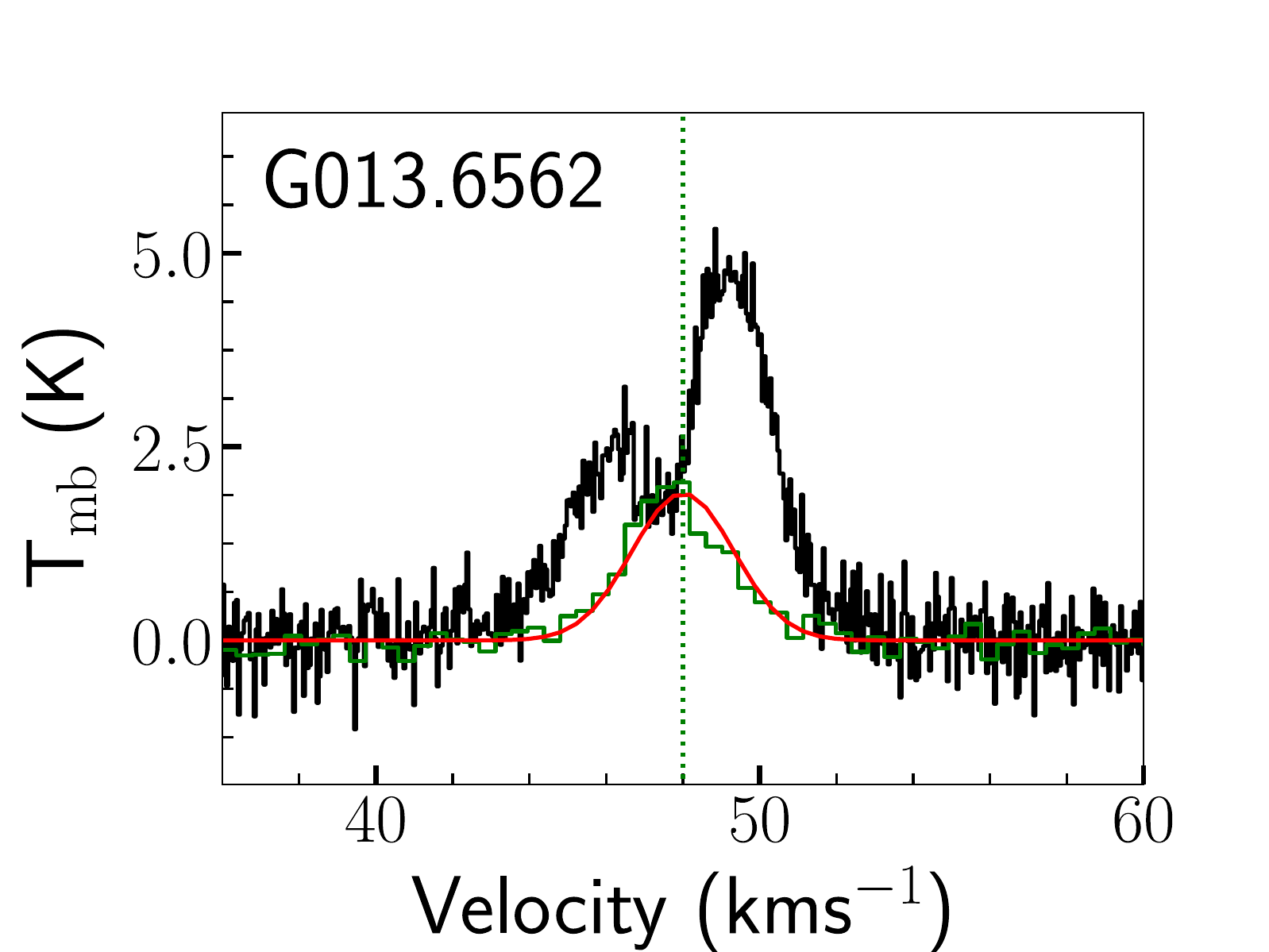}
\includegraphics[width=0.21\textwidth]{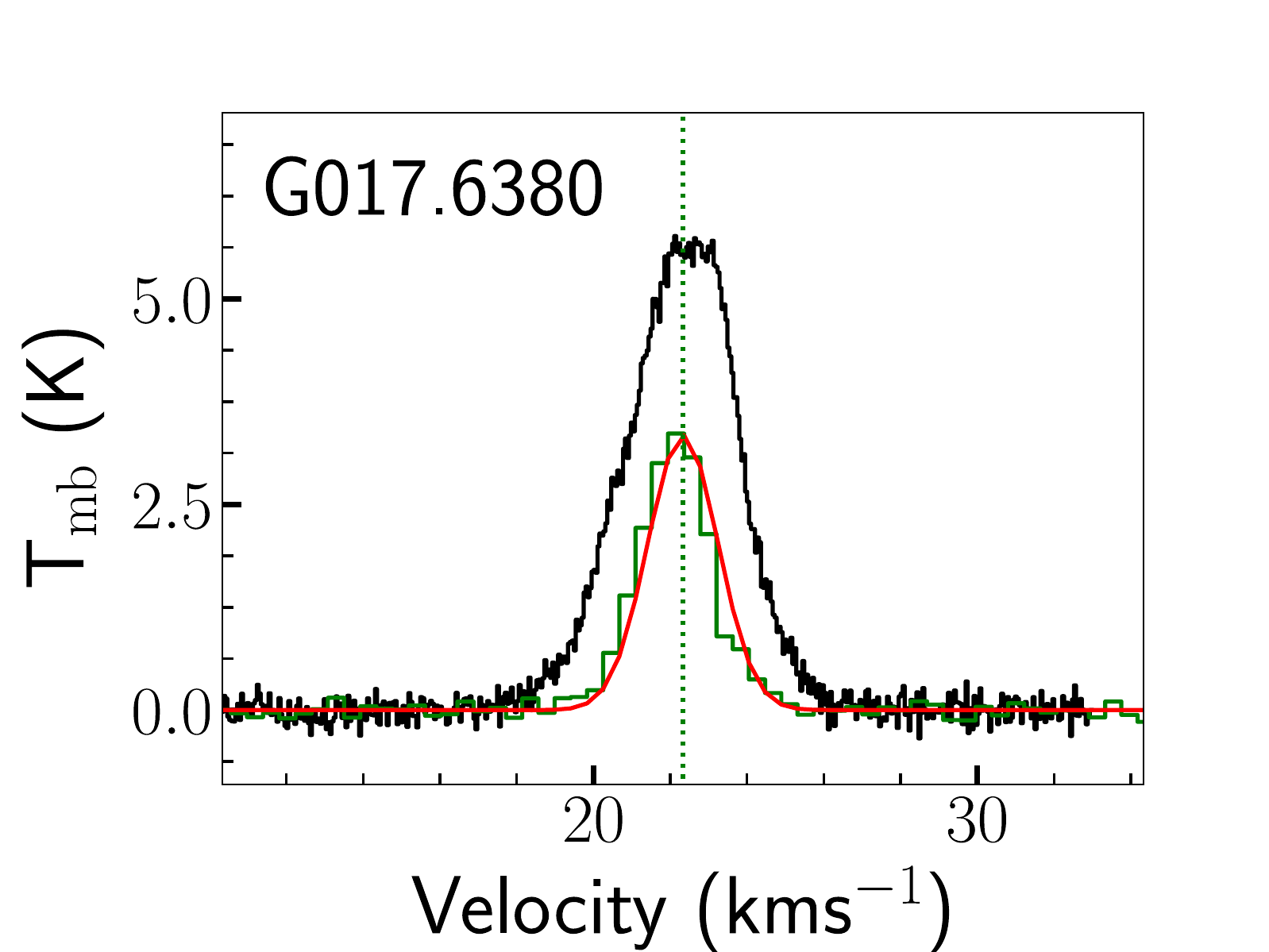}
\includegraphics[width=0.21\textwidth]{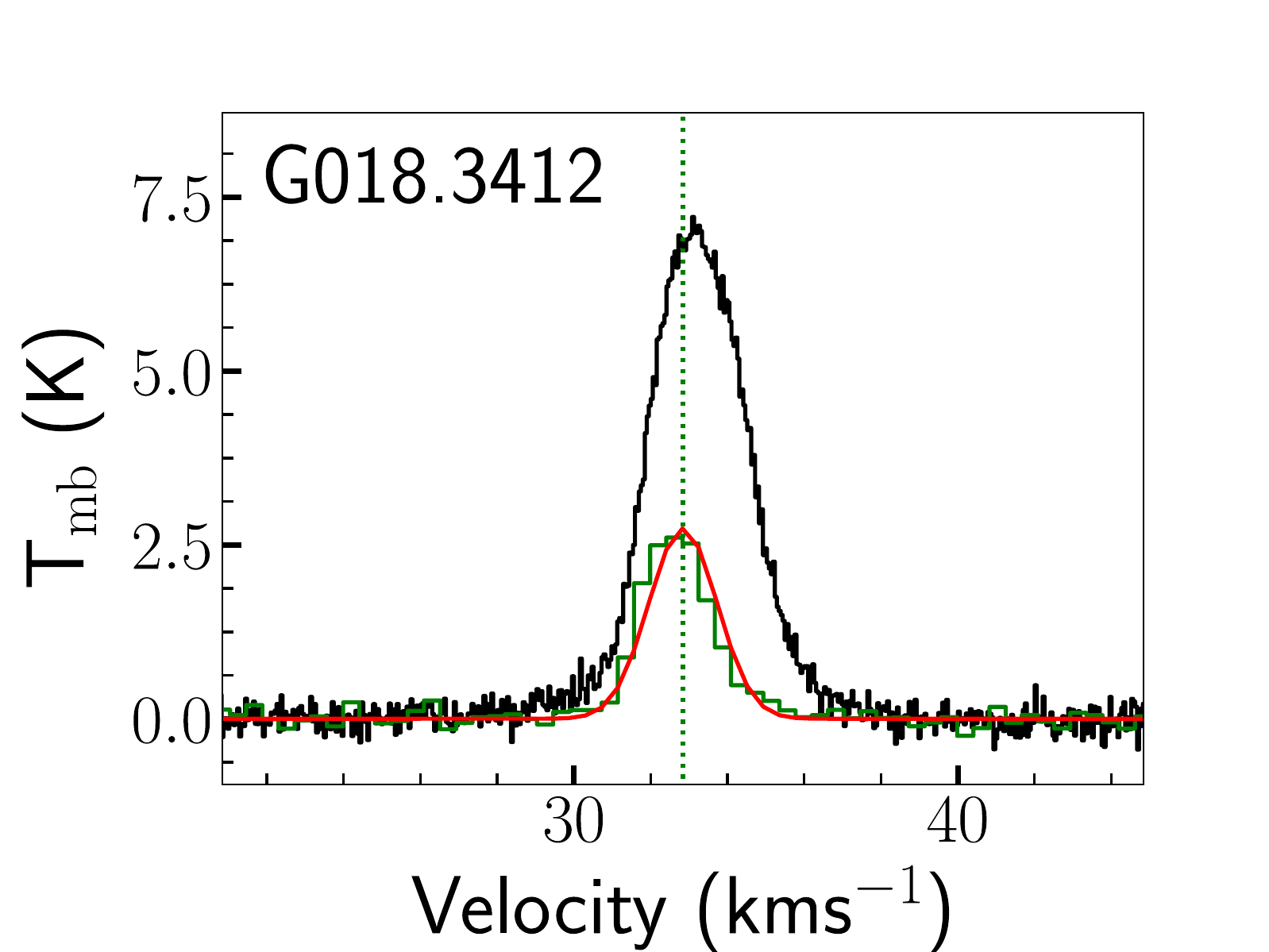}
\includegraphics[width=0.21\textwidth]{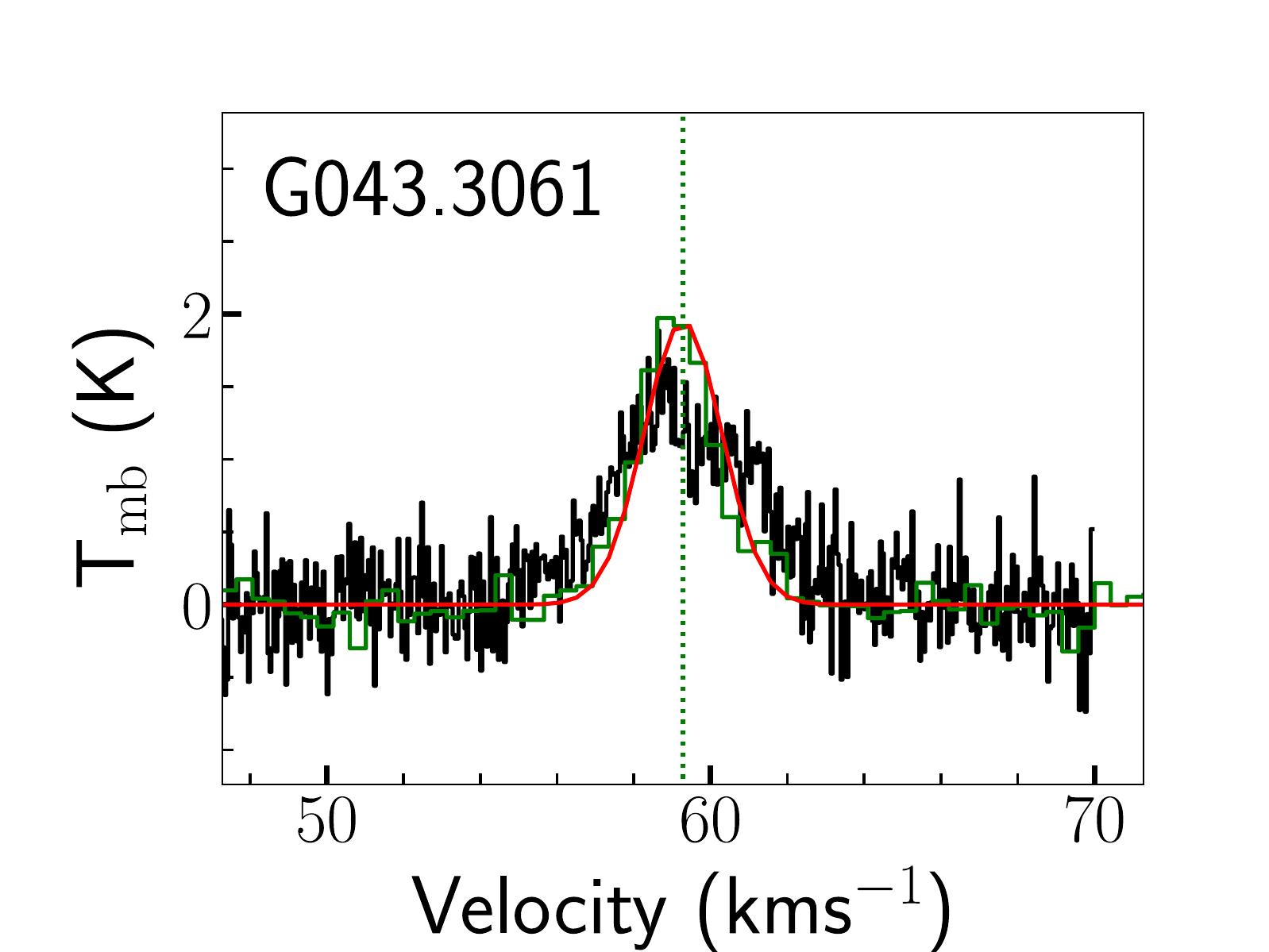}
\includegraphics[width=0.21\textwidth]{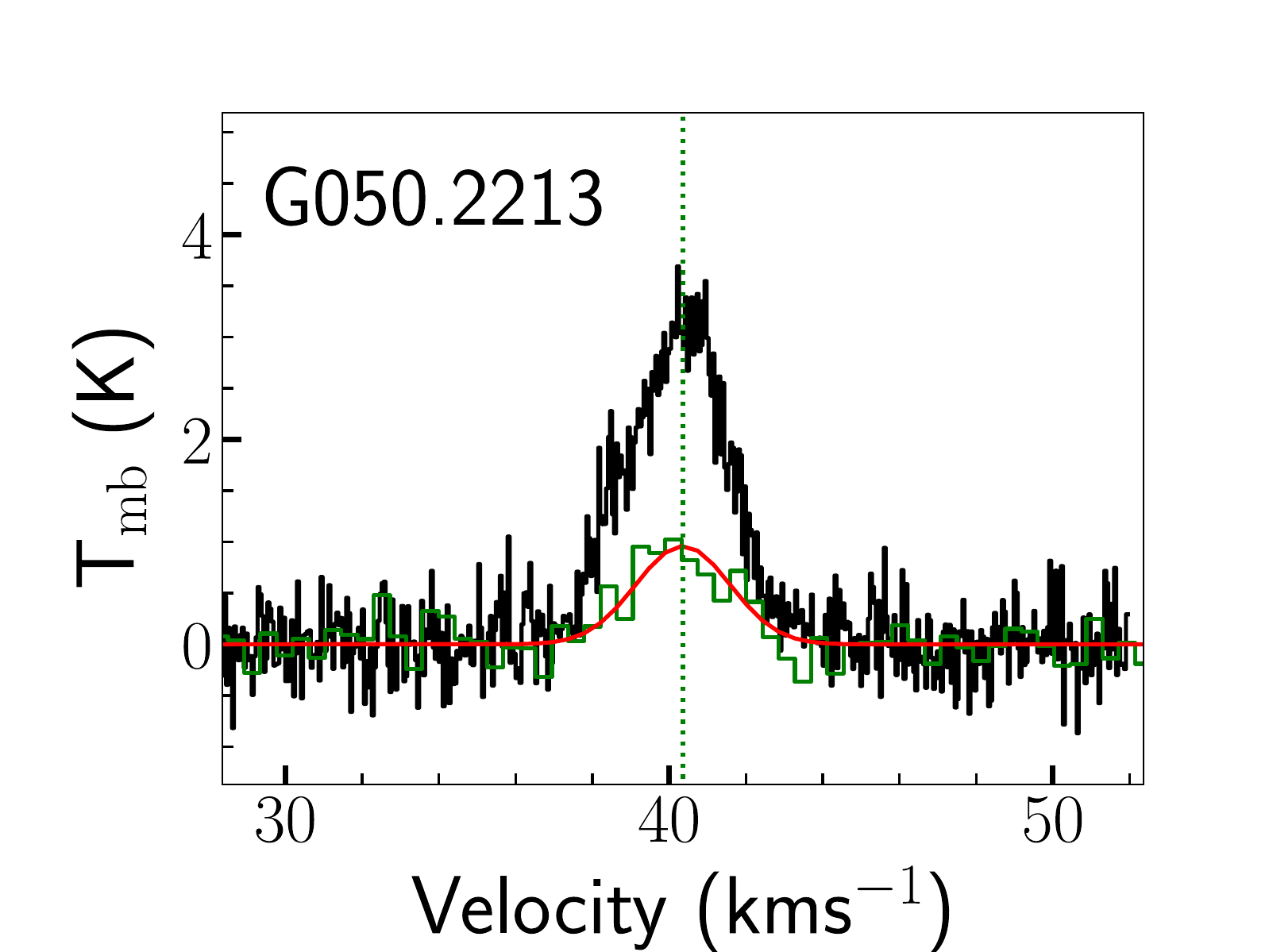}
\includegraphics[width=0.21\textwidth]{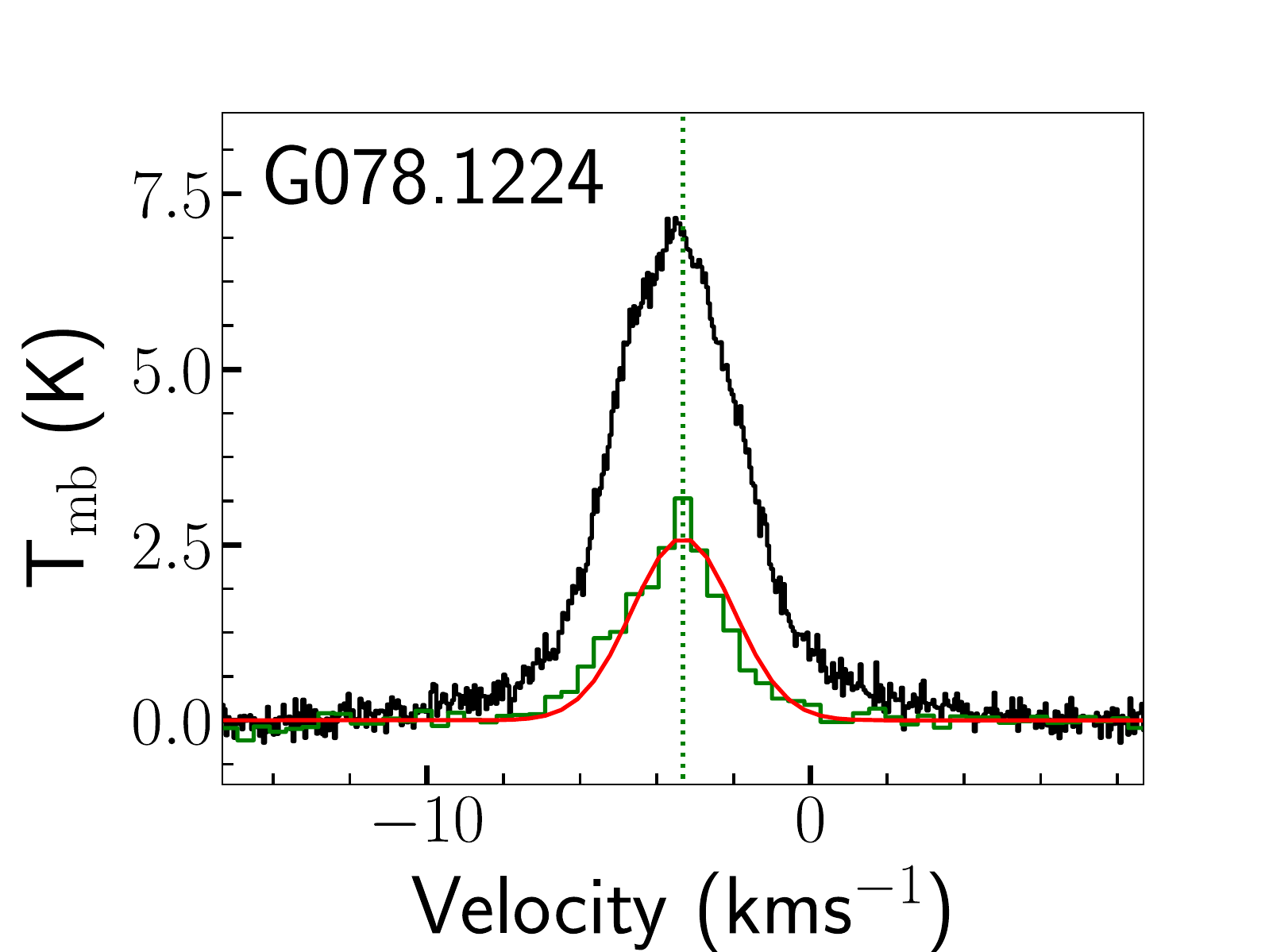}
\includegraphics[width=0.21\textwidth]{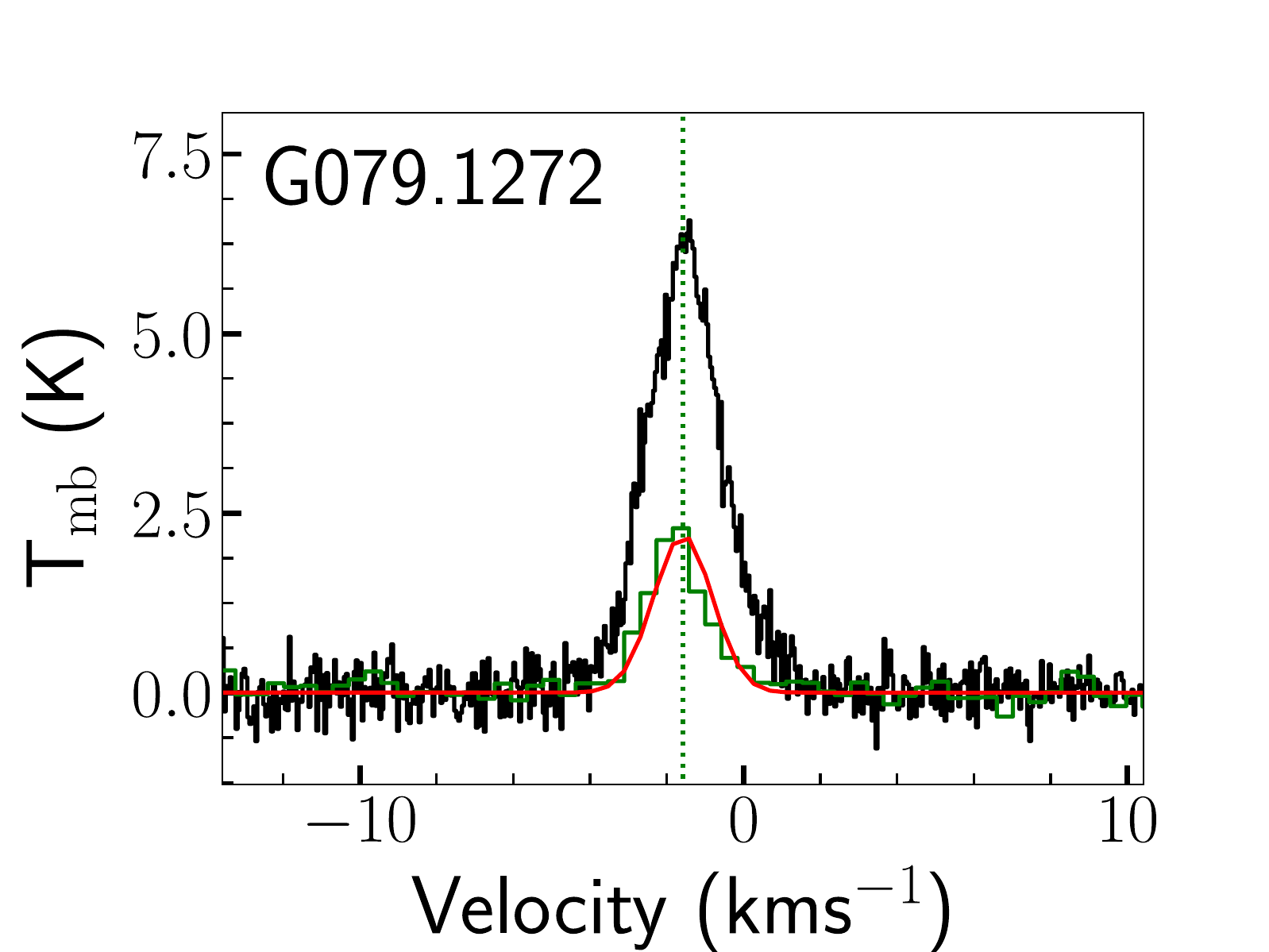}
\includegraphics[width=0.21\textwidth]{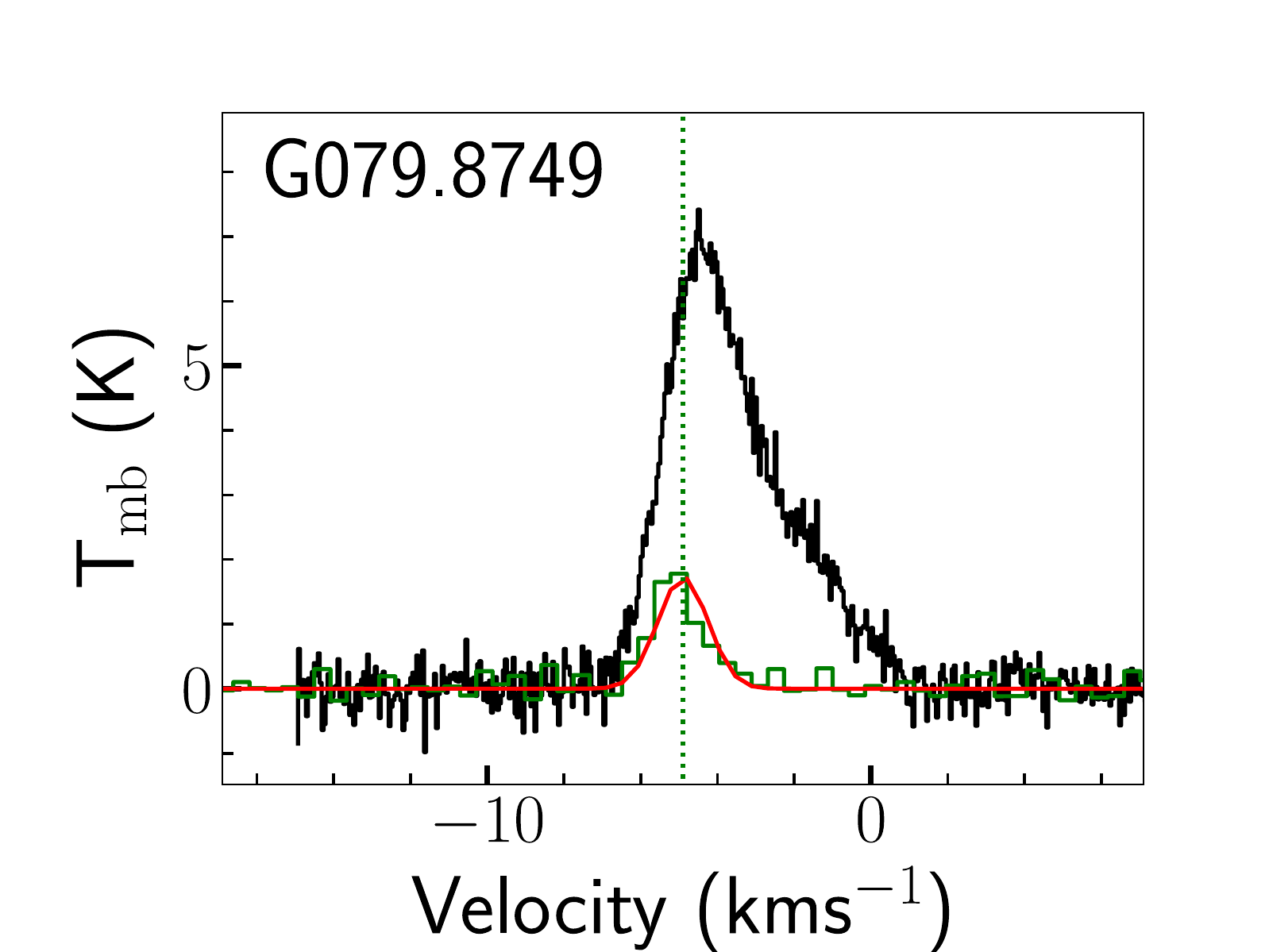}
\includegraphics[width=0.21\textwidth]{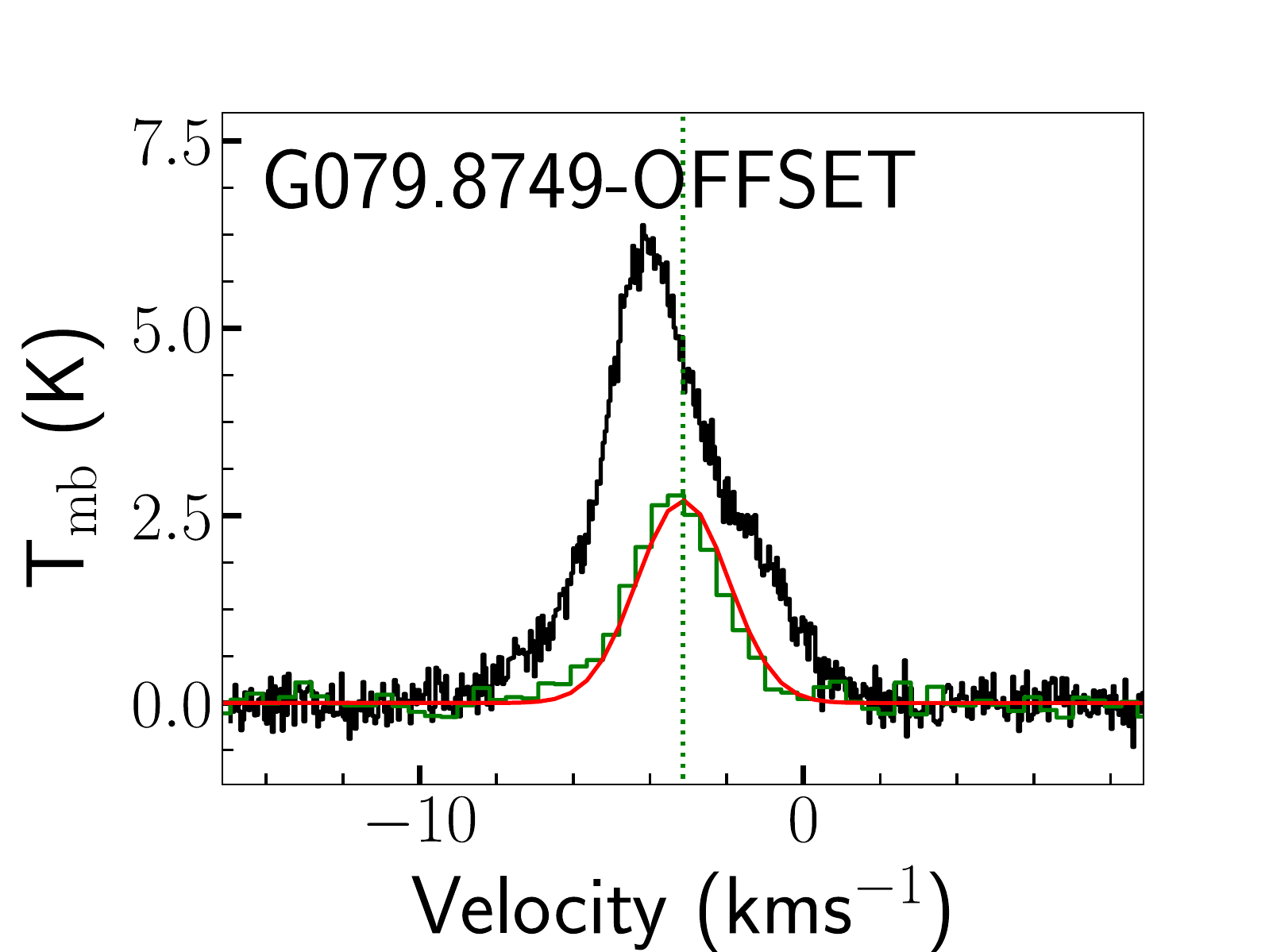}
\includegraphics[width=0.21\textwidth]{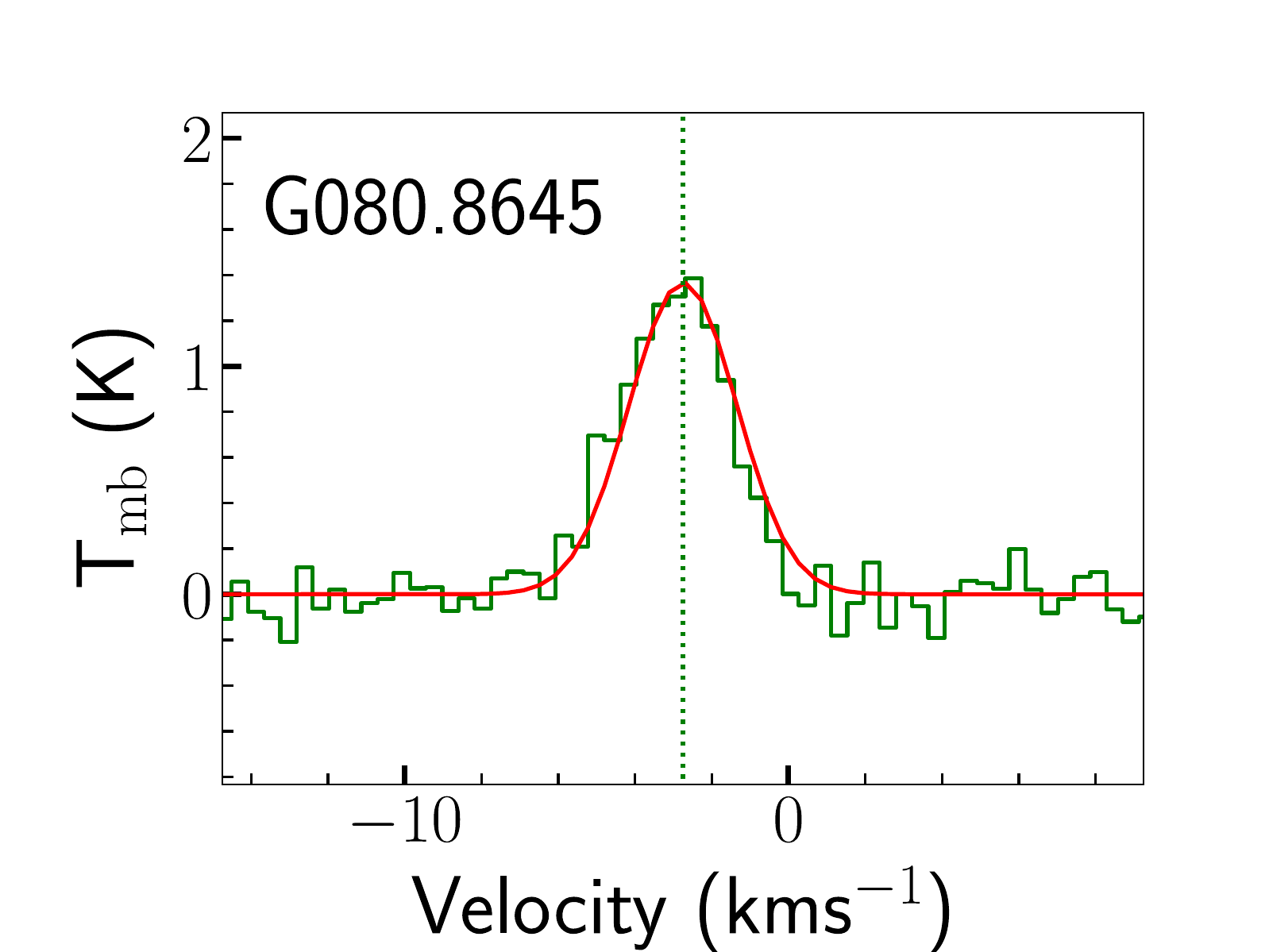}
\includegraphics[width=0.21\textwidth]{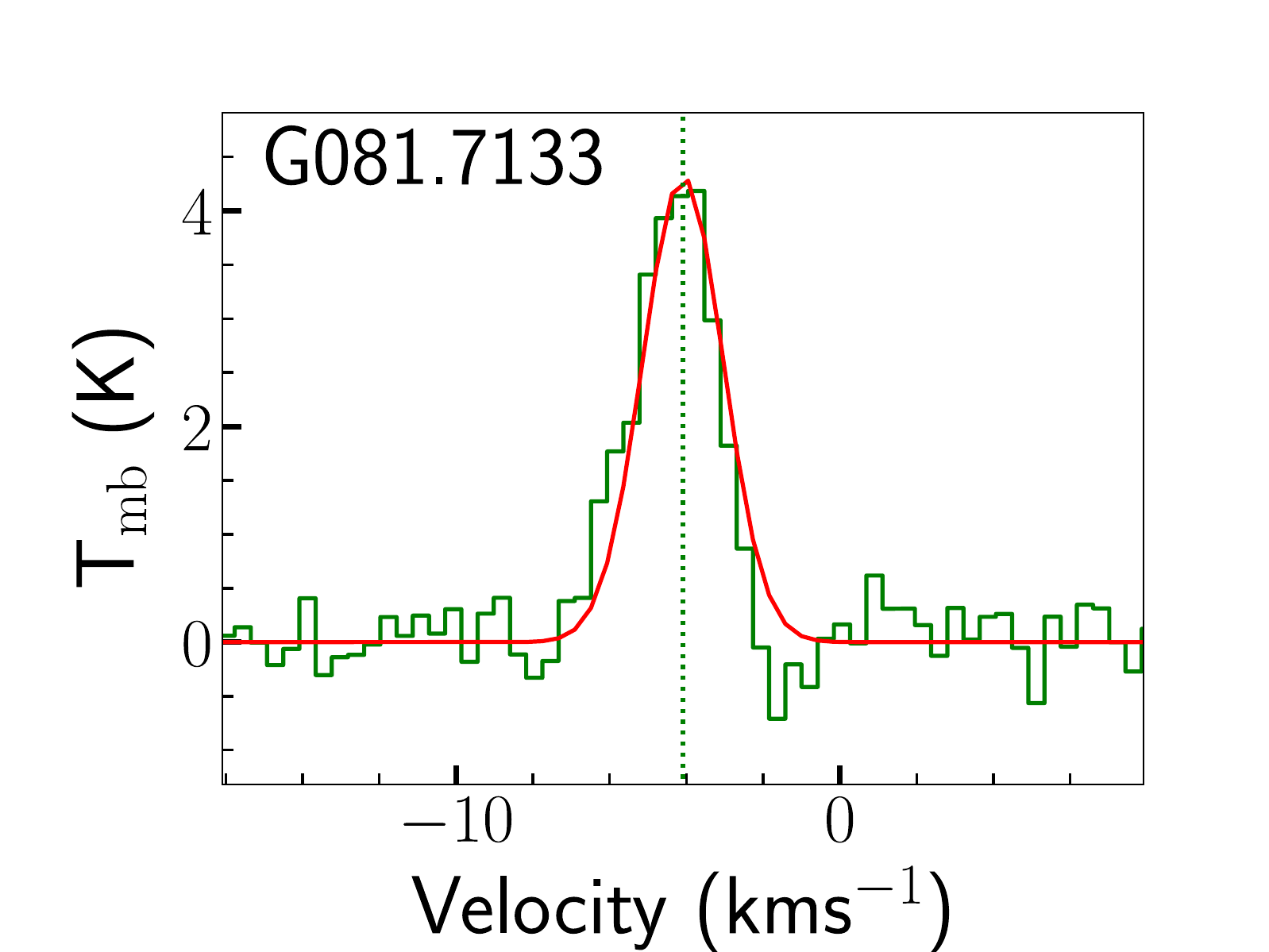}
\includegraphics[width=0.21\textwidth]{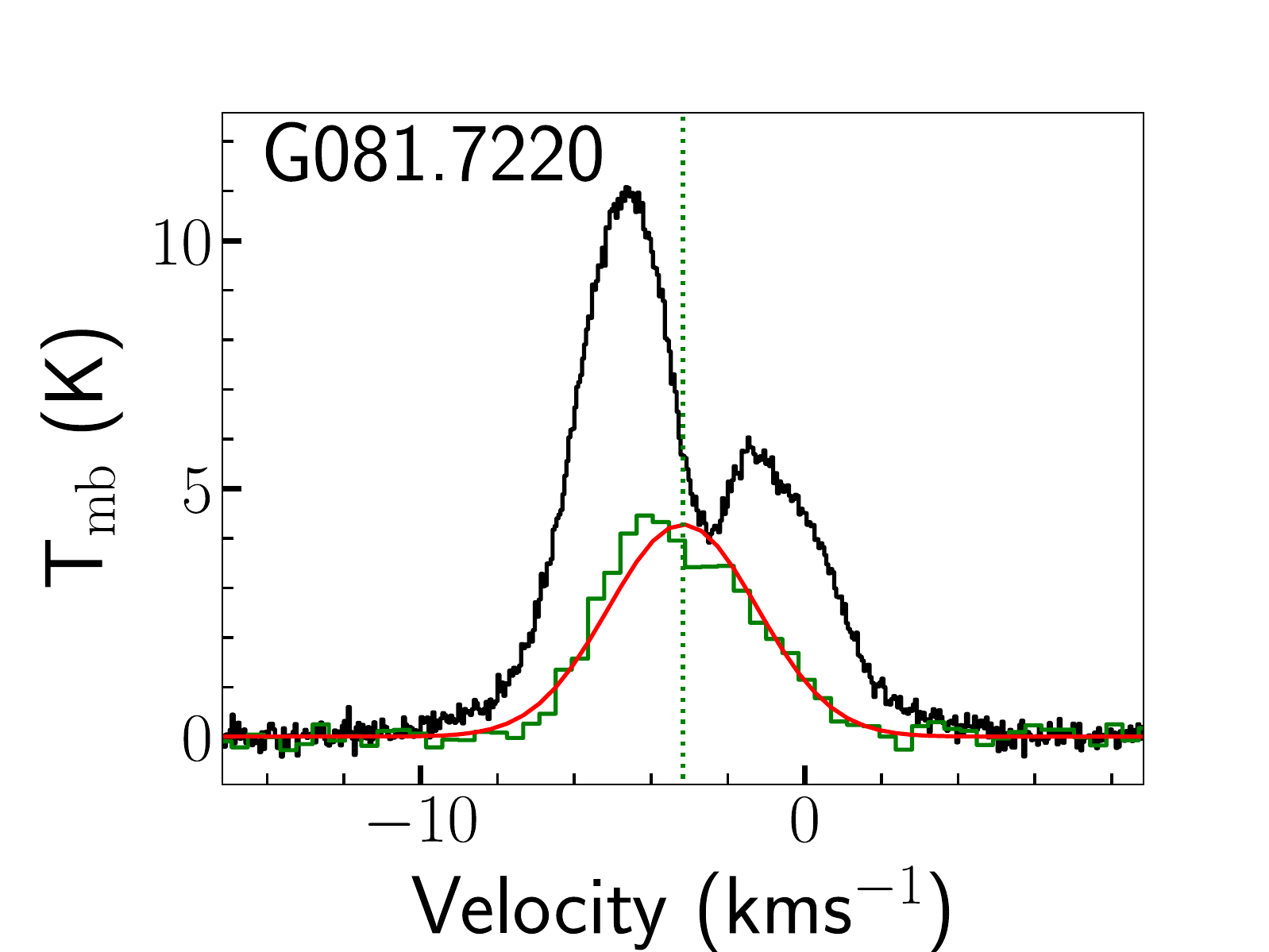}
\includegraphics[width=0.21\textwidth]{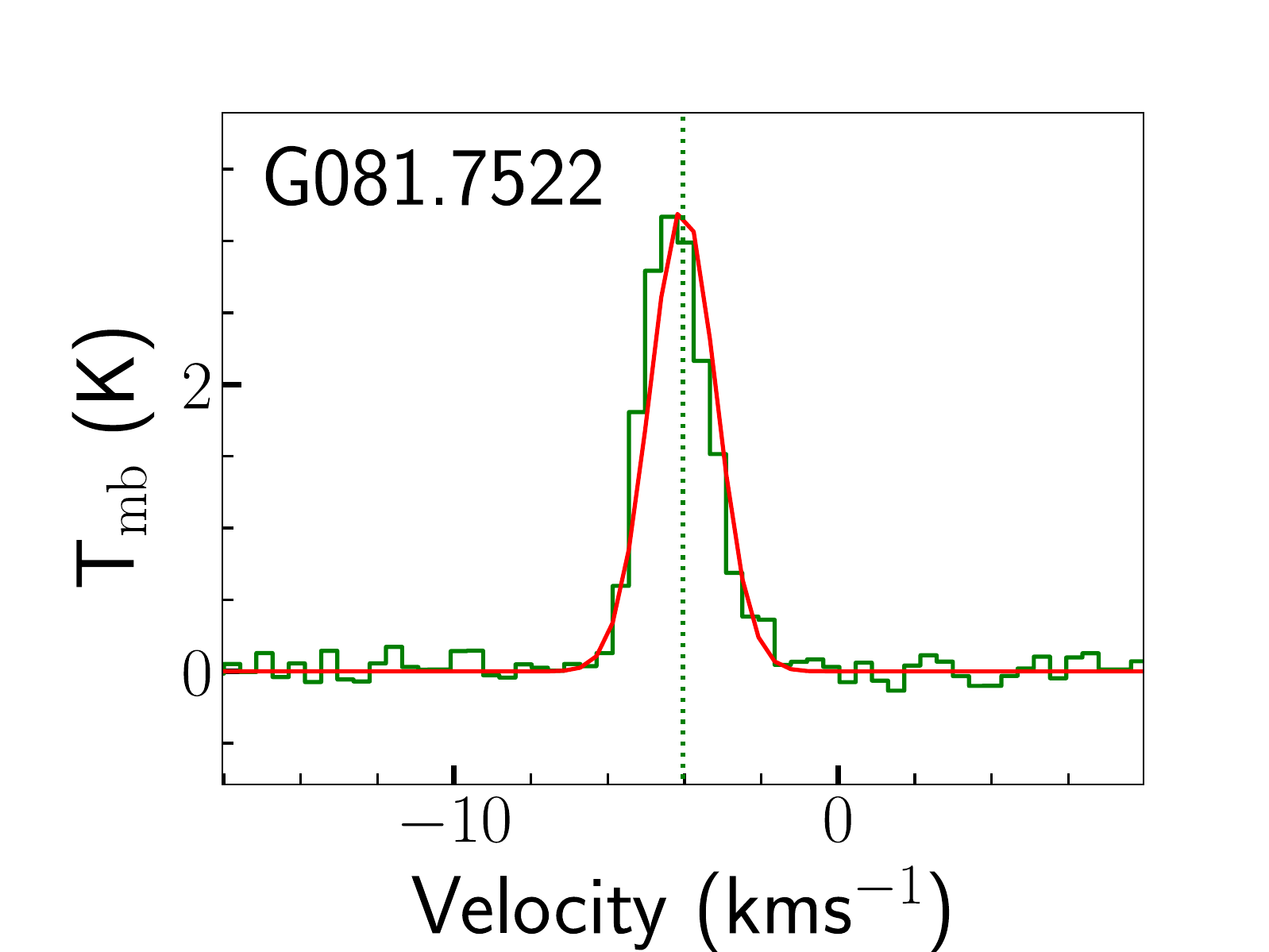}
\includegraphics[width=0.21\textwidth]{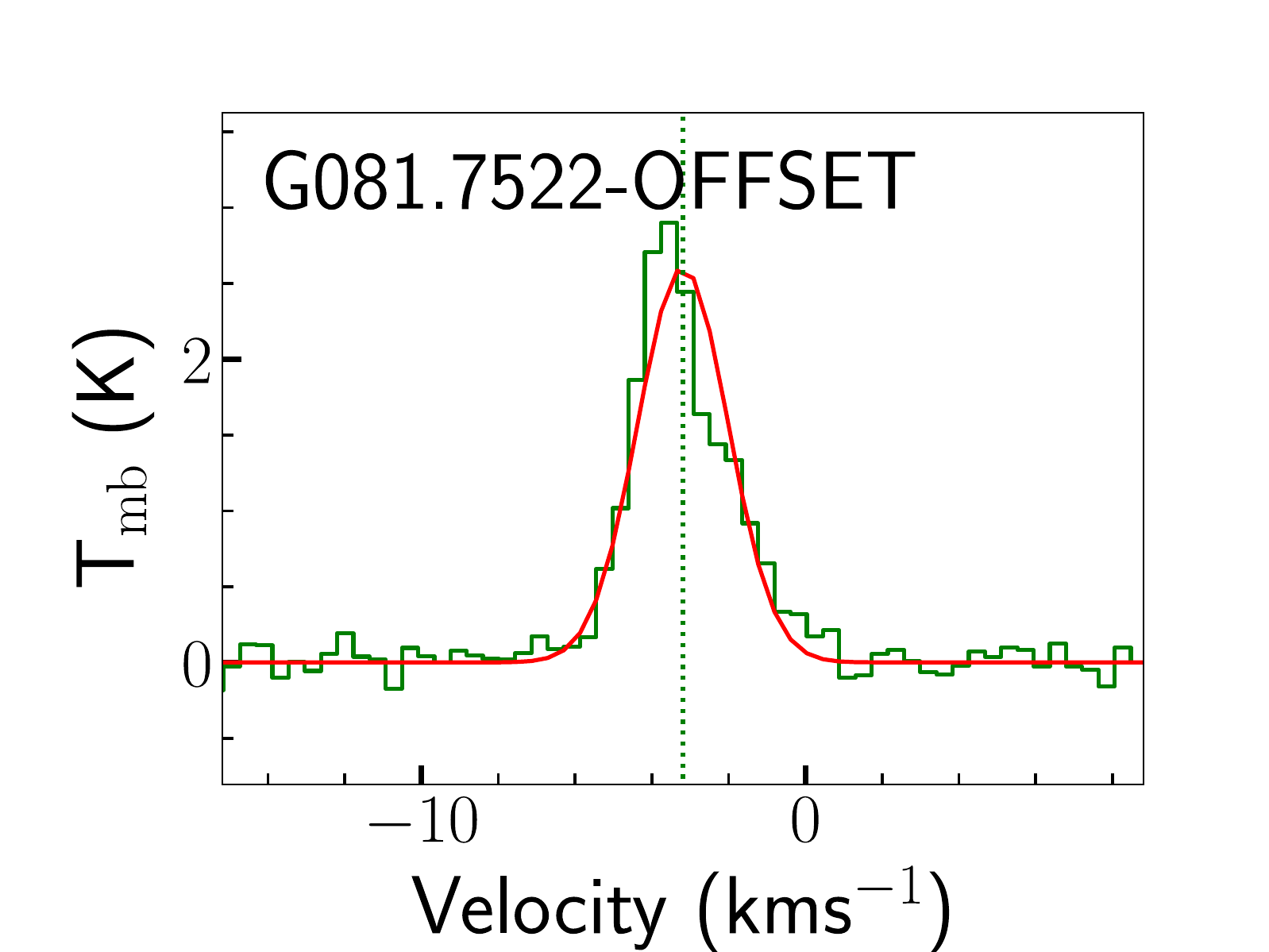}
\includegraphics[width=0.21\textwidth]{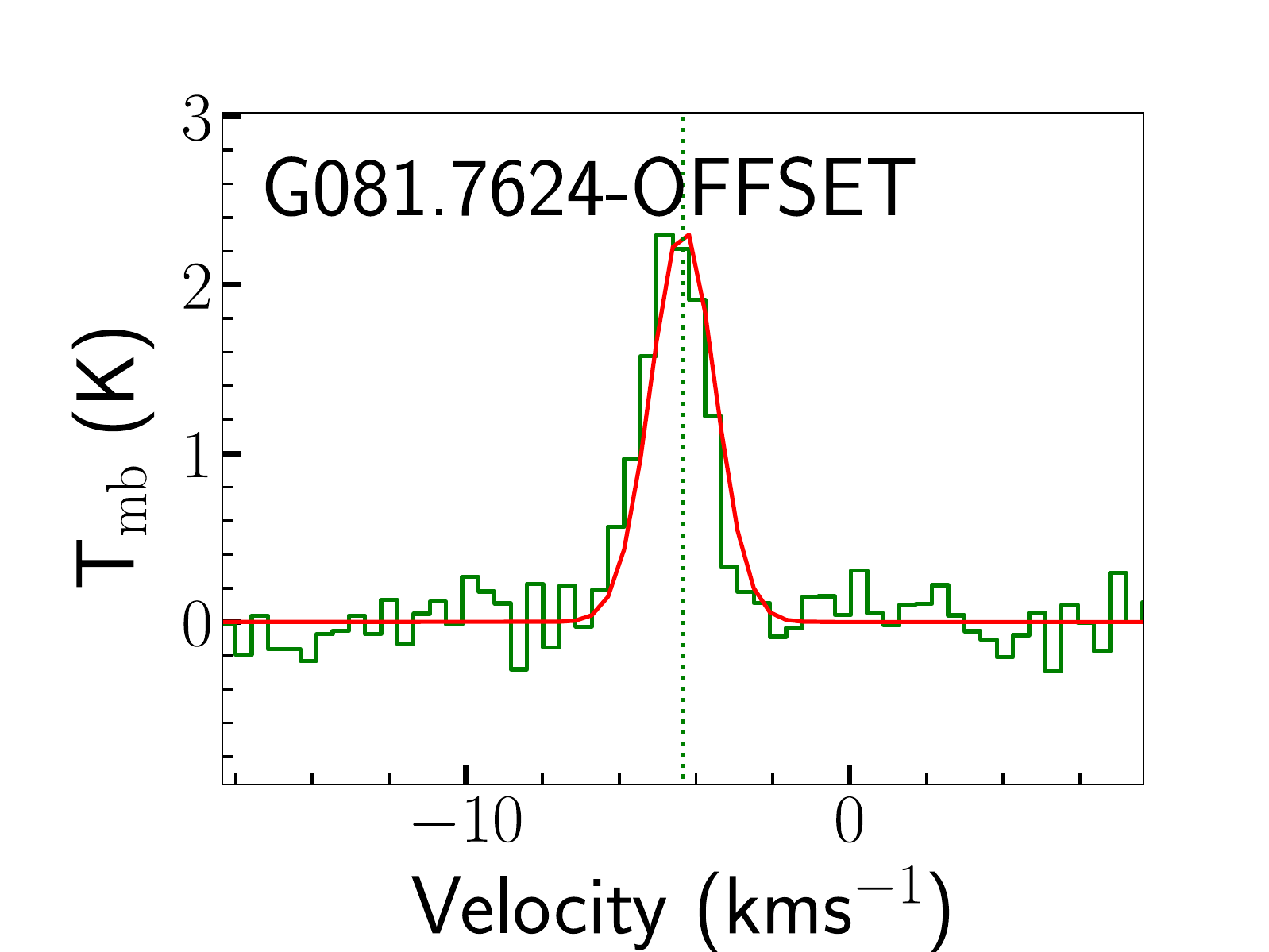}
\includegraphics[width=0.21\textwidth]{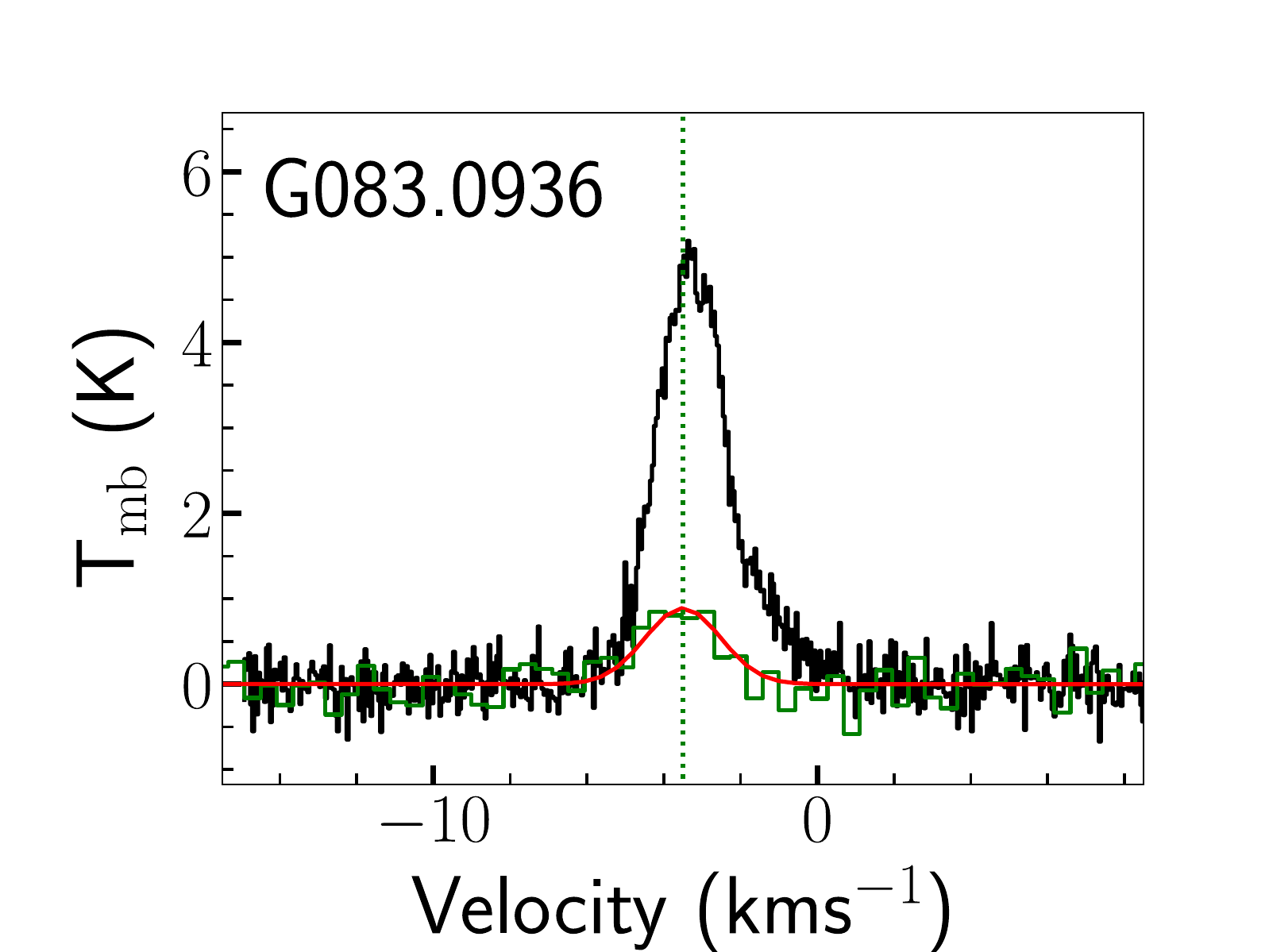}
\includegraphics[width=0.21\textwidth]{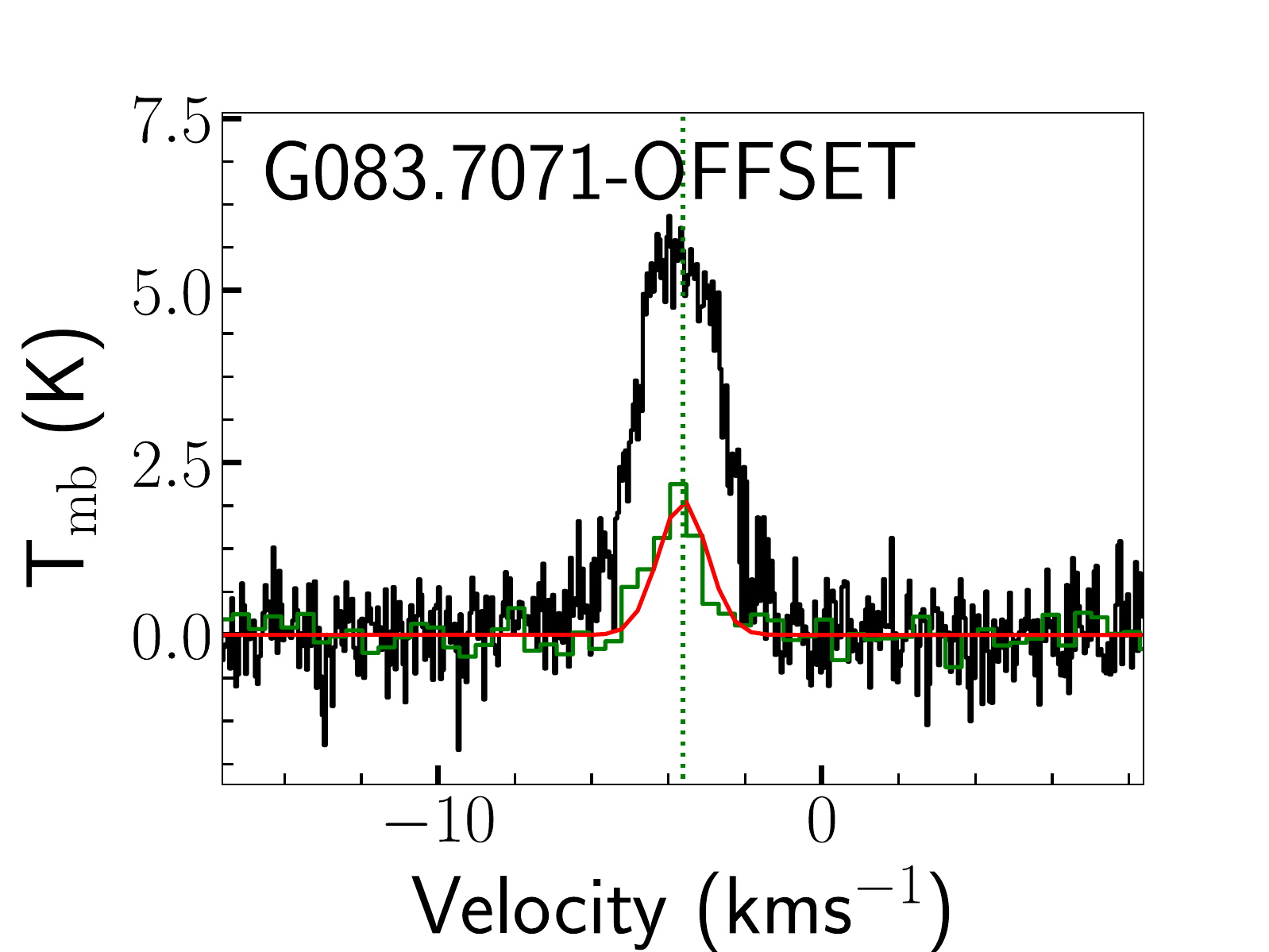}
\includegraphics[width=0.21\textwidth]{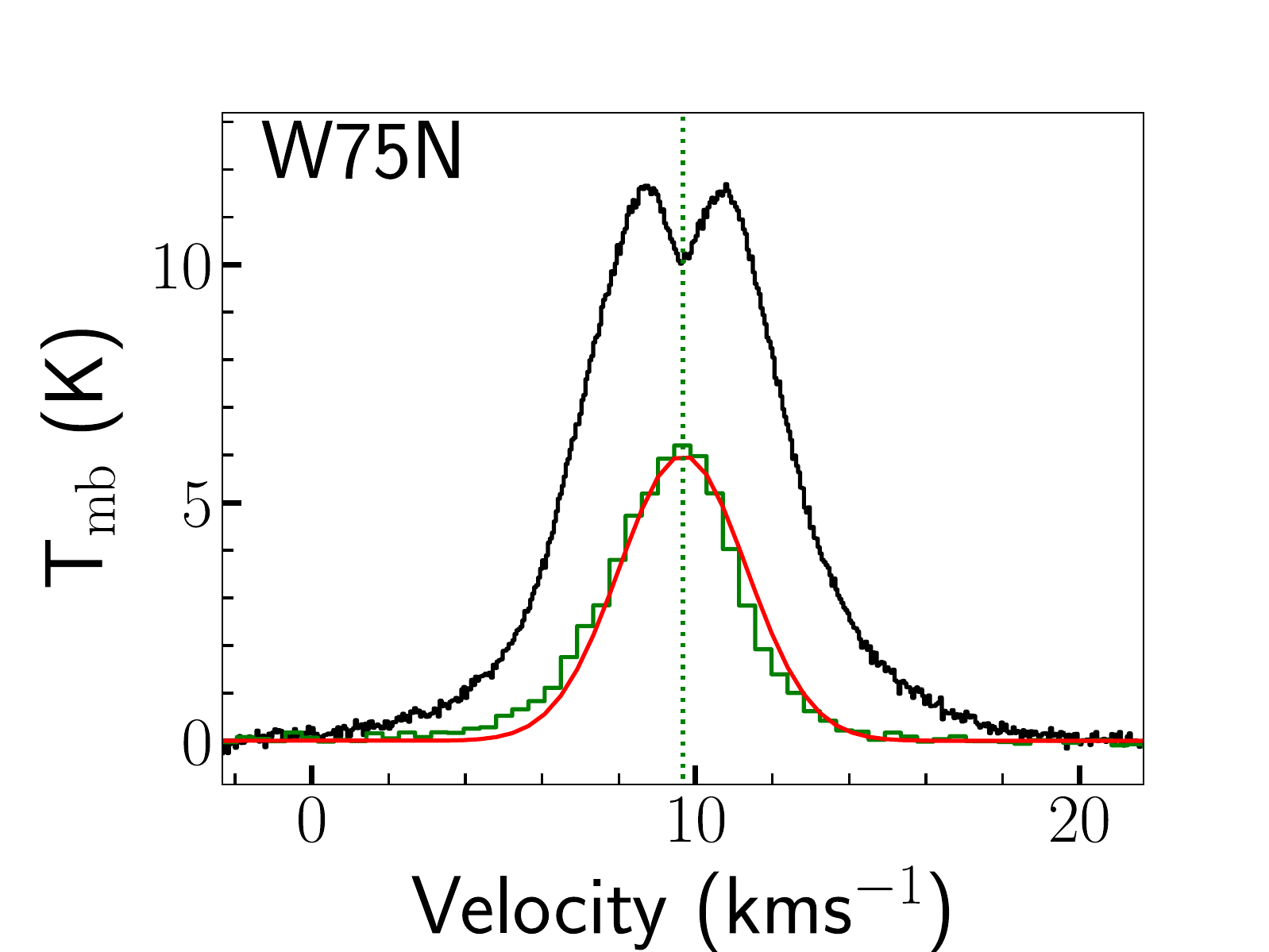}
\includegraphics[width=0.21\textwidth]{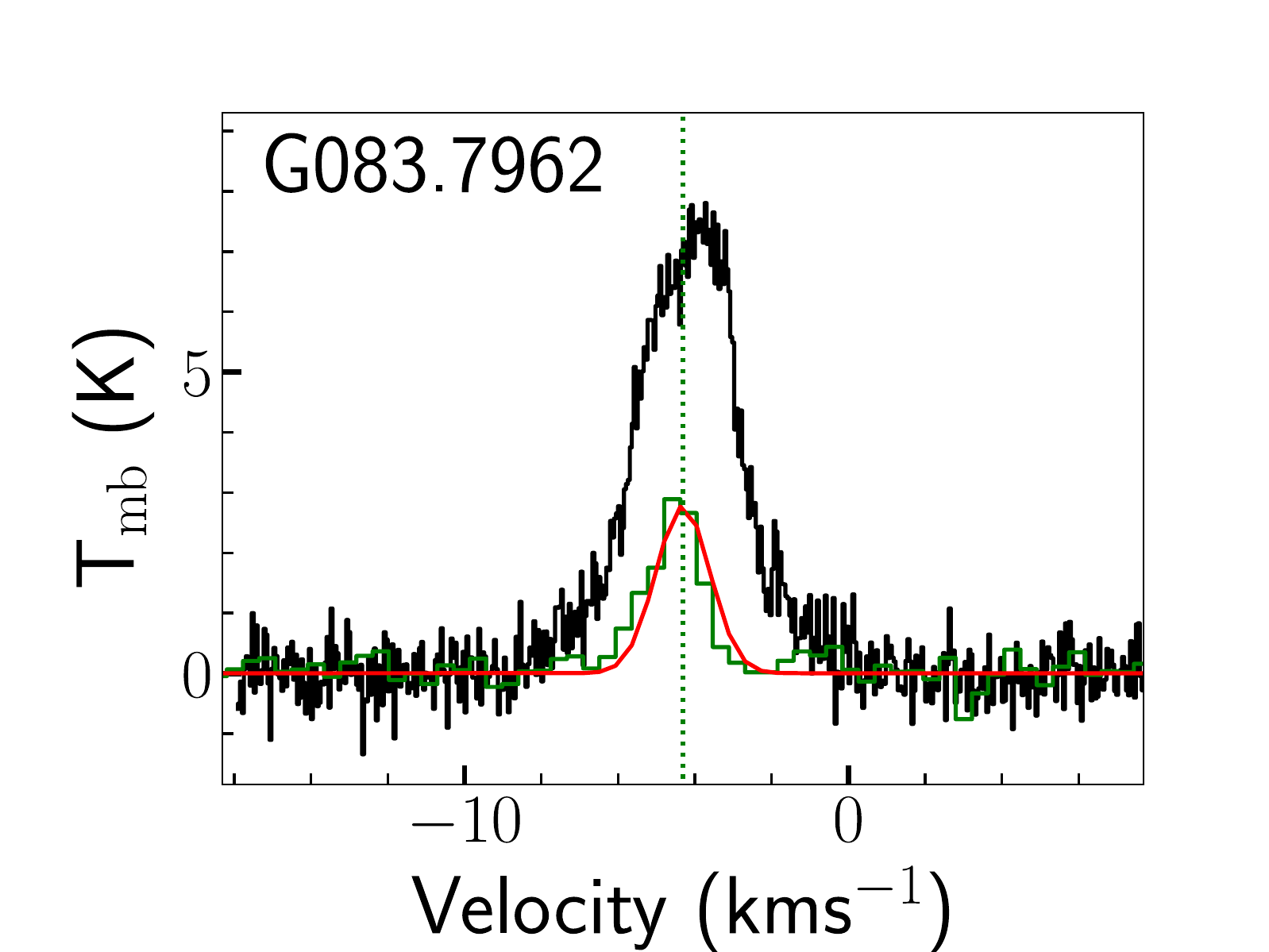}
\includegraphics[width=0.21\textwidth]{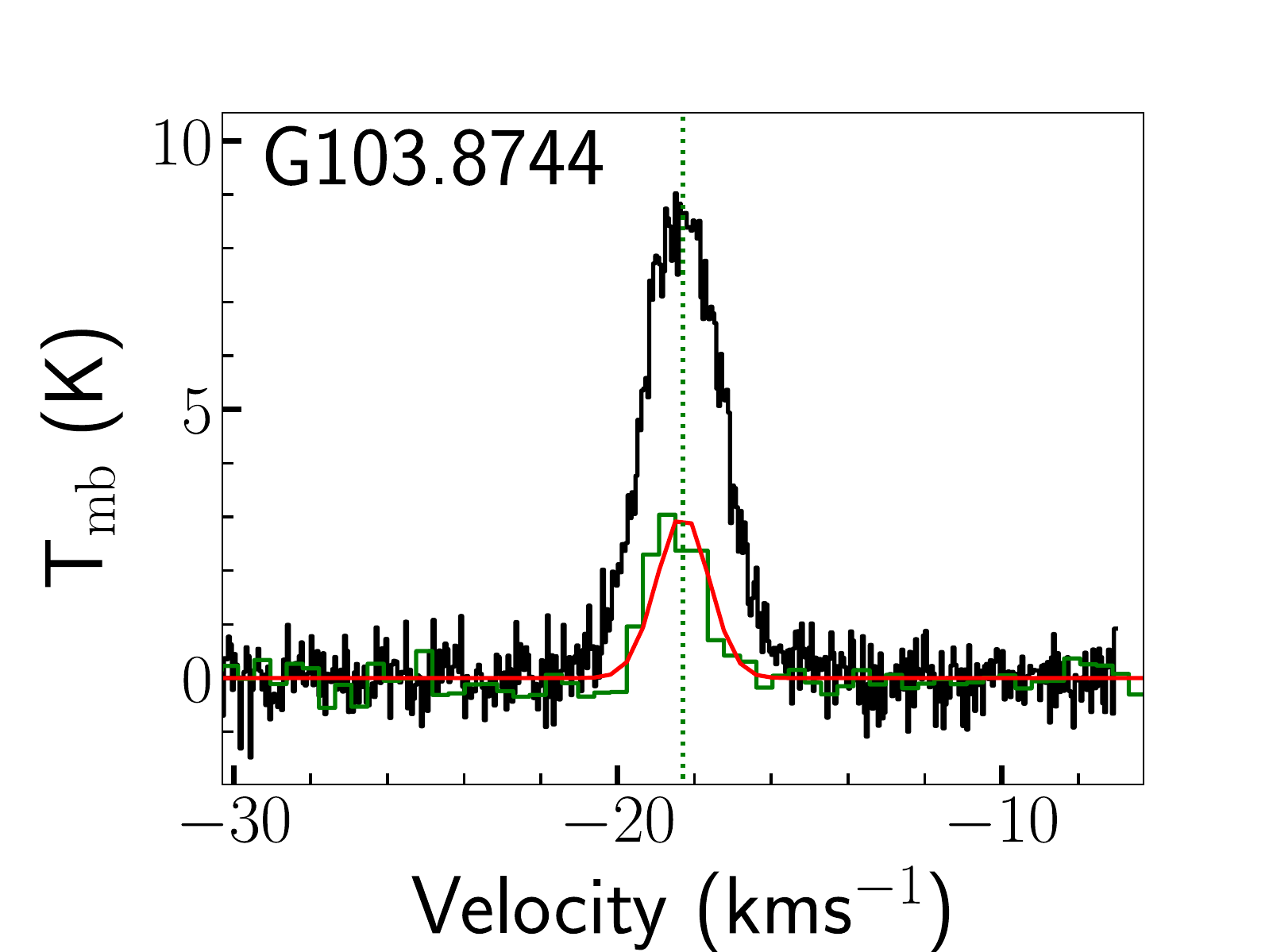}
\includegraphics[width=0.21\textwidth]{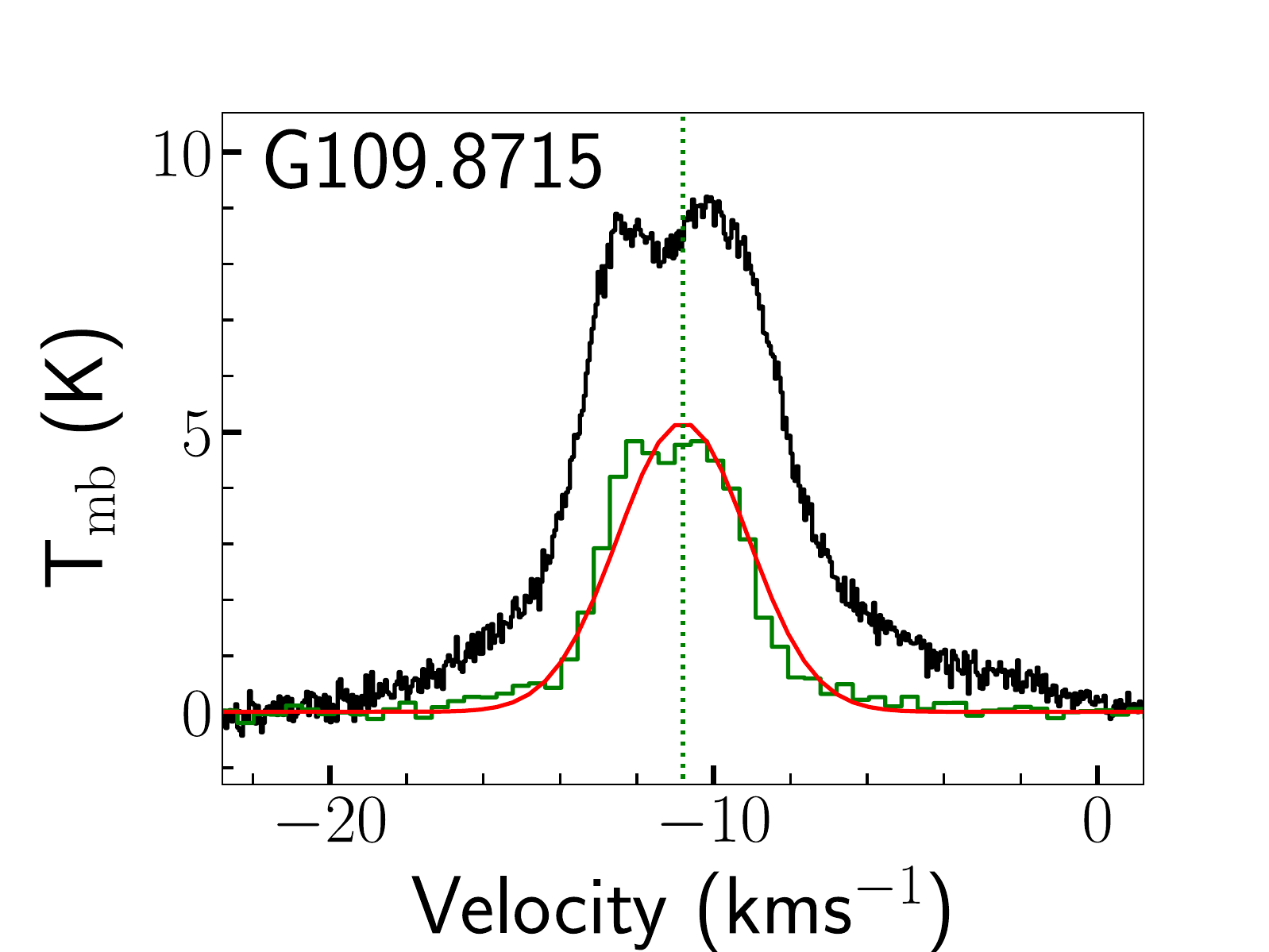}
\includegraphics[width=0.21\textwidth]{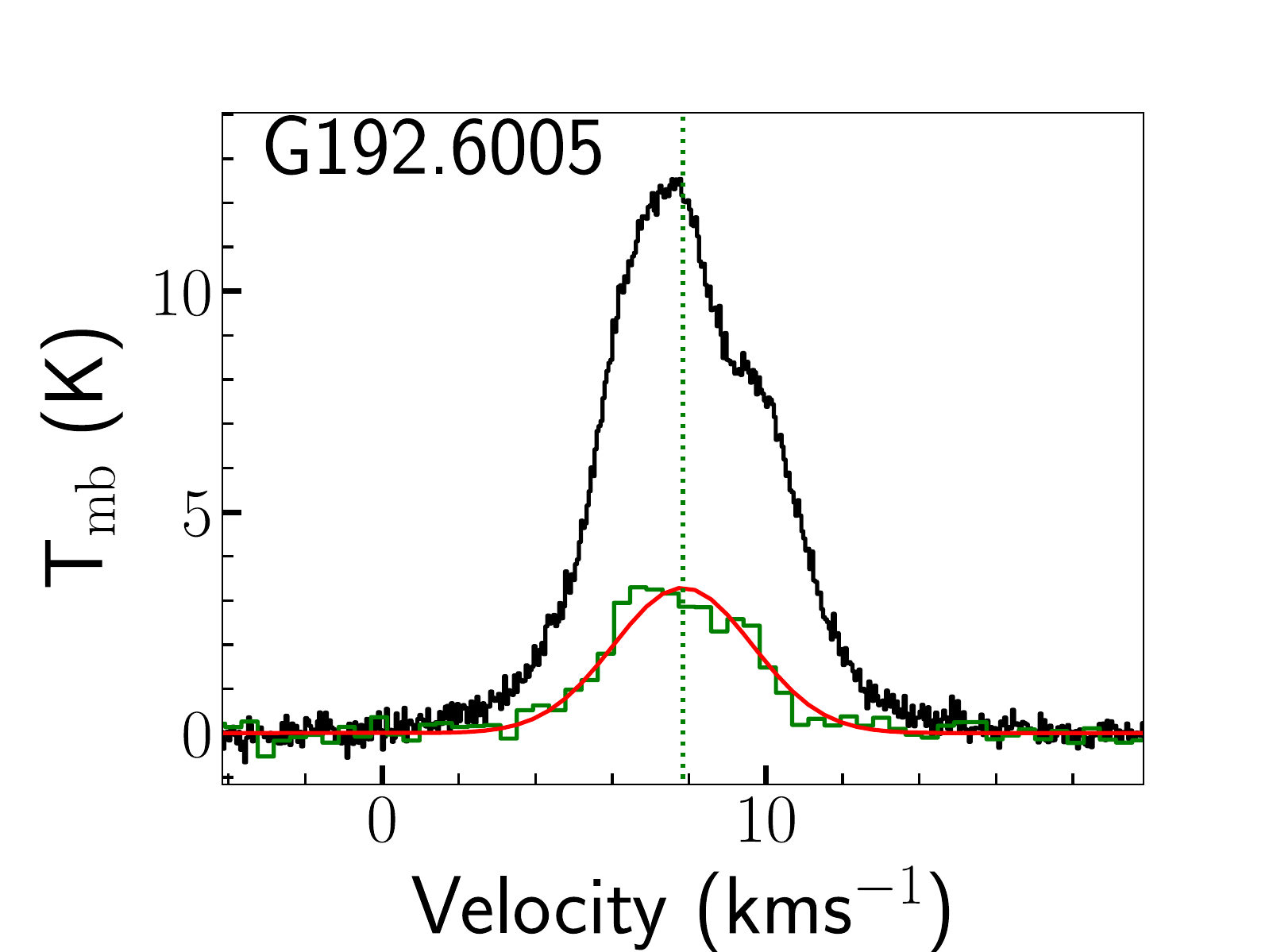}
\includegraphics[width=0.21\textwidth]{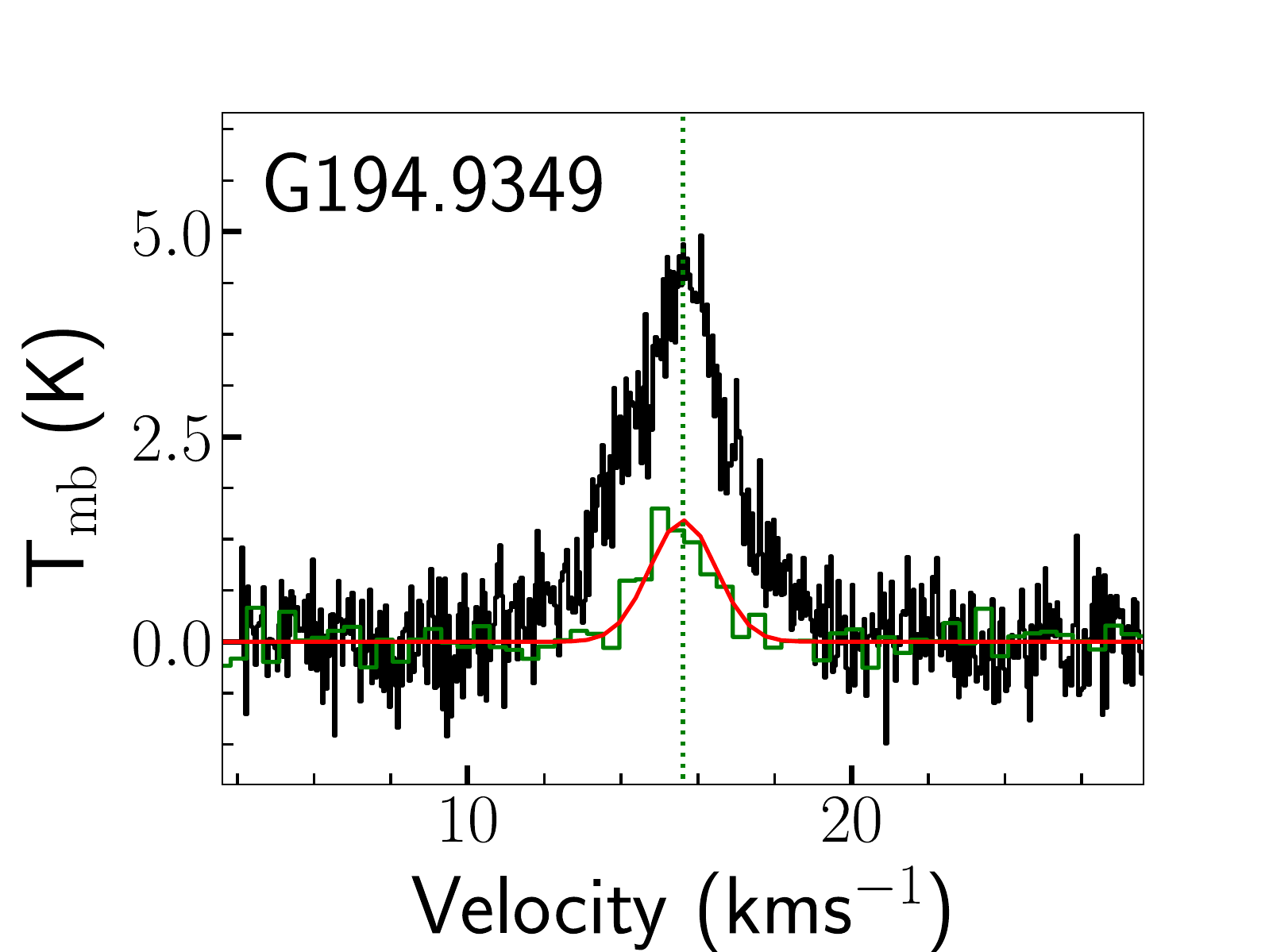}
\includegraphics[width=0.21\textwidth]{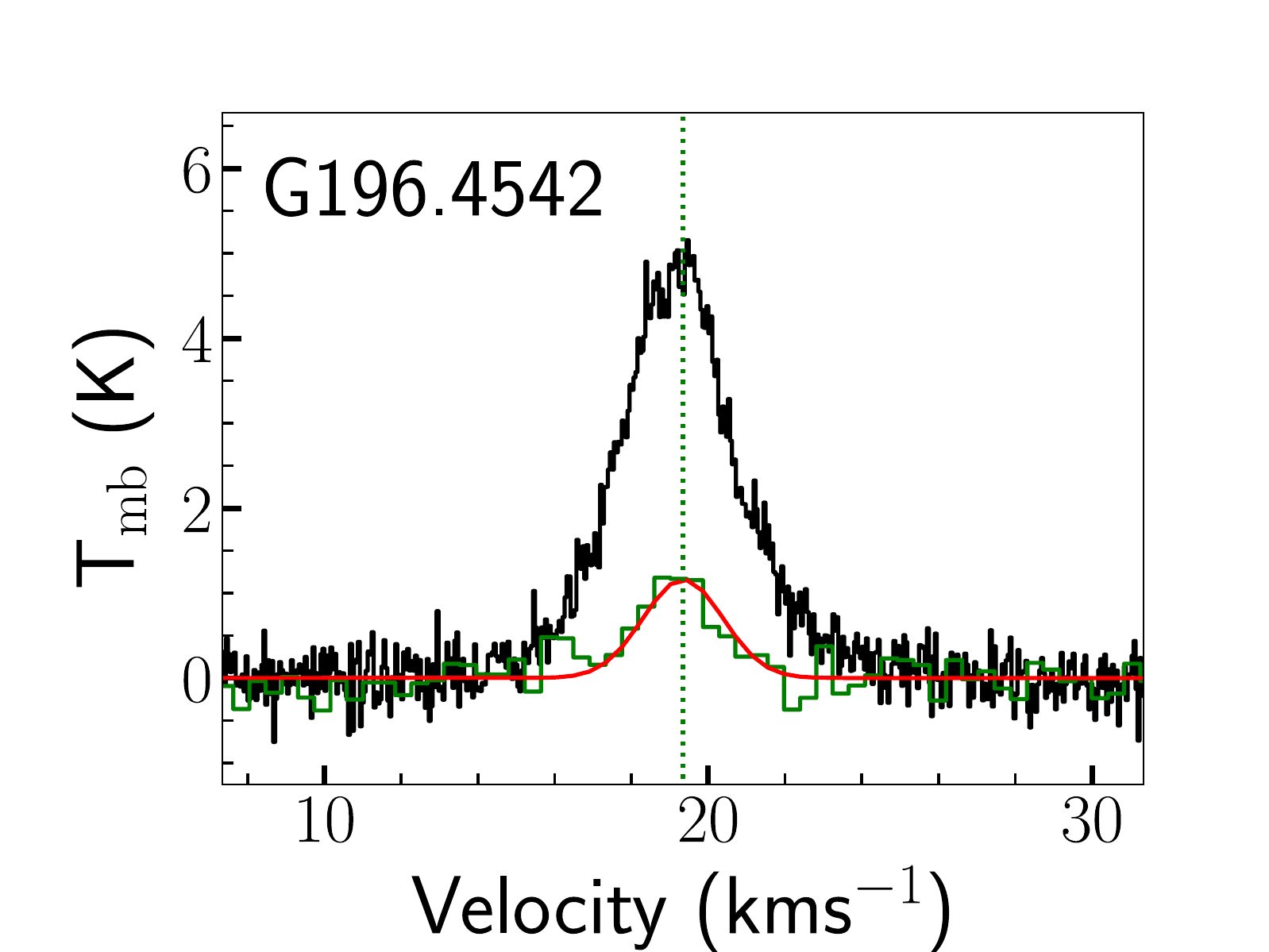}
\includegraphics[width=0.21\textwidth]{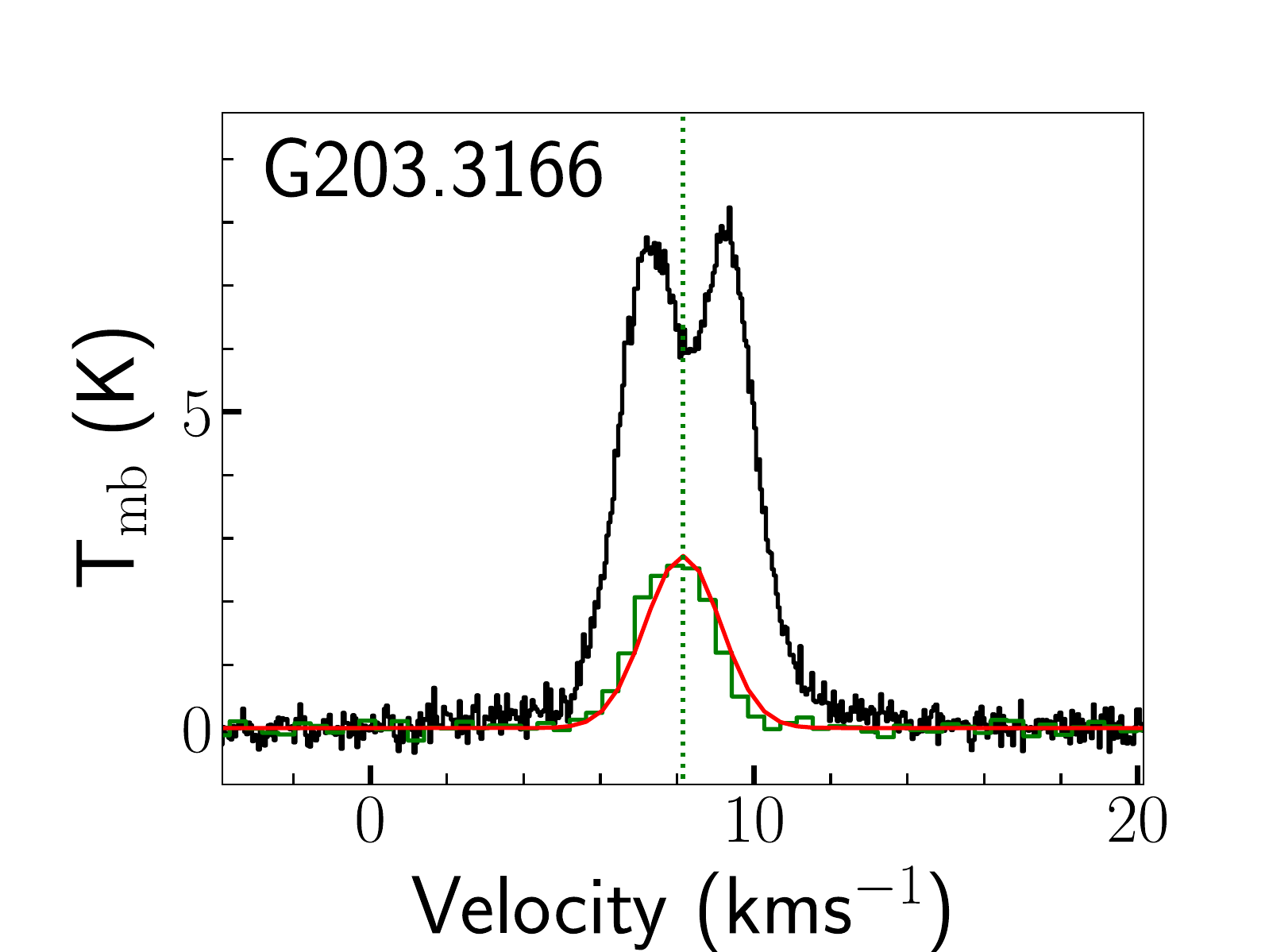}
\includegraphics[width=0.21\textwidth]{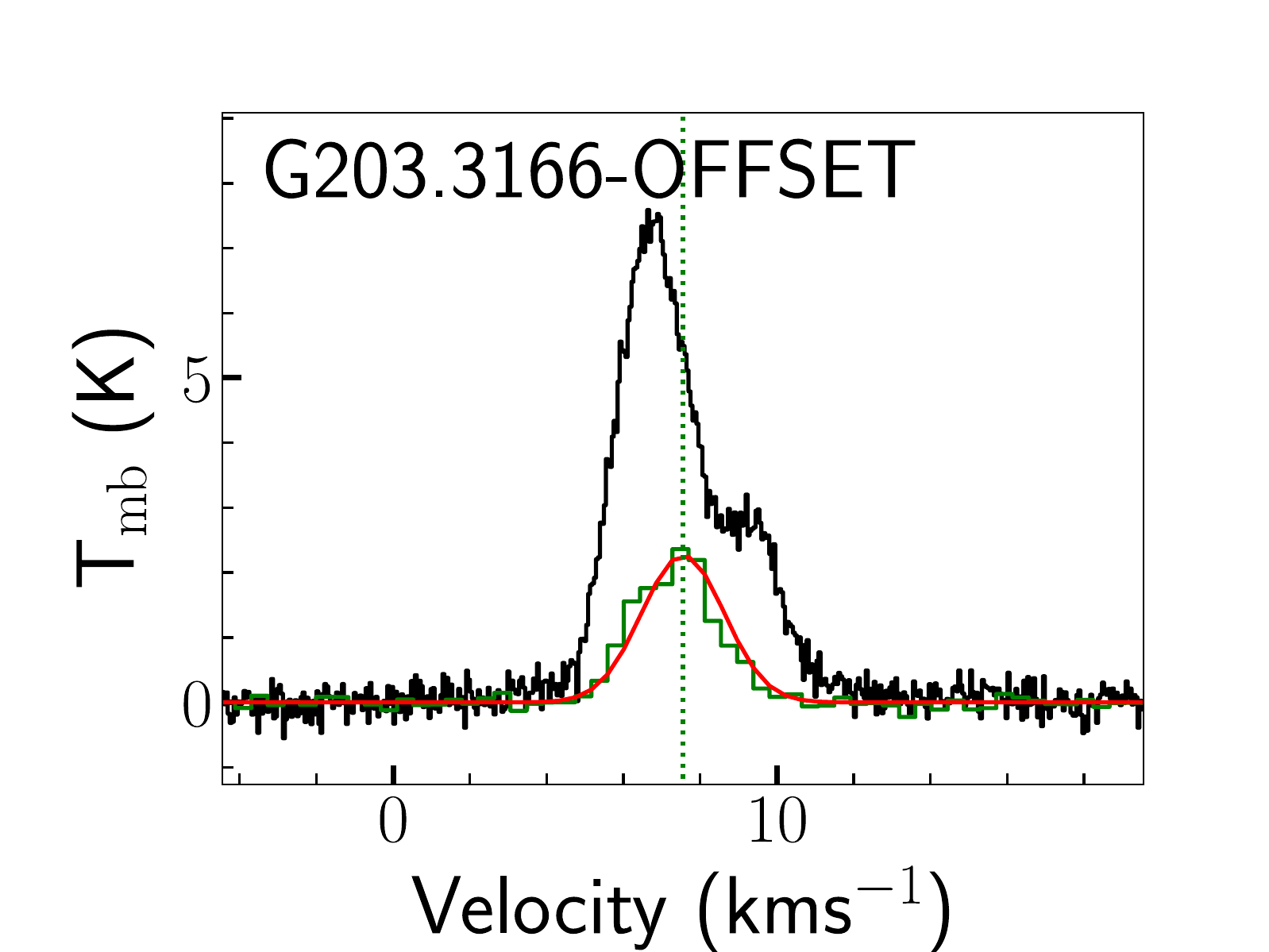}
\includegraphics[width=0.21\textwidth]{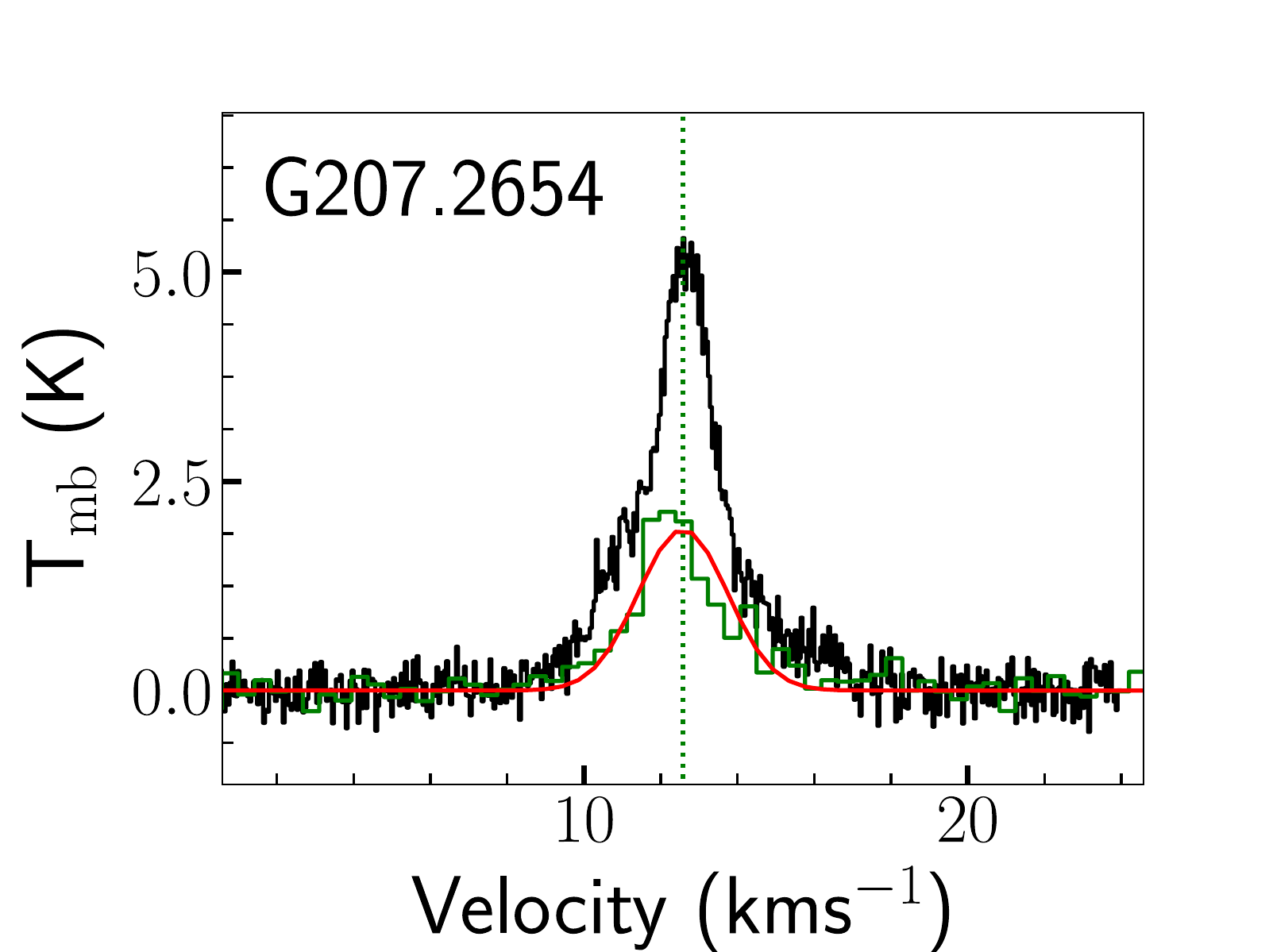}
\includegraphics[width=0.21\textwidth]{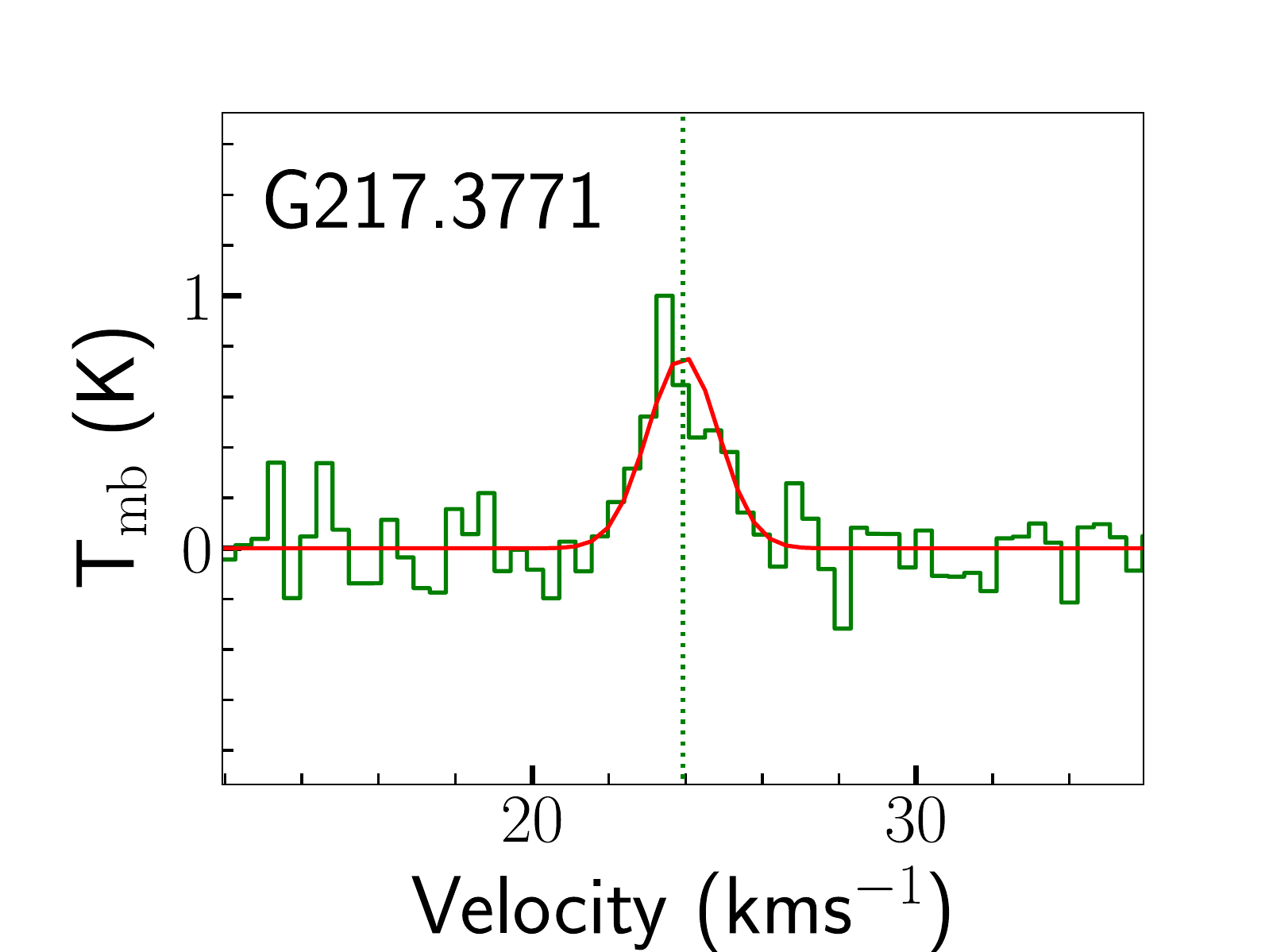}

\caption{\hco (solid black line) and \htco (solid green line) J$=$4-3 spectra averaged over all pixels within the dendrogram fitted masks. The solid red line shows the Gaussian fits to the \htco emission (calculated from Table \ref{aa:tab:htcogaus} of the appendix). The green dotted line is at the position of the \htco V$_{\rm LSR}$ taken from Table \ref{aa:tab:htcogaus}. Both the \htco spectra and respective Gaussian fit have been multiplied by a factor of 4. The velocity scale is the same for all plots, and is the \htco V$_{\rm LSR}$\,$\pm$12\kms. Sources where no \hco observations were undertaken are missing the \hco spectra.  \label{spectra}}
\end{figure*}

The extent of the \htco emission is determined from dendrogram fits made to the \htco zeroth order moment maps, using the python based dendrogram fitting application, {\em astrodendro}\footnote{This research made use of astrodendro, a Python package to compute dendrograms of Astronomical data (http://www.dendrograms.org/)}. An \htco detection is assigned based on a $\geq$5$\sigma$ detection over a minimum of 4 contiguous pixels (approximately equivalent to the beam area of 4.45 pixels). The rms noise per pixel in the integrated intensity maps is obtained using $\Delta\,I=T_{\rm mb(rms)}\,\Delta\,v\sqrt{N_{\rm chan}}$, where T$_{\rm mb(rms)}$ is the rms noise level in K per channel, $\Delta\,v$ is the velocity resolution in \kms (0.42\,\kms for \htco) and N$_{\rm chan}$ is the number of channels used to integrate the emission. The number of channels is determined from the minimum and maximum velocity in the \htco cubes that contain emission above the 3$\sigma$ limit. A sample of the \htco zeroth order moment maps with the \hco emission overlaid are shown in Figure \ref{figure:emission_maps_body} with the remainder provided in the online data. The \hco and \htco spectra (shown in Figure \ref{spectra}) display the average emission per pixel extracted from the sum of all pixels within the \htco dendrogram fitted mask. The Gaussian fits of the \htco spectra, presented in Table \ref{aa:tab:htcogaus} of the Appendix, are extracted from the the sum of the emission over the region.

We detect \htco emission towards 28 of the 31 distance limited RMS sources observed. For the three undetected sources (G233.8306, G081.7131, and G083.9383) no \htco emission is detected in a single pixel above the 3$\sigma$ limit. Towards several sources we find the peak of the \htco emission is offset by more than the FWHM of the JCMT beam ($>$14.5\arcsec) from the RMS position. Furthermore, towards two sources, G081.7522 and G203.3166, two \htco features are identified in the dendrogram fit. One component is associated with the RMS source position and a second structure is located in an offset position ($>$14.5\arcsec from the RMS position). We discuss the offset components in more detail below.

\subsubsection{{\rm \htco} offset components}

An offset component is identified if the centre of the pixel containing the peak of the \htco integrated intensity emission is spatially offset by more than a beam FWHM (14.5\arcsec) from the RMS source position (see Figure B1 in the online data for the HCO$^+$ and H$^{13}$CO$^+$ zeroth order moment maps towards the offset sources) .
In total six \htco offset components are identified:
\begin{enumerate}
\item Towards G079.8749, only one \htco component is identified and is offset from the RMS source position by $\sim$24\arcsec, located at R.A. (J2000)  20$^h$30$^m$29.5$^s$,  Dec. (J2000)   $+$41$^{\circ}$15$^{'}$51.4$^{''}$. However, there is a clear enhancement in the \htco emission towards the RMS source position in agreement with previous ammonia VLA observations \citep{Lu2014}. 
We therefore split the \htco emission into two separate components, a smaller one associated with the RMS source position, and a second larger component located in the offset position now labelled G079.8749-OFFSET. 
\item Towards G081.7522, two \htco components are identified; one feature coincident with the RMS source position and a second offset by $\sim$30\arcsec\ to the south of the RMS position located at  R.A. (J2000)  20$^h$39$^m$00.3$^s$,  Dec. (J2000)   $+$42$^{\circ}$24$^{'}$36.4$^{''}$. We label the second offset component G081.7522-OFFSET. This source, located in the northern part of the DR\,21 filament, was identified as a mm-continuum source (N43) by \citet{Motte2007} and does not have an outflow association in either SiO (2-1) \citep{Motte2007} or CO\,(2-1) \citet{Schneider2010}. 
\item Towards G081.7624, only one \htco component is present, located $\sim$22\arcsec\, to the north of the RMS source position at R.A. (J2000)  20$^h$39$^m$03.3$^s$,  Dec. (J2000)  $+$42$^{\circ}$25$^{'}$50.6$^{''}$, and is now labelled G081.7624-OFFSET. This component also resides in the northern part of the DR\,21 filament and was identified as a mm-continuum source (N53) by \citet{Motte2007}. This \htco feature has associated SiO\,(2-1) emission \citep{Motte2007} and a CO\,(2-1) outflow \citep{Schneider2010}. Furthermore, the $^{12}$CO (3-2) emission in \citet{Maud2015outflows} is also coincident with the offset component and we associate the outflow properties to G081.7624-OFFSET in this work. 
\item For G081.8789 and G081.8652, which were observed in the same JCMT map, the dendrogram fit reveals only one \htco feature,  located between the two RMS positions, $\sim$40\arcsec\, from G081.8789, and $\sim$20\arcsec\, from G081.8652 at R.A. (J2000)  20$^h$38$^m$36.3$^s$,  Dec. (J2000)  $+$42$^{\circ}$37$^{'}$30.3$^{''}$. While \htco emission extends over both sources, there appears to be no obvious enhancement towards either RMS source, this is consistent with the C$^{18}$O emission in this region \citep{Maud2015cores}. The peak of the H$^{13}$CO$^+$ emission is coincident with W75N, which hosts multiple mm continuum peaks and outflow emission (e.g., \citealt{Minh2010}). Furthermore, the $^{12}$CO (3-2) emission in \citet{Maud2015outflows} is also coincident with W75N and we associate the outflow properties to W75N in this work. Given the source confusion in this field, W75N was not listed as an MSX point source. We label this \htco component as W75N but class it as an offset source for the remainder of the analysis as it does not coincide with a listed RMS point source position. 
\item Towards G083.7071, only one \htco component is identified, offset from the RMS position by $\sim$16\arcsec. The peak of the \htco emission is located at R.A. (J2000)  20$^h$33$^m$35.2$^s$,  Dec. (J2000)   $+$45$^{\circ}$35$^{'}$36.5$^{''}$. This component  is not coincident with any previously known source and is labelled G083.7071-OFFSET.
\item Towards G203.3166, two \htco components are identified in the dendrogram fit; the first feature is coincident with the RMS source position and the second component is offset by $\sim$37\arcsec\, to the south east of the RMS position. The offset component, labeled G203.3166-OFFSET, is located at R.A. (J2000)  06$^h$41$^m$12.1$^s$,  Dec. (J2000)   $+$09$^{\circ}$29$^{'}$11.3$^{''}$, and is coincident with the position of C-MM3 (see \citealt{Cunningham2016} and references therein). 
\end{enumerate}

\subsubsection{\hco column density estimates}

We estimate the \hco column density, assuming that the \htco emission is optically thin, following, 
\begin{equation}
\label{c2:eq:hcocolumn}
N_{\rm H^{13}CO^+}=\frac{8\pi\kappa\nu^2}{hc^3}\frac{1}{g_uA_{ul}}Q(T_{\rm ex})e^{\frac{E_u}{kT_{\rm ex}}}\int{T_{\rm mb}dv},
\end{equation}
where $\int{T_{\rm mb}dv}$ is either the average of the \htco integrated intensity (where the emission is the average over all pixels in the dendrogram fit) or the peak \htco integrated intensity extracted at the peak of the \htco emission, and N$_{\rm H^{13}CO^+}$ is then the average or peak column density. Q(T$_{\rm ex}$) is the partition function, and is well approximated by Q(T$_{\rm ex}$)$=$({\it k}T$_{\rm ex}$)/(hB) for linear rotators, where B is the rotational constant and T$_{\rm ex}$ is the excitation temperature. A$_{\rm ul}$ is the Einstein A coefficient in s$^{-1}$, g$_u$ is the degeneracy of the upper energy state and E$_u$ is the energy of the upper state. We assume a value of 44\,K for the excitation temperature as used by \citet{Klaassen2007}. Furthermore, towards a similar sample of RMS selected young massive star forming regions \citep{Cunningham2015PhD} an average rotational temperature of 44\,K was derived from the CH$_3$CN (J$=$5-4) ladder. The \hco column densities are estimated assuming an abundance ratio between \htco and \hco to be 65 \citep{Rygl2013} and are given in Table \ref{c2:table:parameters}.  
In addition, we also provide mass estimates for individual sources in Table \ref{c2:table:parameters}. The masses are taken from \citet{Maud2015cores}, derived using the 850$\mu$m SCUBA fluxes. For sources not listed in \citet{Maud2015cores}, we follow the same procedure and extract the 850$\mu$m fluxes from \citet{DiFrancesco2008} checking that the SCUBA positions are coincident with the position of the offset emission. The mass estimates are used in Sections \ref{section:sio1}, \ref{section:sio2}, and \ref{section:sio3} for comparison with the SiO luminosities, and are used in the bolometric-luminosity-to-mass ratio in Figure \ref{image:Lsio_lbym}.

\begin{table*}
\begin{minipage}{180mm}
\begin{small}
\begin{center}
\label{table:detections_work}
\caption{\label{table:detections_work}Summary of the molecular-line detections towards the sources surveyed. Column 1 gives the Galactic name, sources where a \htco component was detected more than 14.5\arcsec offset from the RMS source position are labelled Galactic name-OFFSET. Column 2 gives the RMS classification of the source. Columns 3, 4, and 5 give the corresponding detection (Y), non-detection (N) or were not observed (-) of SiO, \htco and \hco respectively. Columns 6 and 7 are the asymmetries estimated from the \htco and \hco spectra extracted from both the average emission over the whole source and peak of the \htco emission, where N, R and B represent no asymmetry, red asymmetry and blue asymmetry, respectively. Column 8 is the detection (Y), and non-detection (N) of {\em Herschel} 70$\mu$m, within 14.5\arcsec of the peak of the \htco component. Sources not available or present in the online data by either survey are noted by (-).}
\begin{tabular}{l c c c c c c c}
\hline\hline

Source & RMS    & SiO & H$^{13}$CO$^+$ & HCO$^{+}$  & \multicolumn{2}{c}{Line Asymmetry$^{a}$} &{\em Herschel}  \\
\cline{6-7}\
Name   & Type   & (8-7)     &  (4-3)              & (4-3) &   Average          &   Peak          &           70$\mu$m flux \\

\hline

 &  &&\multicolumn{4}{c}{CO outflow detection} & \\

\hline
G010.8411   & YSO     & N  &  Y &  Y & N & N & -- \\
G012.9090   & YSO     & Y  &  Y &  Y & N & N & Y  \\
G013.6562   & YSO     & Y  &  Y &  Y & R & N & Y  \\
G017.6380   & YSO     & N  &  Y &  Y & N & B & Y  \\
G018.3412   & YSO     & Y  &  Y &  Y & N & N&Y\\
G043.3061   & H{\scriptsize II}     & Y  &  Y & Y &  B &R &Y \\
G050.2213   & YSO     & Y  &  Y & Y &  N &N &Y   \\
G078.1224   & YSO     & Y  &  Y & Y &  N &N&Y \\
G079.1272   & YSO     & Y  &  Y & Y &  N& N& Y  \\
G079.8749   & H{\scriptsize II}     & N  &Y&  Y & R  &  R & Y\\
G079.8749-OFFSET& --    & N  &  Y & Y &  N&B & Y\\
G081.7133   &  H{\scriptsize II}    & Y  & Y  &  -- & -- & --&Y  \\
G081.7220   &  H{\scriptsize II}    & Y  & Y  &  Y & B &B &Y \\
G081.7522   & YSO     & Y  & Y  &  --& --&--& Y \\
G081.7522-OFFSET&--    & N  & Y  &  --& --& --&Y \\
G081.7624   & YSO     & N  & N  &  --& --& --&Y \\
G081.7624-OFFSET & -- & Y  & Y  &  --& --& --&Y \\

G081.8652$^b$ & YSO   & N  & N  & Y & -- &--& N$^{c}$\\
W75N$^b$& --& Y  & Y  & Y & R &N& Y \\
G081.8789$^b$  & H{\scriptsize II}  & N  & N  & Y & --& --&N$^{c}$ \\
G083.0936   & H{\scriptsize II}     & N  & Y  & Y & N &N & --\\
G083.7071   & YSO     & N  & N  & Y & N & N & -- \\
G083.7071-OFFSET& --    & N  & Y  & Y & N & B & -- \\
G083.7962   & H{\scriptsize II}     & N  & Y  & Y & R &N & -- \\
G103.8744   & YSO     & N  & Y  & Y & N &B & Y \\
G109.8715   & YSO     & Y  & Y  & Y & N &B& Y \\
G192.6005   & YSO     & Y  & Y  & Y & N &N& Y\\
G194.9349   & YSO     & N  & Y  & Y & N  &N& Y \\
G203.3166   & YSO     & Y  & Y  & Y & R &R& Y\\
G203.3166-OFFSET& --    & Y  & Y  & Y & B &N& Y\\
G207.2654   & H{\scriptsize II}/YSO & Y  & Y  & Y & N &N& Y \\
\hline 
& & &\multicolumn{4}{c}{No CO outflow detection} &  \\
%No CO Outflow     &   &      &    &      &     &   \\
\hline
G080.8645 & H{\scriptsize II}  &    N  & Y  &  --  &  --  &--&  Y \\
G080.9383 & H{\scriptsize II}  &    N  & N  &  --  &  --  & --& Y  \\
G081.7131 & YSO  &    N  & N  &  --  & -- &--  &  Y  \\
G196.4542 & YSO  &    N  & Y  &  Y   & N   &N &  -- \\
G217.3771 & H{\scriptsize II}  &    N  & Y  &  --  &   -- &--&  Y \\
G233.8306 & YSO  &    N  & N  &  --  &  -- &-- &  Y \\
\hline

\end{tabular}
\end{center}
\end{small}

{\bf Notes}\\
{\it ($a$)} The line asymmetry is given for both the average and peak emission, and is denoted by a B for a blue asymmetry where $\delta$V$\leq$\,$-$0.25, R for red asymmetry where $\delta$V$\geq$\,0.25, and N for no asymmetry. \\
{\it ($b$)} These sources are all spatially located within $\sim$1\,arcminute. The \htco emission peaks between the two RMS sources, $\sim$ 40\arcsec \,from G081.8789, and $\sim$20\arcsec\, from G081.8652. While \htco emission does extend over the whole region, there appears to be no obvious extension or enhancement towards either RMS source, therefore the \htco component is associated with the offset position, W75N, and G081.8789 and G081.8652 are classed as non-detections and their \hco properties are not estimated.\\
{\it ($c$)} Herschel 70$\mu$m emission extends over both sources, however the dendrogram fit cannot separate the emission from the dominant 70$\mu$m component in the field which is associated with the offset position, W75N.\\
\end{minipage}
\end{table*}

\subsection{Detecting active outflow signatures with SiO}

A source is determined to have an SiO detection if a minimum 3$\sigma$ detection is obtained in at least one pixel in the SiO integrated intensity maps. The integrated SiO intensity is extracted from the zeroth order moment maps, where the velocity range is determined using either the velocity of the upper and lower channels above 3$\sigma$ in the SiO channel maps (where possible) or from the $^{12}$CO linewidths taken from \citet{Maud2015outflows}. Furthermore, only pixels situated within the respective \htco integrated intensity mask are considered (this was done to eliminate the possibility of a spurious detection that may appear towards the edge of the maps being identified as a detection). The SiO luminosity (in units of K\,\kms\,kpc$^2$) is calculated from L$_{\rm SiO}=$\,$\int$\,T$_{\rm mb(SiO)}$dv$\times$4$\pi$d$^{2}$, where $\int$\,T$_{\rm(SiO)}$dv is the SiO integrated intensity extracted from the sum of the pixels in the zeroth order moment maps (see Table \ref{c2:table:parameters} for individual source values and Figure B2 of the online data for individual SiO integrated intensity maps), and d is the distance to the respective source. For sources without an SiO detection, we estimate the 3$\sigma$ upper limits using the rms in a single pixel. For the 6 sources without a CO outflow detection a velocity interval of 26\kms (the average $^{12}$CO linewidth of the SiO non-detections) is used to estimate the upper limit of the SiO luminosity.

SiO J$=$\,8-7 is detected towards 14 ($\sim$45\%) of the 31 RMS sources observed (excluding both sources, G045.0711 and G020.7617, that fall outside of the distance limits), see Table \ref{table:detections_work} for a list of detections towards individual sources. We do not detect SiO emission towards the 6 sources without a confirmed CO outflow in \citet{Maud2015outflows}. We detect SiO emission towards 3 of the 6 defined OFFSET sources, G203.3166-OFFSET, G081.7624-OFFSET and W75N. Therefore, we detect SiO emission, including the offset sources, towards 17(46\%) of the distance limited sample (see Table \ref{c2:table:detection_stats}). 

For completeness, we also provide an estimate of the average SiO column density (N$_{\rm SiO}$) and average abundance (X$_{\rm SiO}$) in Table \ref{c2:table:parameters}. We calculate the SiO column density using Equation \ref{c2:eq:hcocolumn} substituting the values for SiO (8-7). For the SiO abundance, we derive the N$_{H_2}$ column density using the $^{12}$CO\,(3-2) column densities given in \citet{Maud2015outflows}, averaged over the blue- and red-shifted outflow lobes. The CO\,(3-2) column densities were used because the SiO emission is likely to be produced as a result of shocks in the jet/outflow and is not expected to be associated with the compact continuum emission tracing the core. However, in doing this we also assume that SiO arises from the same component in the outflow as the CO emission, which may not be the case.

\begin{table*}
%\begin{minipage}{180mm}
\begin{small}
\begin{center}
\caption[Physical Properties Estimated for the Sources]{Physical properties estimated for the sources. The 3$\sigma$ upper limits (represented by $<$) are provided for sources that have no detected emission. Where  (-) represents sources for which no estimate of the property was possible.\label{c2:table:parameters}}
\begin{tabular}{lcccccc}
\hline
Source & Peak N$_{\rm HCO^+}$ &Average N$_{\rm HCO^+}$&Average N$_{\rm SiO}$ & Average X$_{\rm SiO}$ & Mass &$\int{T_{\rm SiO}\,dv}$\\
Name &(cm$^{-2}$)& (cm$^{-2}$) & (cm$^{-2}$) & &(M$_{\odot}$) &(K\,km\,s$^{-1}$)\\
&($\times$10$^{13}$) &($\times$10$^{13}$) &($\times$10$^{12}$) &($\times$10$^{-9}$) & &\\
\hline

 &\multicolumn{4}{c}{SiO Detection}  &\\

\hline
G012.9090  & 24.2$\pm$8.6&5.4$\pm$2.4 & 3.4$\pm$0.7 & 1.1$\pm$0.2 & 1167& 25.4\\
G013.6562  & 5.1$\pm$0.7 & 4.5$\pm$0.4&2.0$\pm$0.4  & 0.5$\pm$0.2 & 1385 &5.2\\   
G018.3412  & 12.1$\pm$0.7 & 4.3$\pm$0.3 &1.6$\pm$0.4  &1.1$\pm$0.2 & 224& 1.0\\   
G043.3061  & 4.8$\pm$1.3 & 3.5$\pm$0.4 & 0.8$\pm$0.2  & --& 595 & 3.2 \\   
%G045.0711   &9.8& 43.4$\pm$6.7 & 28.2$\pm$4.4 &3.0$\pm$0.3  & 1.3 & 21.3$\pm$2.3 &0.6$\pm$0.1 & 5833$\pm$352 & 0.18 &35672$\pm$270 \\   
G050.2213  & 2.7$\pm$0.9 & 2.1$\pm$0.5& 2.0$\pm$0.4 &2.5$\pm$0.7 & 397 & 1.2 \\   
G078.1224  & 17.6$\pm$0.8 & 6.1$\pm$0.3& 4.3$\pm$0.7 &6.7$\pm$1.1 & 90 & 53.1 \\ 
G079.1272  & 4.1$\pm$0.7 & 3.0$\pm$0.4& 1.8$\pm$0.2 &5.4$\pm$1.1 & 24 & 2.1 \\   
G081.7133  & 11.5$\pm$1.3 & 8.0$\pm$0.8& 3.4$\pm$0.7 & 1.8$\pm$0.2& 367 & 35.0\\   
G081.7220  & 69.9$\pm$2.0 & 15.2$\pm$0.5& 10.0$\pm$1.3 & 18.0$\pm$2.9 & 312 & 132.0\\   
G081.7522  & 10.7$\pm$0.8 & 4.9$\pm$0.3& 2.0$\pm$0.7 &1.1$\pm$0.2 & 272 & 1.2 \\   
G081.7624-OFFSET & 6.12$\pm$0.9 & 3.4$\pm$0.5& 4.0$\pm$0.7 & 3.6$\pm$0.7 & 201 & 44.6\\  
W75N       & 102.1$\pm$1.5 & 17.5$\pm$0.3& 6.7$\pm$0.2 & 2.5$\pm$0.2 & 647 & 98.3\\
G109.8715  & 62.3$\pm$1.4 & 15.4$\pm$0.4& 6.5$\pm$0.9 &1.8$\pm$0.2 & 112 & 85.8\\ 
G192.6005  & 14.7$\pm$1.0 & 10.5$\pm$0.6 & 2.5$\pm$0.7 &3.4$\pm$0.9 & 130 & 17.2\\ 
G203.3166  & 11.4$\pm$0.8 & 4.7$\pm$0.2& 1.8$\pm$0.5 & 19.8$\pm$4.3& 61 &19.6\\
G203.3166-OFFSET & 8.9$\pm$0.6 & 4.3$\pm$0.2&  2.2$\pm$0.4 & -- & 33  & 11.8\\ 
G207.2654  & 7.1$\pm$1.5 & 4.0$\pm$0.5&2.2$\pm$0.7  &5.6$\pm$1.4 & 172  & 4.4\\

\hline

 &\multicolumn{4}{c}{SiO Non-detection}& \\

\hline
G010.8411  & 12.2$\pm$1.5 & 4.3$\pm$0.4& $<$1.3  & -- & 139 & $<$0.9 \\
G017.6380  & 19.1$\pm$0.7 & 5.4$\pm$0.2& $<$1.3  & -- & 374 & $<$0.8 \\   
G079.8749  & 2.7$\pm$0.7 & 2.0$\pm$0.4& $<$1.1 & -- & -- & $<$0.7 \\  
G079.8749-OFFSET & 16.1$\pm$1.3 & 5.7$\pm$0.3&  $<$1.1 & -- & --& $<$0.7 \\
G081.7522-OFFSET & 10.8$\pm$1.0 & 5.4$\pm$0.3&$<$1.3   & -- & -- &$<$0.9\\ 
G083.0936     & 2.4$\pm$1.3 & 1.6$\pm$0.6& $<$0.9 &-- & -- & $<$0.5 \\
G083.7071-OFFSET  & 2.5$\pm$0.9 & 2.2$\pm$0.6& $<$1.3  & --  & -- & $<$0.8 \\   
G083.7962    & 3.5$\pm$0.9 & 3.5$\pm$0.6& $<$1.3 & --  & -- & $<$0.9 \\ 
G103.8744    & 4.5$\pm$1.1 & 3.7$\pm$0.8&$<$1.6  &-- & 91 & $<$1.0\\ 
G194.9349    & 3.1$\pm$0.9 & 2.3$\pm$0.6& $<$1.3 & --& -- & $<$0.9 \\ 

\hline
&\multicolumn{4}{c}{No SiO or CO Outflow Detected} & \\

\hline
G080.8645 & 6.3$\pm$0.7 & 3.4$\pm$0.3& $<$1.3 & --& 137 &$<$0.8 \\
%G080.9383 & -- & -- &-- & --  &\\
%G081.7131 & -- & -- & -- & -- & -- & $<$5 &  --&\\
G196.4542 & 3.5$\pm$0.9 & 2.1$\pm$0.5&$<$1.6  & --& 167 &$<$1.0  \\
G217.3771 & 1.8$\pm$0.9 & 1.2$\pm$0.4& $<$1.3 &-- &  --& $<$0.8 \\
%G233.8306 & $<$0.9& -- & -- & -- &-- & $<$18 &-- \\

\hline
\end{tabular}
\end{center}
\end{small}
\end{table*}

\begin{table}
%\begin{minipage}{180mm}
\begin{center}
\caption{Summary of the outflow and infall detections\label{c2:table:detection_stats}}
\centering
\begin{tabular}{cccccc}
\hline\hline
Source   & Total    & SiO       &  Total & \multicolumn{2}{c}{Blue Asymmetric$^{a}$}  \\
Type              & & 3$\sigma$ &  HCO$^+$& \multicolumn{2}{c}{Profile}  \\
\cline{5-6}\
              &    &   & Observed &  Ave &Peak         \\
\hline
YSO           & 20  & 10 &  16 & 0 &3\\
H{\scriptsize II}           & 10  & 3  &  6  & 2 & 1\\
H{\scriptsize II}/YSO       &  1  & 1  &  1  & 0 &0\\
OFFSET        &  6  & 3  &  4  & 1 &2 \\
\hline
Total (\%)        &  37  & 17(46\%)  &  27(73\%)  & 3(12\%) & 6(24\%)\\
\hline
\hline
\end{tabular}
%\label{table:Continuum}
\end{center}
{\bf Notes}\\
{\it ($a$)} The asymmetry is derived for 25 sources using the \htco and \hco emission extracted from both the average (Ave) and peak (Peak) spectra. Where the average spectra are taken from the emission averaged over all pixels taken from the dendrogram fits and the peak spectra are taken from the pixel at the peak of the \htco emission. 
\end{table}

%---------------------------------------Infall Signatures------------------------------------------------------------------
\subsection{Infall signatures determined from the HCO$^+$ and H$^{13}$CO$^+$ emission}

Both \hco (4-3) and \htco (4-3) are dense-gas tracers (n$_{\mbox{crit}}$$\sim$8$\times$10$^{6}$cm$^{-3}$) and, as such, their emission can be used to probe the dynamics of the dense-gas, such as infall or expansion. Infall  is typically interpreted if a blue asymmetry, either from a double-peaked line profile with a brighter blue peak or a single-peak profile, is observed in the optically thick HCO$^{+}$ transition, and is offset from the optically thin isotopologue, H$^{13}$CO$^{+}$, which shows only a single peaked component at rest velocity (e.g. \citealt{Myers1996}). A single-peak in the optically thin \htco\,(4-3) line allows us to distinguish between self absorption and multiple line of sight components in the optically thick HCO$^{+}$ profile. The predominance of either a blue or red asymmetry is quantified by the skewness parameter \citep{Mardones1997} which is estimated from,

\begin{equation}
\delta\,V=\frac{V_{\rm thick}-V_{\rm thin}}{\Delta\,V_{\rm thin}},
\end{equation}
where V$_{\rm thick}$ and V$_{\rm thin}$ are the LSR velocities at line peak for the optically thick \hco\,(4-3) and optically thin \htco\,(4-3) transitions, respectively. The velocity difference is then normalized by the FWHM of the optically thin H$^{13}$CO$^{+}$ line ($\Delta$\,V$_{\rm thin}$). The \htco FWHM and V$_{\rm LSR}$ are taken from the Gaussian fits presented in Table \ref{aa:tab:htcogaus} of the Appendix and V$_{\rm thick}$ is taken from the position of the brightest emission peak in the \hco spectrum. To explore the presence of global infall in these regions, the spectra shown in Figure \ref{spectra} are extracted from the average of the emission over all pixels within the dendrogram-fitted masks. The result is then the dimensionless skewness parameter $\delta$\,V. A significant blue or red excess is defined as $\delta$V\,$\leq$\,$-$0.25 or $\delta$V\,$\geq$0.25, respectively \citep{Mardones1997}. Of the sources where it was possible to determine the asymmetries, 3 objects show a blue excess indicative of infall and 5 show a red excess (expansion) and 17 show no red or blue excess. All 3 sources with a blue excess have a corresponding SiO detection and 3 of the 5 sources with a red excess have an SiO detection (see Table \ref{table:detections_work} for individual sources). The number of sources with an infall detection is consistent with the number of sources without an infall detection given the Poisson errors of 3$\pm$1.7 and 5$\pm$2.25 respectively. However, as the emission is extracted from the full source extent and likely encompasses multiple protostars, this may add noise and mask the signs of global infall. Therefore, we also assess the asymmetry considering the spectra from the \htco peak position, finding a total of 6 sources (see Table \ref{table:detections_work}) with a blue asymmetry and 3 with a red asymmetry. Only a single source G081.7220 displays a blue asymmetry in both the averaged and peak spectra. Furthermore, the Poisson errors are again consistent for sources displaying a blue and red asymmetry of 6$\pm$2.5 and 3$\pm$1.7, respectively. This suggests that the majority of sources in our sample show no preference for global infall motions. However, an important consideration is the sensitivity for infall asymmetry to line optical depth, excitation temperature, and density. \citet{Smith2013} find an increase in the blue asymmetry of the optically thick line with decreasing beam size, suggesting that matching the beam size with the energy of line transition will increase the detection of infall signatures. Furthermore, as noted we are likely sensitive to multiple sources within the JCMT beam which can add noise to the observations. Future higher spatial resolution observations, resolving individual protostars, will be able to directly test this. It should also be noted that \hco is a known tracer of outflow emission in massive star forming regions (e.g. \citealt{WalkerSmith2014}), and several of the regions display broad line wings in the \hco spectra which can be clearly seen in Figure \ref{spectra} (e.g. G109.8715). In addition, several sources show an offset between the red- and blue-shifted \hco emission in Figure B1 of the online data (e.g. G050.2213, G192.6005, and G207.2654), again suggesting that the \hco emission is influenced by the outflow in several regions. Furthermore, sources that display no asymmetric line profile in the \hco spectra, taken over the whole source extent, but show a blue asymmetric profile in the spectra taken from the peak, tend to show multiple components in the red- and blue-shifted \hco emission maps (e.g. G017.6562 and G083.7071-OFFSET). This may add to the lack of consistency between the presence of asymmetry in the peak and average spectral line profiles.

\subsection{Far-IR associations}

We obtain 70$\mu$m
{\em Herschel} PACS fluxes for 25 of the 31 RMS sources with available data in the
archive. The 70$\mu$m flux was extracted using dendrogram
fits, again using the python package {\em astrodendro}. The minimum number of
contiguous pixels was set to the beam area of the {\em Herschel} map and the minimum
detection was set to 5$\sigma$. The rms noise for each source was
determined from an aperture local to that source and not from the entire map,
therefore larger regions with higher levels of emission may have higher noise
estimates. We assign {\em Herschel} 70$\mu$m emission to an \htco component if the
peak of the 70$\mu$m emission is within 14.5\arcsec\, of the peak of the \htco
component (see Table \ref{table:detections_work} for the association of a
70$\mu$m component with a respective \htco component). Of the 25 RMS sources
with available {\em Herschel} 70$\mu$m data, 23 have an associated {\em Herschel}
peak. Only G081.8652, and G081.8789 do not have an associated 70$\mu$m
component. However, as with the \htco emission towards these sources, 70$\mu$m
emission extends over both RMS source positions but there is no obvious
enhancement towards either RMS source and the peak of the 70$\mu$m emission
coincides with the offset \htco component, W75N. All of the OFFSET
sources have an associated 70$\mu$m component within 14.5\arcsec\, of the
identified \htco peak\footnote{It should be noted that for G203.3166-OFFSET when observed at higher spatial resolution (e.g. \citealt{Cunningham2016}) the 70$\mu$m emission is not directly associated with the offset position C-MM3}. The sum of the 70$\mu$m flux, within the dendrogram
mask, is converted to a luminosity through
L$_{70\mu m}$\,=\,4$\pi$d$^2$.\,F$_{70\mu m}$, using a 25$\mu$m bandwidth for the {\em Herschel} 70$\mu$m PACS filter.

\begin{table*}
\begin{minipage}{180mm}
\begin{center}
\caption{Summary of properties between SiO-detected and non-detected sources.}
\centering
\begin{tabular}{l c c c c c c c c c c c}
\hline\hline
Source properties  &\multicolumn{2}{c}{SiO-detected$^a$}& &\multicolumn{2}{c}{No SiO Detected$^a$} & &\multicolumn{2}{c}{No SiO or CO Detected$^a$}& &\multicolumn{2}{c}{ KS-test$^b$} \\
\cline{2-3}\cline{5-6}\cline{8-9}\cline{11-12}
   &   Mean     & Median & & Mean & Median & & Mean & Median& & No SiO & No SiO\\
         & & &     &       &  &  &  &  & & & or CO\\

\hline
Distance (kpc)  & 1.9$\pm$1.1  &  1.4 & & 1.6$\pm$0.3 & 1.4 & &  2.1$\pm$1.1  & 1.4 & &0.45 &0.63 \\
L$_{*}$ (L$_{\odot}$ $\times$10$^{3}$)&13$\pm$12&11 & &15$\pm$27 & 5& &20$\pm$18 & 11 & &0.52 &0.51\\ 
\htco FWHM$^c$ (km\,s$^{-1}$)& 2.9$\pm$0.8 & 2.7 & &  2.0$\pm$0.5 & 1.9 & & 2.7$\pm$0.5 & 2.5 & &0.05 &0.11 \\
C$^{18}$O FWHM (km\,s$^{-1}$)& 2.9$\pm$0.7 &2.6 && 2.6$\pm$0.7 & 2.7& &-- & --& &0.48 & --\\
L$_{70\mu m}$( L$_{\odot}$ $\times$10$^{2}$) & 27$\pm$30& 18 & & 17$\pm$34  &  2  & & 18$\pm$15 & 10 & &0.21 &0.33\\    
Mass (M$_{\odot}$) &   364$\pm$378 &  224 & &  201$\pm$124  &  139  & &  107$\pm$64 &  137 & & 0.91 & 0.18\\
Average N$_{\rm HCO^+}$( cm$^{-2}$ $\times$10$^{13}$)& 6.9$\pm$4.7&4.7 & & 3.6$\pm$1.5&3.6 & &2.3$\pm$0.9&2.1& &0.29&0.06 \\
Peak N$_{\rm HCO^+}$( cm$^{-2}$ $\times$10$^{13}$)& 22.1$\pm$27.4&11.4 & & 7.7$\pm$6.0&4.2 & &3.9$\pm$1.9&3.5& &0.12&0.04 \\
$^{12}$CO linewidth (km\,s$^{-1}$)& 48$\pm$19  &  56  & & 26$\pm$6 & 25  & &  -- &  --  & & 0.009 &--\\

\hline
\hline
\end{tabular}
\label{table:mean_ks}
\end{center}
{\bf Notes}\\
{\it ($a$)} The mean is given with $\pm$ standard deviation. For the No SiO Detected sample the mean and median values are estimated considering sources without an SiO-detection that have a CO outflow detection. For the No SiO or CO Detected column the mean and median values are estimated considering only those sources with no SiO and no CO outflow. \\. 
{\it ($b$)} The results of the KS test for the No SiO column is considering sources with and without and SiO detection that have a CO outflow detection. The No SiO or CO is considering all sources without an SiO detection, including those sources with no CO outflow in \citet{Maud2015outflows}.\\
{\it ($c$)} The \htco FWHM presented here is extracted from the dendrogram fits to the full source extent.\\
\end{minipage}
\end{table*}

%%%%%%%%%%%%%%%%%%%%%%%%%%%%%%%%%%%%%%%%%%%%%%%%%%%%%%%%%%%%%%%%%%%%%%%%%%%%%%%%%%%%%%%%%%%%%%%%%%%%%%%%%%%%%%%%%%%%%%%%%%%%%%%%%%%%%5
%%%%%%%%%%%%%%%%%%%%%%%%%%%%%%%%%%%%%%%%%%%%%%%%%%%%%%%%%%%%%%%%%%%%%%%%%%%%%%%%%%%%%%%%%%%%%%%%%%%%%%%%%%%%%%%%%%%%%%%%%%%%%%%%%%%%%
\section{Discussion}           %%%%%%%%%%%%%%%%%%%%%%%%%% Discussion %%%%%%%%%%%%%%%%%%%%
%%%%%%%%%%%%%%%%%%%%%%%%%%%%%%%%%%%%%%%%%%%%%%%%%%%%%%%%%%%%%%%%%%%%%%%%%%%%%%%%%%%%%%%%%%%%%%%%%%%%%%%%%%%%%%%%%%%%%%%%%%%%%%%%%%%%%
%%%%%%%%%%%%%%%%%%%%%%%%%%%%%%%%%%%%%%%%%%%%%%%%%%%%%%%%%%%%%%%%%%%%%%%%%%%%%%%%%%%%%%%%%%%%%%%%%%%%%%%%%%%%%%%%%%%%%%%%%%%%%%%%%%%%%
\subsection{Comparison of SiO-detected and non-detected source properties}

SiO emission is detected towards approximately 46\% of the sources. Table \ref{table:mean_ks} presents the average, median, and standard deviation of the source properties (e.g., bolometric luminosity, distance, and \hco column density) for the SiO-detected and non-detected samples. For completeness we also include sources without an SiO or a CO outflow detection. We perform Kolmogorov-Smirnoff (KS) tests to determine if the source properties of the SiO-detected and non-detected sources are drawn from the same underlying distribution. The returned {\it p-}value from the KS test gives the confidence level at which the null hypothesis (i.e. that the two samples originate from the same underlying distribution) can be rejected. A value of $\leq$0.01 is associated with a high confidence that the two populations originate from different underlying distributions. We find no difference in the distance to sources with or without an SiO detection. The median distance is 1.4\,kpc for both samples. If the emission traced by SiO is considerably smaller than the beam, suggesting a very young outflow, then beam dilution may be responsible for the remainder of the SiO non-detections. However, this would need to be tested with higher angular resolution observations. We note that several of the SiO non-detections have the weakest \htco emission in the sample, but there is no obvious difference in the masses or bolometric luminosities between populations. Therefore, the lack of an SiO detection towards these sources should not be due to sensitivity limitations in the sample. 

We find no significant differences between the source properties of the SiO-detected and SiO non-detected populations (see Table \ref{table:mean_ks} for a list of all returned {\it p-}values). If we compare the outflow properties taken from \citet{Maud2015outflows}, such as the outflow velocity, momentum, force, mass and energy, we find only the CO outflow velocity has a {\it p}-value $\leq$0.01 between the SiO-detected and SiO non-detected sample. Furthermore, only sources with an SiO detection have a $^{12}$CO\,(3-2) total linewidth $>$35\,\kms, suggesting SiO emission is a more efficient tracer of high-velocity outflows. This is consistent with the CO outflow velocity ranges observed by \citet{Gibb2007} towards a sample of young massive stars, where sources with detected SiO (5-4) have associated outflows with a total maximum CO velocity of $>$36\,\kms. This suggests that the detection of the higher J transitions of SiO is an indication of the presence of a high-velocity outflow and is consistent with the expected shock velocities ($>$25\,\kms) required to disrupt dust grains (e.g. \citealt{Schilke1997}). For the remaining CO outflow properties we find no difference between the SiO-detected and non-detected samples. However, we find the total outflow-mass between the two populations has a {\it p}-value$\geq$0.9999, thus the outflow-masses estimated from the CO emission are drawn from the same distribution.

\subsection{SiO luminosity as a function of source properties\label{section:sio1}}

\begin{figure*}

\includegraphics[width=0.48\textwidth]{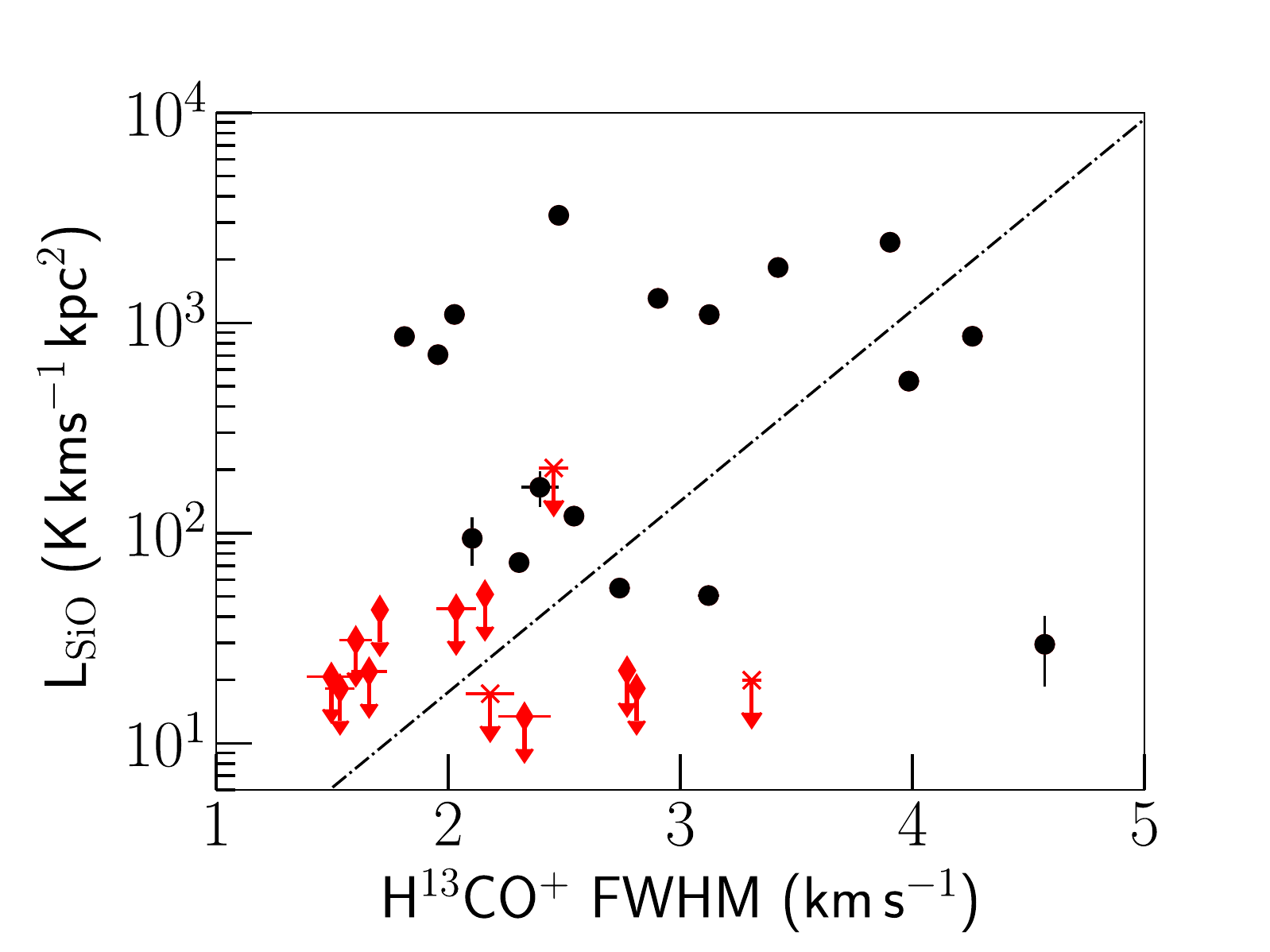}
\includegraphics[width=0.48\textwidth]{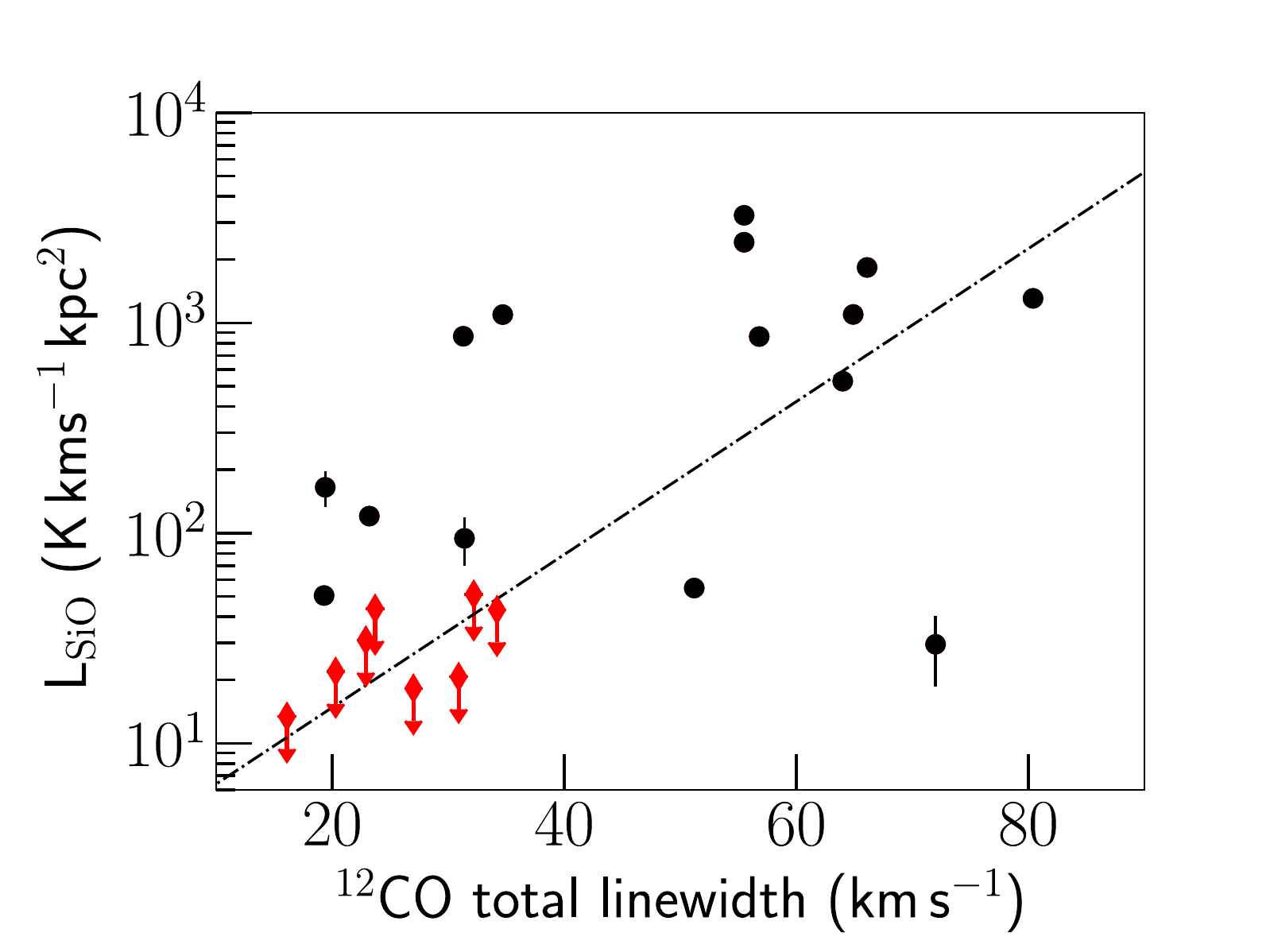}
\includegraphics[width=0.48\textwidth]{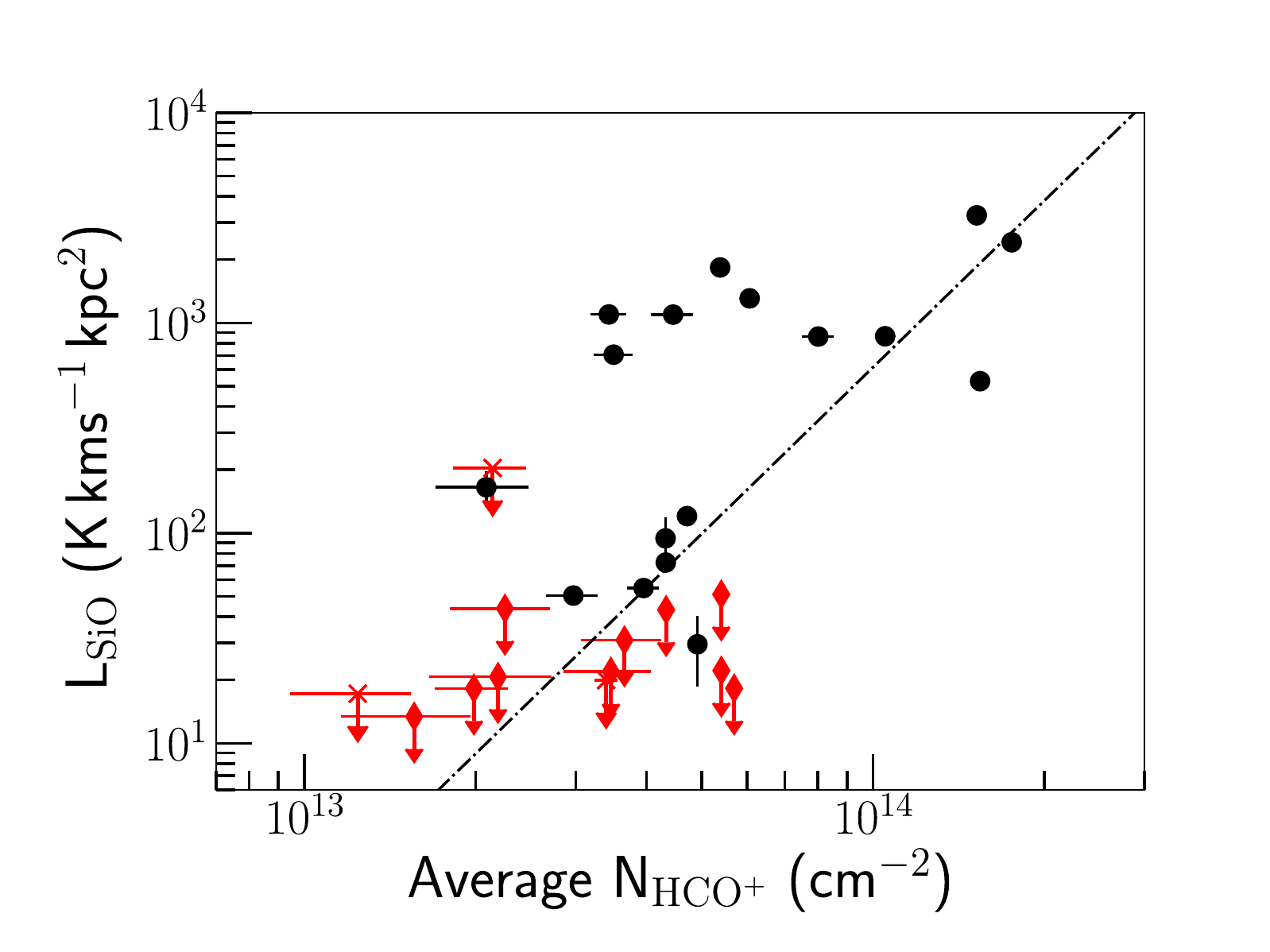}
\includegraphics[width=0.48\textwidth]{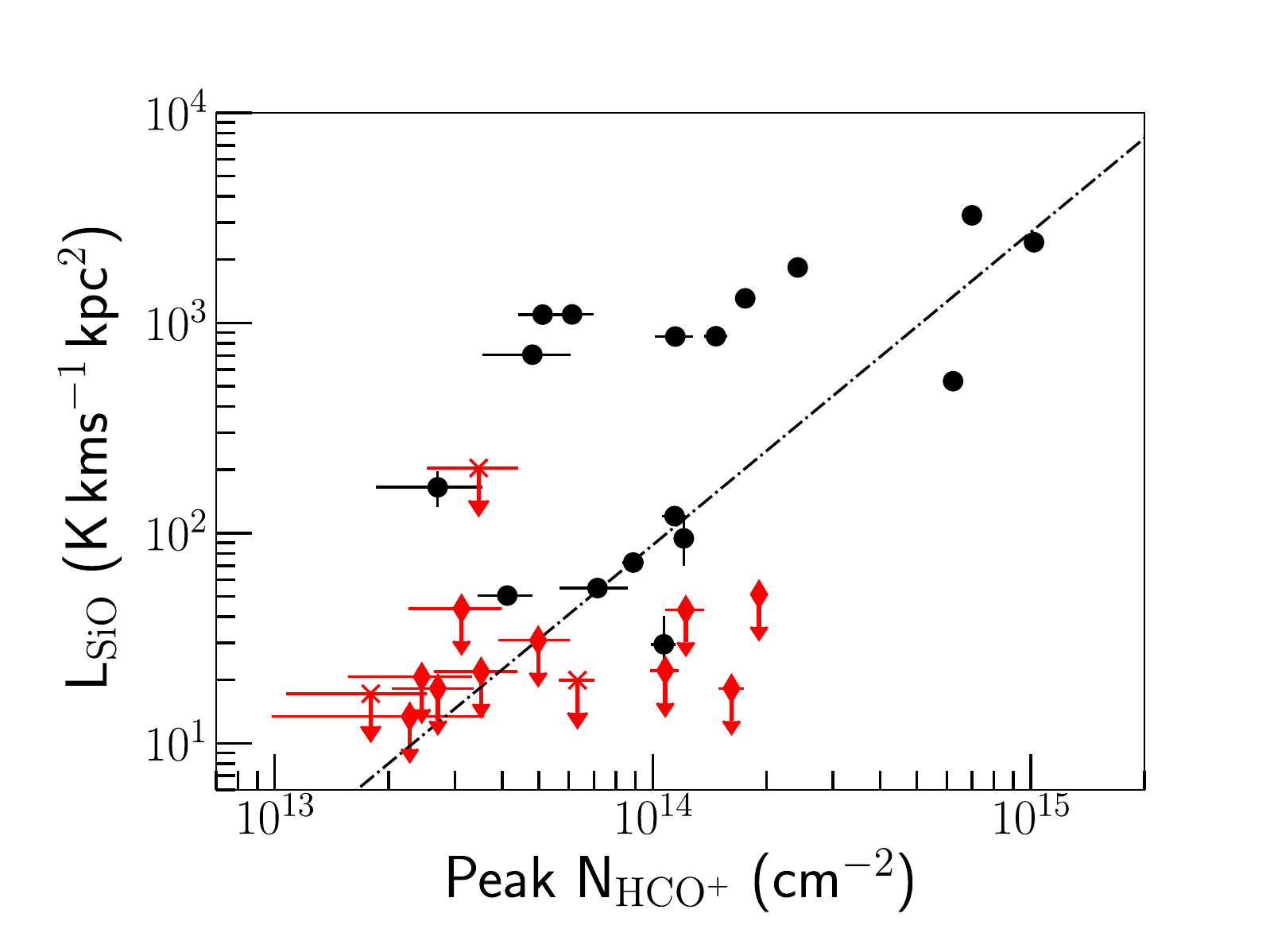}
\includegraphics[width=0.48\textwidth]{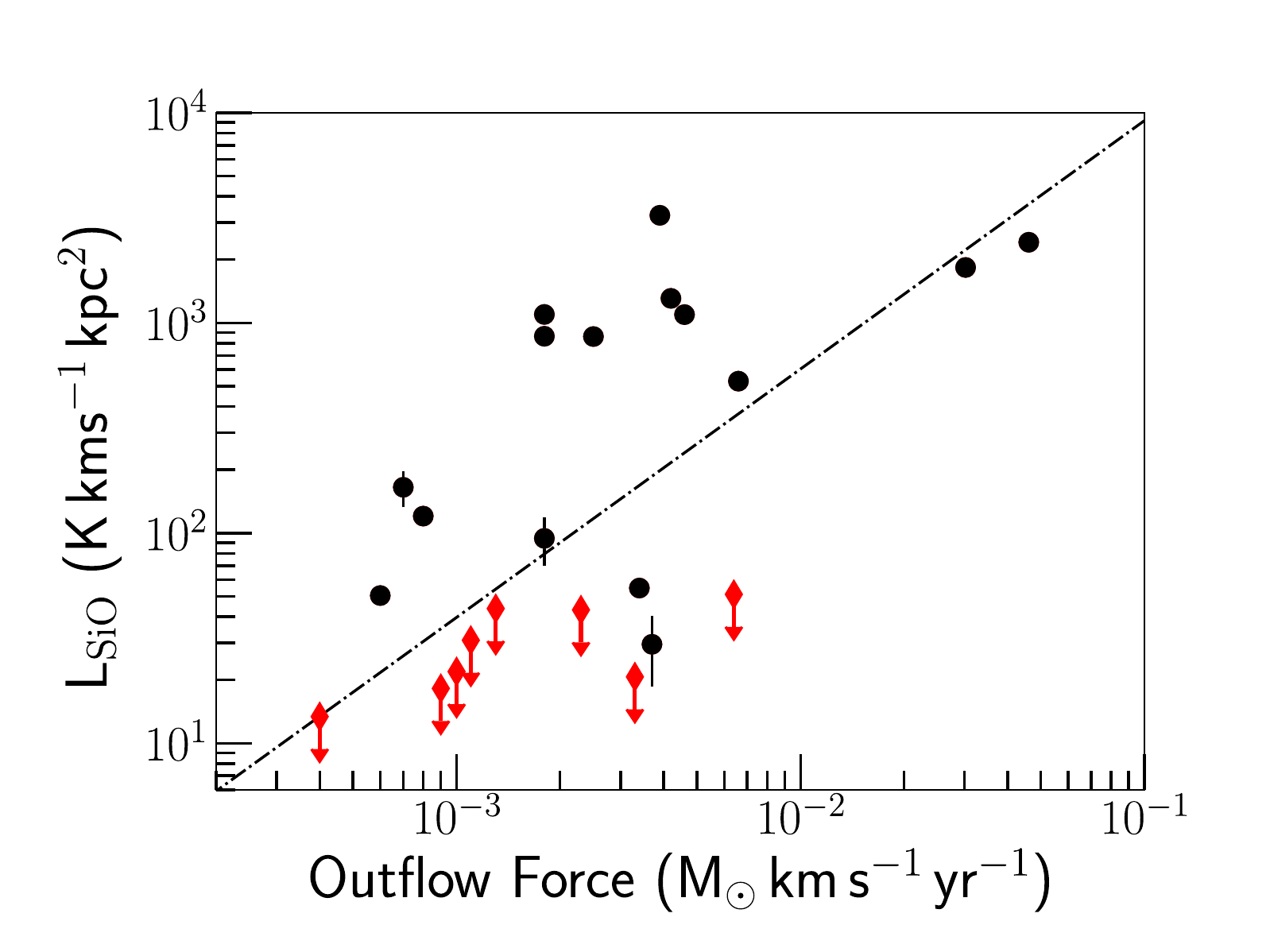}
\includegraphics[width=0.48\textwidth]{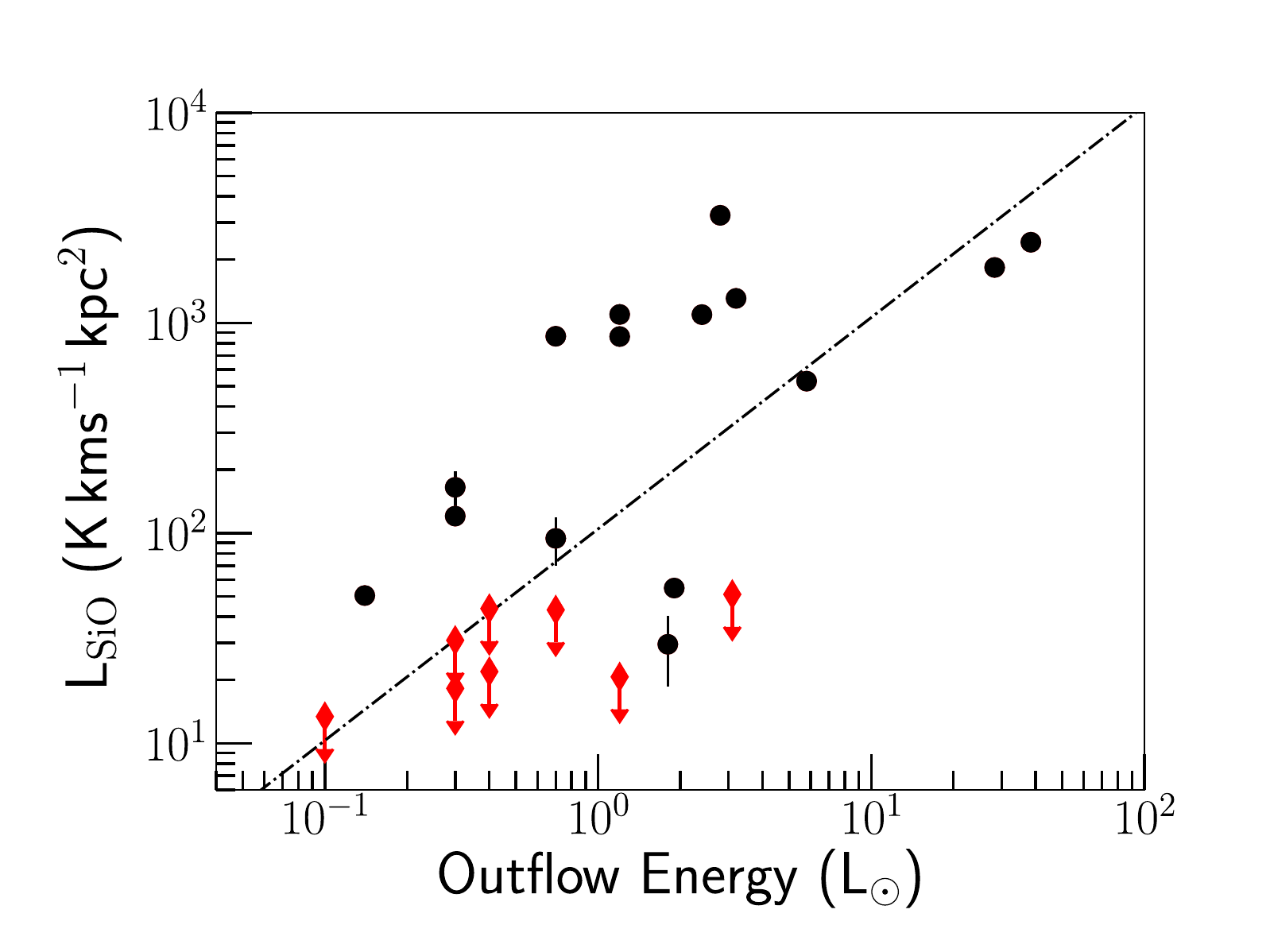}

\caption{SiO luminosity as a function of source properties, for those parameter relationships given in Table \ref{c2:table:sio_detected_correlation_coeffs} with a P-value $\leq$0.01 and a correlation coefficient $>$0.53. The plots include the \htco FWHM, $^{12}$CO\,(3-2) total linewidth or outflow velocity, average and peak \hco column density, outflow force and outflow energy. The black circles represent those sources with an SiO detection above 3$\sigma$ that have a confirmed CO outflow detection in \citet{Maud2015outflows}. The red diamonds and red crosses represent the 3$\sigma$ upper limits for non-detected SiO sources with and without a CO outflow respectively. The black dashed lines are linear regression fits (provided in Table \ref{c2:table:sio_detected_correlation_coeffs}) to the data. For sources that show no L$_{\rm SiO}$ error bars the errors are smaller than the symbols. It should be noted that the errors in the SiO luminosity do not account for uncertainties in source distance, and should be seen as minimal errors.  
\label{c2:image:correclation_siodet}}
\end{figure*}

We perform Spearman rank correlations along with linear regression fits to the estimated source properties as a function of the SiO luminosity with the outcomes presented in Table \ref{c2:table:sio_detected_correlation_coeffs}. 
We assume that a correlation is significant, given the small sample sizes, if a P-value of $\leq$0.01 is found and a correlation coefficient (R) of $>$0.53 is obtained. The \htco FWHM, \hco column density (both the peak and average), $^{12}$CO\,(3-2) total linewidth or outflow velocity, outflow force and energy are all found to correlate with the SiO luminosity and are presented in Figure \ref{c2:image:correclation_siodet}. We find no correlation between the bolometric luminosity, source mass, outflow-mass, or momentum with the SiO luminosity.

The correlation of the SiO luminosity with the \htco FWHM was also observed by \citet{Klaassen2012} in their sample of high mass star forming regions. Furthermore, for both the SiO-detected and non-detected sources, the \htco FWHM is greater than would be estimated considering only the linewidth-size relation \citep{Larson1981}, and may then be a measure of the turbulence in these regions. This would suggest an increase in turbulence with increasing SiO luminosity, which may be a product of the shocks associated with the production of SiO. 
However, no correlation is observed between the C$^{18}$O FWHM from \citet{Maud2015cores} and the SiO luminosity. Towards several sources (e.g. G050.2213, G192.6005, and G207.2654) the \hco red- and blue-shifted emission (see Figure B1 in the online data) appears to be offset, indicating the \hco emission is tracing the outflows in these sources. Thus, it may also be the case that the \htco emission is sensitive to the outflows in these regions. A correlation is observed between the SiO luminosity and the \hco column density, which suggests a preference for increased SiO emission towards sources with higher densities, as seen in the low mass regime \citep{Gibb2004}. 
Furthermore, we find the SiO luminosity is correlated with the CO outflow velocity, again showing the association of SiO emission with high velocity outflows as seen in previous works (e.g. \citealt{Gibb2007}).

\begin{table*}
%\begin{minipage}{180mm}
\begin{center}
\caption[Linear Fit Values for SiO Luminosity Relationships]{Spearman rank correlation statistics for source properties as a function of the SiO luminosity. The P-value represents the probability of a correlation arising by chance, R is the resultant correlation coefficient, and N is the number of sources in each sample and the linear fit for SiO luminosity relationship is also given for properties showing a correlation. The Spearman rank probability of false correlation, the linear correlation coefficient and the resulting linear regression fit were derived using the ASURV package (\citealt{Feigelson1985}; \citealt{Isobe1986}; \citealt{Lavalley1992}) considering the 3$\sigma$ upper limits only. The linear fits are for the log$_{10}$ of the SiO luminosity and the log$_{10}$ of the source properties (excluding the \htco FWHM and $^{12}$CO\,(3-2) linewdith). The \htco FWHM is extracted from the dendrogram fits to the full source extent.\\ \label{c2:table:sio_detected_correlation_coeffs}}
\centering
\begin{tabular}{lccccl}
\hline\hline
Correlation with L$_{SiO}$        & &N &P-value & R  &Linear fit\\
%\cline{2-4}
\hline
L$_{*}$ (L$_{\odot}$) & & 32& 0.509& --& --\\
\htco FWHM (km\,s$^{-1}$)& & 30& $\leq$0.003 &0.66& Log$_{10}$(L$_{\rm SiO}$)\,=\,(0.91$\pm$0.18)$\times$\htco FWHM $-$ 0.58\\
C$^{18}$O FWHM (km\,s$^{-1}$)& & 23& 0.121 &--& --\\
L$_{70\mu m}$ (L$_{\odot}$)& & 29 & 0.049 &--& --\\
Mass (M$_{\odot}$) & & 23& 0.033&--&--\\
N$_{\rm HCO^+}$ Peak (cm$^{-2}$)&& 30&0.003&0.59& Log$_{10}$(L$_{\rm SiO}$)\,$=$\,(1.49$\pm$0.38)$\times$Log$_{10}$(N$_{\rm HCO^+}$)  $-$ 18.89\\
N$_{\rm HCO^+}$ Average (cm$^{-2}$)&& 30&0.003&0.64& Log$_{10}$(L$_{\rm SiO}$)\,$=$\,(2.63$\pm$0.62)$\times$Log$_{10}$(N$_{\rm HCO^+}$) $-$ 34.05\\
$^{12}$CO linewidth(km\,s$^{-1}$) &&23 & $\leq$0.004&0.62& Log$_{10}$(L$_{\rm SiO}$)\,$=$\,(0.04$\pm$0.01)$\times$$^{12}$CO linewidth + 0.44\\
M$_{\rm total}$ (M$_{\odot}$)& &23 &0.54 & --& --\\
P$_{\rm total}$(M$_{\odot}$km\,s$^{-1}$) & &23& 0.08& -- & --\\
E$_{\rm total}$ (ergs)& &23& 0.02& -- & --\\
$\dot{\rm M}_{\rm total}$ (M$_{\odot}$yr$^{-1}$) &&23 &0.11 &-- & --\\
$\dot{\rm P}_{\rm total}$ (M$_{\odot}$kms$^{-1}$yr$^{-1}$)&&23 & 0.01& 0.60&  Log$_{10}$(L$_{\rm SiO}$)\,$=$\,(1.18$\pm$0.32)$\times$Log$_{10}$($\dot{\rm P}_{\rm total}$) + 5.14\\
$\dot{\rm E}_{\rm total}$(L$_{\odot}$) &&23 & 0.004& 0.67&  Log$_{10}$(L$_{\rm SiO}$)\,$=$\,(1.00$\pm$0.25)$\times$Log$_{10}$($\dot{\rm E}_{\rm total}$) + 2.02\\
\hline
\hline
\end{tabular}
%\label{table:Continuum}
\end{center}

\end{table*}

\subsection{SiO luminosity with source evolution \label{section:sio2}} 

SiO emission was predominantly detected towards Class 0 sources in the low-mass regime \citep{Gibb2004}  suggesting a preference for SiO emission towards younger, denser sources with faster outflows. However, in the high-mass regime the evolutionary sequence and outflow properties are less constrained, and previous works have found the observed SiO luminosity and integrated intensity to both increase and decrease as a function of evolution (e.g.  \citealt{Klaassen2012}; \citealt{SanchezMonge2013}; \citealt{Leurini2014}). In this work we aim to establish if an evolutionary trend, as seen in the low-mass regime, does transfer to the high-mass regime. With this in mind, we purposely selected a sample of massive star forming regions from the RMS survey with a range of luminosity and evolutionary stage (categorised into two evolutionary stages in the RMS survey; massive YSOs (MYSOs) and compact H{\scriptsize II} regions, see \citealt{Lumsden2013}). A KS test shows no significant difference in the SiO luminosity between the MYSOs and H{\scriptsize II} regions. Moreover, the SiO luminosities of the OFFSET and H{\scriptsize II}/YSO sources also show no obvious differences compared with the MYSO sample.
\begin{figure}

\includegraphics[width=0.49\textwidth]{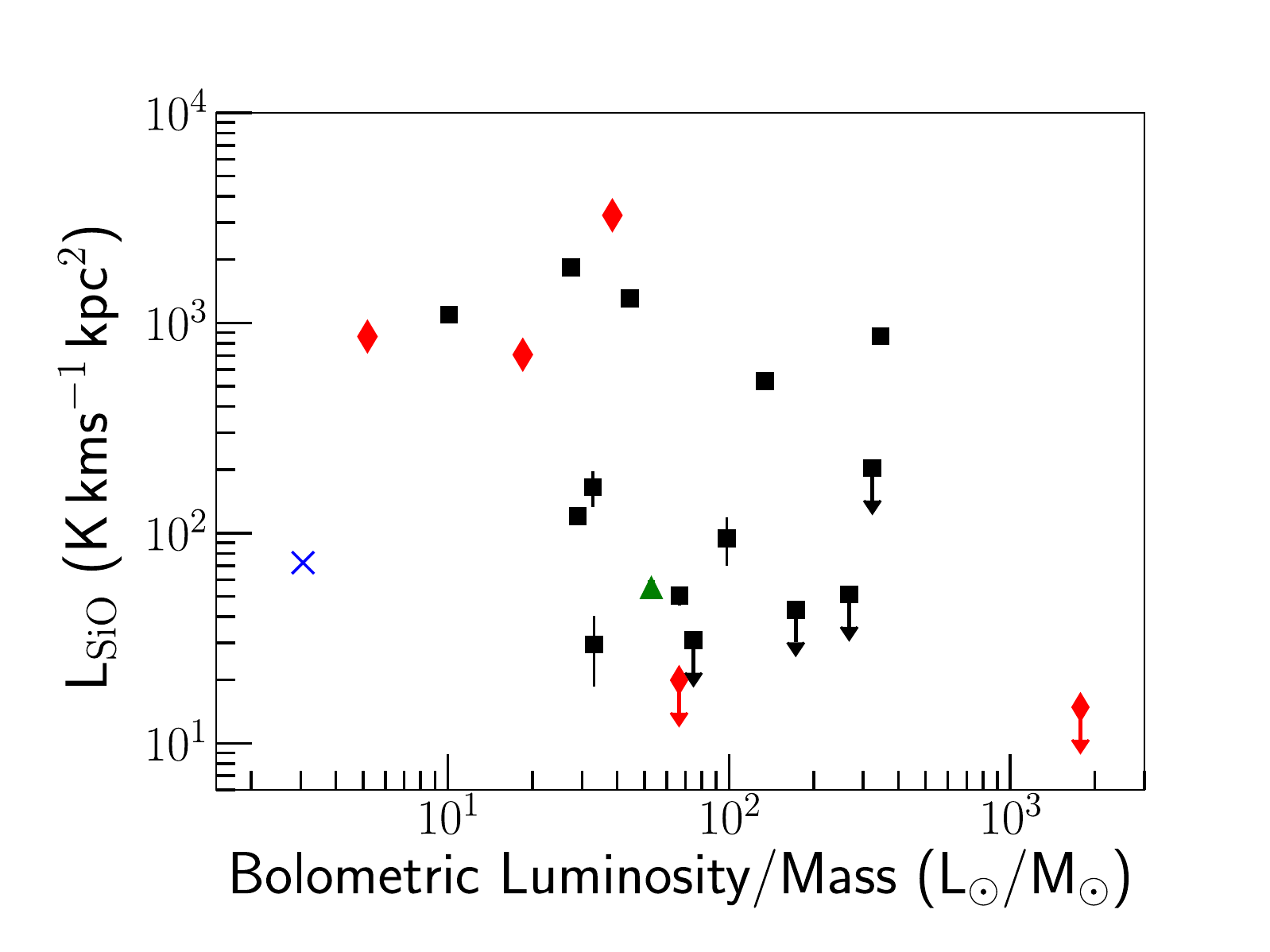}
\caption[SiO Luminosity as a Function of L$_{\odot}$/M$_{\odot}$]{SiO luminosity as a function of bolometric luminosity-to-mass ratio (L$_{\odot}$/M$_{\odot}$). The black squares, red diamonds, and green triangle represent YSOs, H{\scriptsize II}, and H{\scriptsize II}/YSO sources with an SiO detection, respectively. Sources without an SiO detection are represented as upper limits using the same colours/symbols as for the SiO-detected sources. The blue cross represents the offset source G203.3166-OFFSET, where the bolometric luminosity estimate is taken from \citet{Cunningham2016}. We do not have bolometric luminosity estimates for the remaining OFFSET sources, and as such, are subsequently missing from the figure.\label{image:Lsio_lbym}} %The lower panel 
\end{figure}

\begin{figure}

\includegraphics[width=0.49\textwidth]{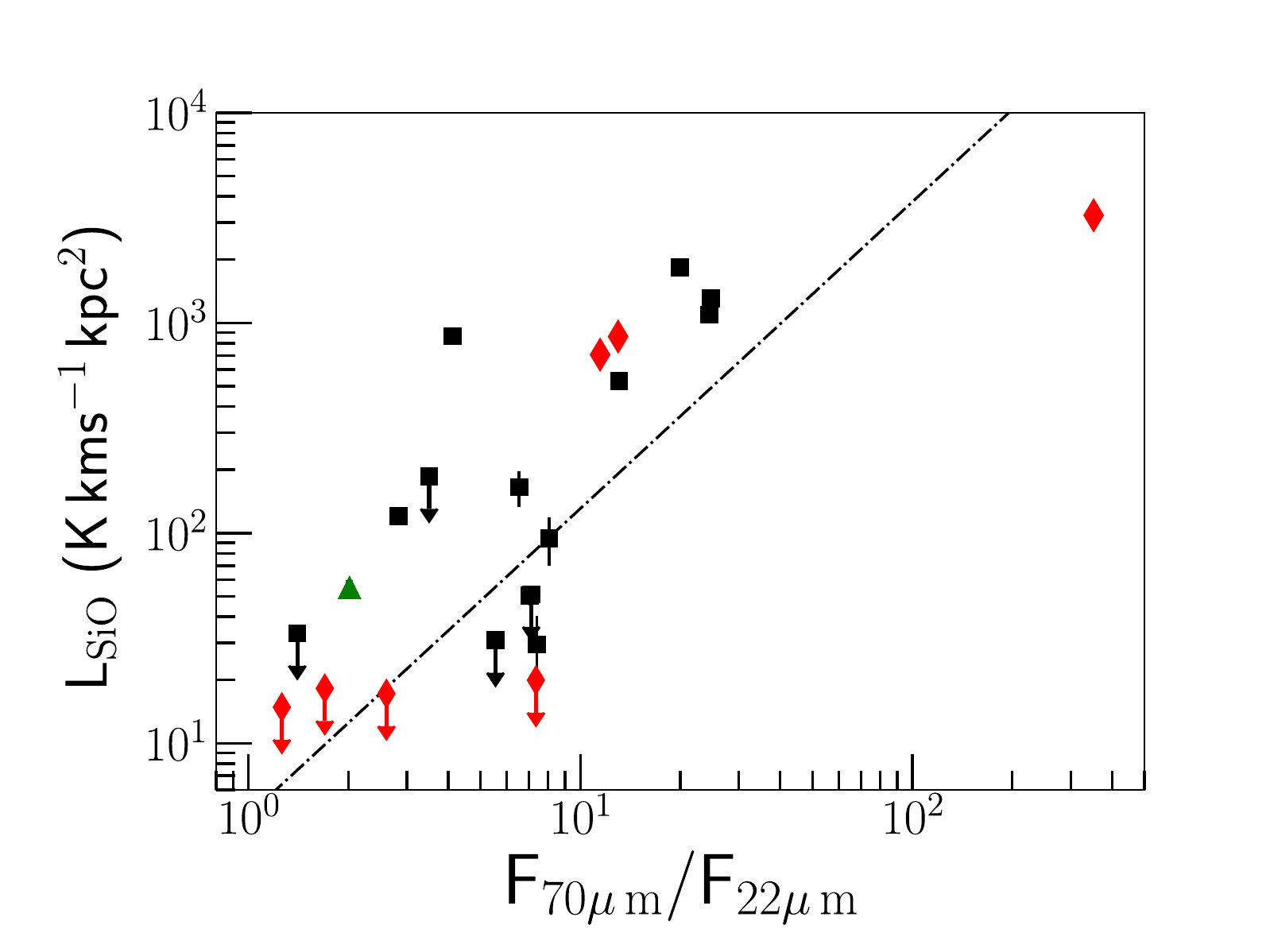}
\caption[SiO Luminosity as a Function of F70/F22 colour]{SiO luminosity as a function of the F70/F22 colour ratio. The symbols and colours are the same as in Figure \ref{image:Lsio_lbym}. The offset sources are not included as their 22$\mu$m fluxes were not available. The linear fit is shown by the black dashed line and has a correlation coefficient of R$=$0.76  and a linear fit given by Log$_{10}$(L$_{\rm SiO}$)\,$=$(1.46$\pm$0.30)$\times$Log$_{10}$(F70/F22)+0.66.\label{image:Lsio_f70f22ratio}}

\end{figure}

\subsubsection{Evolutionary indicators \label{section:sio3}}

We adopt the approach used by \citet{SanchezMonge2013} and \citet{LopezSepulcre2011} and compare the bolometric luminosity-to-mass ratio (L$_{bol}$/M$_{\odot}$), suggested as an indicator of the age of a given source, to the SiO luminosity in Figure \ref{image:Lsio_lbym}. \citet{Molinari2008} showed that as a source evolves the luminosity is expected to increase more rapidly than the core envelope mass, which is expected to decrease only slightly due to mass loss from winds and jets. An increase in the bolometric luminosity-to-mass ratio would potentially indicate a more advanced evolutionary stage. 
We find that all of the SiO non-detected sources have higher ($>$50 L$_{\odot}$/M$_{\odot}$) bolometric-to-luminosity ratios. The result of the KS test returns a {\it p}-value of 0.009  between the SiO-detected and SiO non-detected samples, suggesting they are drawn from different populations. However, it should be noted that the sample size for the non-detected sources is low and includes 3 of those sources without either an SiO detection or an associated CO outflow. As with the bolometric luminosity, we find no difference between the RMS classifications. This was previously observed by both \citet{Urquhart2014} and \citet{Maud2015cores}, where no indistinguishable differences in the bolometric luminosity-to-mass ratios between the MYSOs and compact H{\scriptsize II} regions from the RMS survey were found. This suggests that these sources are either likely to be at a similar evolutionary stage or it may be the case that the luminosity for this IR-bright stage has stopped rapidly increasing and the bolometric luminosity-to-mass ratio may not be sensitive enough to distinguish the evolutionary stages in this sample.  
For the offset source G203.3166-OFFSET or CMM-3\footnote{The bolometric luminosity for this source is taken from \citet{Cunningham2016} and the mass is taken from the SCUBA mass estimated here.} this source shows the smallest bolometric luminosity-to-mass ratio in the sample, in agreement with it being a young protostar (e.g. \citealt{Cunningham2016}; \citealt{Watanabe2015}). 

In addition, we use the ratio of mid- to far-IR colours as a potential indication of age in these sources. In Figure \ref{image:Lsio_f70f22ratio} we plot the ratio of the 70$\mu$m flux estimated from the {\em Herschel} data with the 22$\mu$m \textsc{WISE} flux (F70/F22)\footnote{The \textsc{WISE} flux is extracted directly from the RMS survey database, for one source, G012.9090, which does not have a WISE flux, the MSX 21$\mu$m flux is used.}. As a source evolves and the emission moves to shorter wavelengths, we would expect the F70/F22 colour ratio to decrease. A Spearman rank correlation test gives a correlation coefficient of 0.76 and probability of a false correlation given by $<$0.001 between the SiO luminosity and the F70/F22 colour ratio. Thus, the SiO luminosity is stronger in the redder, potentially younger, more embedded sources. This may indicate that the colour ratio is more sensitive to evolution in these sources. However, a KS test between the SiO-detected and SiO non-detected sources only gives a {\it p}-value of $\sim$0.06 and is therefore not significant between the populations, which may again be a result of these sources being at a similar evolutionary stage. \citealt{Csengeri2016} noted that the bolometric luminosity-to-mass ratio can be dominated by the most massive IR-bright source in the region, the F70/F22 colour ratio is also likely to suffer from this. Furthermore, as suggested by \citet{Maud2015outflows} there is evidence that several of these regions are likely to host multiple outflows at this spatial resolution (e.g. G081.7220/DR21 OH; \citealt{Girart2013}, CMM-3 in NGC2264-C; \citealt{Watanabe2017}). Additionally, we are assuming that the SiO emission is associated with the IR-bright RMS source in all regions, which was not the case towards G203.3166/NGC2264-C \citep{Cunningham2016}. This is likely to add scatter to the statistics.

\section{Conclusions}

We present the results of a JCMT SiO\,(8-7), \htco and \hco\,(4-3) survey towards a distance limited sample of 31 massive star-forming regions drawn from the RMS survey. The presence of a young, active outflow is associated with the detection of SiO\,(8-7) emission and we use previous $^{12}$CO\,(3-2) data \citep{Maud2015outflows} to determine outflow properties and identify potential fossil outflows. We explore the presence of possible global infall from the \hco and \htco\,(4-3) emission. Our results are summarized below.
\begin{enumerate}
\item We detect SiO\,(8-7) emission towards $\sim$46\% of the sources, where the lack of an SiO detection does not appear to be due to sensitivity limitations or distance to the sources. We find only the CO outflow velocity shows a significant difference between SiO-detected (i.e. a potentially active outflow) and SiO non-detected (i.e. a potentially fossil remnant driven outflow) sources. Thus, the detection of SiO is an indication of the presence of a high velocity, likely active outflow and is consistent with the expected shock velocities required to disrupt dust grains. In addition, correlations between the SiO luminosity and the \htco FWHM, \hco column density, $^{12}$CO\,(3-2) total outflow velocity, outflow force and energy are found. Thus, the production and strength of the SiO emission is increased towards potentially more turbulent regions with increased column densities. Similarly, regions with faster and more powerful outflows are more likely to produce stronger SiO emission, as observed in the low mass regime. However, it is possible the \htco is also tracing the outflow emission in these sources.
\item  We find tentative evidence from the bolometric-luminosity-to-mass ratio and F70/F22 colour ratios that sources with an SiO detection are associated with potentially younger, more embedded regions. However, if multiple outflows are present or the SiO emission is not associated with the RMS source this would likely add scatter to the statistics. Higher resolution observations are required to fully explore this.

\item We do not find a significant number of blue asymmetric profiles, indicative of global infall, towards these sources. However, \hco\,(4-3) may not be best suited tracer, at this spatial resolution, to detect global infall signatures in these sources. Higher spatial resolution observations, where infall motions onto individual protostars can be resolved, will be able to directly probe this.  
\end{enumerate}

\section*{Acknowledgments}
We thank the referee for their useful comments which improved the clarity of this manuscript. NC thanks the Green Bank Observatory which is a facility of the National Science Foundation operated under cooperative agreement with Associated Universities, Inc. IM thanks Comunidad Aut\'onoma de Madrid for funding his research through a "Talento-CAM" fellowship (2016-T1/TIC-1890). This paper has made use of information from the RMS survey database at \url{http://rms.leeds.ac.uk/cgi-bin/public/RMS_DATABASE.cgi} which was constructed with support from the Science and Technology Facilities Council of the United Kingdom.

\section*{Supporting Information}
Additional supporting information may be found in the online version of this article:
{\bf Appendix B}. HCO$^{+}$, H$^{13}$CO$^+$ and SiO emission maps.
Please note: Oxford University Press is not responsible for the
content or functionality of any supporting materials supplied by
the authors. Any queries (other than missing material) should be
directed to the corresponding author for this article.
%%%%%%%%%%%%%%%%%%%%%%%%%%%%%%%%%%%%%%%%%%%%%%%%%%



\bibliographystyle{mnras}
\bibliography{JCMT_reference} % if your bibtex file is called example.bib

\begin{thebibliography}{}
\makeatletter
\relax
\def\mn@urlcharsother{\let\do\@makeother \do\$\do\&\do\#\do\^\do\_\do\%\do\~}
\def\mn@doi{\begingroup\mn@urlcharsother \@ifnextchar [ {\mn@doi@}
  {\mn@doi@[]}}
\def\mn@doi@[#1]#2{\def\@tempa{#1}\ifx\@tempa\@empty \href
  {http://dx.doi.org/#2} {doi:#2}\else \href {http://dx.doi.org/#2} {#1}\fi
  \endgroup}
\def\mn@eprint#1#2{\mn@eprint@#1:#2::\@nil}
\def\mn@eprint@arXiv#1{\href {http://arxiv.org/abs/#1} {{\tt arXiv:#1}}}
\def\mn@eprint@dblp#1{\href {http://dblp.uni-trier.de/rec/bibtex/#1.xml}
  {dblp:#1}}
\def\mn@eprint@#1:#2:#3:#4\@nil{\def\@tempa {#1}\def\@tempb {#2}\def\@tempc
  {#3}\ifx \@tempc \@empty \let \@tempc \@tempb \let \@tempb \@tempa \fi \ifx
  \@tempb \@empty \def\@tempb {arXiv}\fi \@ifundefined
  {mn@eprint@\@tempb}{\@tempb:\@tempc}{\expandafter \expandafter \csname
  mn@eprint@\@tempb\endcsname \expandafter{\@tempc}}}

\bibitem[\protect\citeauthoryear{{Asaki}, {Imai}, {Sobolev}  \&
  {Parfenov}}{{Asaki} et~al.}{2014}]{Asaki2014}
{Asaki} Y.,  {Imai} H.,  {Sobolev} A.~M.,   {Parfenov} S.~Y.,  2014, \mn@doi
  [\apj] {10.1088/0004-637X/787/1/54}, \href
  {http://adsabs.harvard.edu/abs/2014ApJ...787...54A} {787, 54}

\bibitem[\protect\citeauthoryear{{Bally}, {Reipurth}, {Lada}  \&
  {Billawala}}{{Bally} et~al.}{1999}]{Bally1999}
{Bally} J.,  {Reipurth} B.,  {Lada} C.~J.,   {Billawala} Y.,  1999, \mn@doi
  [\aj] {10.1086/300672}, \href
  {http://adsabs.harvard.edu/abs/1999AJ....117..410B} {117, 410}

\bibitem[\protect\citeauthoryear{{Bonnell}, {Bate}, {Clarke}  \&
  {Pringle}}{{Bonnell} et~al.}{2001}]{Bonnell2001}
{Bonnell} I.~A.,  {Bate} M.~R.,  {Clarke} C.~J.,   {Pringle} J.~E.,  2001,
  \mn@doi [\mnras] {10.1046/j.1365-8711.2001.04270.x}, \href
  {http://adsabs.harvard.edu/abs/2001MNRAS.323..785B} {323, 785}

\bibitem[\protect\citeauthoryear{{Bontemps}, {Andre}, {Terebey}  \&
  {Cabrit}}{{Bontemps} et~al.}{1996}]{Bontemps1996}
{Bontemps} S.,  {Andre} P.,  {Terebey} S.,   {Cabrit} S.,  1996, \aap, 311, 858

\bibitem[\protect\citeauthoryear{{Buckle} et~al.,}{{Buckle}
  et~al.}{2009}]{Buckle2009}
{Buckle} J.~V.,  et~al., 2009, \mn@doi [\mnras]
  {10.1111/j.1365-2966.2009.15347.x}, 399, 1026

\bibitem[\protect\citeauthoryear{{Csengeri} et~al.,}{{Csengeri}
  et~al.}{2016}]{Csengeri2016}
{Csengeri} T.,  et~al., 2016, \mn@doi [\aap] {10.1051/0004-6361/201425404},
  586, A149

\bibitem[\protect\citeauthoryear{{Cunningham}}{{Cunningham}}{2015}]{Cunningham%
2015PhD}
{Cunningham} N.,  2015, PhD thesis, University of Leeds

\bibitem[\protect\citeauthoryear{{Cunningham}, {Lumsden}, {Cyganowski}, {Maud}
  \& {Purcell}}{{Cunningham} et~al.}{2016}]{Cunningham2016}
{Cunningham} N.,  {Lumsden} S.~L.,  {Cyganowski} C.~J.,  {Maud} L.~T.,
  {Purcell} C.,  2016, \mn@doi [\mnras] {10.1093/mnras/stw359}, 458, 1742

\bibitem[\protect\citeauthoryear{{Di Francesco}, {Johnstone}, {Kirk},
  {MacKenzie}  \& {Ledwosinska}}{{Di Francesco} et~al.}{2008}]{DiFrancesco2008}
{Di Francesco} J.,  {Johnstone} D.,  {Kirk} H.,  {MacKenzie} T.,
  {Ledwosinska} E.,  2008, \mn@doi [\apjs] {10.1086/523645}, 175, 277

\bibitem[\protect\citeauthoryear{{Duarte-Cabral}, {Bontemps}, {Motte},
  {Gusdorf}, {Csengeri}, {Schneider}  \& {Louvet}}{{Duarte-Cabral}
  et~al.}{2014}]{DuarteCabral2014}
{Duarte-Cabral} A.,  {Bontemps} S.,  {Motte} F.,  {Gusdorf} A.,  {Csengeri} T.,
   {Schneider} N.,   {Louvet} F.,  2014, \mn@doi [\aap]
  {10.1051/0004-6361/201423677}, \href
  {http://adsabs.harvard.edu/abs/2014A\%26A...570A...1D} {570, A1}

\bibitem[\protect\citeauthoryear{{Feigelson} \& {Nelson}}{{Feigelson} \&
  {Nelson}}{1985}]{Feigelson1985}
{Feigelson} E.~D.,  {Nelson} P.~I.,  1985, \mn@doi [\apj] {10.1086/163225},
  \href {http://adsabs.harvard.edu/abs/1985ApJ...293..192F} {293, 192}

\bibitem[\protect\citeauthoryear{{Flower} \& {Pineau des For{\^e}ts}}{{Flower}
  \& {Pineau des For{\^e}ts}}{2012}]{Flower2012}
{Flower} D.~R.,  {Pineau des For{\^e}ts} G.,  2012, \mn@doi [\mnras]
  {10.1111/j.1365-2966.2012.20481.x}, \href
  {http://adsabs.harvard.edu/abs/2012MNRAS.421.2786F} {421, 2786}

\bibitem[\protect\citeauthoryear{{Gibb}, {Richer}, {Chandler}  \&
  {Davis}}{{Gibb} et~al.}{2004}]{Gibb2004}
{Gibb} A.~G.,  {Richer} J.~S.,  {Chandler} C.~J.,   {Davis} C.~J.,  2004,
  \mn@doi [\apj] {10.1086/381309}, 603, 198

\bibitem[\protect\citeauthoryear{{Gibb}, {Davis}  \& {Moore}}{{Gibb}
  et~al.}{2007}]{Gibb2007}
{Gibb} A.,  {Davis} C.,   {Moore} T.,  2007, \mn@doi [\mnras]
  {10.1111/j.1365-2966.2007.12455.x}, 382, 1213

\bibitem[\protect\citeauthoryear{{Girart}, {Frau}, {Zhang}, {Koch}, {Qiu},
  {Tang}, {Lai}  \& {Ho}}{{Girart} et~al.}{2013}]{Girart2013}
{Girart} J.~M.,  {Frau} P.,  {Zhang} Q.,  {Koch} P.~M.,  {Qiu} K.,  {Tang}
  Y.-W.,  {Lai} S.-P.,   {Ho} P.~T.~P.,  2013, \mn@doi [\apj]
  {10.1088/0004-637X/772/1/69}, 772, 69

\bibitem[\protect\citeauthoryear{{Guillet}, {Jones}  \& {Pineau Des
  For{\^e}ts}}{{Guillet} et~al.}{2009}]{Guillet2009}
{Guillet} V.,  {Jones} A.~P.,   {Pineau Des For{\^e}ts} G.,  2009, \mn@doi
  [\aap] {10.1051/0004-6361/200811115}, 497, 145

\bibitem[\protect\citeauthoryear{{Gusdorf}, {Cabrit}, {Flower}  \& {Pineau Des
  For{\^e}ts}}{{Gusdorf} et~al.}{2008}]{Gusdorf2008b}
{Gusdorf} A.,  {Cabrit} S.,  {Flower} D.~R.,   {Pineau Des For{\^e}ts} G.,
  2008, \mn@doi [\aap] {10.1051/0004-6361:20078900}, 482, 809

\bibitem[\protect\citeauthoryear{{Isobe}, {Feigelson}  \& {Nelson}}{{Isobe}
  et~al.}{1986}]{Isobe1986}
{Isobe} T.,  {Feigelson} E.~D.,   {Nelson} P.~I.,  1986, \mn@doi [\apj]
  {10.1086/164359}, \href {http://adsabs.harvard.edu/abs/1986ApJ...306..490I}
  {306, 490}

\bibitem[\protect\citeauthoryear{{Jenness}, {Currie}, {Tilanus}, {Cavanagh},
  {Berry}, {Leech}  \& {Rizzi}}{{Jenness} et~al.}{2015}]{Jenness2015}
{Jenness} T.,  {Currie} M.~J.,  {Tilanus} R.~P.~J.,  {Cavanagh} B.,  {Berry}
  D.~S.,  {Leech} J.,   {Rizzi} L.,  2015, \mn@doi [\mnras]
  {10.1093/mnras/stv1545}, \href
  {http://adsabs.harvard.edu/abs/2015MNRAS.453...73J} {453, 73}

\bibitem[\protect\citeauthoryear{{Klaassen} \& {Wilson}}{{Klaassen} \&
  {Wilson}}{2007}]{Klaassen2007}
{Klaassen} P.~D.,  {Wilson} C.~D.,  2007, \mn@doi [\apj] {10.1086/518760},
  \href {http://adsabs.harvard.edu/abs/2007ApJ...663.1092K} {663, 1092}

\bibitem[\protect\citeauthoryear{{Klaassen}, {Testi}  \& {Beuther}}{{Klaassen}
  et~al.}{2012}]{Klaassen2012}
{Klaassen} P.,  {Testi} L.,   {Beuther} H.,  2012, \mn@doi [\aap]
  {10.1051/0004-6361/201118350}, 538, A140

\bibitem[\protect\citeauthoryear{{Kutner} \& {Ulich}}{{Kutner} \&
  {Ulich}}{1981}]{Kutner1981}
{Kutner} M.~L.,  {Ulich} B.~L.,  1981, \mn@doi [\apj] {10.1086/159380}, 250,
  341

\bibitem[\protect\citeauthoryear{{Larson}}{{Larson}}{1981}]{Larson1981}
{Larson} R.~B.,  1981, \mn@doi [\mnras] {10.1093/mnras/194.4.809}, 194, 809

\bibitem[\protect\citeauthoryear{{Lavalley}, {Isobe}  \&
  {Feigelson}}{{Lavalley} et~al.}{1992}]{Lavalley1992}
{Lavalley} M.,  {Isobe} T.,   {Feigelson} E.,  1992, in {Worrall} D.~M.,
  {Biemesderfer} C.,   {Barnes} J.,  eds,  Astronomical Society of the Pacific
  Conference Series Vol. 25, Astronomical Data Analysis Software and Systems I.
  p.~245

\bibitem[\protect\citeauthoryear{{Leurini}, {Codella}, {L{\'o}pez-Sepulcre},
  {Gusdorf}, {Csengeri}  \& {Anderl}}{{Leurini} et~al.}{2014}]{Leurini2014}
{Leurini} S.,  {Codella} C.,  {L{\'o}pez-Sepulcre} A.,  {Gusdorf} A.,
  {Csengeri} T.,   {Anderl} S.,  2014, \mn@doi [\aap]
  {10.1051/0004-6361/201424251}, 570, A49

\bibitem[\protect\citeauthoryear{{L{\'o}pez-Sepulcre}
  et~al.,}{{L{\'o}pez-Sepulcre} et~al.}{2011}]{LopezSepulcre2011}
{L{\'o}pez-Sepulcre} A.,  et~al., 2011, \mn@doi [\aap]
  {10.1051/0004-6361/201015827}, 526, L2

\bibitem[\protect\citeauthoryear{{Lu}, {Zhang}, {Liu}, {Wang}  \& {Gu}}{{Lu}
  et~al.}{2014}]{Lu2014}
{Lu} X.,  {Zhang} Q.,  {Liu} H.~B.,  {Wang} J.,   {Gu} Q.,  2014, \mn@doi
  [\apj] {10.1088/0004-637X/790/2/84}, 790, 84

\bibitem[\protect\citeauthoryear{{Lumsden}, {Hoare}, {Urquhart}, {Oudmaijer},
  {Davies}, {Mottram}, {Cooper}  \& {Moore}}{{Lumsden}
  et~al.}{2013}]{Lumsden2013}
{Lumsden} S.~L.,  {Hoare} M.~G.,  {Urquhart} J.~S.,  {Oudmaijer} R.~D.,
  {Davies} B.,  {Mottram} J.~C.,  {Cooper} H.~D.~B.,   {Moore} T.~J.~T.,  2013,
  \mn@doi [\apjs] {10.1088/0067-0049/208/1/11}, 208, 11

\bibitem[\protect\citeauthoryear{{Mardones}, {Myers}, {Tafalla}, {Wilner},
  {Bachiller}  \& {Garay}}{{Mardones} et~al.}{1997}]{Mardones1997}
{Mardones} D.,  {Myers} P.~C.,  {Tafalla} M.,  {Wilner} D.~J.,  {Bachiller} R.,
    {Garay} G.,  1997, \apj, 489, 719

\bibitem[\protect\citeauthoryear{{Maud}, {Lumsden}, {Moore}, {Mottram},
  {Urquhart}  \& {Cicchini}}{{Maud} et~al.}{2015a}]{Maud2015cores}
{Maud} L.~T.,  {Lumsden} S.~L.,  {Moore} T.~J.~T.,  {Mottram} J.~C.,
  {Urquhart} J.~S.,   {Cicchini} A.,  2015a, \mn@doi [\mnras]
  {10.1093/mnras/stv1334}, \href
  {http://adsabs.harvard.edu/abs/2015MNRAS.452..637M} {452, 637}

\bibitem[\protect\citeauthoryear{{Maud}, {Moore}, {Lumsden}, {Mottram},
  {Urquhart}  \& {Hoare}}{{Maud} et~al.}{2015b}]{Maud2015outflows}
{Maud} L.~T.,  {Moore} T.~J.~T.,  {Lumsden} S.~L.,  {Mottram} J.~C.,
  {Urquhart} J.~S.,   {Hoare} M.~G.,  2015b, \mn@doi [\mnras]
  {10.1093/mnras/stv1635}, 453, 645

\bibitem[\protect\citeauthoryear{{McKee} \& {Tan}}{{McKee} \&
  {Tan}}{2003}]{McKeeandTan2003}
{McKee} C.,  {Tan} J.,  2003, \mn@doi [\apj] {10.1086/346149}, 585, 850

\bibitem[\protect\citeauthoryear{{Minh}, {Su}, {Chen}, {Liu}, {Yan}  \&
  {Kim}}{{Minh} et~al.}{2010}]{Minh2010}
{Minh} Y.~C.,  {Su} Y.-N.,  {Chen} H.-R.,  {Liu} S.-Y.,  {Yan} C.-H.,   {Kim}
  S.-J.,  2010, \mn@doi [\apj] {10.1088/0004-637X/723/2/1231}, \href
  {http://adsabs.harvard.edu/abs/2010ApJ...723.1231M} {723, 1231}

\bibitem[\protect\citeauthoryear{{Molinari}, {Pezzuto}, {Cesaroni}, {Brand},
  {Faustini}  \& {Testi}}{{Molinari} et~al.}{2008}]{Molinari2008}
{Molinari} S.,  {Pezzuto} S.,  {Cesaroni} R.,  {Brand} J.,  {Faustini} F.,
  {Testi} L.,  2008, \mn@doi [\aap] {10.1051/0004-6361:20078661}, \href
  {http://adsabs.harvard.edu/abs/2008A\%26A...481..345M} {481, 345}

\bibitem[\protect\citeauthoryear{{Molinari} et~al.,}{{Molinari}
  et~al.}{2010}]{Molinari2010}
{Molinari} S.,  et~al., 2010, \mn@doi [\aap] {10.1051/0004-6361/201014659},
  518, L100

\bibitem[\protect\citeauthoryear{{Motte}, {Bontemps}, {Schilke}, {Schneider},
  {Menten}  \& {Brogui{\`e}re}}{{Motte} et~al.}{2007}]{Motte2007}
{Motte} F.,  {Bontemps} S.,  {Schilke} P.,  {Schneider} N.,  {Menten} K.~M.,
  {Brogui{\`e}re} D.,  2007, \mn@doi [\aap] {10.1051/0004-6361:20077843}, 476,
  1243

\bibitem[\protect\citeauthoryear{{Motte} et~al.,}{{Motte}
  et~al.}{2010}]{Motte2010}
{Motte} F.,  et~al., 2010, \mn@doi [\aap] {10.1051/0004-6361/201014690}, 518,
  L77

\bibitem[\protect\citeauthoryear{{Motte}, {Bontemps}  \& {Louvet}}{{Motte}
  et~al.}{2017}]{Motte2017}
{Motte} F.,  {Bontemps} S.,   {Louvet} F.,  2017, preprint, \href
  {http://adsabs.harvard.edu/abs/2017arXiv170600118M} {} (\mn@eprint {arXiv}
  {1706.00118})

\bibitem[\protect\citeauthoryear{{Mottram} et~al.,}{{Mottram}
  et~al.}{2017}]{Mottram2017}
{Mottram} J.~C.,  et~al., 2017, \mn@doi [\aap] {10.1051/0004-6361/201628682},
  \href {http://adsabs.harvard.edu/abs/2017A%26A...600A..99M} {600, A99}

\bibitem[\protect\citeauthoryear{{Myers}, {Mardones}, {Tafalla}, {Williams}  \&
  {Wilner}}{{Myers} et~al.}{1996}]{Myers1996}
{Myers} P.~C.,  {Mardones} D.,  {Tafalla} M.,  {Williams} J.~P.,   {Wilner}
  D.~J.,  1996, \mn@doi [\apjl] {10.1086/310146}, 465, L133

\bibitem[\protect\citeauthoryear{{Peretto} et~al.,}{{Peretto}
  et~al.}{2014}]{Peretto2014}
{Peretto} N.,  et~al., 2014, \mn@doi [\aap] {10.1051/0004-6361/201322172},
  \href {http://adsabs.harvard.edu/abs/2014A\%26A...561A..83P} {561, A83}

\bibitem[\protect\citeauthoryear{{Pilbratt} et~al.,}{{Pilbratt}
  et~al.}{2010}]{Pilbratt2010}
{Pilbratt} G.~L.,  et~al., 2010, \mn@doi [\aap] {10.1051/0004-6361/201014759},
  518, L1

\bibitem[\protect\citeauthoryear{{Poglitsch}, {Waelkens}, {Geis},
  {Feuchtgruber}, {Vandenbussche}, {Rodriguez}, {Krause}  \&
  {Renotte}}{{Poglitsch} et~al.}{2010}]{Poglitsch2010}
{Poglitsch} A.,  {Waelkens} C.,  {Geis} N.,  {Feuchtgruber} H.,
  {Vandenbussche} B.,  {Rodriguez} L.,  {Krause} O.,   {Renotte} E.,  2010,
  \mn@doi [\aap] {10.1051/0004-6361/201014535}, 518, L2

\bibitem[\protect\citeauthoryear{{Rygl}, {Wyrowski}, {Schuller}  \&
  {Menten}}{{Rygl} et~al.}{2013}]{Rygl2013}
{Rygl} K.~L.~J.,  {Wyrowski} F.,  {Schuller} F.,   {Menten} K.~M.,  2013,
  \mn@doi [\aap] {10.1051/0004-6361/201219574}, 549, A5

\bibitem[\protect\citeauthoryear{{S{\'a}nchez-Monge}, {L{\'o}pez-Sepulcre},
  {Cesaroni}, {Walmsley}, {Codella}, {Beltr{\'a}n}, {Pestalozzi}  \&
  {Molinari}}{{S{\'a}nchez-Monge} et~al.}{2013}]{SanchezMonge2013}
{S{\'a}nchez-Monge} {\'A}.,  {L{\'o}pez-Sepulcre} A.,  {Cesaroni} R.,
  {Walmsley} C.~M.,  {Codella} C.,  {Beltr{\'a}n} M.~T.,  {Pestalozzi} M.,
  {Molinari} S.,  2013, \mn@doi [\aap] {10.1051/0004-6361/201321589}, 557, A94

\bibitem[\protect\citeauthoryear{{Schilke}, {Walmsley}, {Pineau des Forets}  \&
  {Flower}}{{Schilke} et~al.}{1997}]{Schilke1997}
{Schilke} P.,  {Walmsley} C.~M.,  {Pineau des Forets} G.,   {Flower} D.~R.,
  1997, \aap, 321, 293

\bibitem[\protect\citeauthoryear{{Schneider}, {Csengeri}, {Bontemps}, {Motte},
  {Simon}, {Hennebelle}, {Federrath}  \& {Klessen}}{{Schneider}
  et~al.}{2010}]{Schneider2010}
{Schneider} N.,  {Csengeri} T.,  {Bontemps} S.,  {Motte} F.,  {Simon} R.,
  {Hennebelle} P.,  {Federrath} C.,   {Klessen} R.,  2010, \mn@doi [\aap]
  {10.1051/0004-6361/201014481}, 520, A49

\bibitem[\protect\citeauthoryear{{Smith}, {Shetty}, {Beuther}, {Klessen}  \&
  {Bonnell}}{{Smith} et~al.}{2013}]{Smith2013}
{Smith} R.~J.,  {Shetty} R.,  {Beuther} H.,  {Klessen} R.~S.,   {Bonnell}
  I.~A.,  2013, \mn@doi [\apj] {10.1088/0004-637X/771/1/24}, 771, 24

\bibitem[\protect\citeauthoryear{{Tig{\'e}} et~al.,}{{Tig{\'e}}
  et~al.}{2017}]{Tige2017}
{Tig{\'e}} J.,  et~al., 2017, \mn@doi [\aap] {10.1051/0004-6361/201628989},
  \href {http://adsabs.harvard.edu/abs/2017A%26A...602A..77T} {602, A77}

\bibitem[\protect\citeauthoryear{{Urquhart} et~al.,}{{Urquhart}
  et~al.}{2014}]{Urquhart2014}
{Urquhart} J.~S.,  et~al., 2014, \mn@doi [\mnras] {10.1093/mnras/stu1207},
  \href {http://adsabs.harvard.edu/abs/2014MNRAS.443.1555U} {443, 1555}

\bibitem[\protect\citeauthoryear{{Walker-Smith}, {Richer}, {Buckle}, {Hatchell}
   \& {Drabek-Maunder}}{{Walker-Smith} et~al.}{2014}]{WalkerSmith2014}
{Walker-Smith} S.~L.,  {Richer} J.~S.,  {Buckle} J.~V.,  {Hatchell} J.,
  {Drabek-Maunder} E.,  2014, \mn@doi [\mnras] {10.1093/mnras/stu512}, \href
  {http://adsabs.harvard.edu/abs/2014MNRAS.440.3568W} {440, 3568}

\bibitem[\protect\citeauthoryear{{Watanabe} et~al.,}{{Watanabe}
  et~al.}{2015}]{Watanabe2015}
{Watanabe} Y.,  et~al., 2015, \mn@doi [\apj] {10.1088/0004-637X/809/2/162},
  \href {http://adsabs.harvard.edu/abs/2015ApJ...809..162W} {809, 162}

\bibitem[\protect\citeauthoryear{{Watanabe}, {Sakai}, {L{\'o}pez-Sepulcre},
  {Sakai}, {Hirota}, {Liu}, {Su}  \& {Yamamoto}}{{Watanabe}
  et~al.}{2017}]{Watanabe2017}
{Watanabe} Y.,  {Sakai} N.,  {L{\'o}pez-Sepulcre} A.,  {Sakai} T.,  {Hirota}
  T.,  {Liu} S.-Y.,  {Su} Y.-N.,   {Yamamoto} S.,  2017, \mn@doi [\apj]
  {10.3847/1538-4357/aa88b6}, \href
  {http://adsabs.harvard.edu/abs/2017ApJ...847..108W} {847, 108}

\bibitem[\protect\citeauthoryear{{Williams}, {Peretto}, {Avison},
  {Duarte-Cabral}  \& {Fuller}}{{Williams} et~al.}{2018}]{Williams2018}
{Williams} G.~M.,  {Peretto} N.,  {Avison} A.,  {Duarte-Cabral} A.,   {Fuller}
  G.~A.,  2018, preprint, \href
  {http://adsabs.harvard.edu/abs/2018arXiv180107253W} {} (\mn@eprint {arXiv}
  {1801.07253})

\bibitem[\protect\citeauthoryear{{Wu} et~al.,}{{Wu} et~al.}{2014}]{Wu2014}
{Wu} Y.~W.,  et~al., 2014, \mn@doi [\aap] {10.1051/0004-6361/201322765}, 566,
  A17

\makeatother
\end{thebibliography}

%%%%%%%%%%%%%%%%%%%%%%%%%%%%%%%%%%%%%%%%%%%%%%%%%%

%%%%%%%%%%%%%%%%% APPENDICES %%%%%%%%%%%%%%%%%%%%%

\appendix
\section{H$^{13}$CO$^+$ Gaussian fits}
Presented below are the resulting Gaussian fits to the H$^{13}$CO$^+$ spectra.
%\section{H$^{13}$CO$^+$ Gaussian fits}
% along with the HCO$^+$ and H$^{13}$CO$^+$ integrated intensity maps for each source and the extracted total spectra from the masked regions are shown.
%\vspace{-15mm}
\begin{table*}
%\begin{minipage}{180mm}
\begin{small}
\begin{center}
\caption{Fitted parameters from a single Gaussian fit to the sum of the H$^{13}$CO$^+$\,(4-3) line emission extracted from all pixels within H$^{13}$CO$^+$ (4-3) 5$\sigma$ masked regions. Column 1 is the galactic name, Column 2 is the RMS classification, and Column 3 gives the total number of pixels in the H$^{13}$CO$^+$\,(4-3) masks used to extract the emission. Columns 4, 5, 6 and 7 give peak, central velocity, FWHM and integrated intensity from a single Gaussian fit to the sum of the H$^{13}$CO$^+$ (4-3) emission extracted  from all pixels within the masked region.\label{aa:tab:htcogaus}}
\begin{tabular}{l cccccc}
\hline
Source   &  Type  & No of pixels &summed T$_{mb}$ & V$_{\rm LSR}$   & $\delta$V& summed T$_{mb}$dv\\
Name      & 	  &              &   (K)   & (km\,s$^{-1}$)&  (km\,s$^{-1}$) & (K\,km\,s$^{-1}$) \\
\hline  
 CO  Outflow             &          &	    &                &                 &               &                 \\
\hline
G010.8411-02.5919        & YSO      & 17  & 14.38$\pm$0.76 & 11.98$\pm$0.04  & 1.71$\pm$0.10 & 26.12$\pm$2.11  \\ 
G012.9090-00.2607        & YSO      & 38  & 19.94$\pm$0.58 & 37.32$\pm$0.05  & 3.42$\pm$0.11 & 72.65$\pm$3.21  \\
G013.6562-00.5997        & YSO      &  4  & 1.90$\pm$0.12  & 48.00$\pm$0.10   & 3.13$\pm$0.23 & 6.32$\pm$0.62  \\
G017.6380+00.1566        & YSO      & 33  & 27.57$\pm$0.77 & 22.33$\pm$0.03  & 2.16$\pm$0.07 & 63.38$\pm$2.71  \\
G018.3412+01.7681        & YSO      & 21  & 14.37$\pm$0.56 & 32.84$\pm$0.04  & 2.10$\pm$0.10 & 32.19$\pm$1.91  \\
%G020.7617-00.0638        & HII/YSO  
G043.3061-00.2106        & HII      & 6   & 2.92$\pm$0.21  & 59.28$\pm$0.08  & 2.40$\pm$0.20 & 7.45$\pm$0.81  \\
G045.0711+00.1325        & HII      & 13  & 6.97$\pm$0.14  & 59.10$\pm$0.06  & 6.20$\pm$0.15 & 45.98$\pm$1.43  \\
G050.2213-00.6063        & YSO      & 4   & 0.96$\pm$0.15  & 40.36$\pm$0.23  & 2.90$\pm$0.54 & 2.97$\pm$0.72  \\
G078.1224+03.6320        & YSO      & 22  & 14.26$\pm$0.46 & -3.32$\pm$0.05  & 3.12$\pm$0.12 & 47.39$\pm$2.35  \\
G079.1272+02.2782        & YSO      & 5   & 2.74$\pm$0.26  & -1.58$\pm$0.09  & 1.81$\pm$0.20 & 5.28$\pm$0.77   \\
G079.8749+01.1821        & HII      & 5   & 2.16$\pm$0.29  & -4.90$\pm$0.10  & 1.54$\pm$0.24 & 3.53$\pm$0.71   \\ 
G079.8749+01.1821-OFFSET & --       & 19  & 12.84$\pm$0.49 & -3.14$\pm$0.05  & 2.81$\pm$0.12 & 38.44$\pm$2.25   \\ 
G081.7133+00.5589        & YSO      & 8   & 8.63$\pm$0.53  & -4.09$\pm$0.08  & 2.47$\pm$0.18 & 22.76$\pm$2.14   \\
G081.7220+00.5699        & HII      & 53  & 58.83$\pm$1.32 & -3.17$\pm$0.05  & 4.57$\pm$0.12 & 286.36$\pm$9.84  \\
G081.7522+00.5906        & YSO      & 23  & 18.58$\pm$0.70 & -4.04$\pm$0.04  & 2.03$\pm$0.09 & 40.11$\pm$2.29    \\
G081.7522+00.5906-OFFSET & --       & 20  & 13.02$\pm$0.51 & -3.19$\pm$0.05  & 2.77$\pm$0.12 & 38.43$\pm$2.28    \\
G081.7624+00.5916-OFFSET & YSO      & 19  & 11.11$\pm$0.94 & -4.34$\pm$0.08  & 1.96$\pm$0.19 & 23.2$\pm$3.03 \\ 
G081.8652+00.7800        & YSO      & --  &  --   &   --     &   --     &  --           \\
W75N 			 & --       & 56  & 83.83$\pm$0.75 & 9.67$\pm$0.017  & 3.90$\pm$0.04 & 348.53$\pm$4.78  \\
G081.8789+00.7822        & HII      & --  &  --   &  --      &   --     &  --           \\ 
G083.0936+03.2724        & HII      & 4   & 0.89$\pm$0.22  & -3.50$\pm$0.29  & 2.33$\pm$0.67 & 2.22$\pm$0.84    \\
G083.7071+03.2817        & YSO      & --  & -- &  --   &   --     &   --    \\
G083.7071+03.2817-OFFSET & --       & 4   & 1.95$\pm$0.35  & -3.62$\pm$0.13  & 1.50$\pm$0.31 & 3.11$\pm$0.85   \\
G083.7962+03.3058        & HII      & 4   & 2.78$\pm$0.33  & -4.31$\pm$0.10  & 1.66$\pm$0.23 & 4.91$\pm$0.88   \\
G103.8744+01.8558        & YSO      & 5   & 3.80$\pm$0.55  & -18.30$\pm$0.11 & 1.60$\pm$0.27 & 6.49$\pm$1.44   \\
G109.8715+02.1156        & YSO      & 71  & 91.63$\pm$1.31 & -10.80$\pm$0.03 & 3.99$\pm$0.07 & 388.87$\pm$8.49  \\
G192.6005-00.0479        & YSO      & 5   & 4.11$\pm$0.14  & 7.84$\pm$0.07   & 4.26$\pm$0.17 & 18.65$\pm$0.98   \\
G194.9349-01.2224        & YSO      & 4   & 1.48$\pm$0.27  & 15.61$\pm$0.18  & 2.03$\pm$0.43 & 3.20$\pm$0.89    \\
G203.3166+02.0564        & YSO      & 50  & 34.05$\pm$1.06 & 8.15$\pm$0.04   & 2.31$\pm$0.08 & 83.60$\pm$3.96    \\
G203.3166+02.0564-OFFSET & --       & 28  & 15.87$\pm$0.56 & 7.55$\pm$0.04   & 2.54$\pm$0.10 & 42.96$\pm$2.32   \\
G207.2654-01.8080        & HII/YSO  & 12  &  5.77$\pm$0.45 & 12.58$\pm$0.10  & 2.74$\pm$0.24 & 16.83$\pm$1.98   \\
\hline 
 No CO Outflow       &    &   &             &        &           &             \\
\hline
%G011.9454-00.0373 & HII  & 18:11:53.20  &  -\,18:36:21.8  &   0.10         &  0.05     &   0.75       \\
G080.8645+00.4197        & HII  & 11 &  3.76$\pm$0.22   & -2.75$\pm$0.095  & 3.31$\pm$0.22  & 13.26$\pm$1.17  \\ 
G080.9383-00.1268        & HII  & -- &  --   &   --     &   --     &  --           \\
G081.7131+00.5792        & YSO  & -- &  --   &   --     &   --     &  --           \\
G196.4542-01.6777        & YSO  & 4  & 1.16$\pm$0.18    & 19.34$\pm$0.18 & 2.46$\pm$0.43  & 3.20$\pm$0.89 \\ 
G217.3771-00.0828        & HII  & 4  &  0.76$\pm$0.17   & 23.93$\pm$0.25 & 2.18$\pm$0.58  & 1.76$\pm$0.62 \\
G233.8306-00.1803        & YSO  & -- &  --   &   --     &   --     &  --           \\
\hline

\hline

\end{tabular}
\end{center}
\end{small}
%\end{minipage}{180mm}
\end{table*}
\newpage
\section{H$^{13}$CO$^+$, HCO$^{+}$ and SiO zeroth order moment maps}

\begin{figure*}

\includegraphics[width=0.49\textwidth]{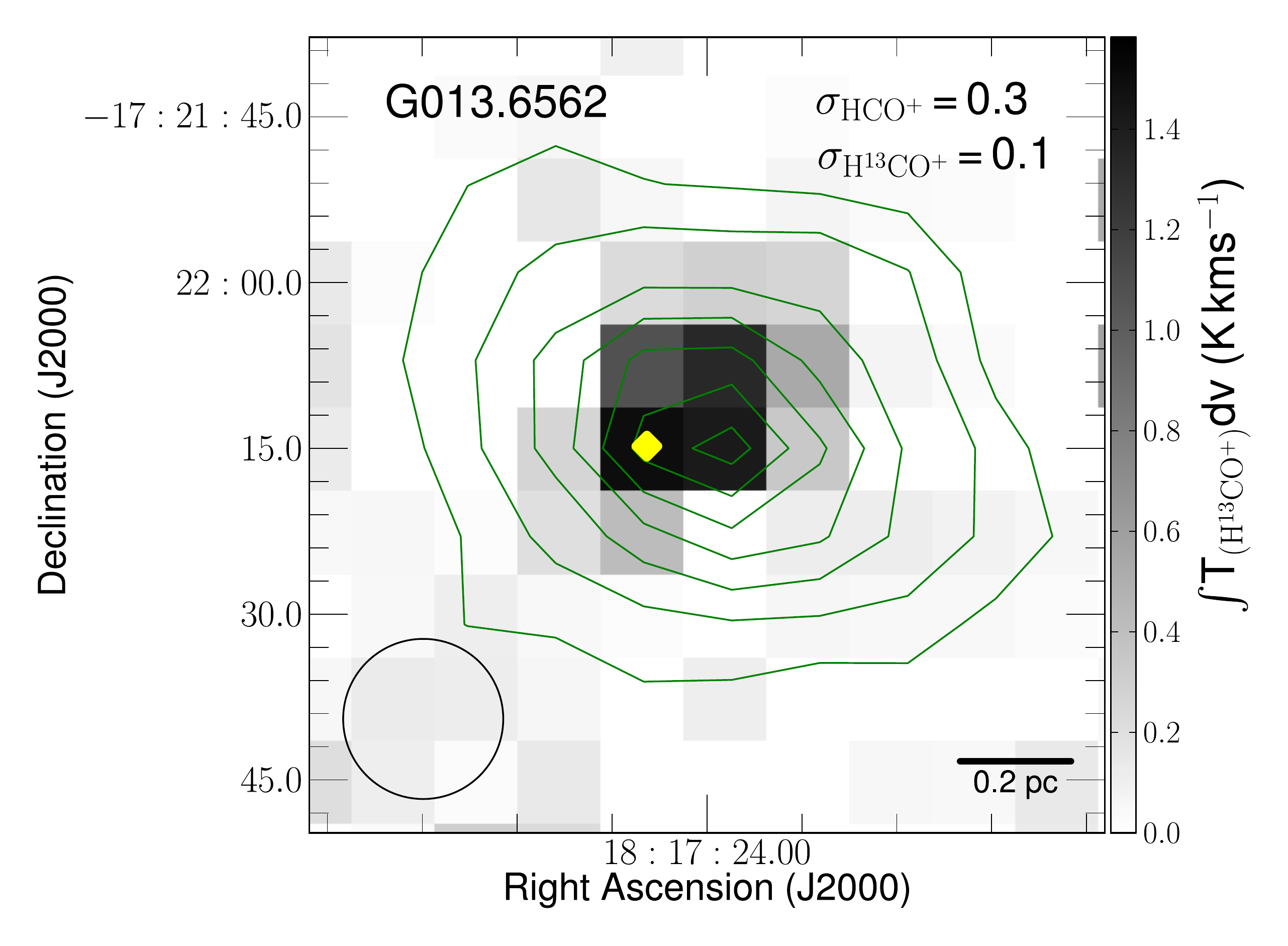}
\includegraphics[width=0.49\textwidth]{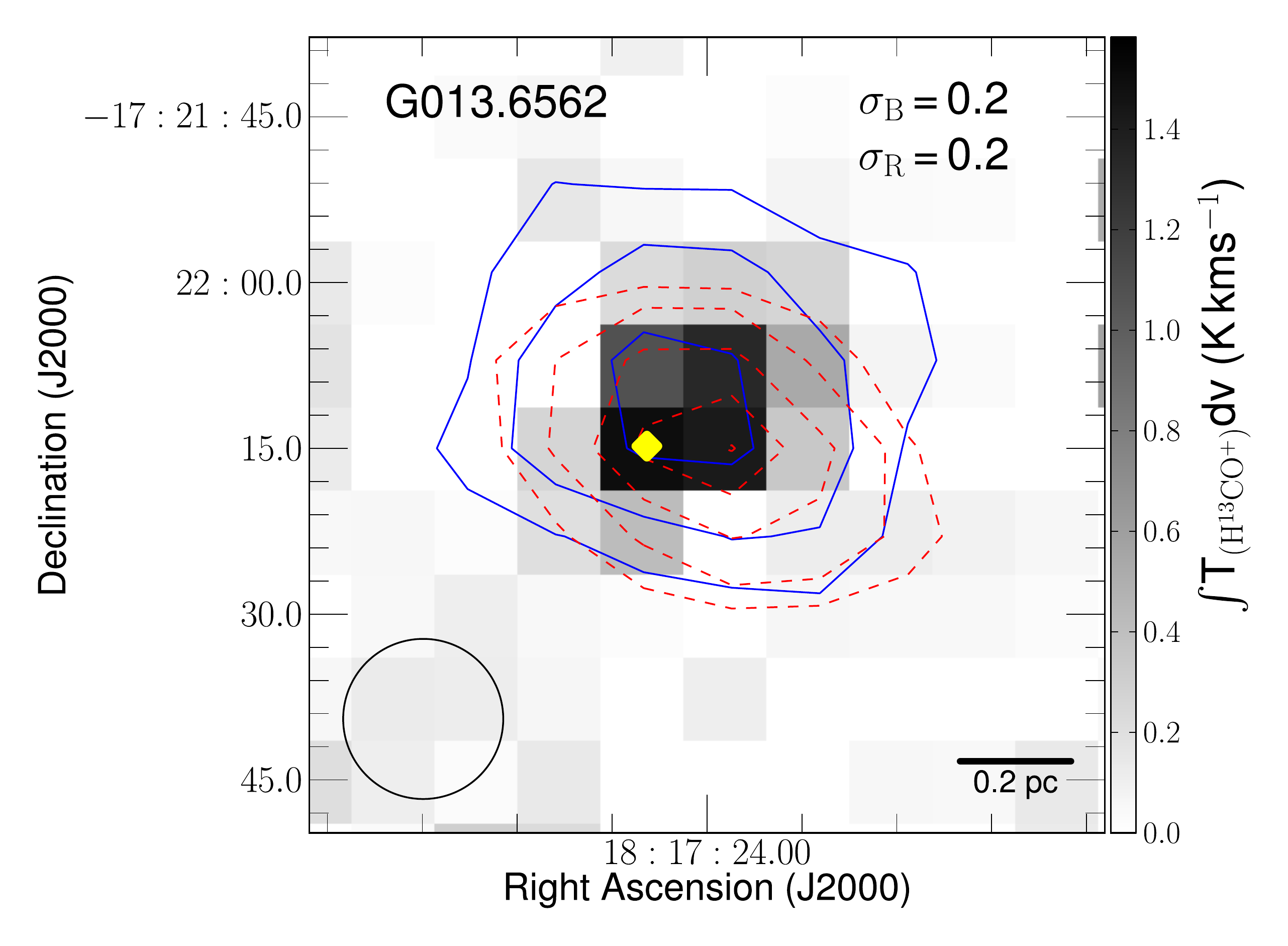}
\includegraphics[width=0.49\textwidth]{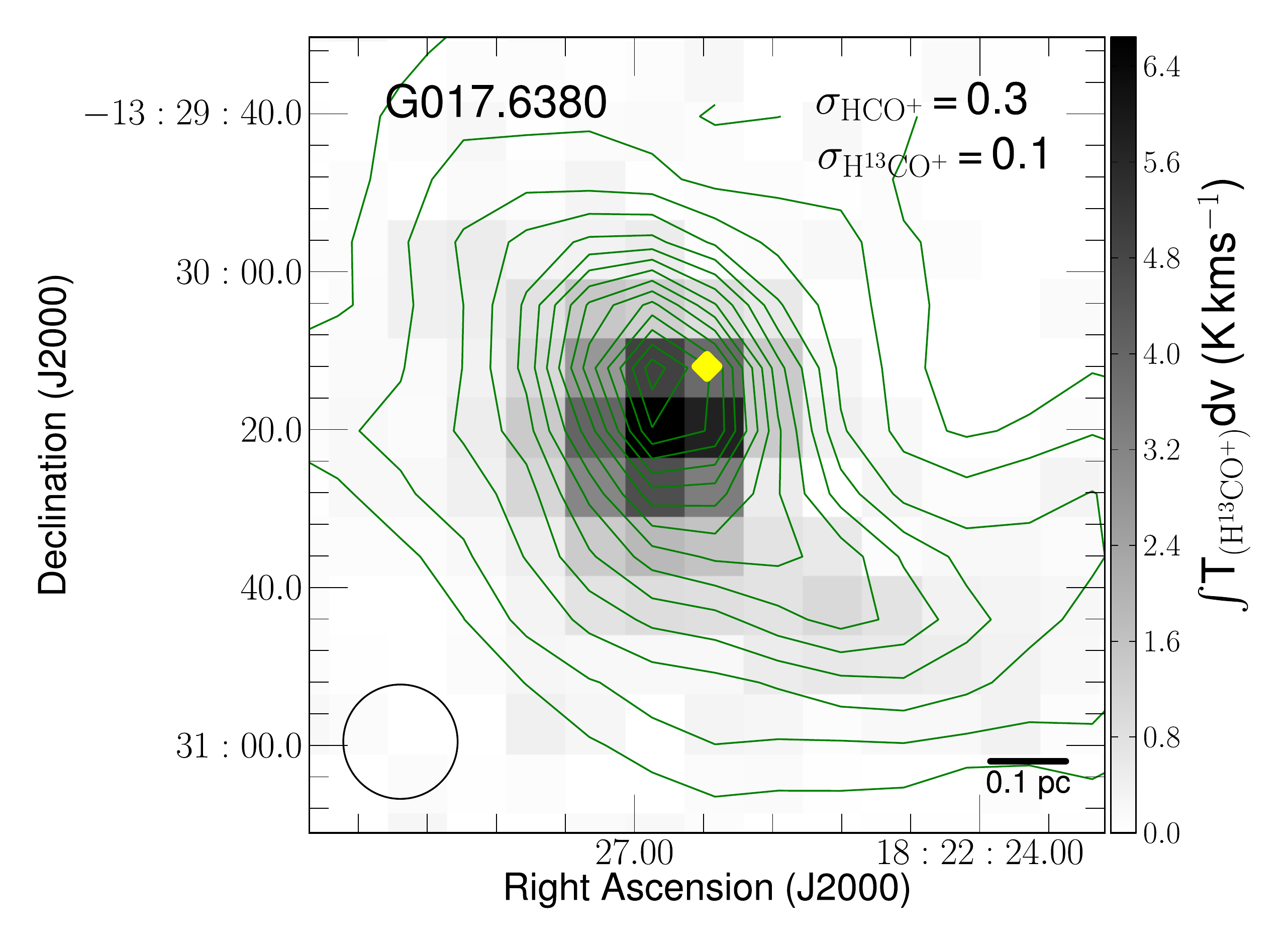}
\includegraphics[width=0.49\textwidth]{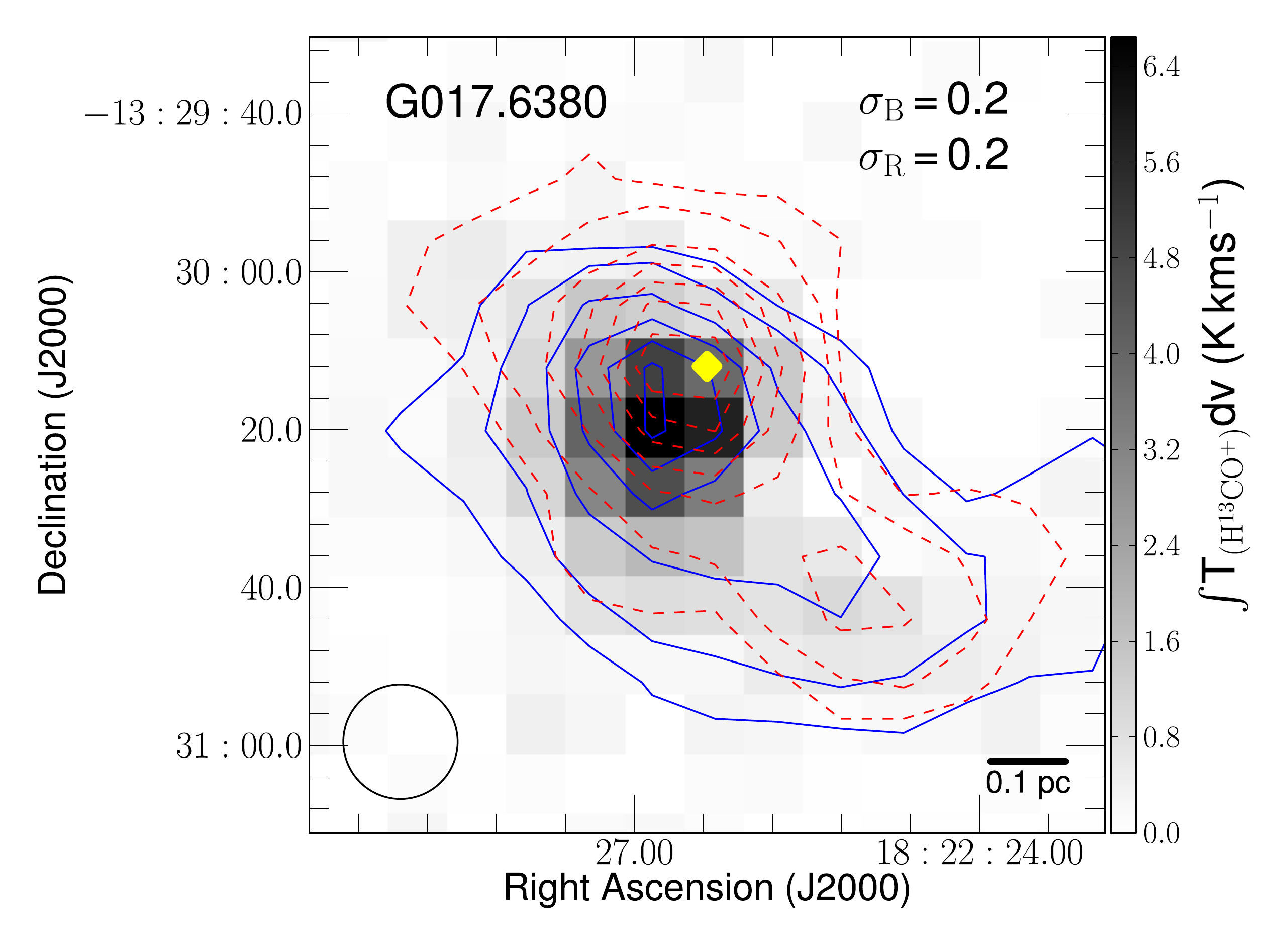}
\includegraphics[width=0.49\textwidth]{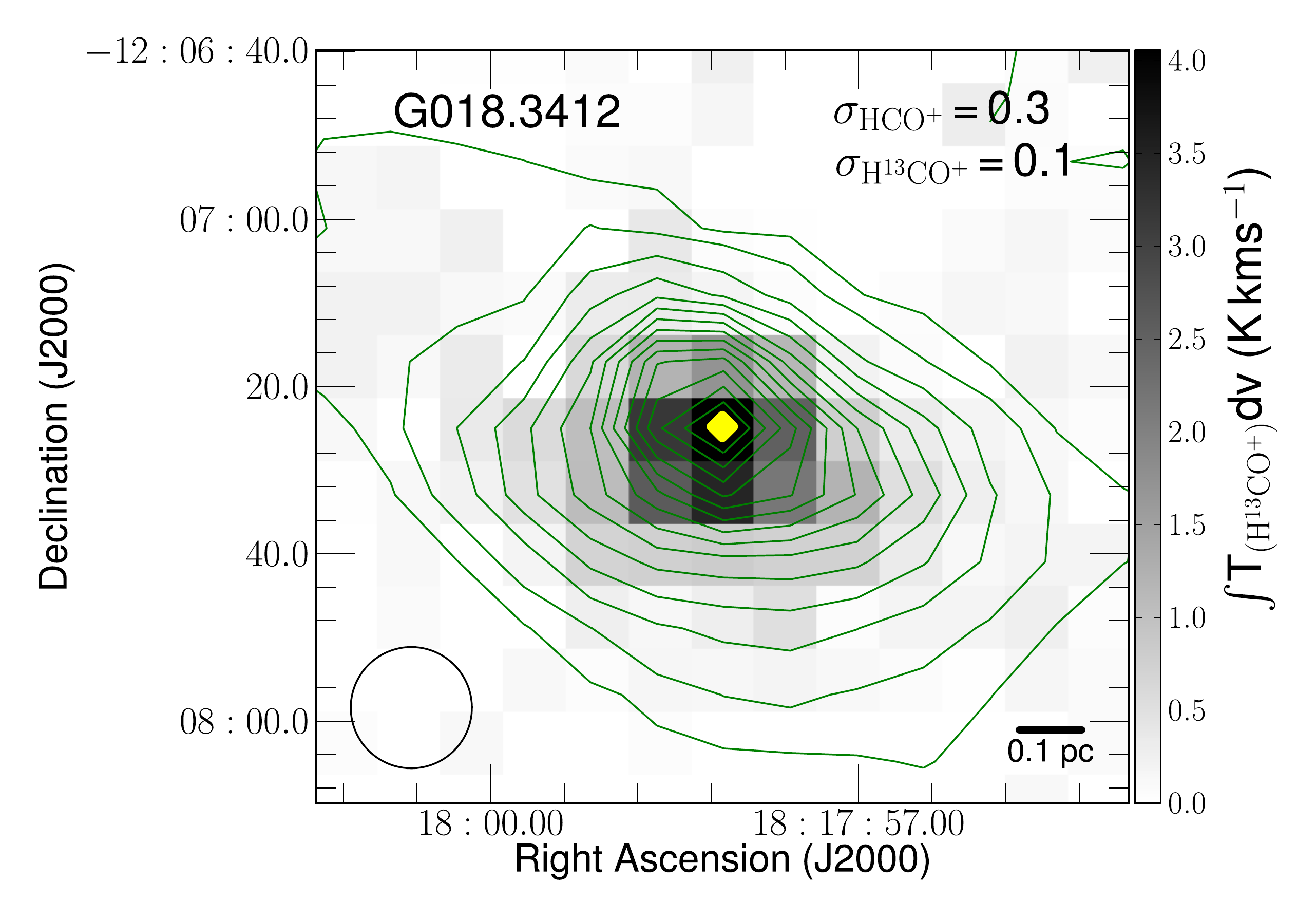}
\includegraphics[width=0.49\textwidth]{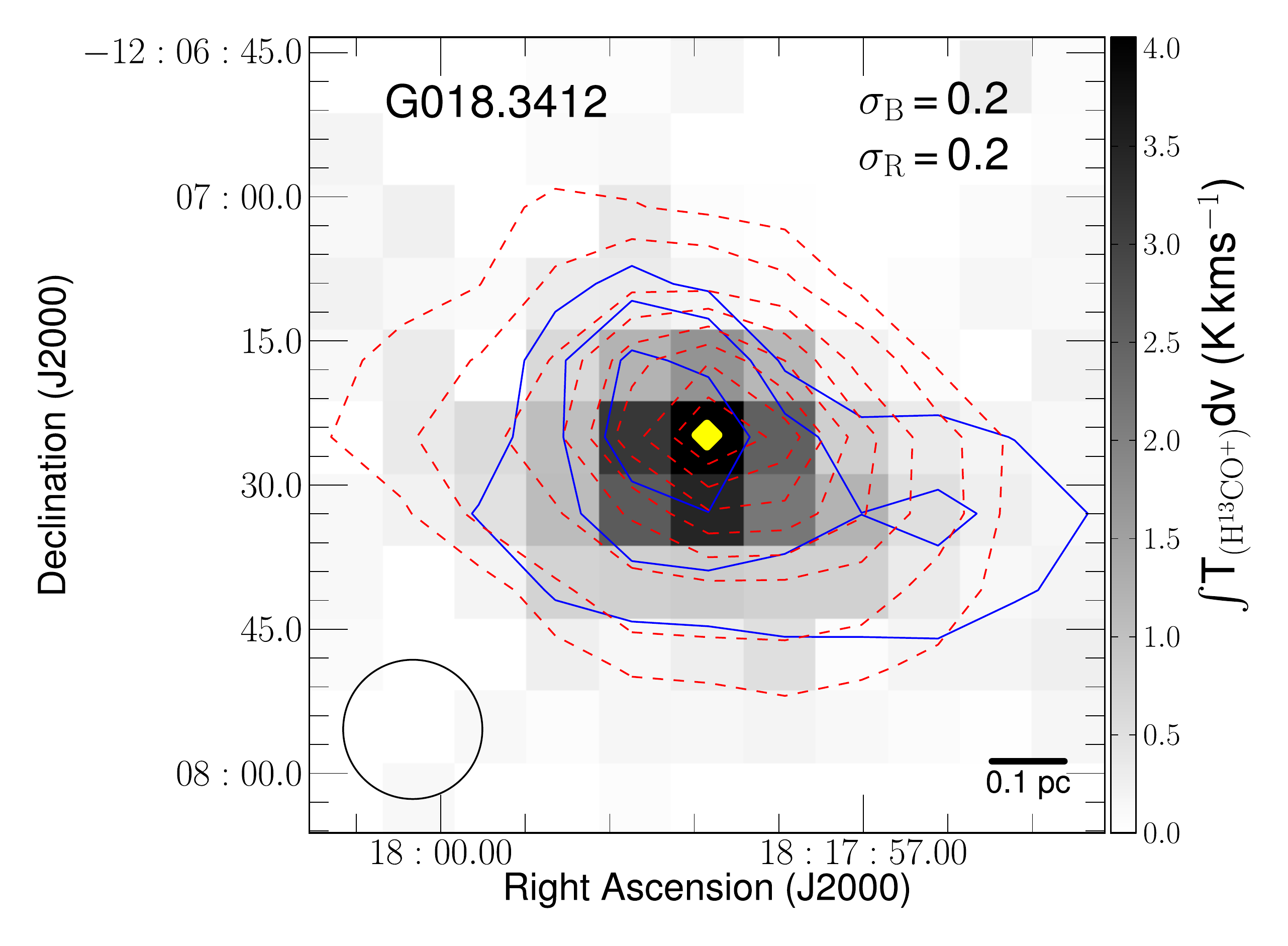}

\caption{\htco and \hco zeroth order moment maps. The total integrated \htco emission is shown in greyscale, integrated from the minimum to maximum channels with a 3$\sigma$ detection. The yellow diamonds and crosses are the RMS source position(s) and offset position, respectively. The JCMT beam is shown in the bottom left corner, and the source name is shown in the top left corner. {\bf Left:} The total integrated \hco emission is overlaid in green solid contours (again integrated from the minimum and maximum channels with a 3$\sigma$ detection). The 1$\sigma$ rms (in units of K.\kms) for both the \hco ($\sigma$$_{\rm HCO^+}$) and \htco ($\sigma$$_{\rm H^{13}CO^+}$) integrated intensity maps are given in the top right corner. The \hco contour levels are from 1$\sigma\times$(5, 10, 20,... to peak in-steps of 10$\sigma$). {\bf Right}: The red- and blue-shifted \hco emission is shown by the red (dashed) and blue (solid) contours, respectively. The blue- and red-shifted contours are taken from the minimum and maximum channels with 3$\sigma$ emission, respectively, excluding the central velocity range defined by the \htco FWHM (see Table \ref{aa:tab:htcogaus} for the \htco FWHM values). The 1$\sigma$ levels for the red- ($\sigma$$_{\rm R}$) and blue-shifted ($\sigma$$_{\rm B}$) emission are given in the top right corner, where the contour levels are from 1$\sigma\times$(5, 10, 20,... to peak in-steps of 10$\sigma$). The velocity ranges used for the total integrated \hco emission are 45.0$-$51.0\,\kms for G013.6562, 18.9$-$25.7\,\kms for G017.6380, and 30.2$-$36.4\,\kms for G018.3412.
\label{figure:emission_maps}}
\end{figure*}

\begin{figure*}

\includegraphics[width=0.49\textwidth]{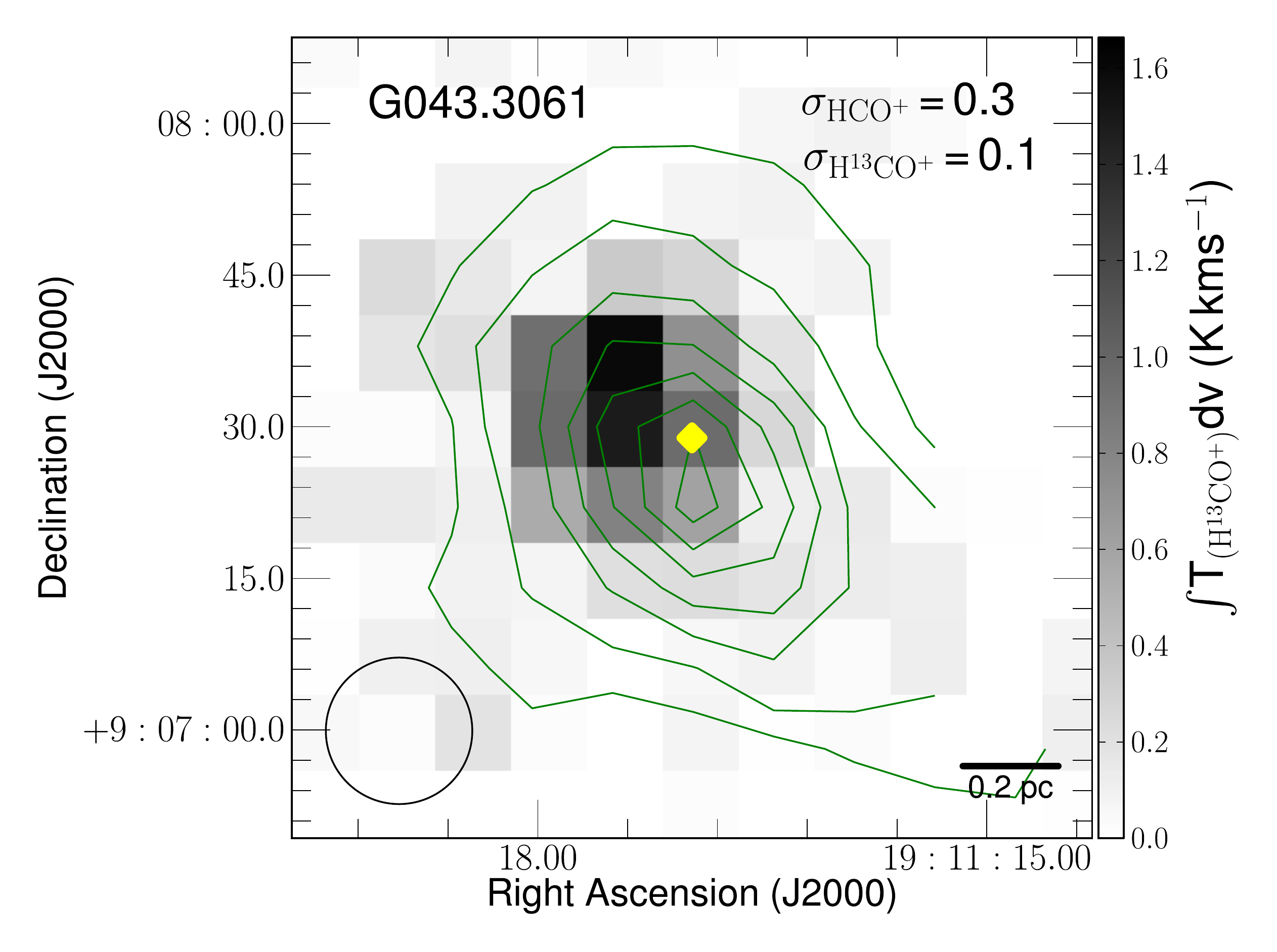}
\includegraphics[width=0.49\textwidth]{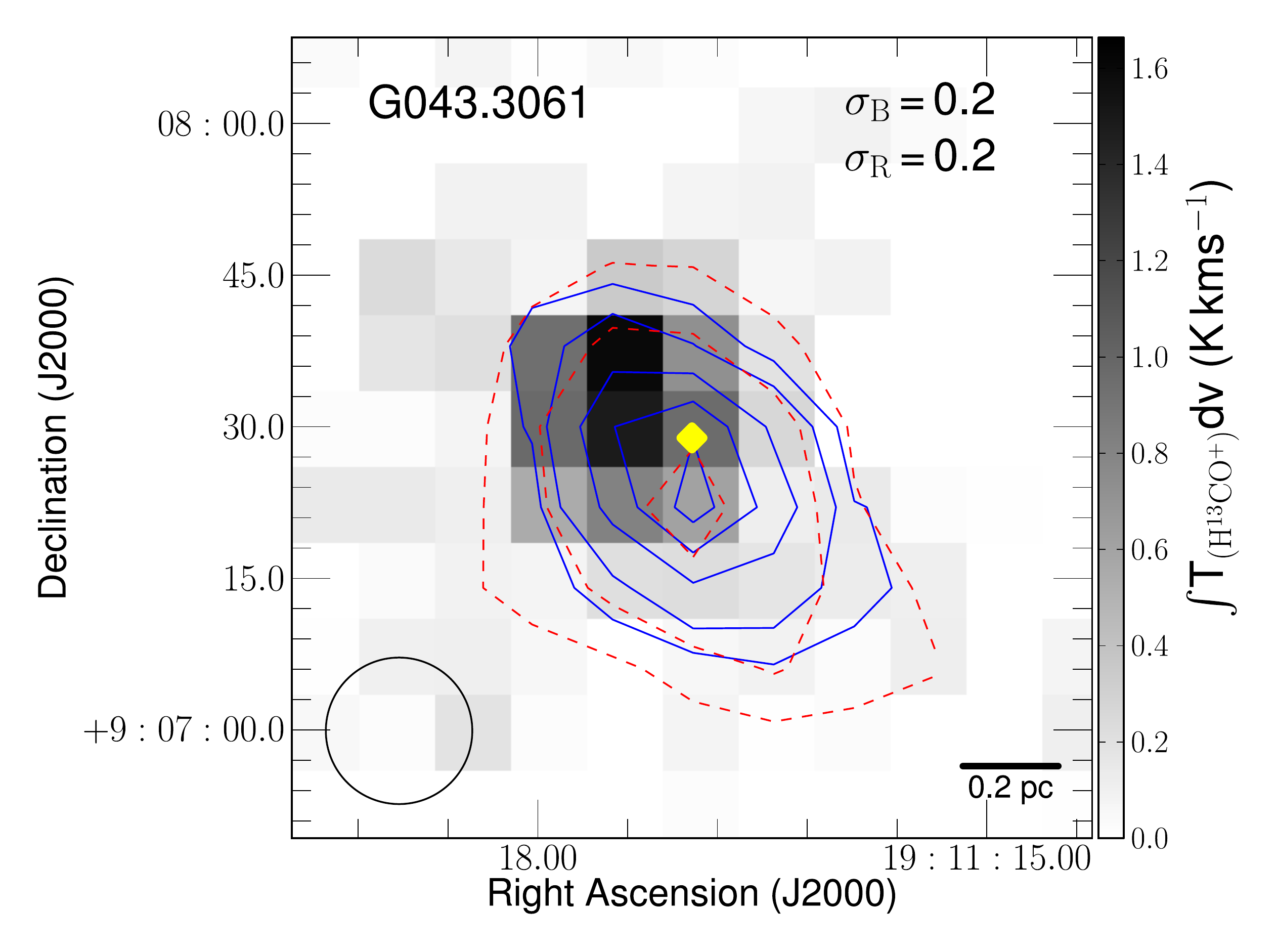}
\includegraphics[width=0.49\textwidth]{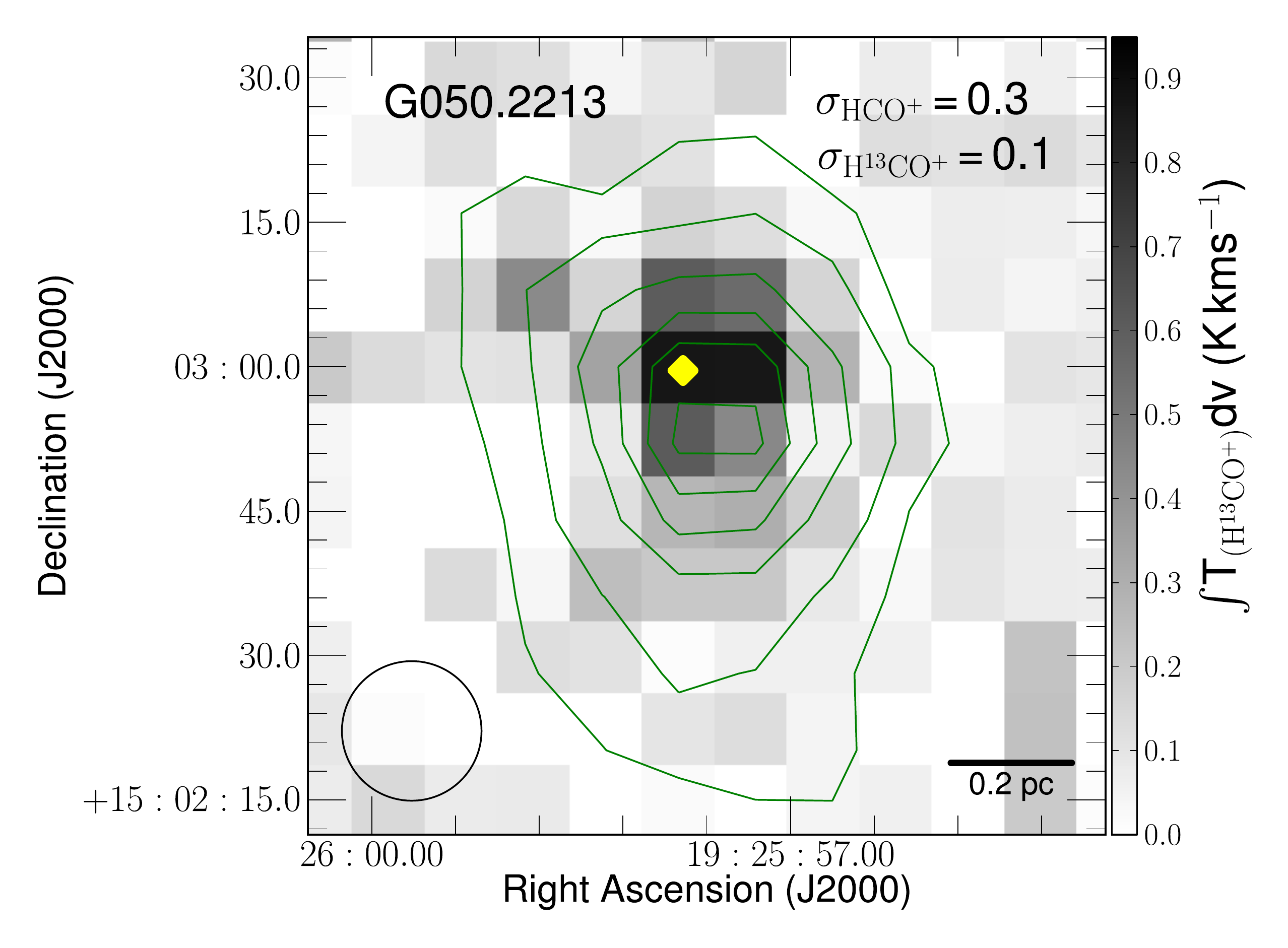}
\includegraphics[width=0.49\textwidth]{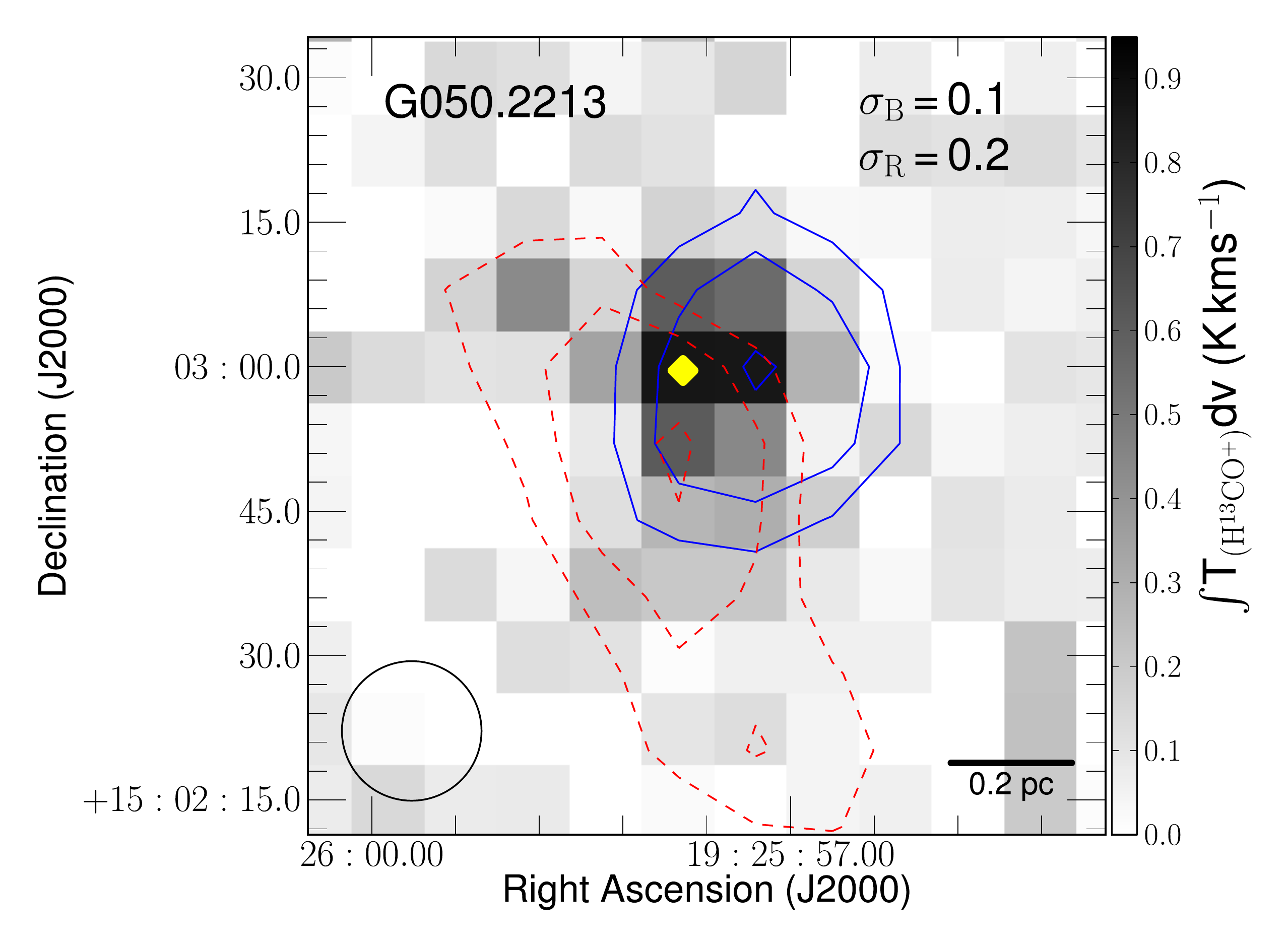}
\includegraphics[width=0.49\textwidth]{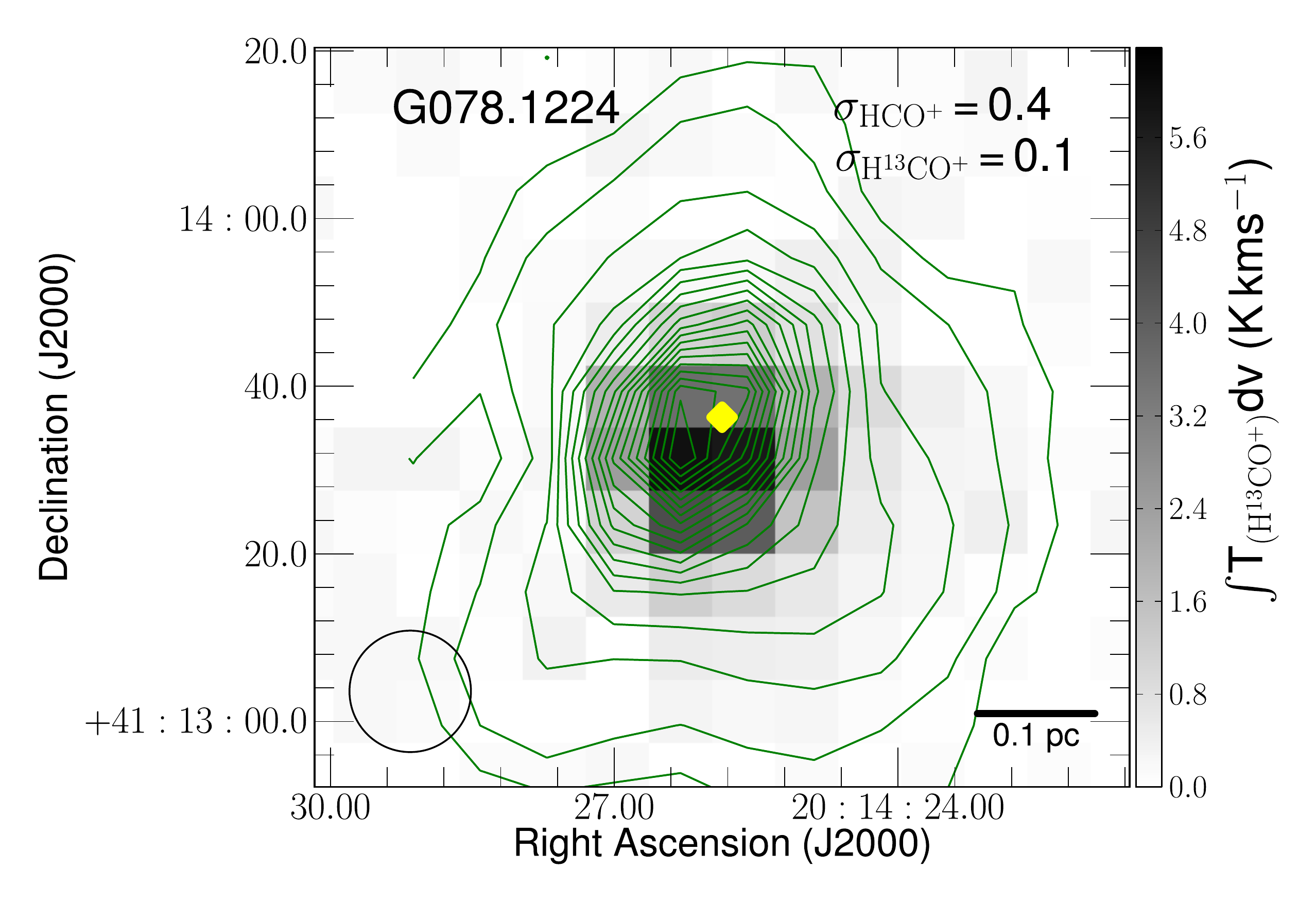}
\includegraphics[width=0.49\textwidth]{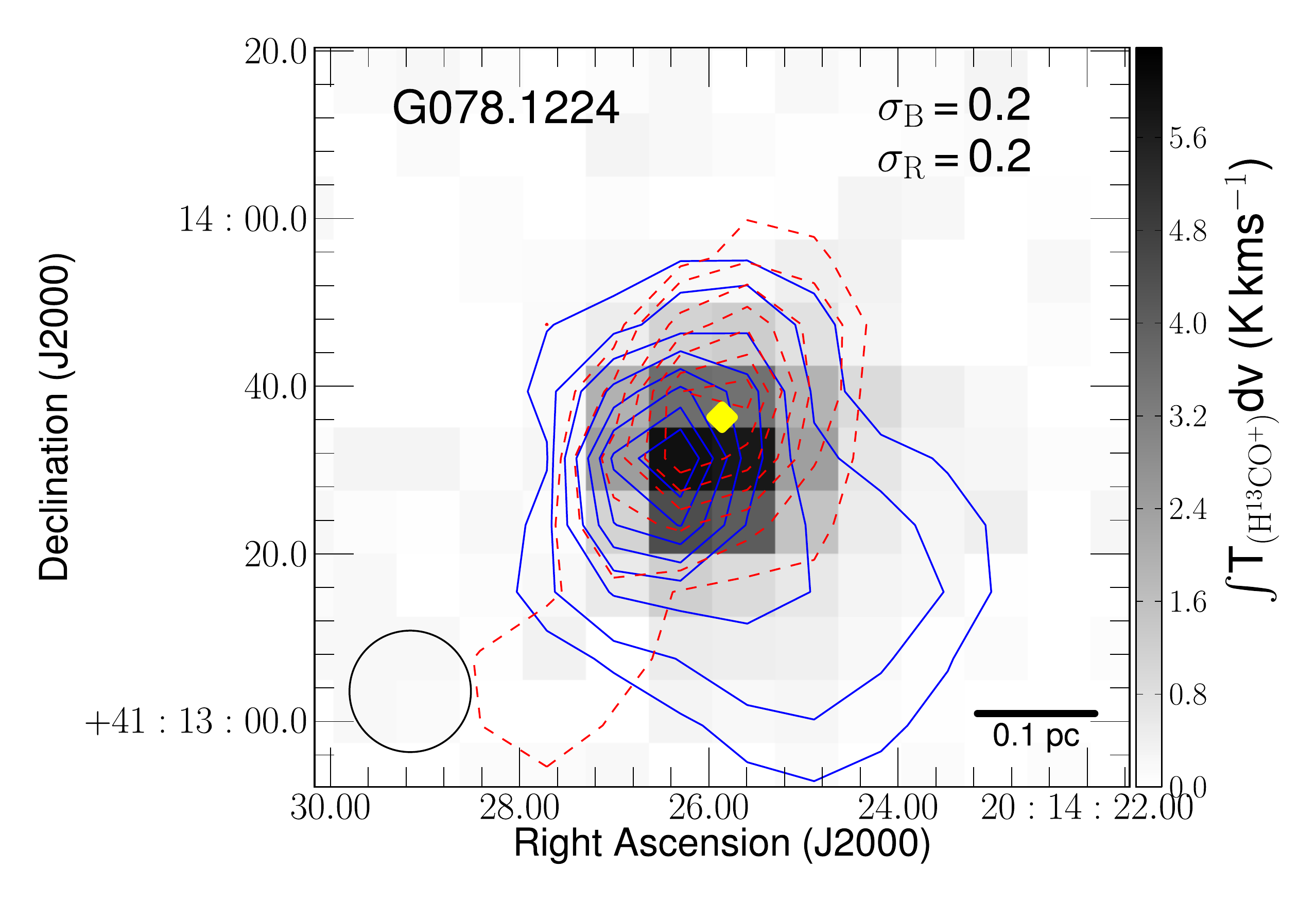}
\contcaption{--\,The velocity ranges used to integrate the total emission are 55.7\,--\,62.7\,\kms for G043.3061, 38.1\,--\,43.3\,\kms for G050.2213, and -8.1\,--\,1.6\,\kms for G078.1224.}
\end{figure*}

\begin{figure*}
\includegraphics[width=0.49\textwidth]{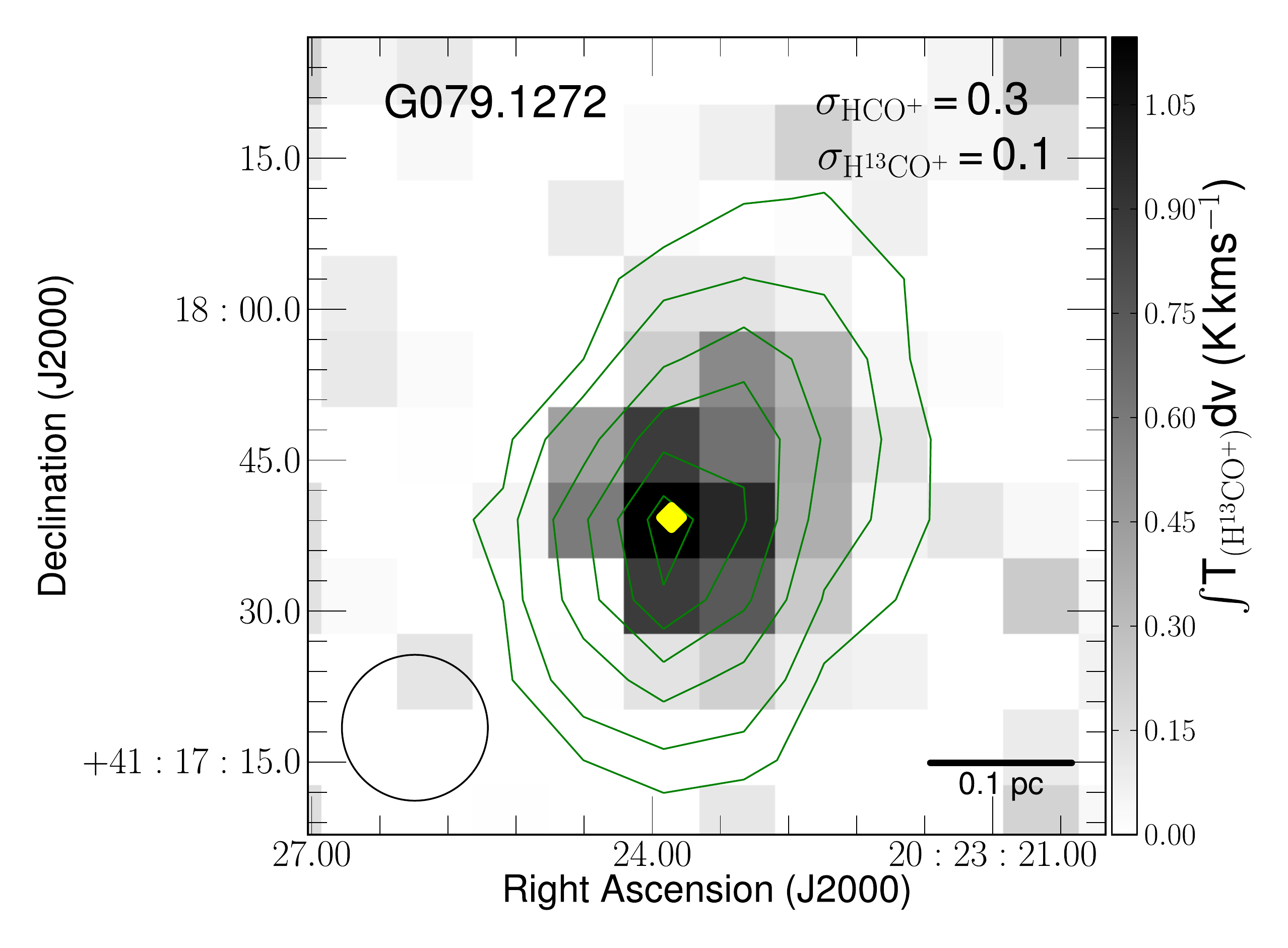}
\includegraphics[width=0.49\textwidth]{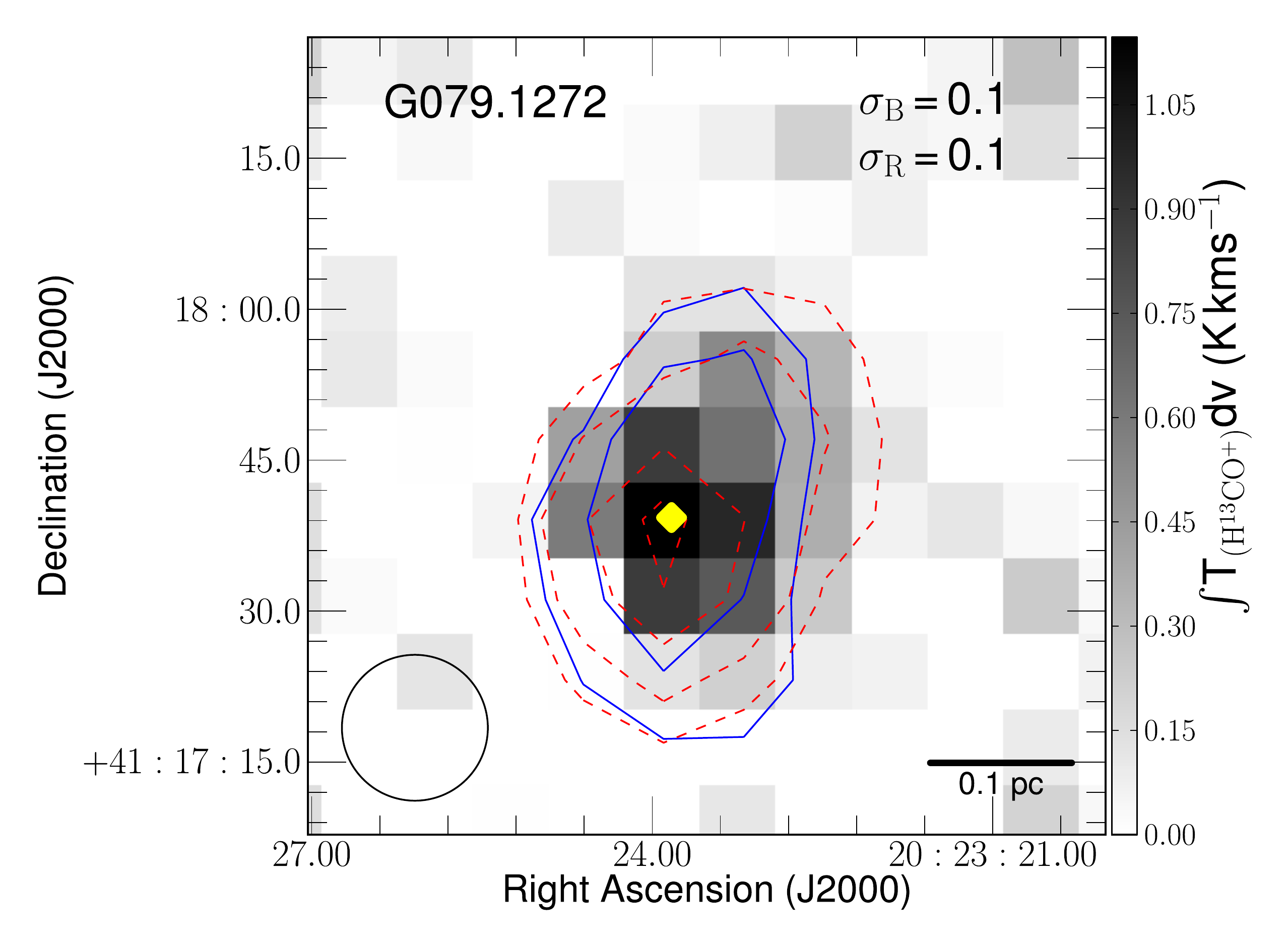}
\includegraphics[width=0.49\textwidth]{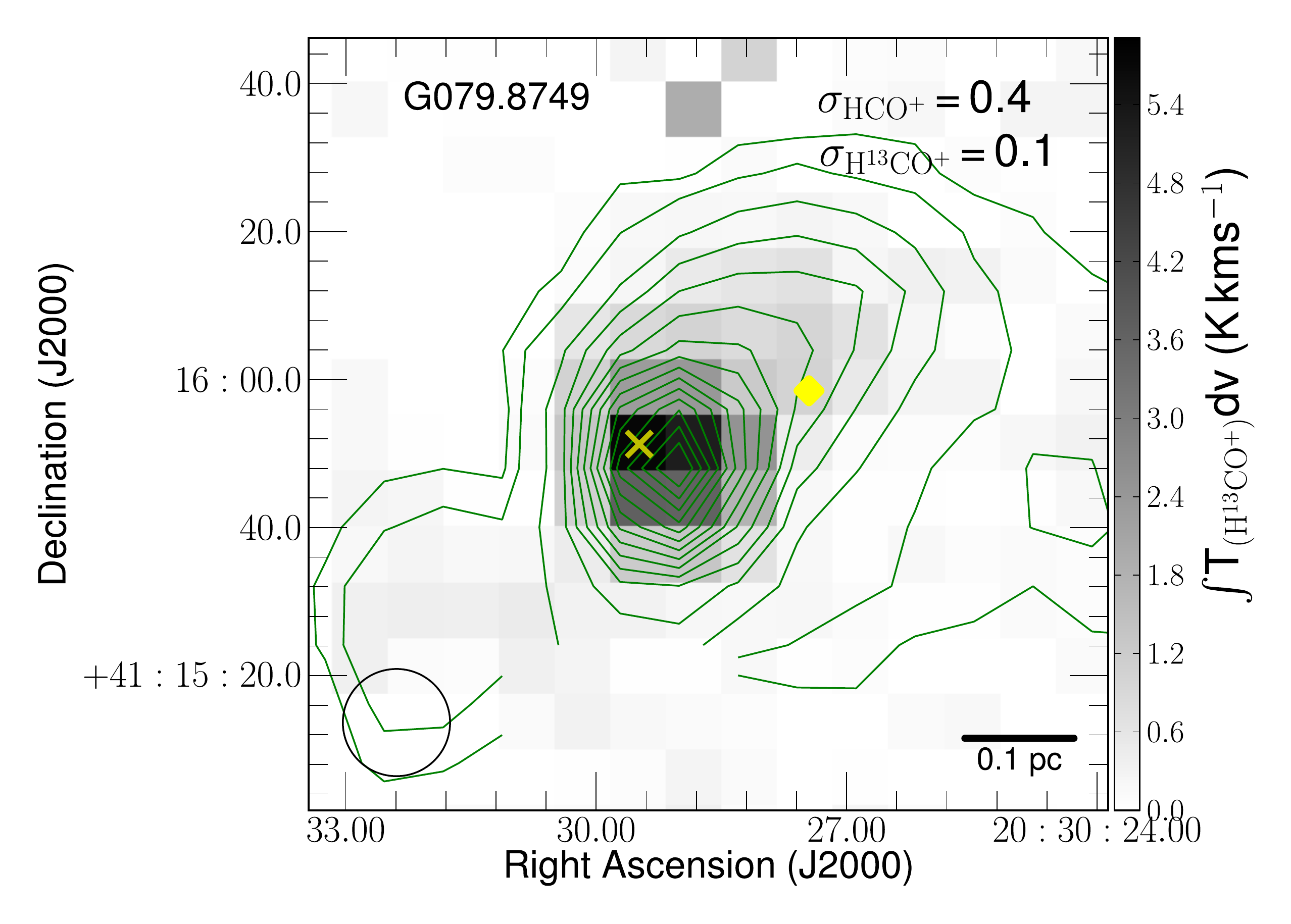}
\includegraphics[width=0.49\textwidth]{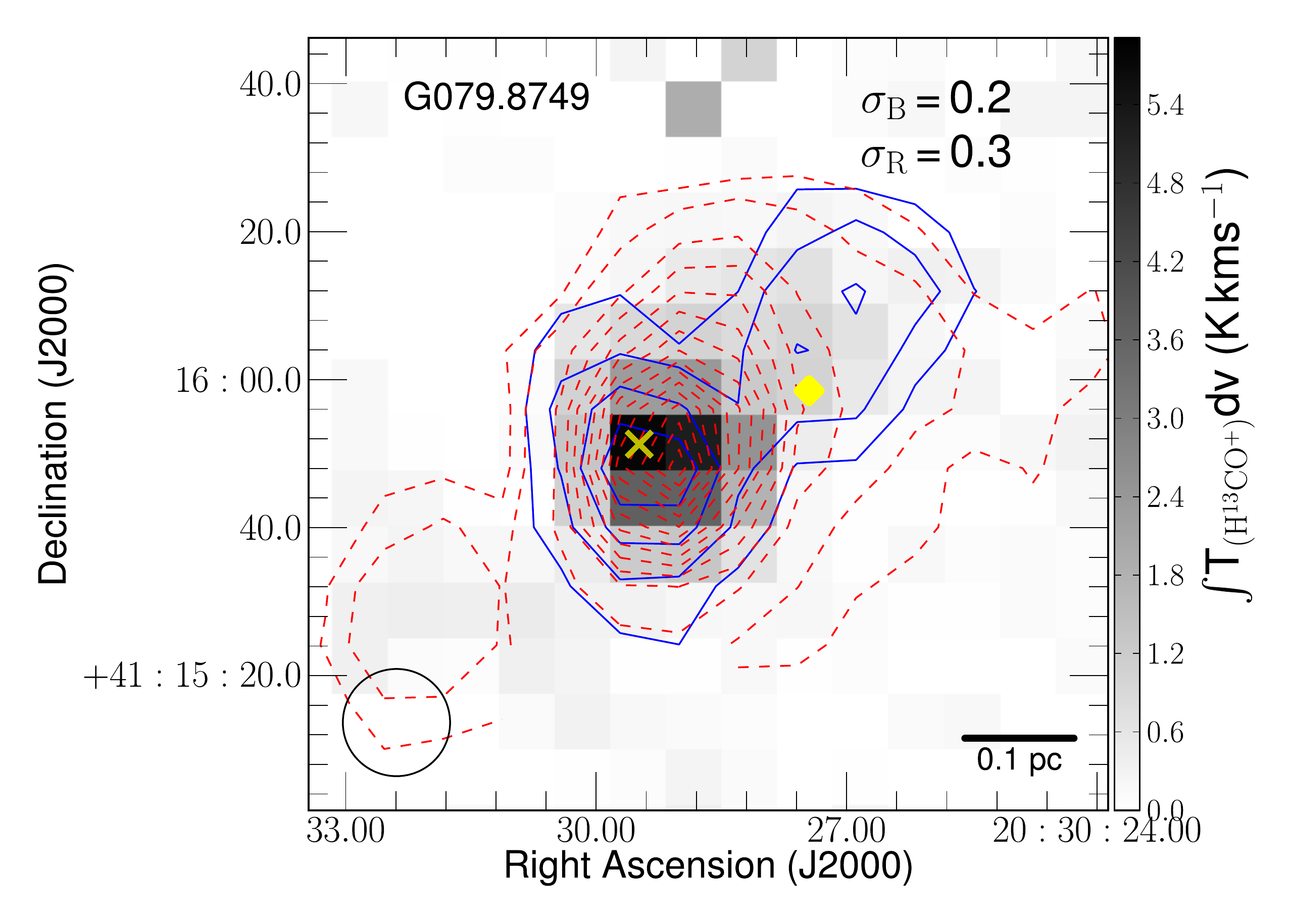}
\includegraphics[width=0.49\textwidth]{G079_8749_test_EMISSION_MAP_TOTAL_REF_COMMENTS_bbox}
\includegraphics[width=0.49\textwidth]{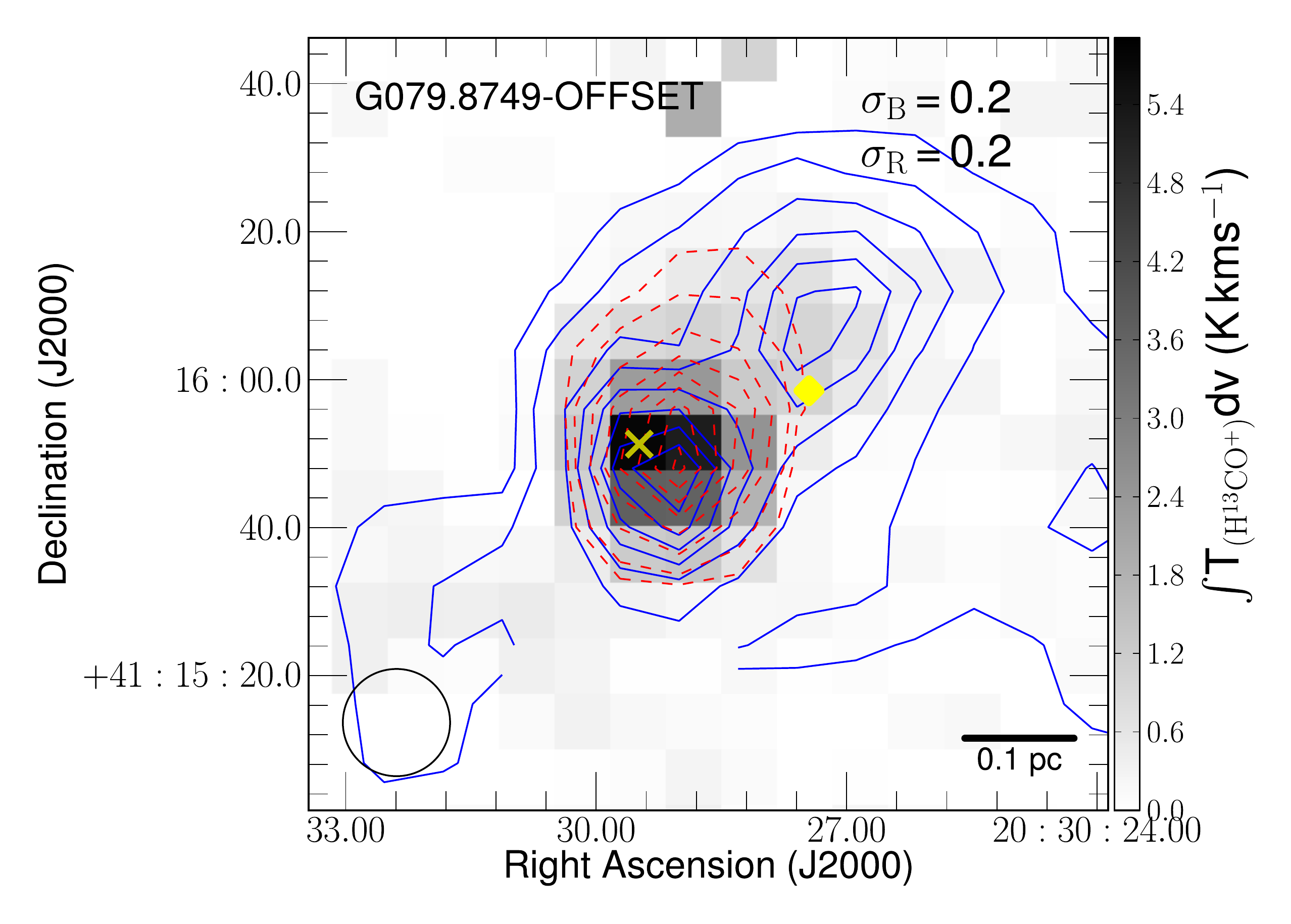}
\contcaption{--\,The velocity ranges used to integrate the total emission are -3.3\,--\,0.9\,\kms for G079.1272, -7.7\,--\,0.9\,\kms for both G079.8749 and G079.8749-OFFSET.}
\end{figure*}

\begin{figure*}
\includegraphics[width=0.49\textwidth]{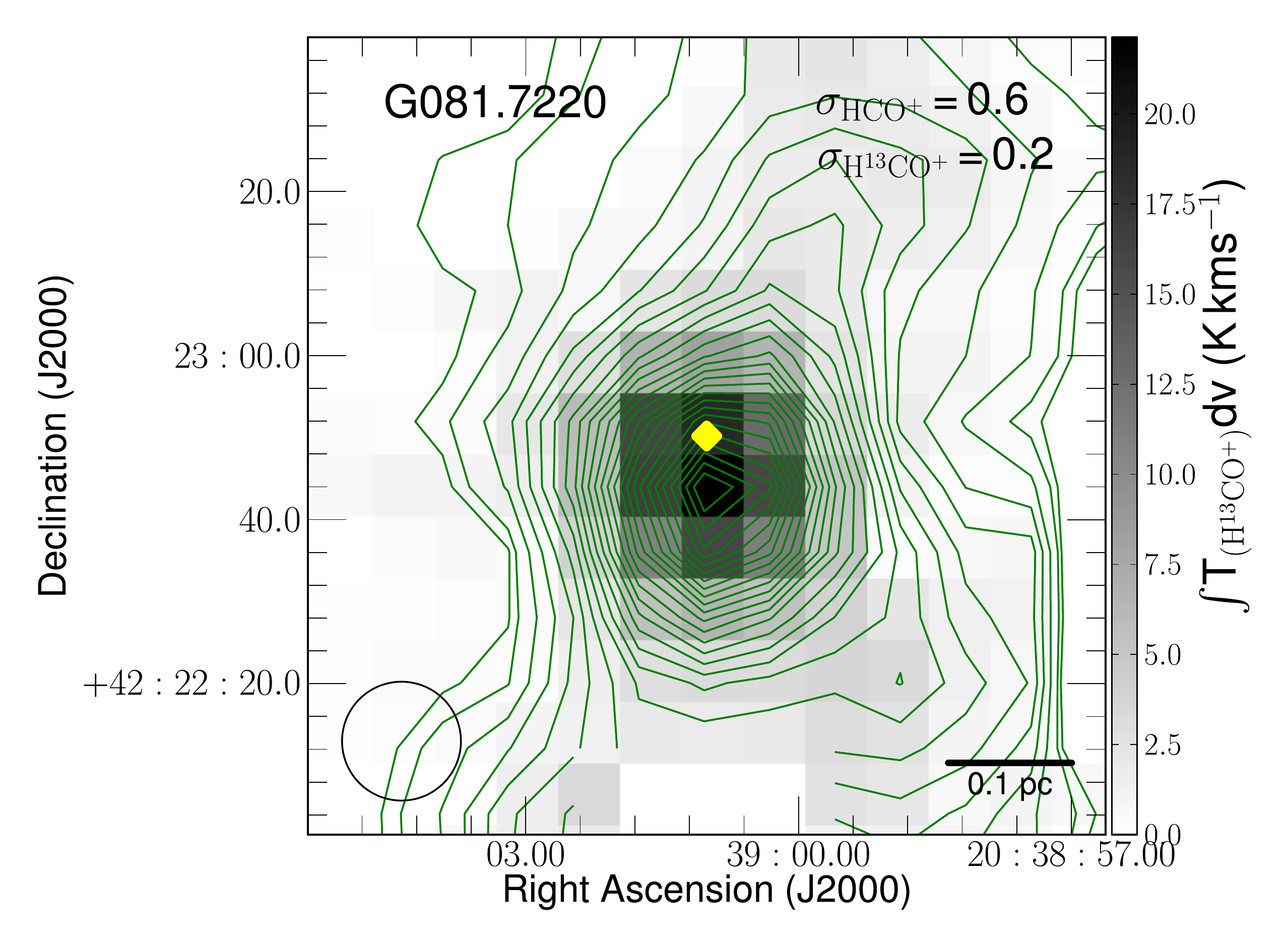}
\includegraphics[width=0.49\textwidth]{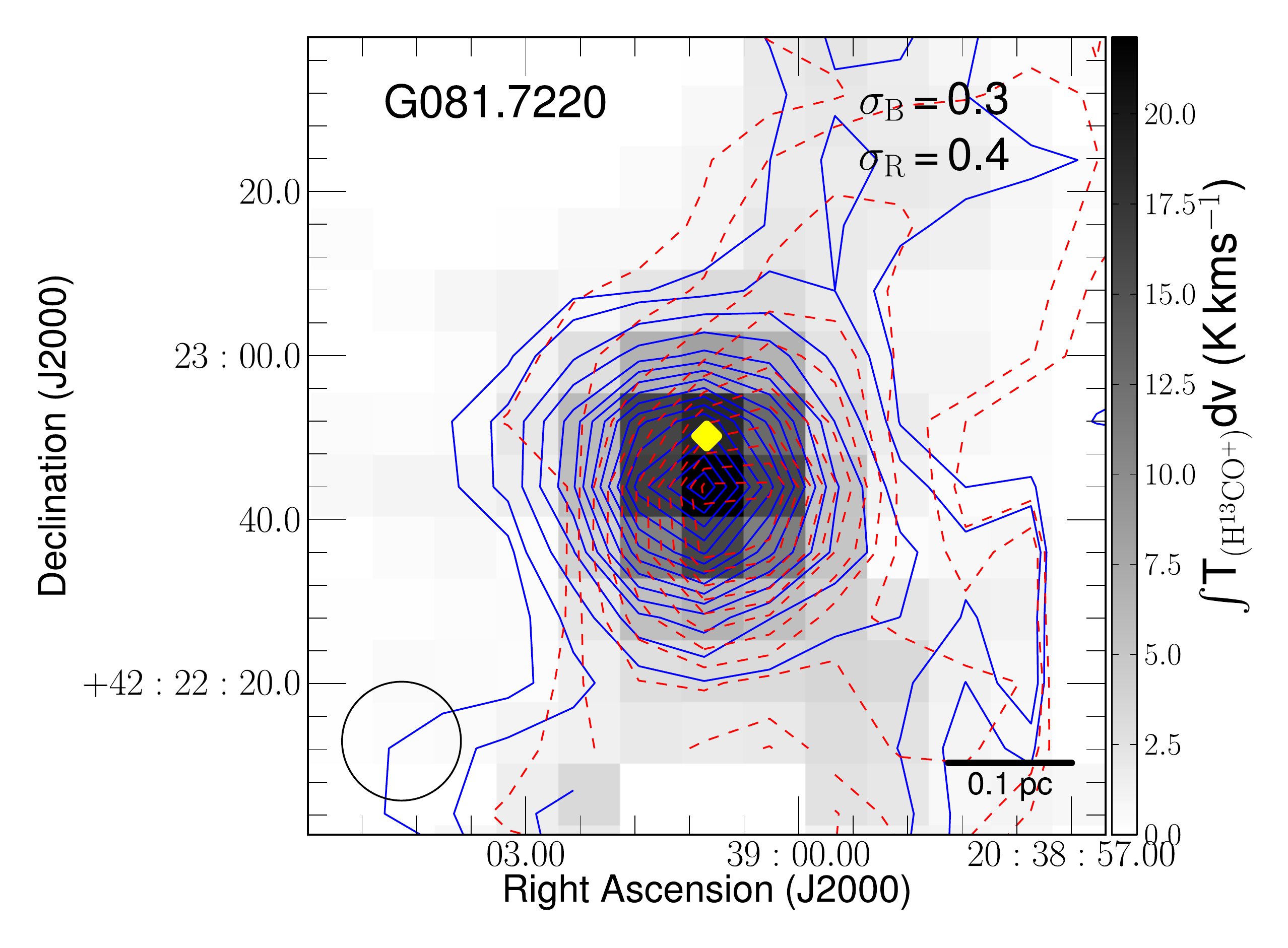}
\includegraphics[width=0.49\textwidth]{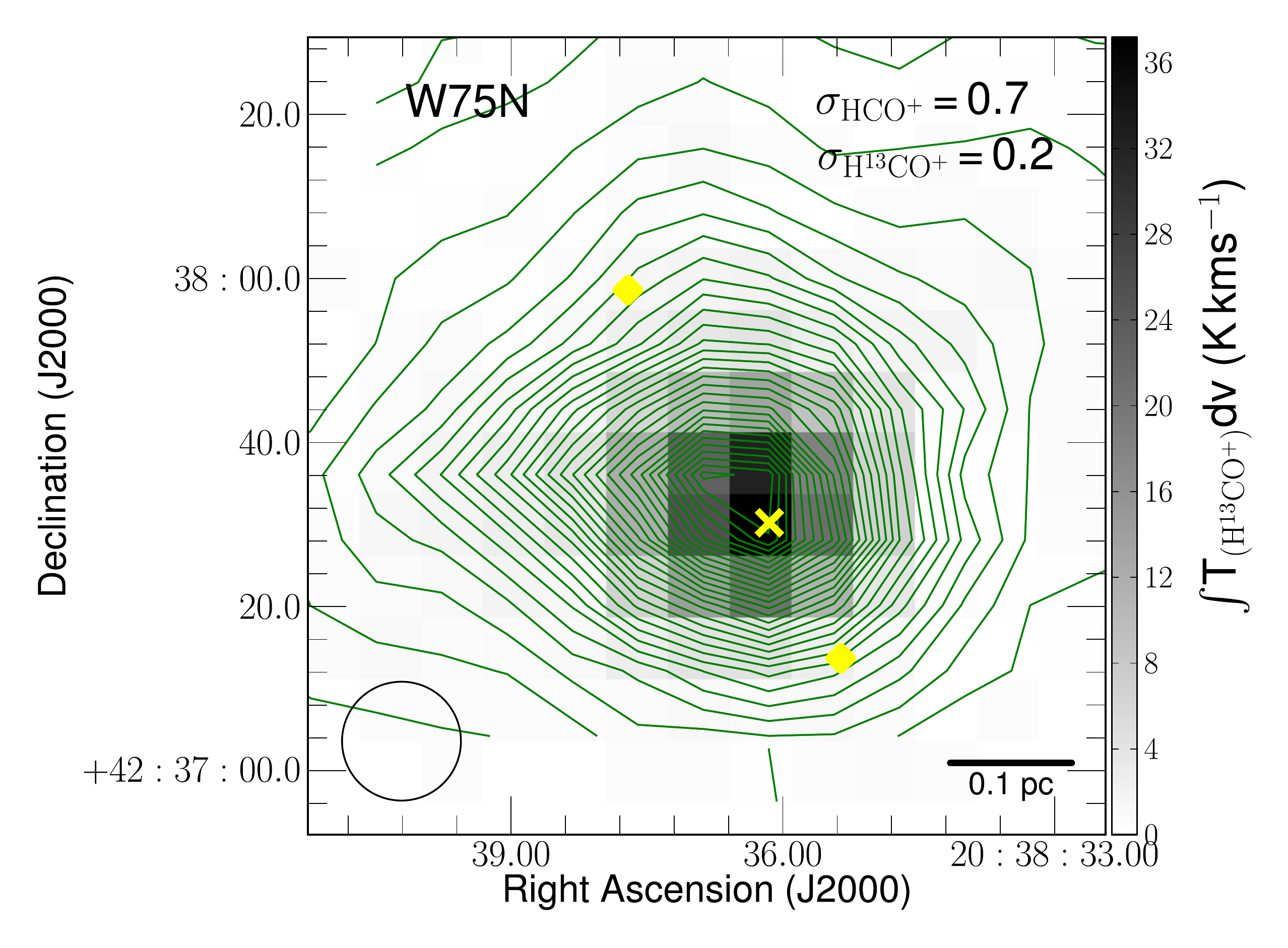}
\includegraphics[width=0.49\textwidth]{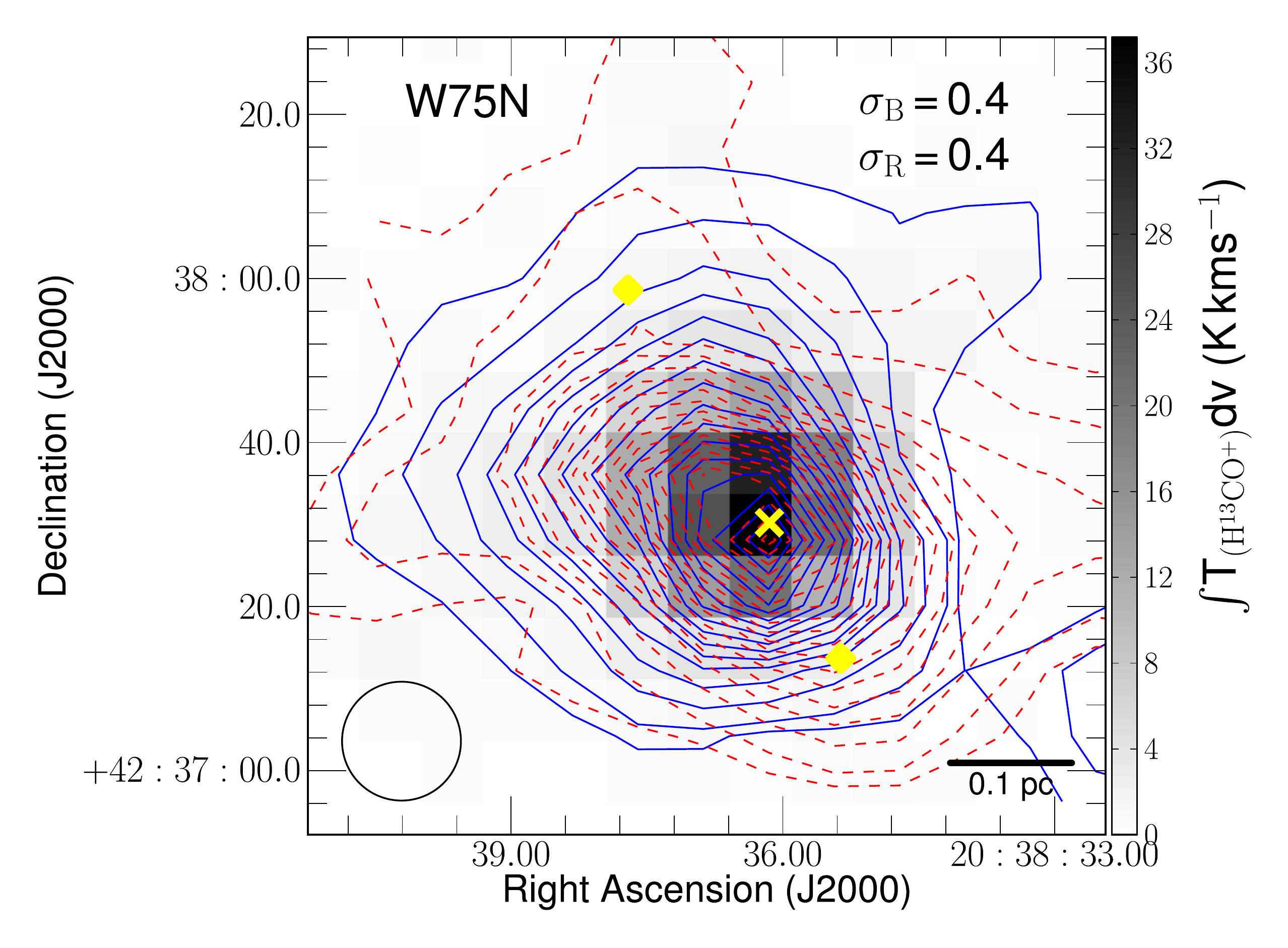}
\includegraphics[width=0.49\textwidth]{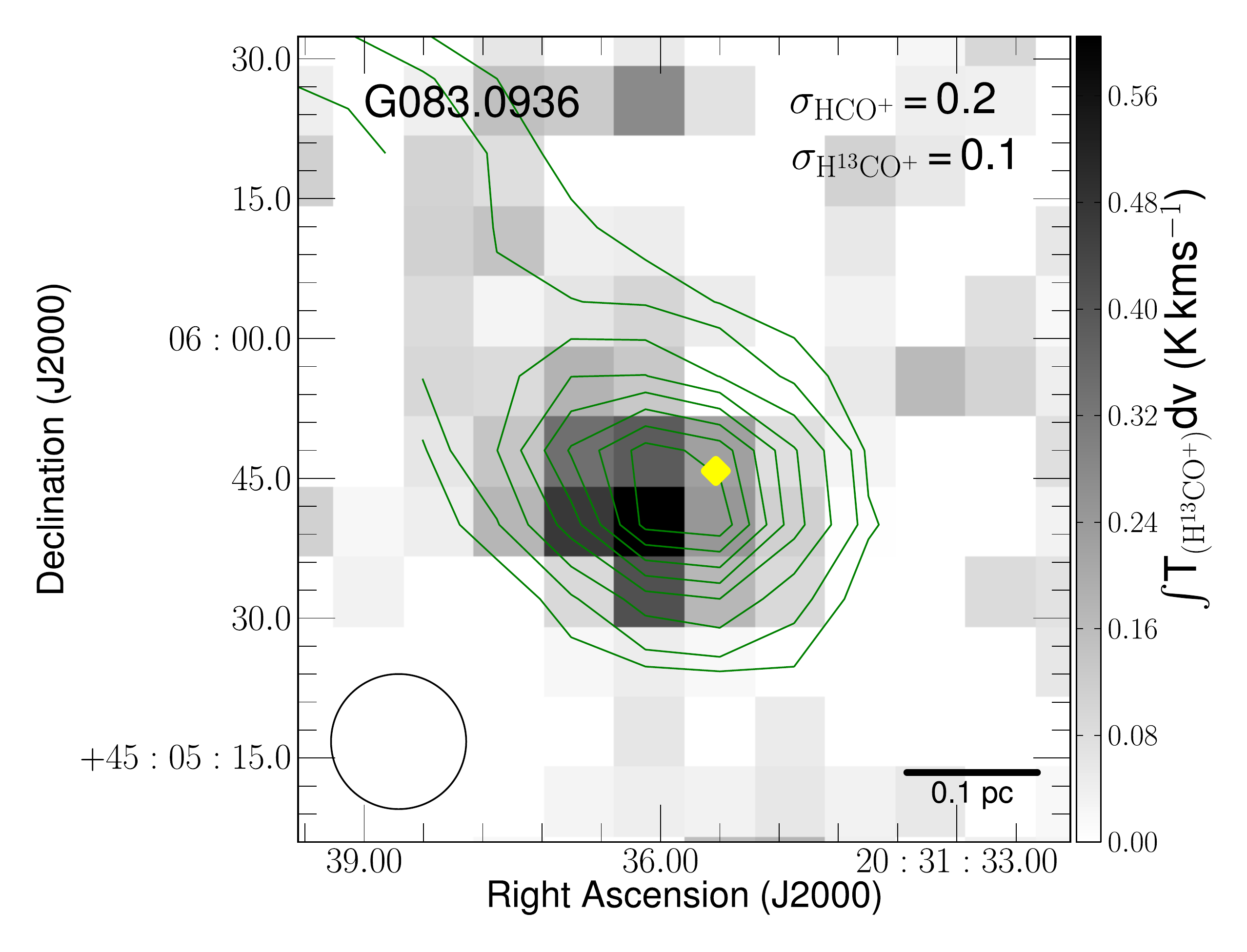}
\includegraphics[width=0.49\textwidth]{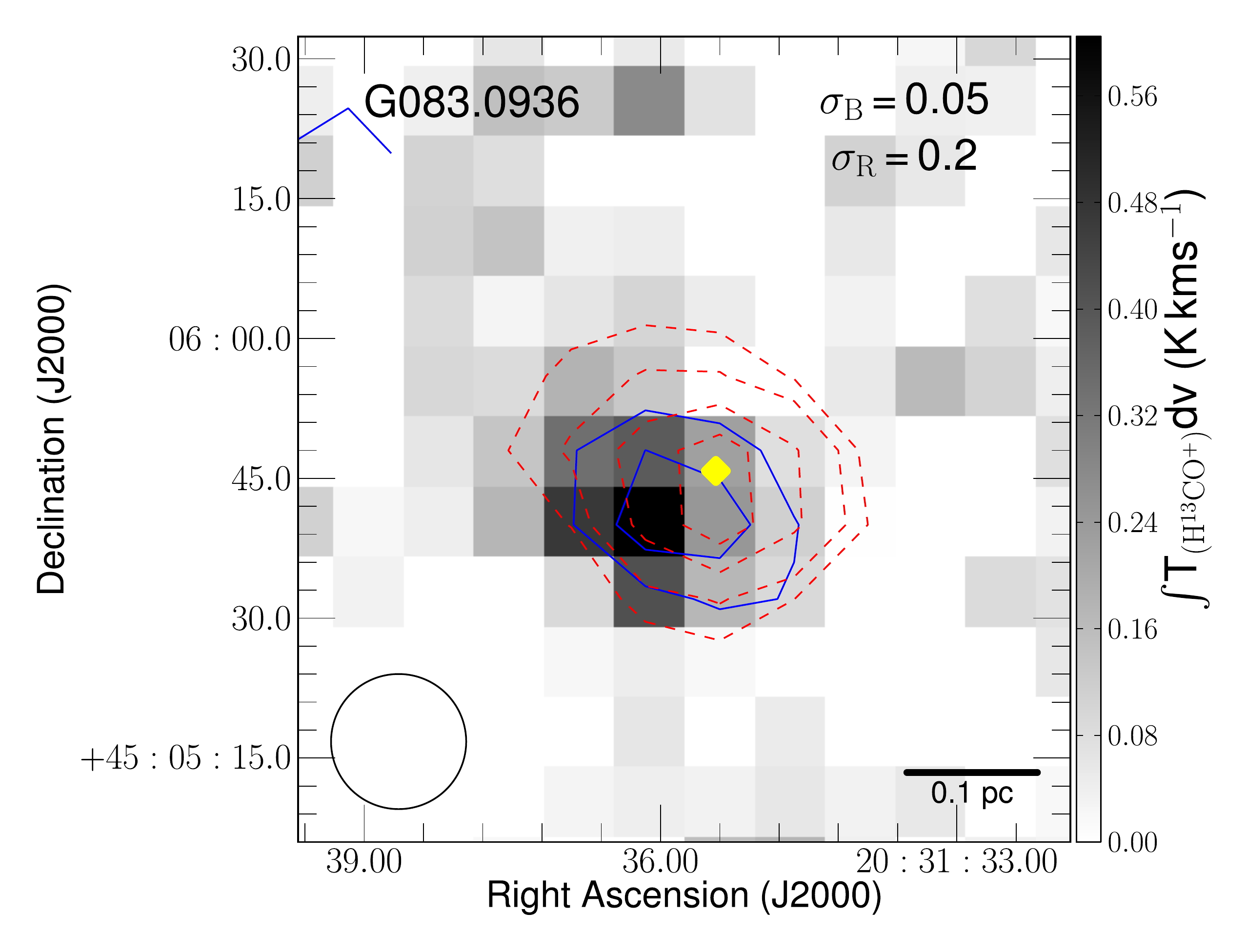}
\contcaption{--\,The velocity ranges used to integrate the total emission are -9.5\,--\,4.8\,\kms for G081.7220, 0.8\,--\,18.1\,\kms for W75N, and -5.1\,--\,-0.1\,\kms for G083.0936.}
\end{figure*}

\begin{figure*}
\includegraphics[width=0.49\textwidth]{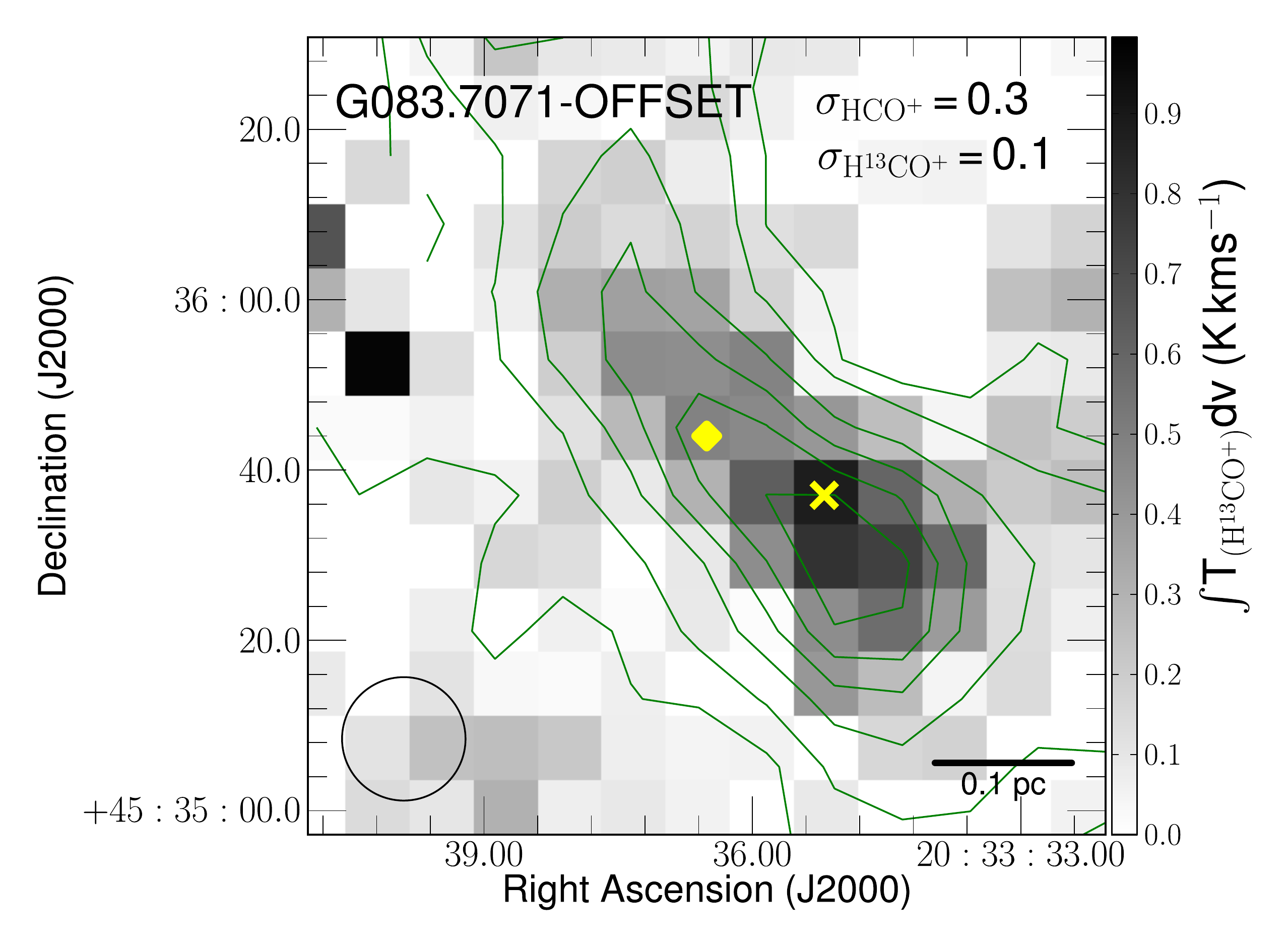}
\includegraphics[width=0.49\textwidth]{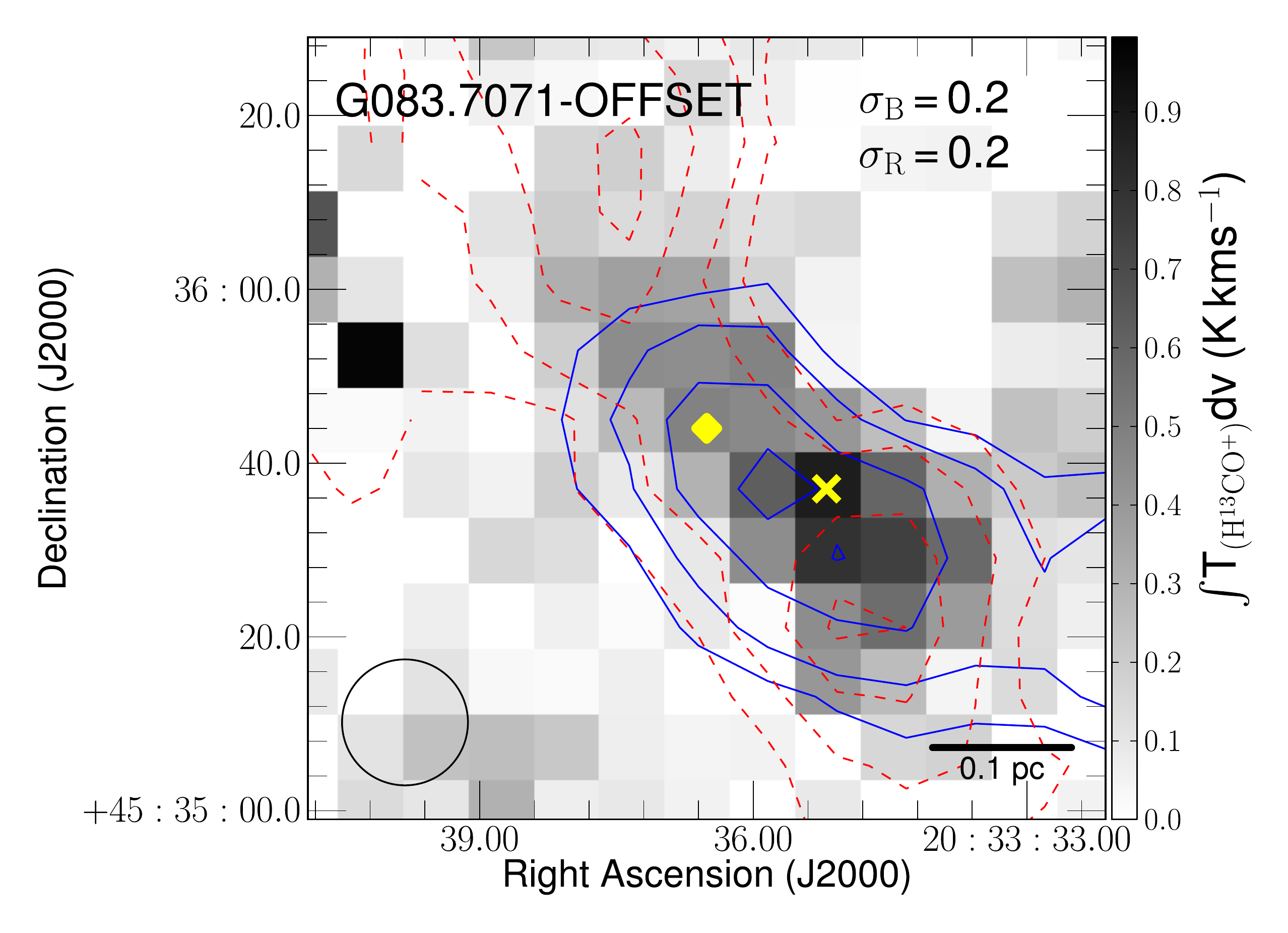}
\includegraphics[width=0.49\textwidth]{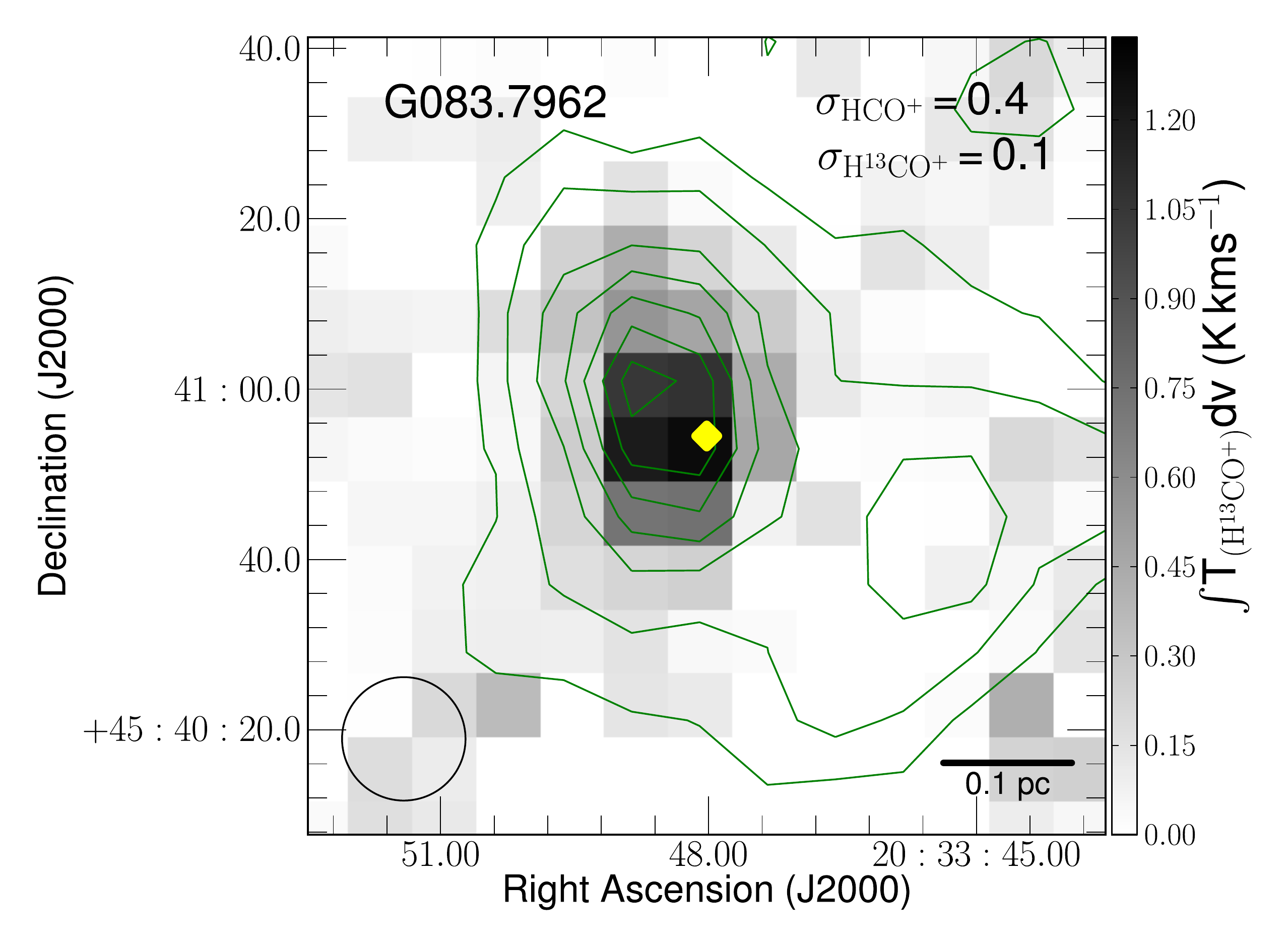}
\includegraphics[width=0.49\textwidth]{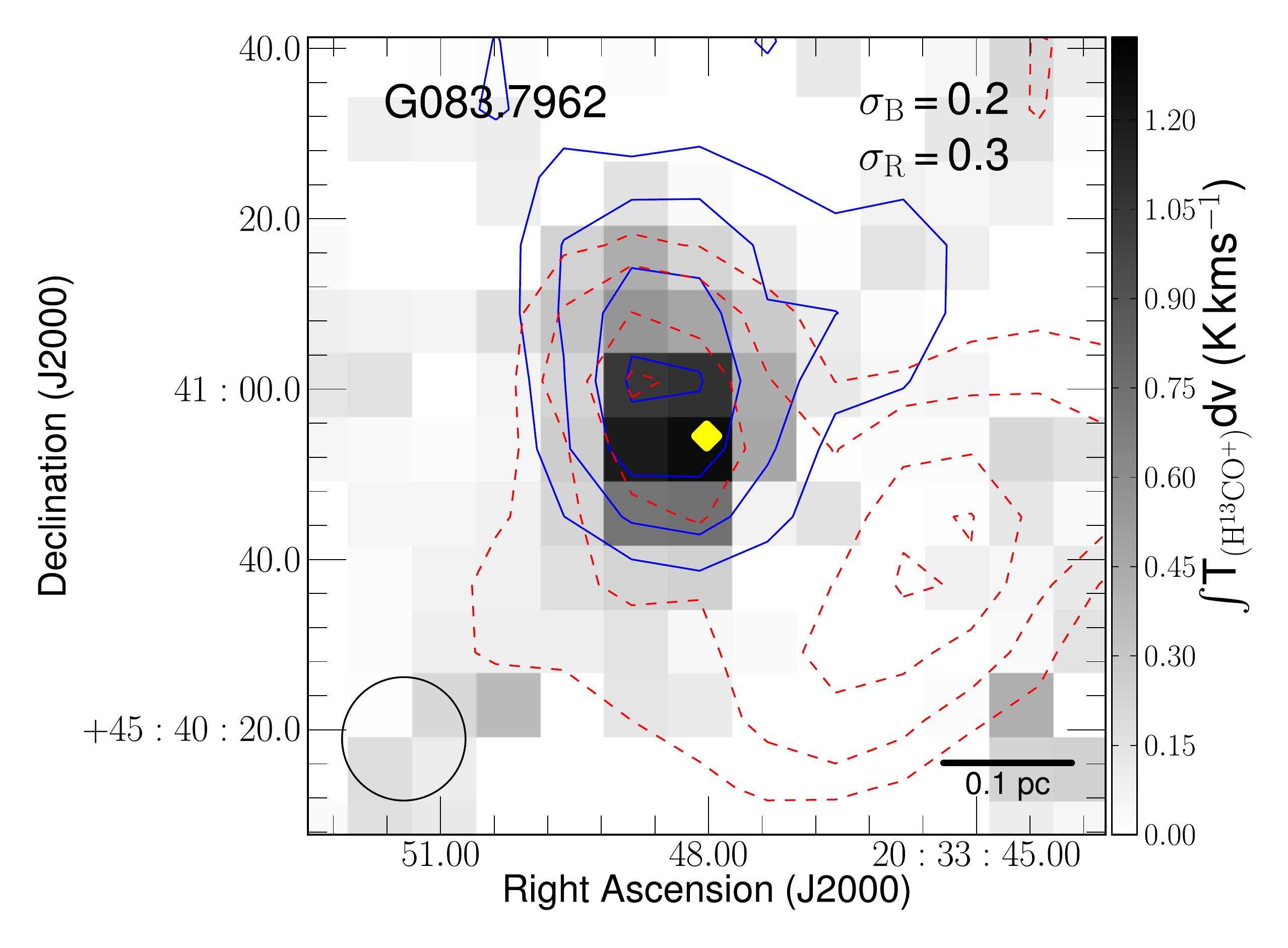}
\includegraphics[width=0.49\textwidth]{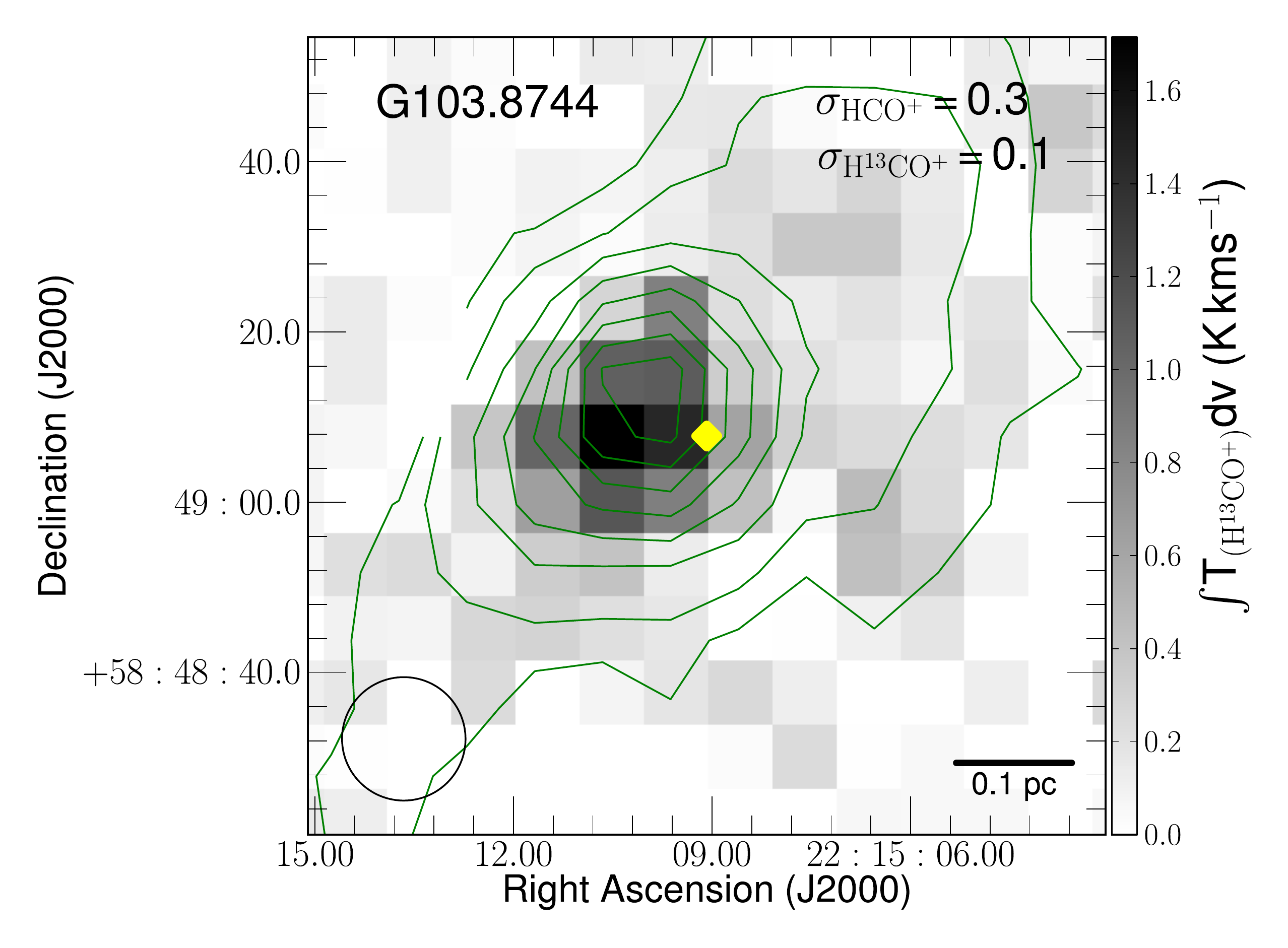}
\includegraphics[width=0.49\textwidth]{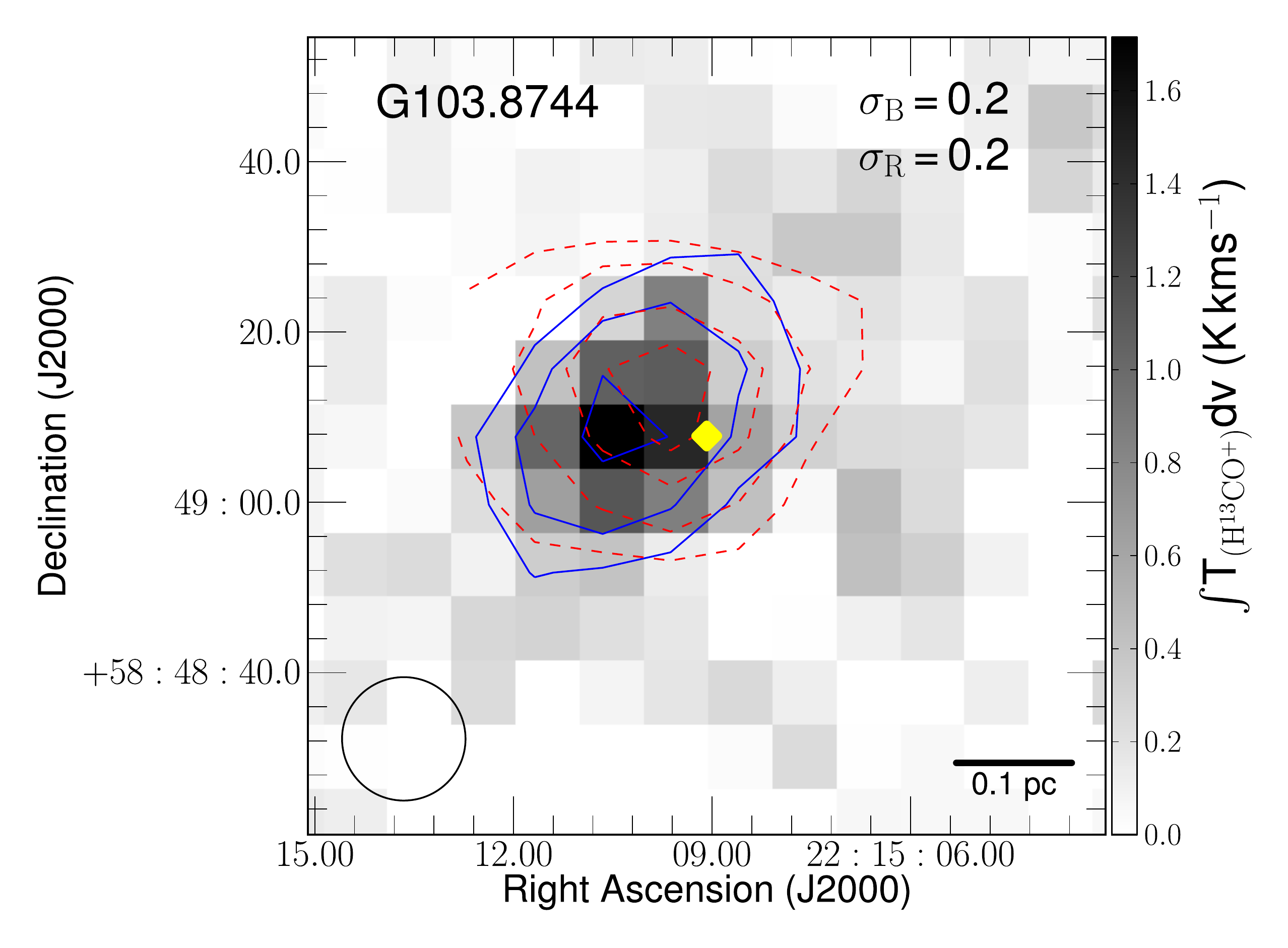}
\contcaption{--\,The velocity ranges used to integrate the total emission are -5.8\,--\,-1.3\,\kms for G083.7071-OFFSET, -6.2\,--\,-0.3\,\kms for G083.7962, and -20.2\,--\,-16.3\,\kms for G103.8744.}
\end{figure*}

\begin{figure*}
\includegraphics[width=0.49\textwidth]{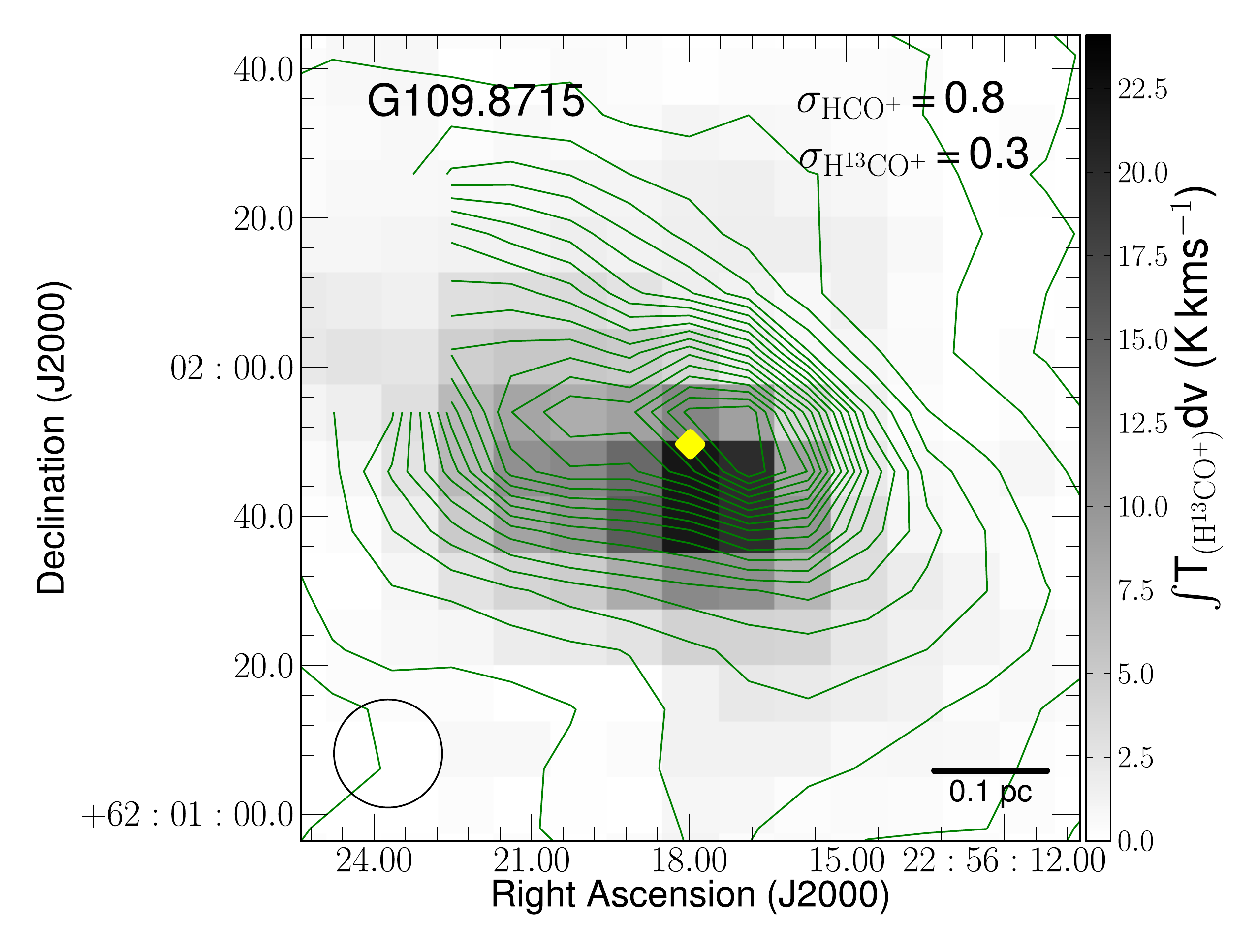}
\includegraphics[width=0.49\textwidth]{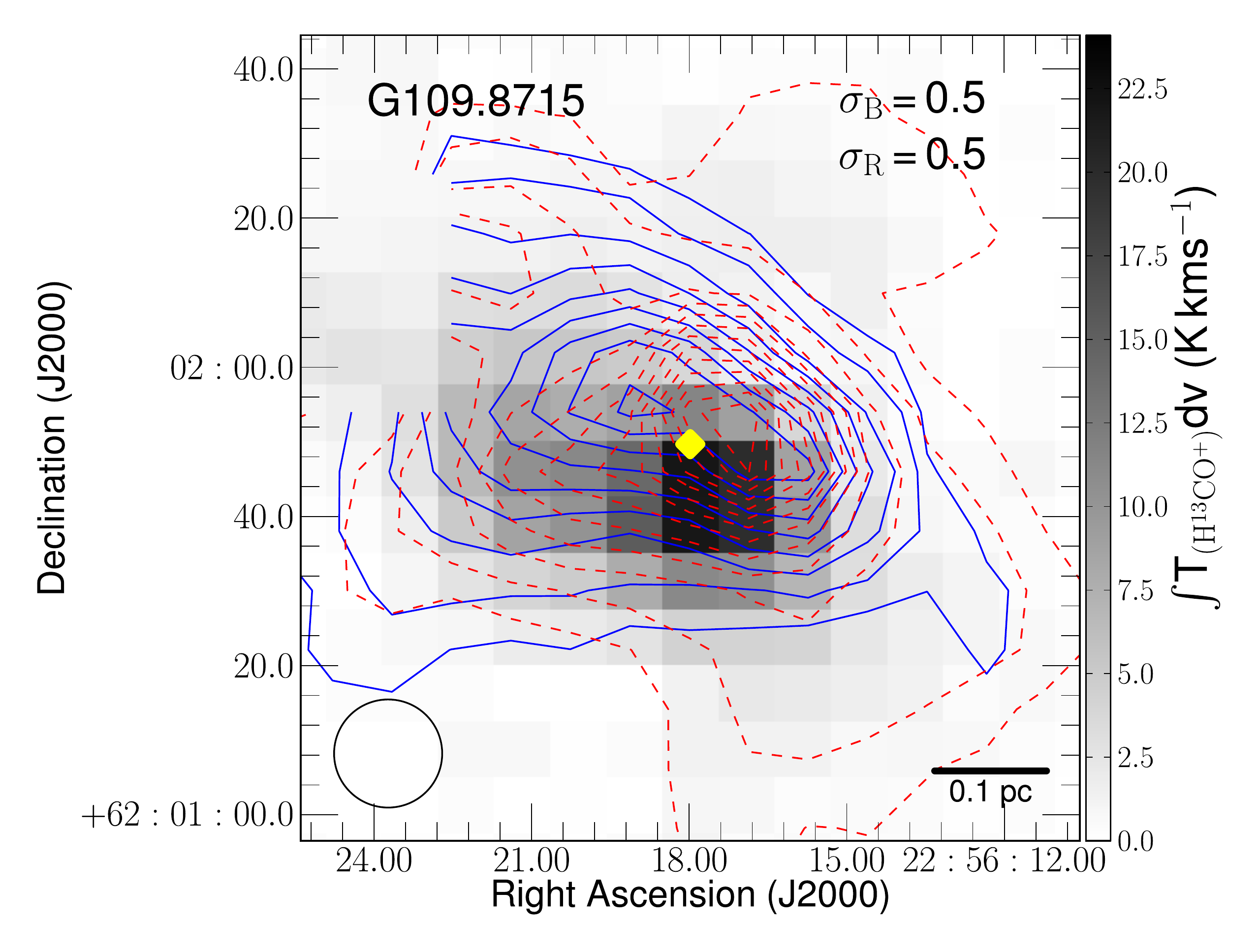}
\includegraphics[width=0.49\textwidth]{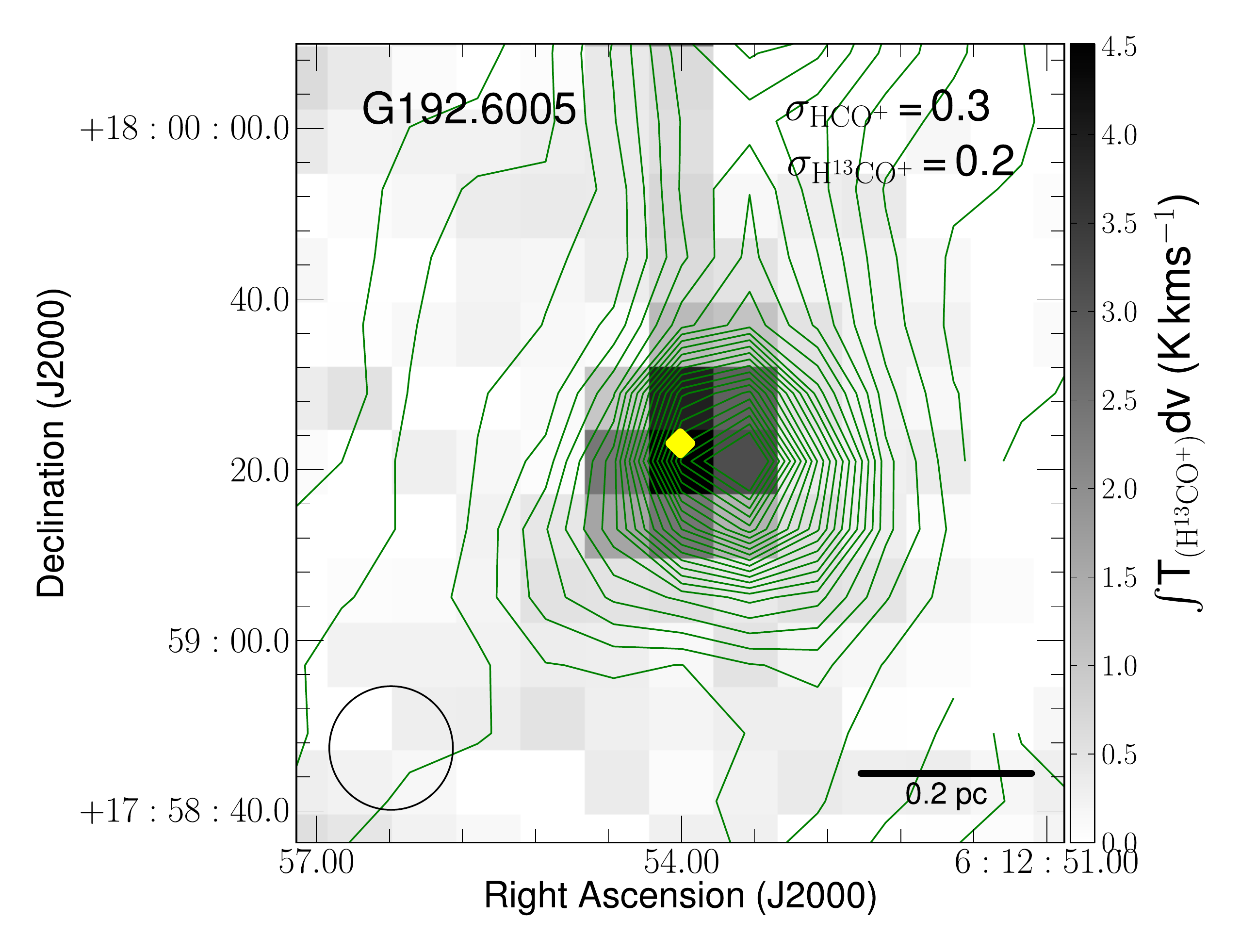}
\includegraphics[width=0.49\textwidth]{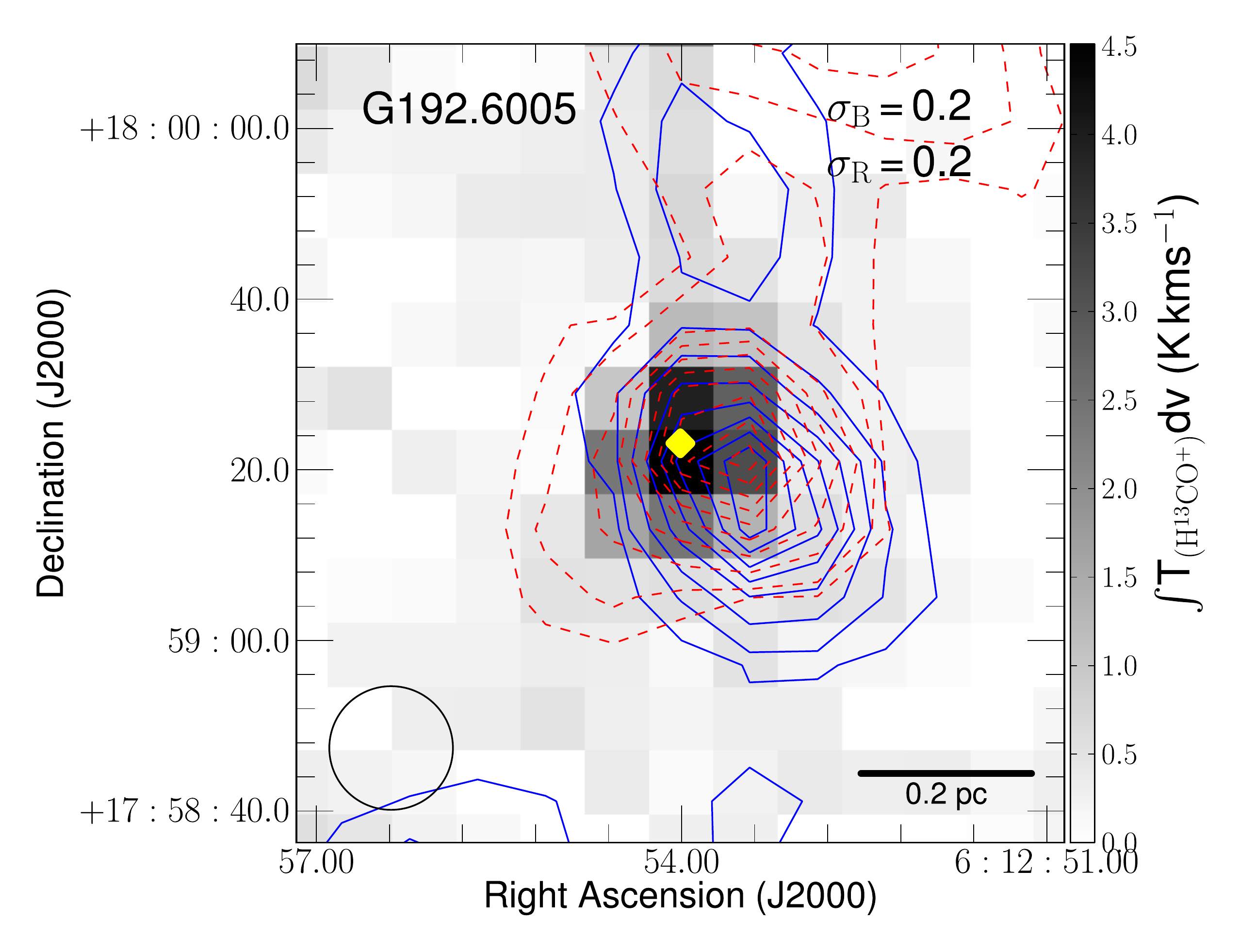}
\includegraphics[width=0.49\textwidth]{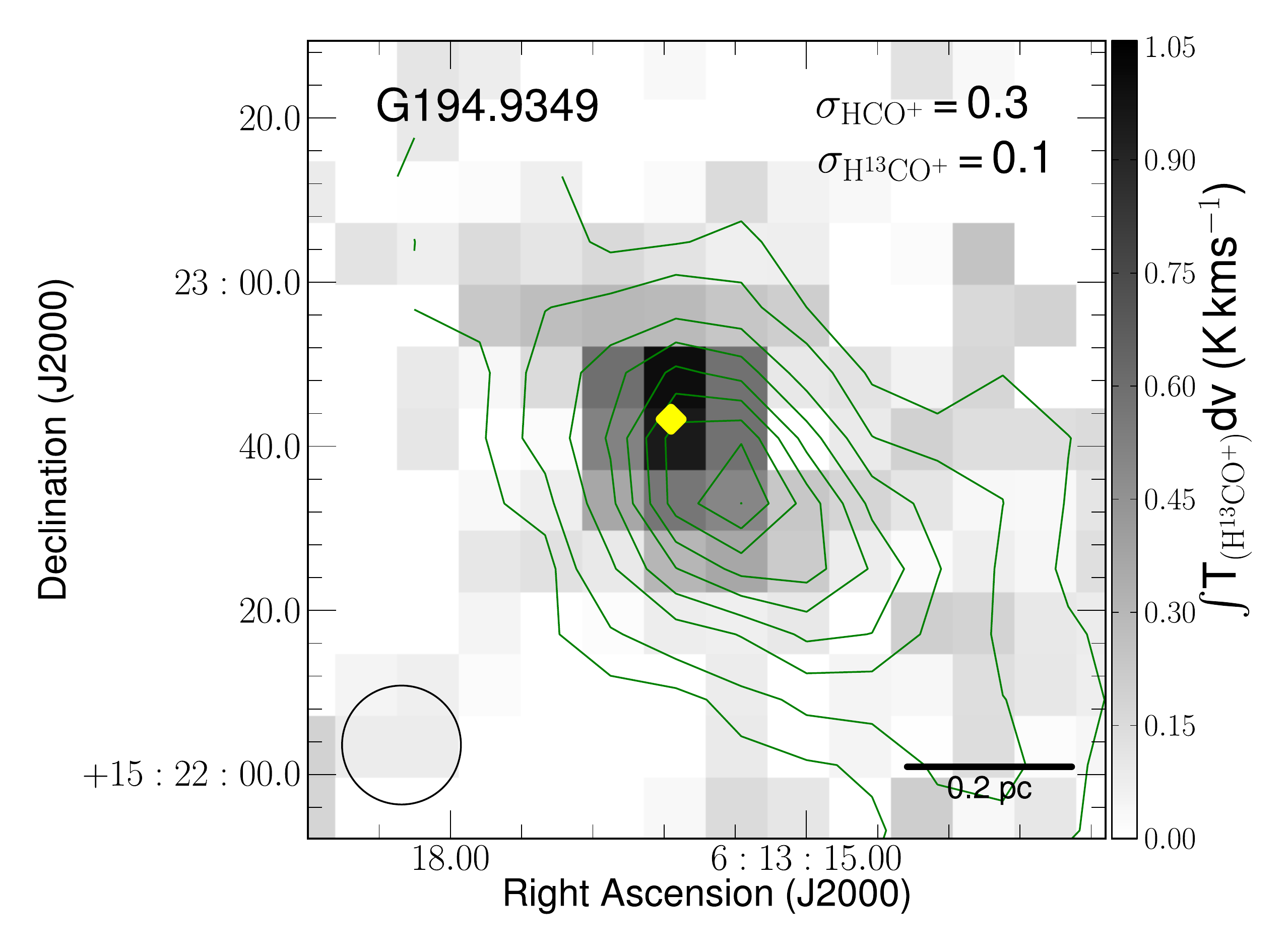}
\includegraphics[width=0.49\textwidth]{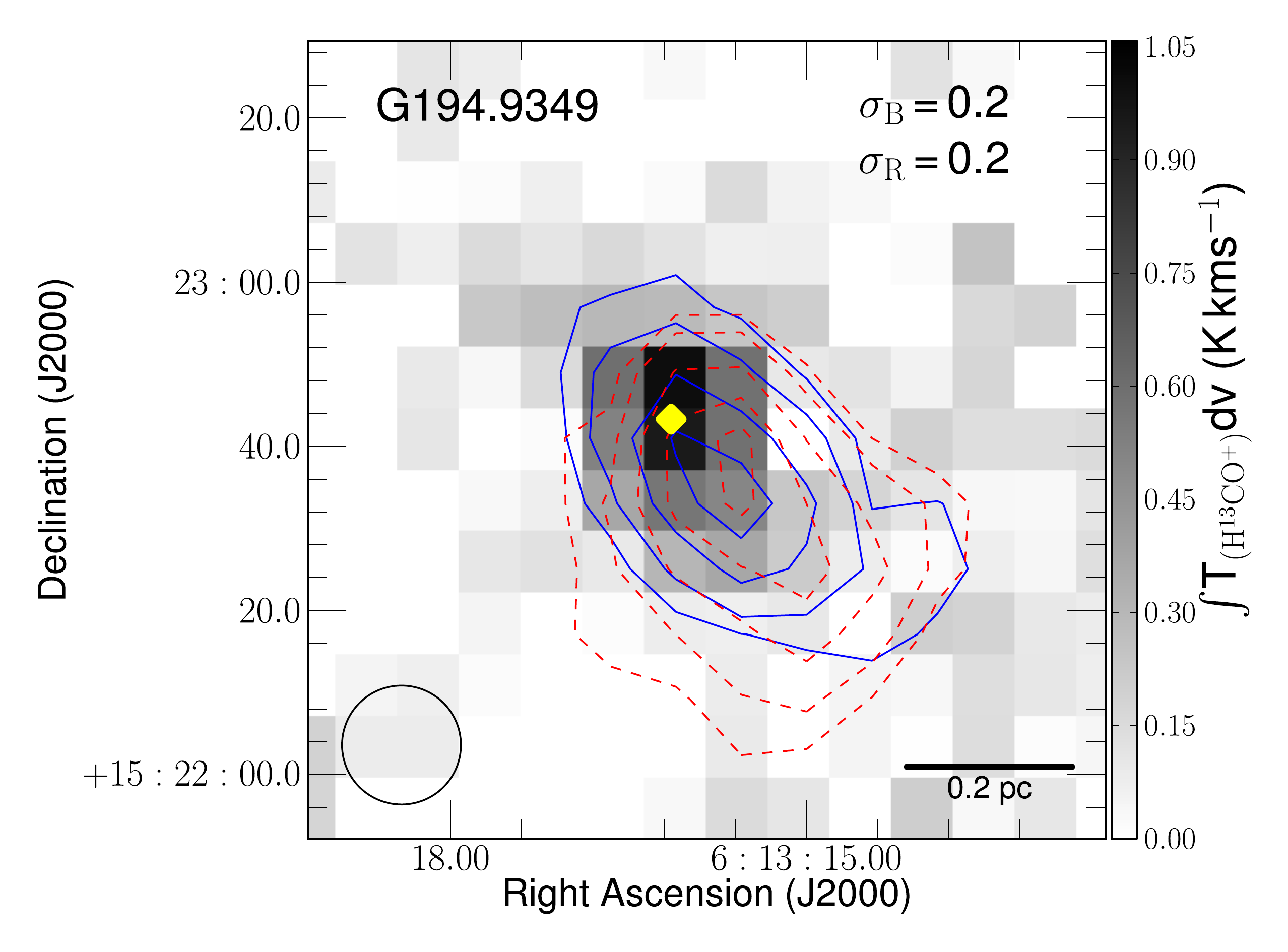}

\contcaption{--\,The velocity ranges used to integrate the total emission are -19.1\,--\,-1.1\,\kms for G109.8715, 2.8\,--\,12.7\,\kms for G192.6005, and 12.2\,--\,18.9\,\kms for G194.9349. }
\end{figure*}

\begin{figure*}
\includegraphics[width=0.49\textwidth]{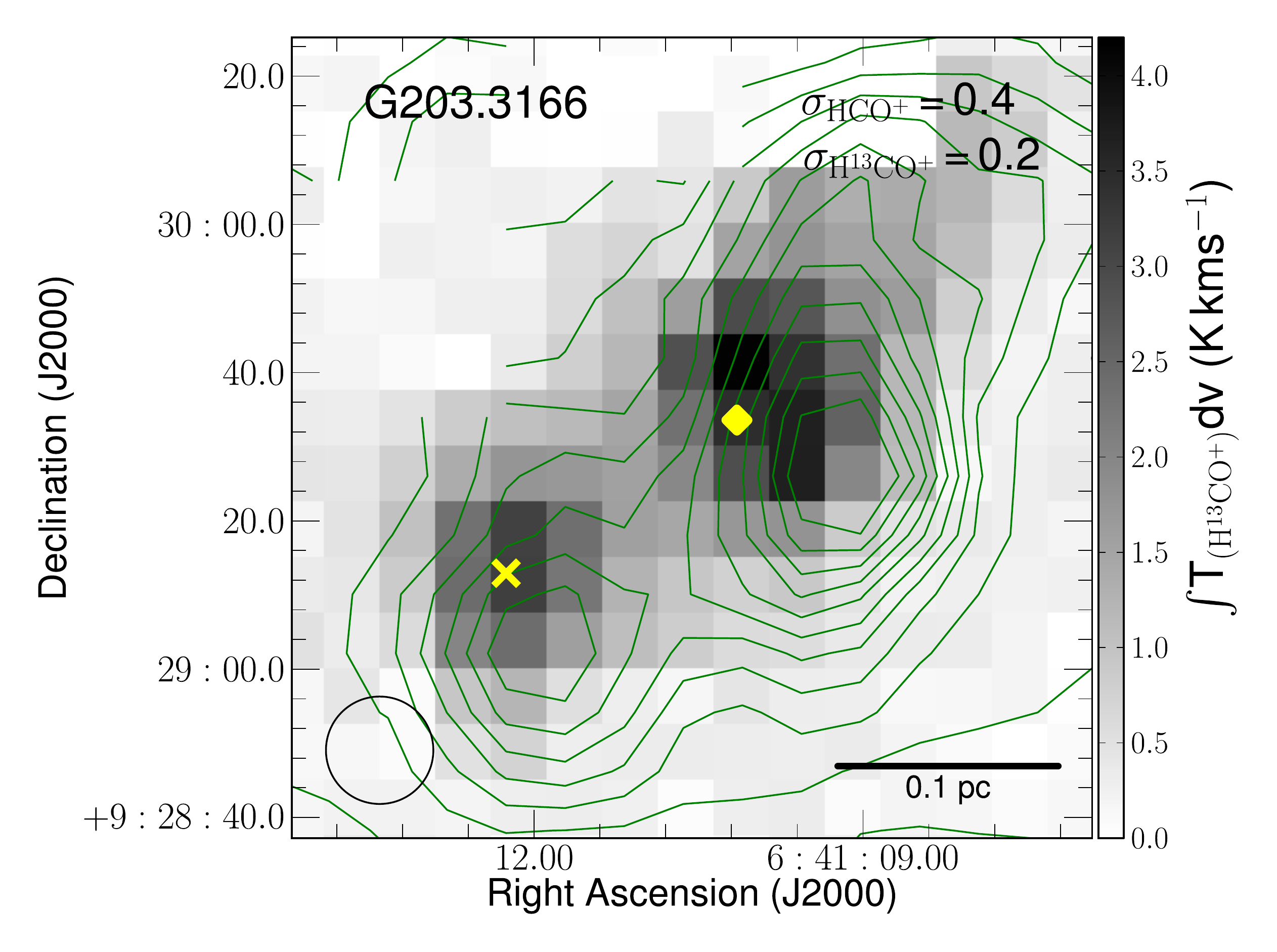}
\includegraphics[width=0.49\textwidth]{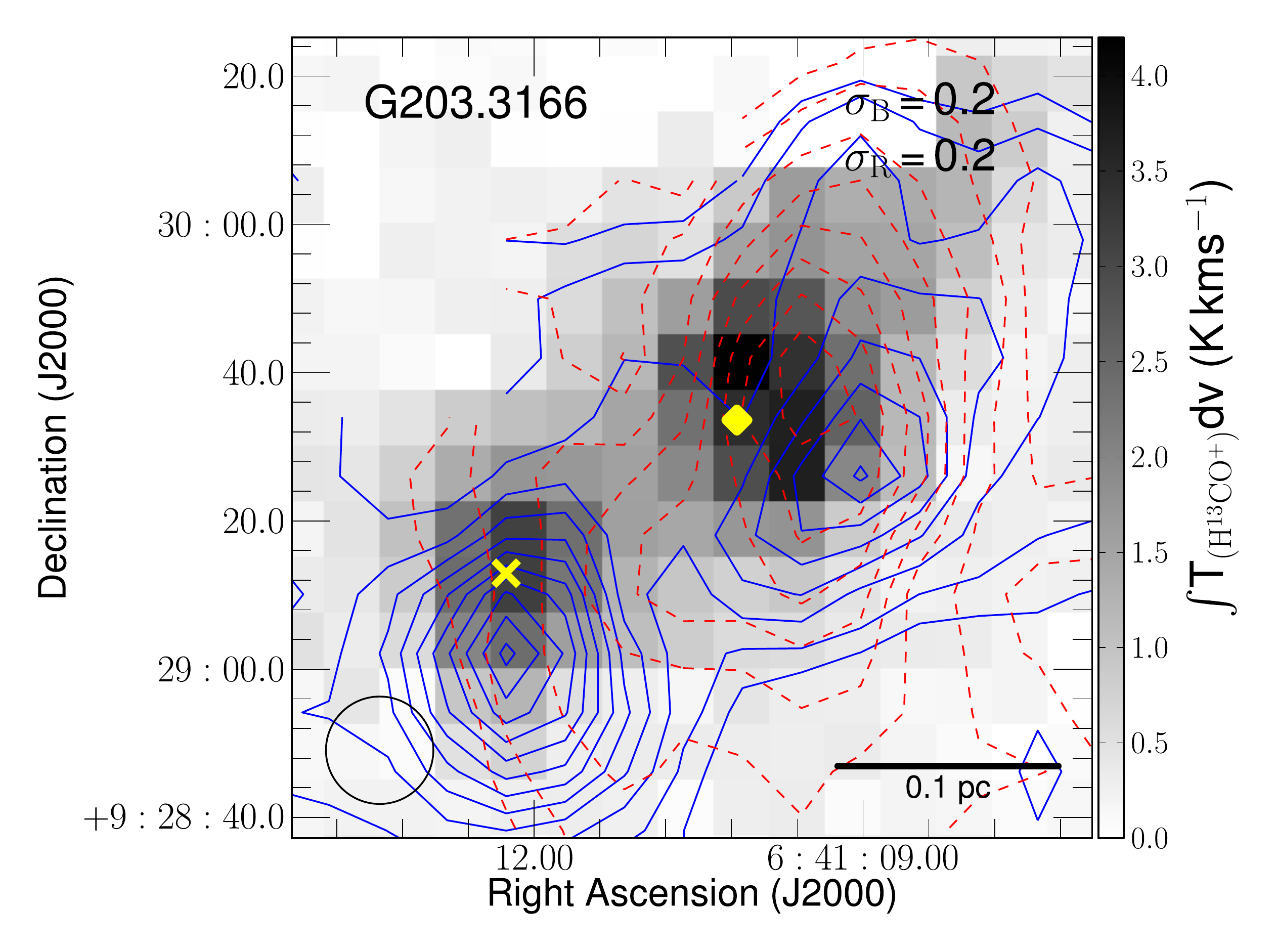}
\includegraphics[width=0.49\textwidth]{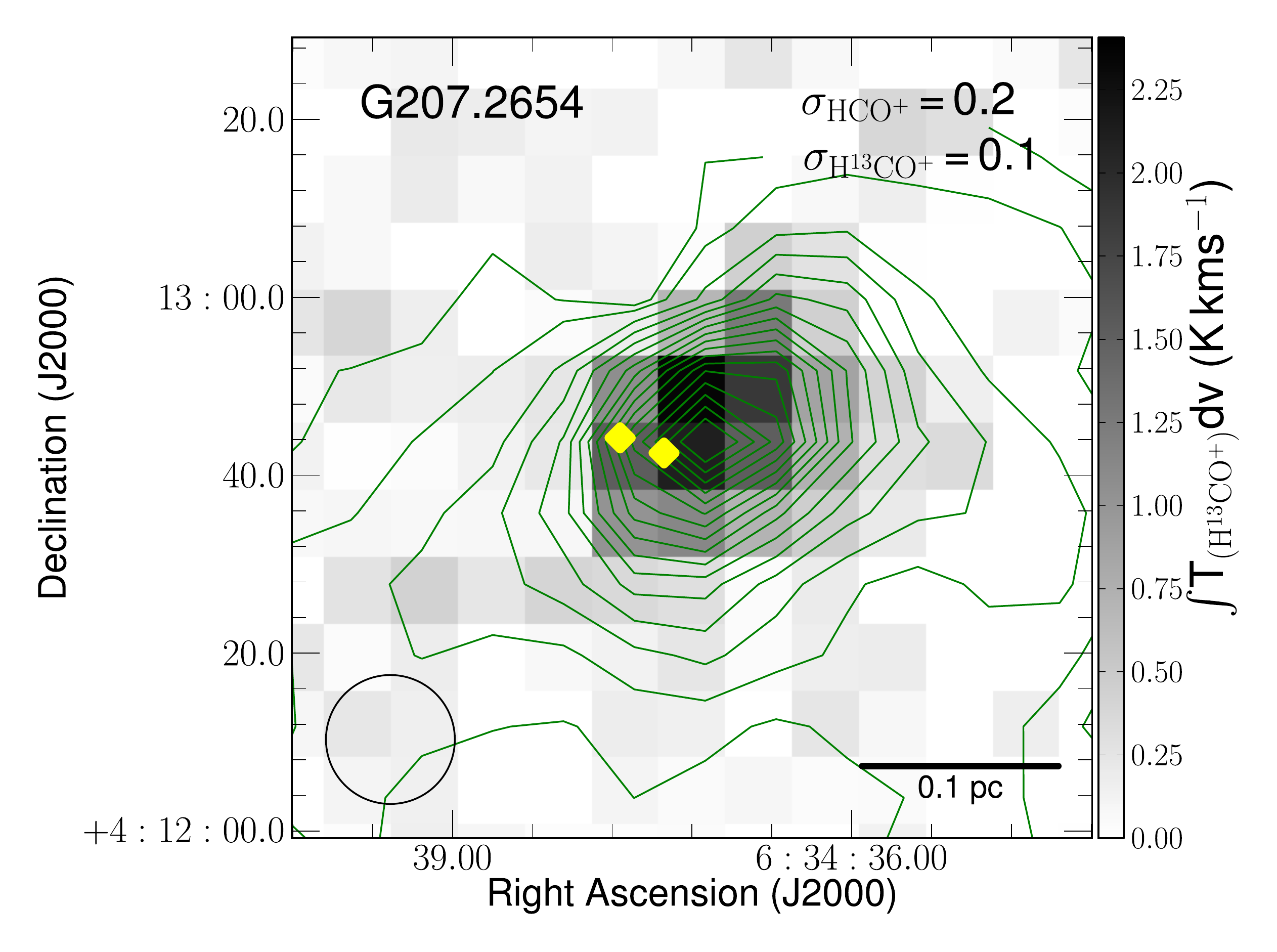}
\includegraphics[width=0.49\textwidth]{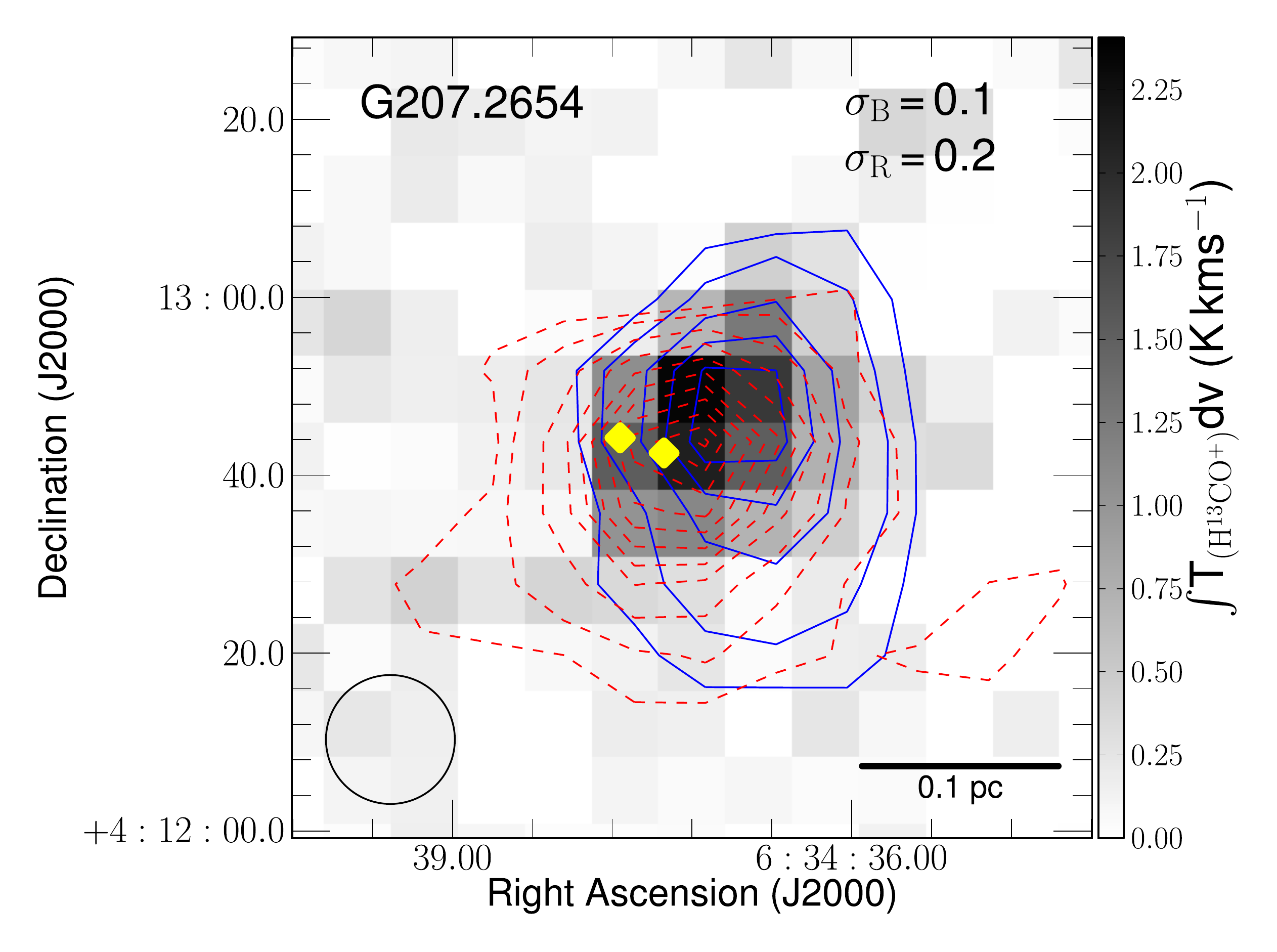}
\includegraphics[width=0.49\textwidth]{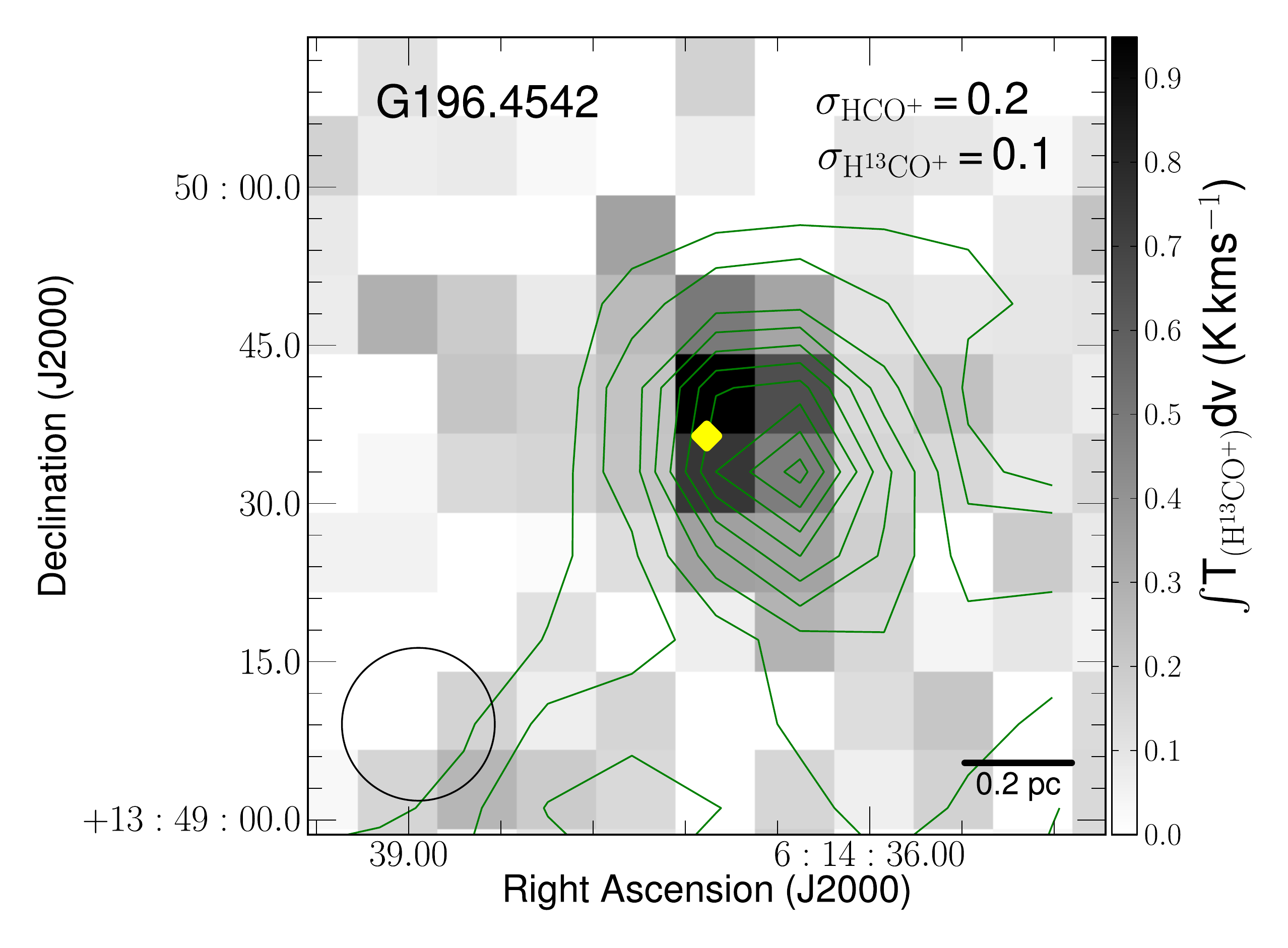}
\includegraphics[width=0.49\textwidth]{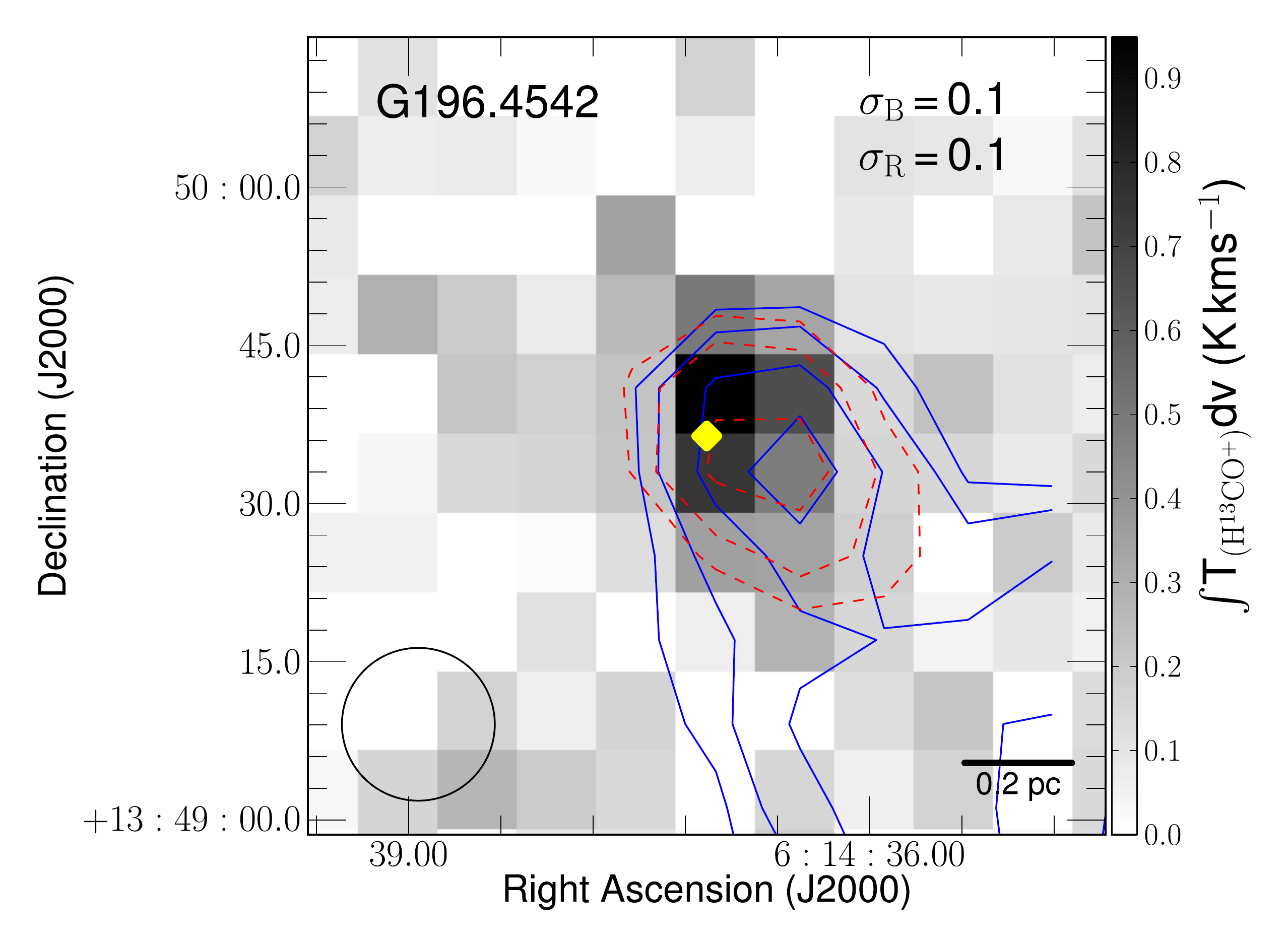}
\contcaption{--\,The velocity ranges used to integrate the total emission are 4.2\,--\,12.6\,\kms for G203.3166, 9.2\,--\,17.7\,\kms for G207.2654, and 16.6\,--\,22.0\,\kms for G196.4542. We do not include a separate map for G203.3166-OFFSET as the emission ranges are similar.}
\end{figure*}

%----------------------------------------------------------------------------------SiO MAPS----------------------------------------------------------------------

\begin{figure*}
\includegraphics[width=0.49\textwidth]{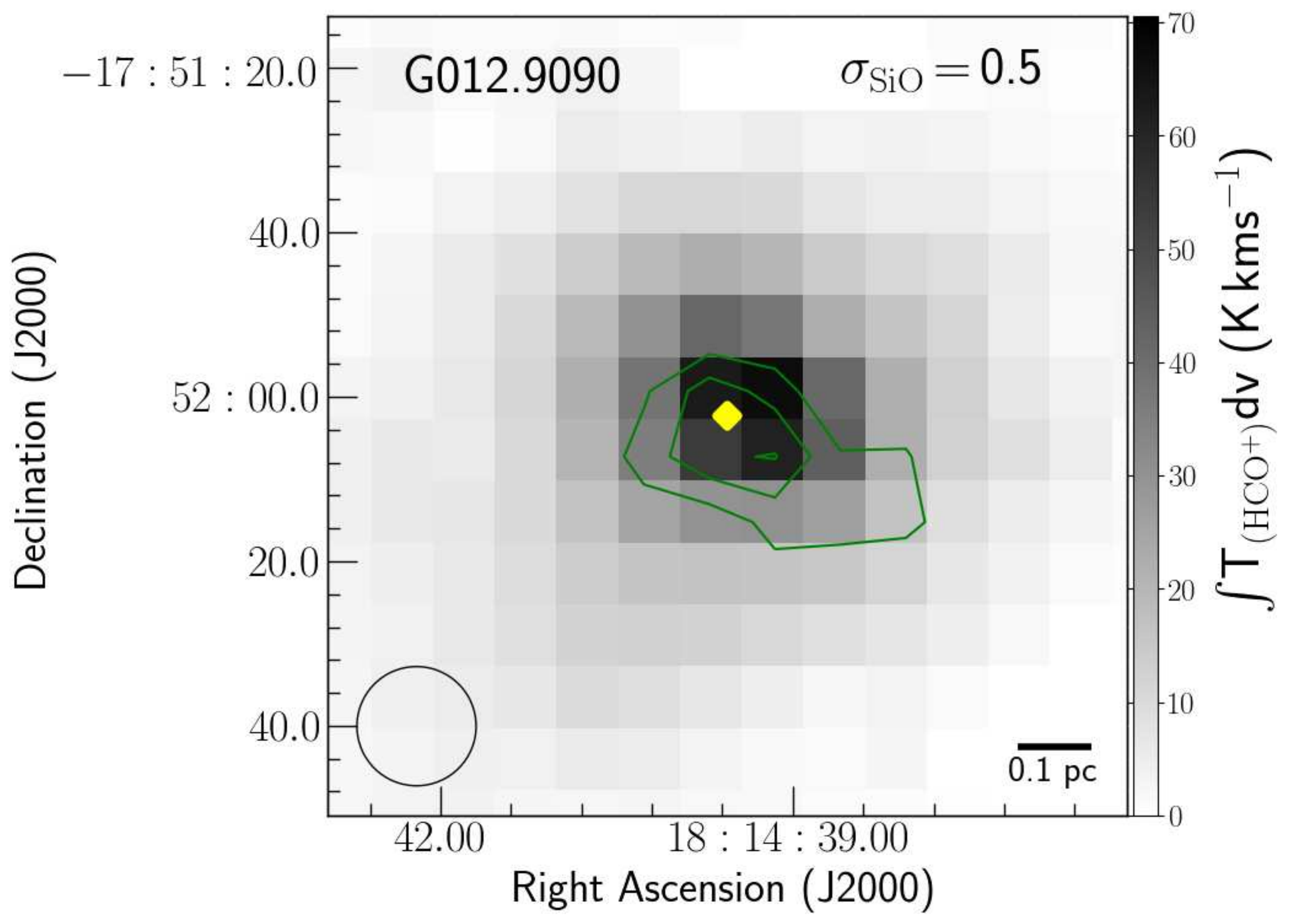}
\includegraphics[width=0.49\textwidth]{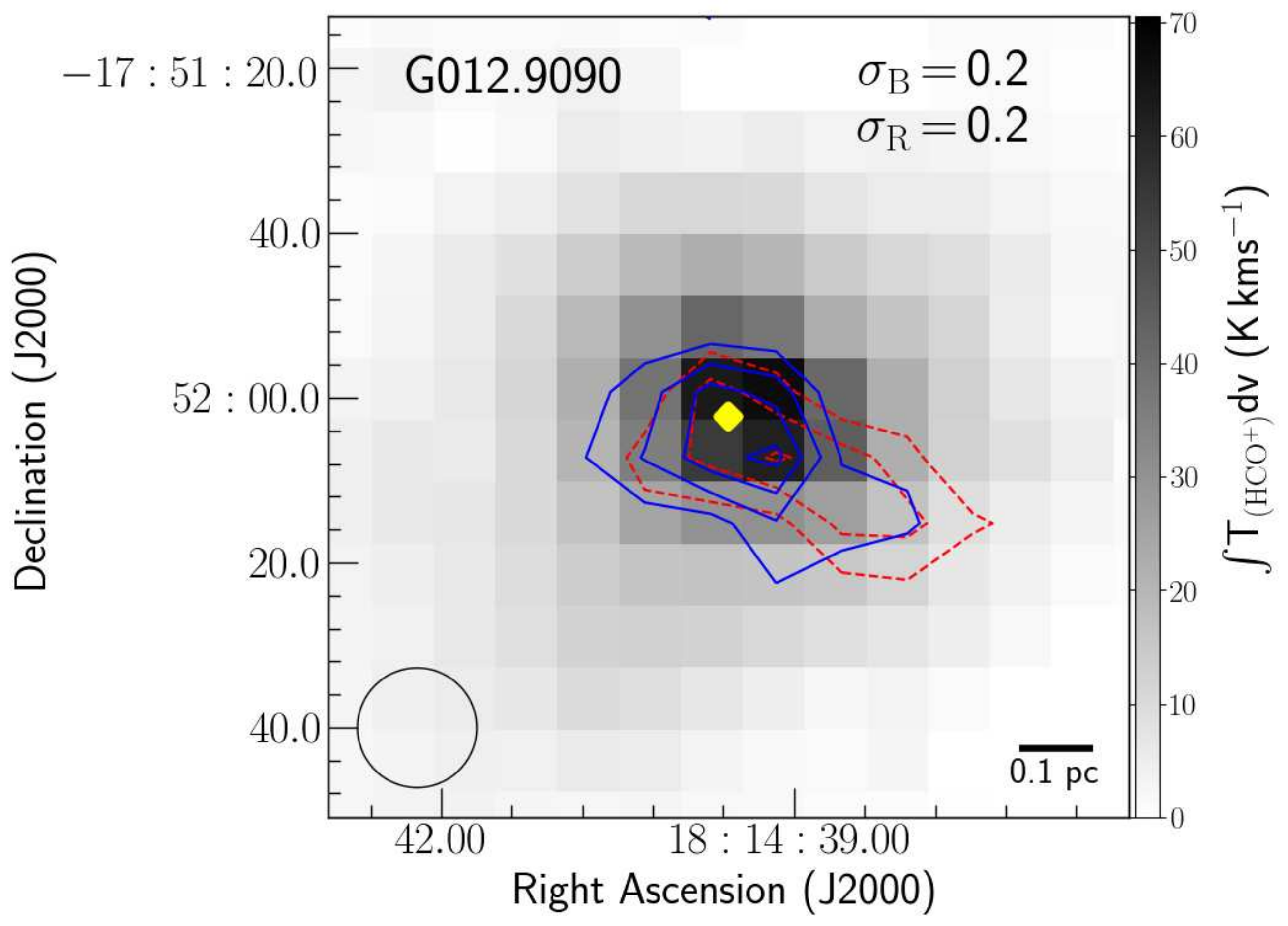}
\includegraphics[width=0.49\textwidth]{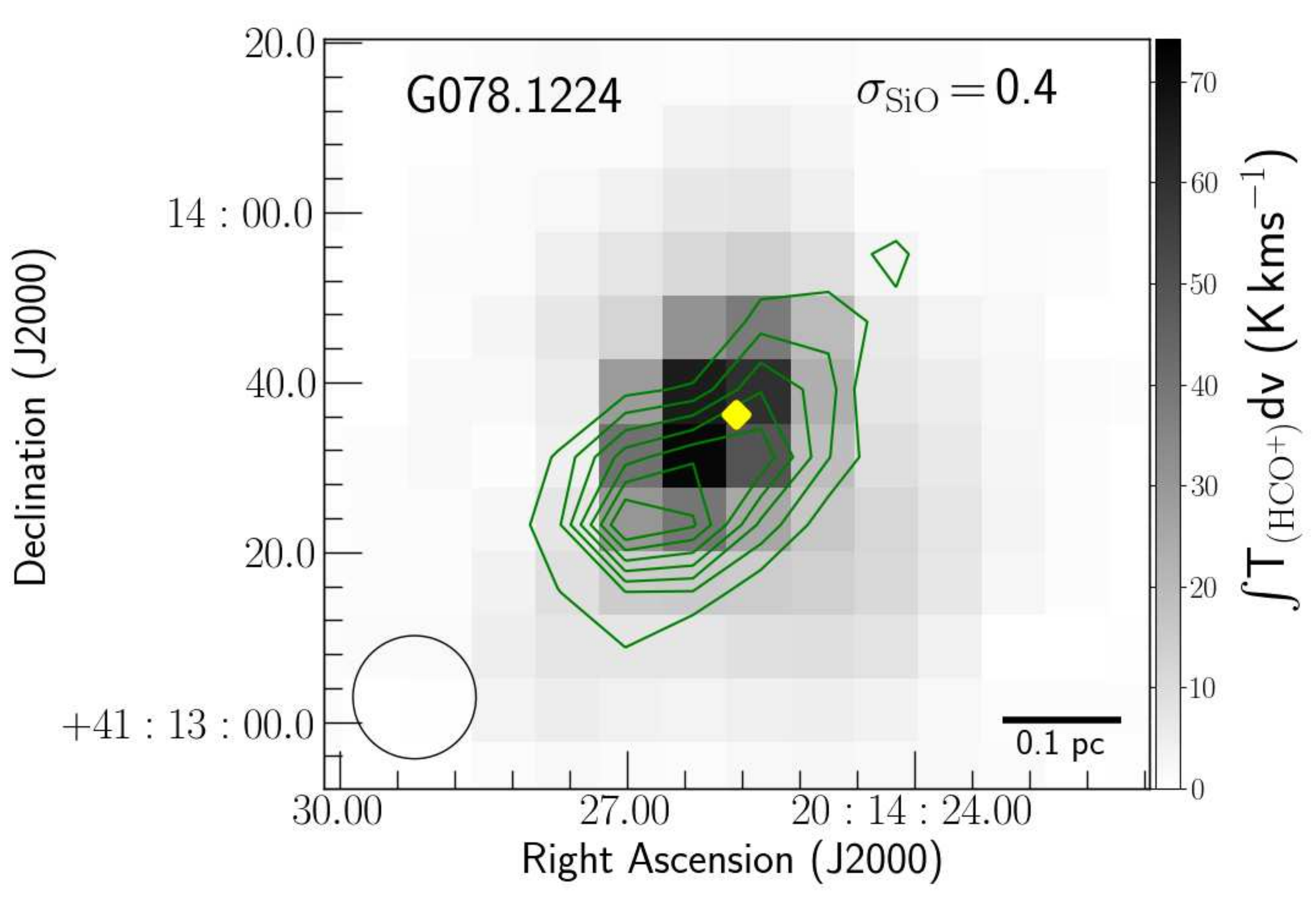}
\includegraphics[width=0.49\textwidth]{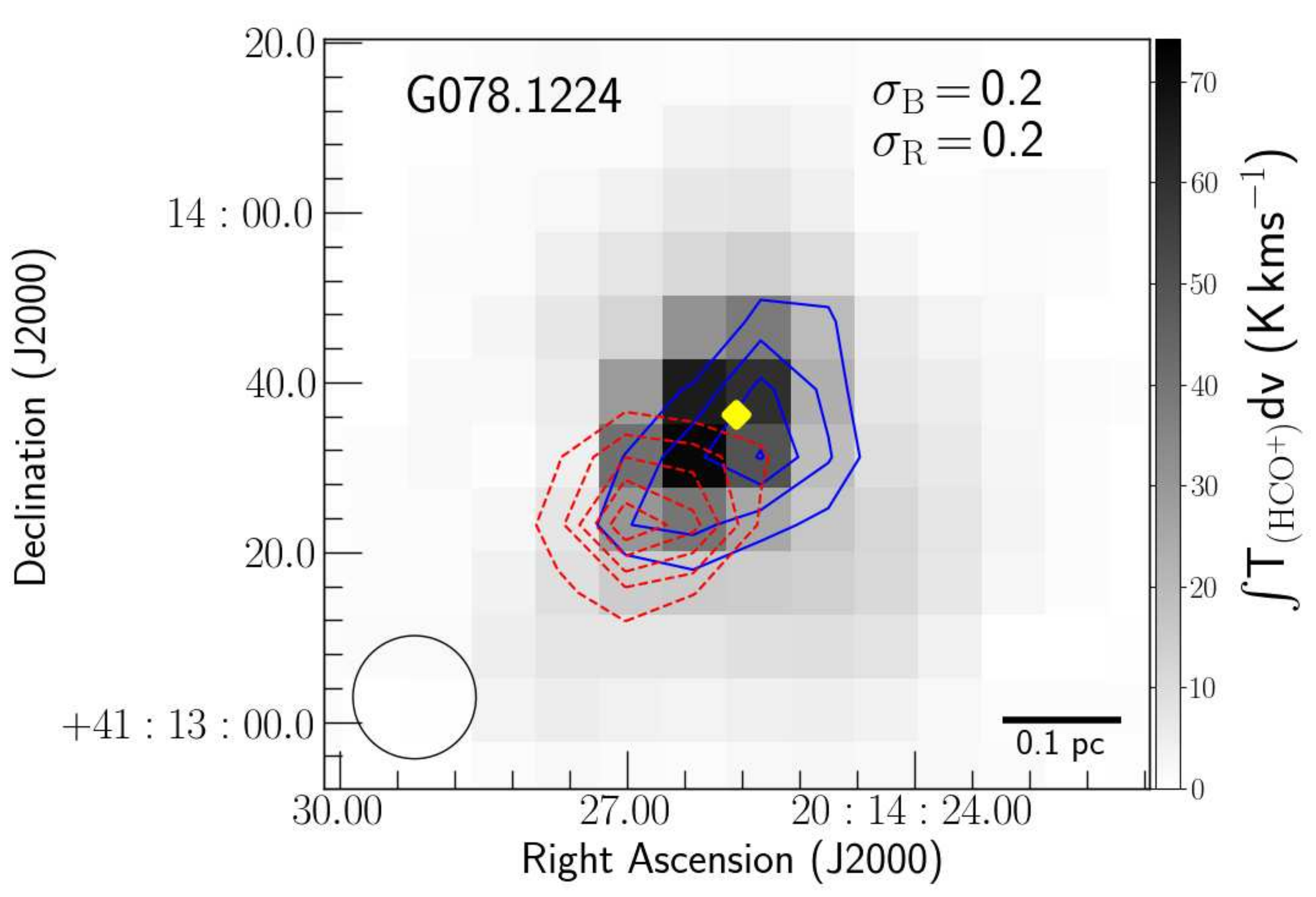}
\includegraphics[width=0.49\textwidth]{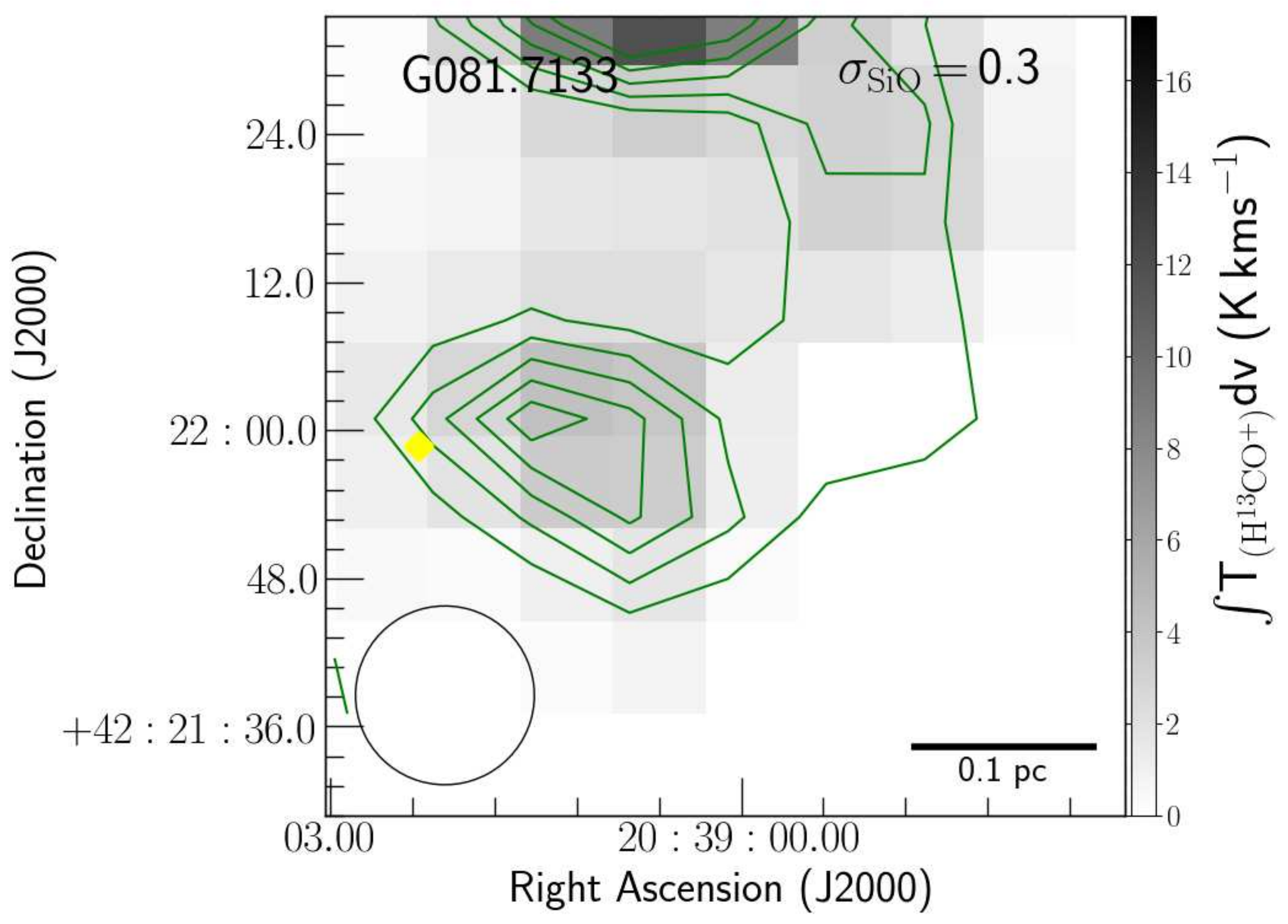}
\includegraphics[width=0.49\textwidth]{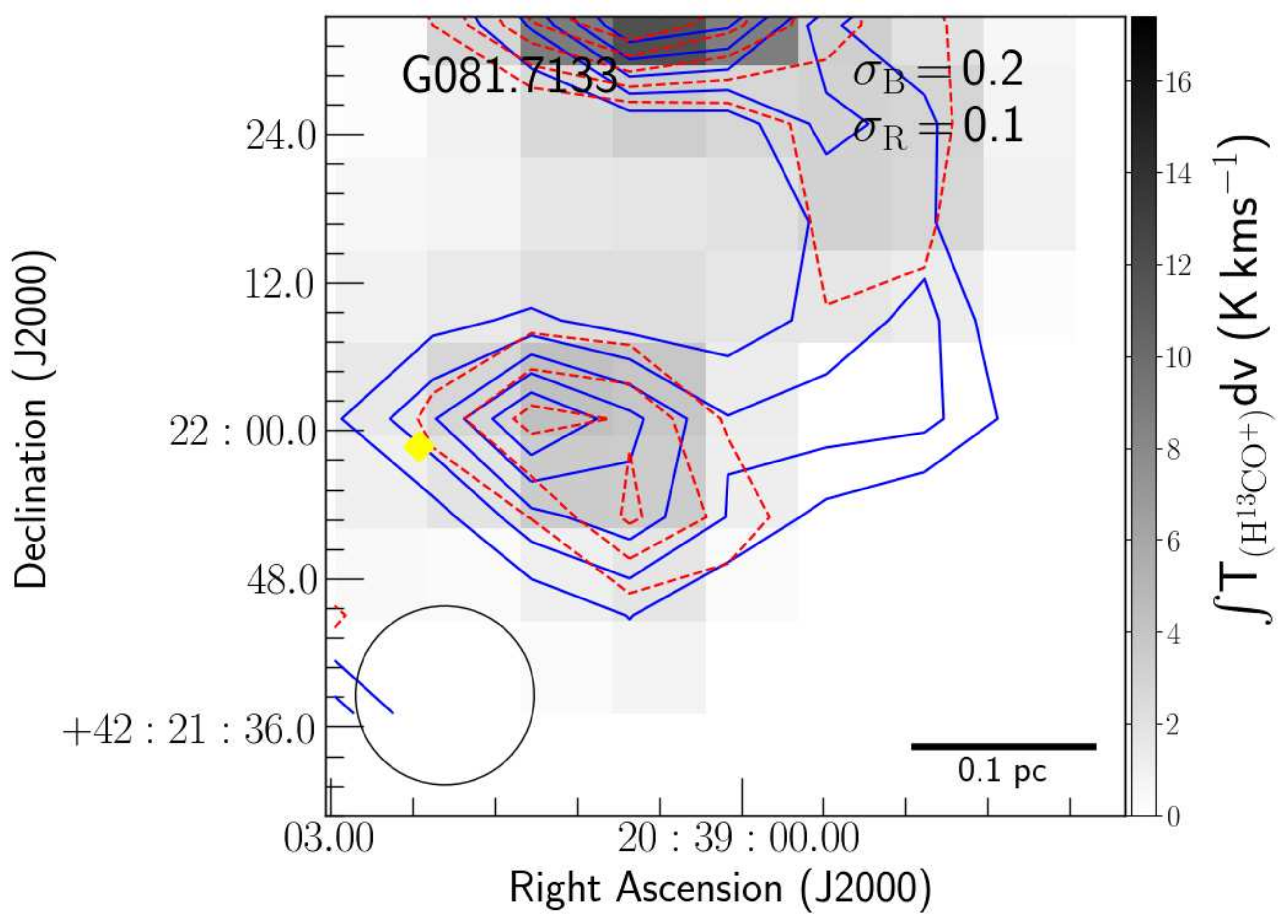}
\caption{SiO and \hco zeroth order moment maps. The total integrated \hco emission is shown in greyscale (apart from sources where no \hco emission was observed and the \htco emission is displayed), integrated from the minimum to maximum channels with a 3$\sigma$ detection (and is the same as given in Figure. B1). The yellow diamonds and crosses are the RMS source position(s) and offset position, respectively. The JCMT beam is shown in the bottom left corner, and the source name is shown in the top left corner. For sources with SiO emission extending over less than four pixels we show only the total integrated emission as shown in the left column, and for sources with emission extending over more than 4 pixels we show the red- and blue-shifted SiO contours as shown in the right column here. {\bf Left:} The total integrated SiO emission is overlaid in green solid contours (again integrated from the minimum and maximum channels with a 3$\sigma$ detection). The 1$\sigma$ rms (in units of K.\kms) for the SiO ($\sigma$$_{\rm SiO}$) integrated intensity is given in the top right corner. The contour levels are from 1$\sigma\times$(3, 5, 7, 9,... to peak in-steps of 2$\sigma$). {\bf Right}: The red- and blue-shifted SiO emission is given by the red (dashed) and blue (solid) contours, respectively. The blue- and red-shifted contours are taken from the minimum and maximum channels with 3$\sigma$ emission to the source V$_{\rm LSR}$ (taken from Table \ref{aa:tab:htcogaus}), respectively. The 1$\sigma$ levels for the red- ($\sigma$$_{\rm R}$) and blue-shifted ($\sigma$$_{\rm B}$) emission are given in the top right corner, where the contour levels are from 1$\sigma\times$(3, 5, 7,... to peak in-steps of 2$\sigma$). The velocity ranges used for the total integrated SiO emission are 25.1.0$-$46.9.0\,\kms for G012.9090, -25.2\,--\,16.8\,\kms for G078.1224, and -13.2\,--\,3.6\,\kms for G081.7133.\label{SiOAppendixmapps}}
\end{figure*}

\begin{figure*}
\includegraphics[width=0.49\textwidth]{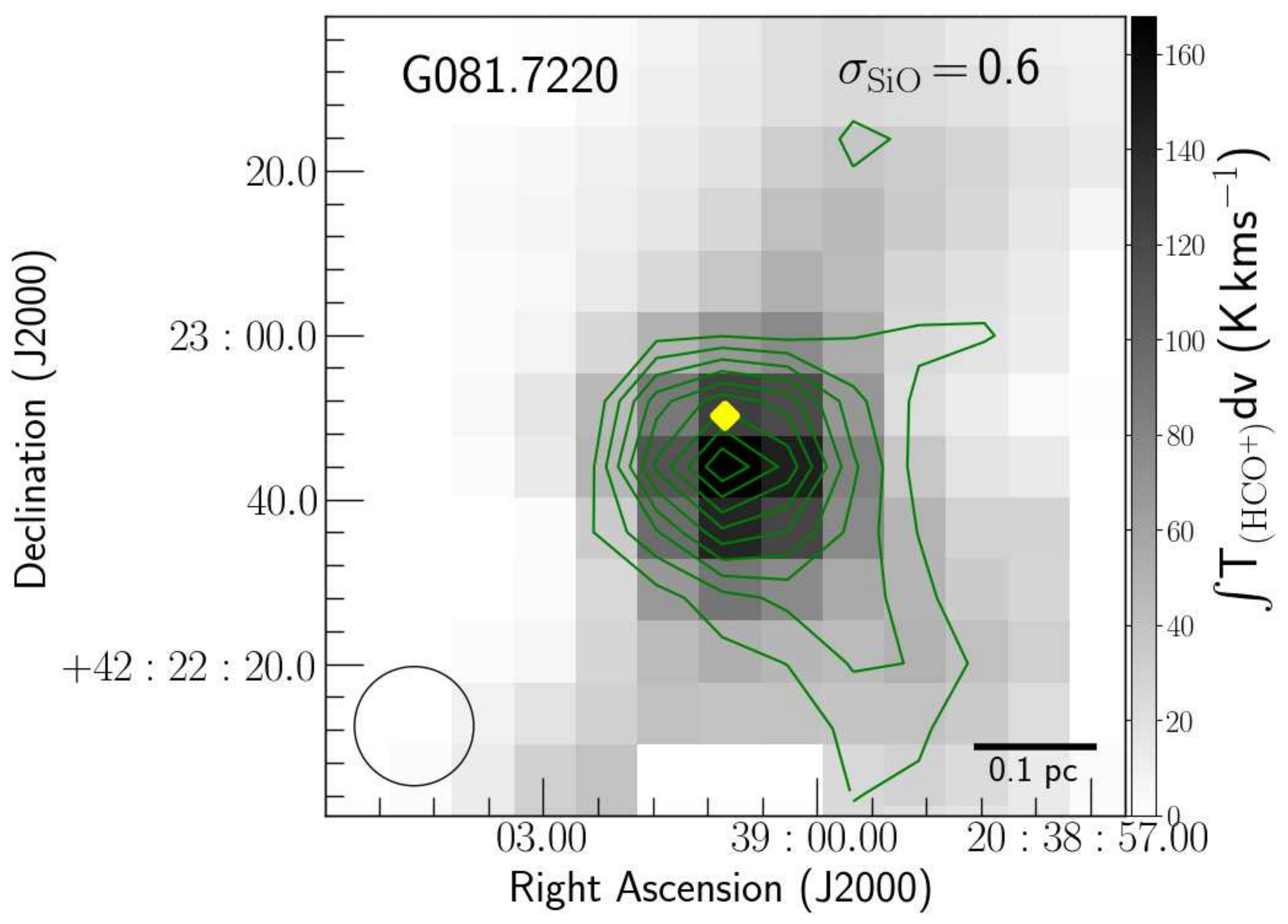}
\includegraphics[width=0.49\textwidth]{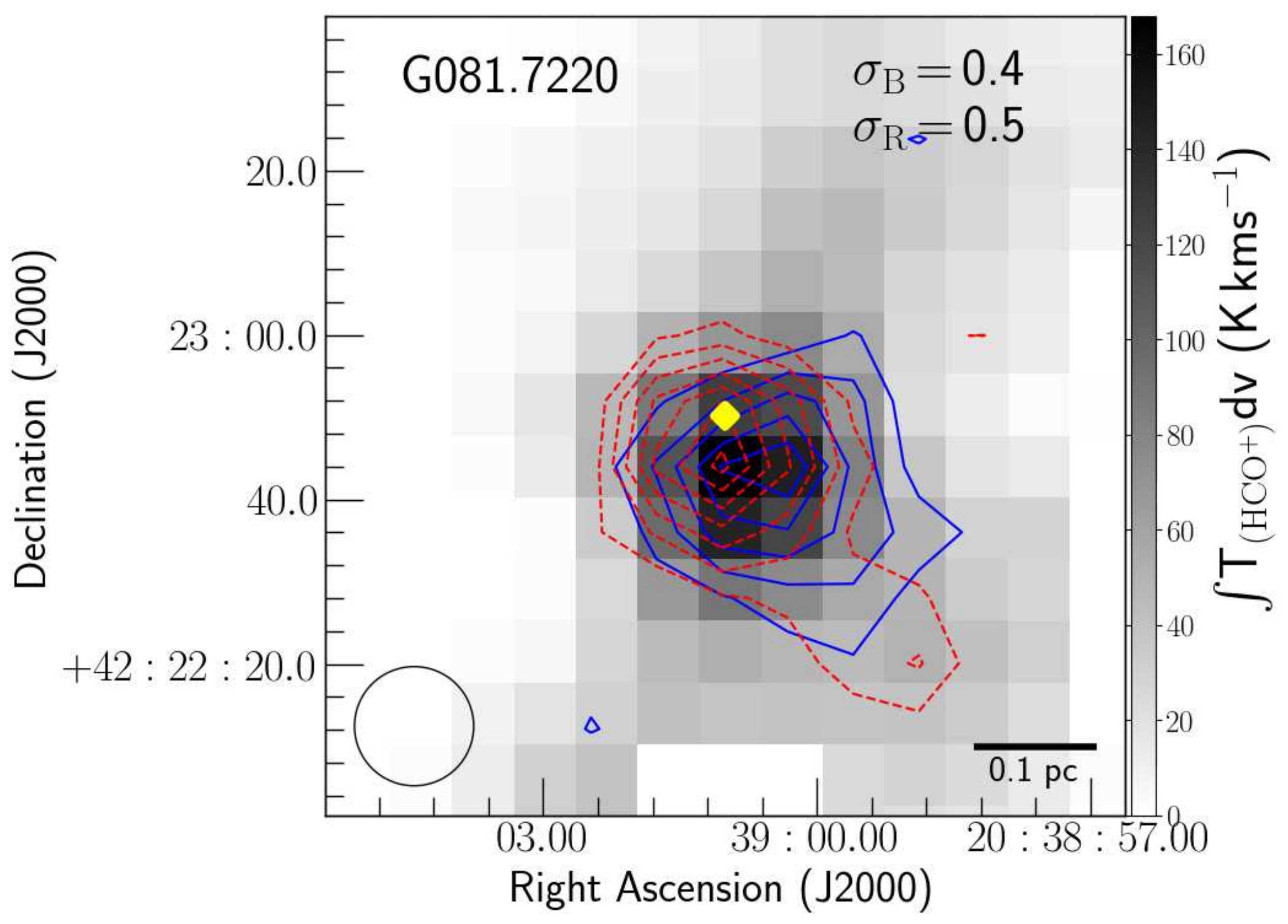}
\includegraphics[width=0.49\textwidth]{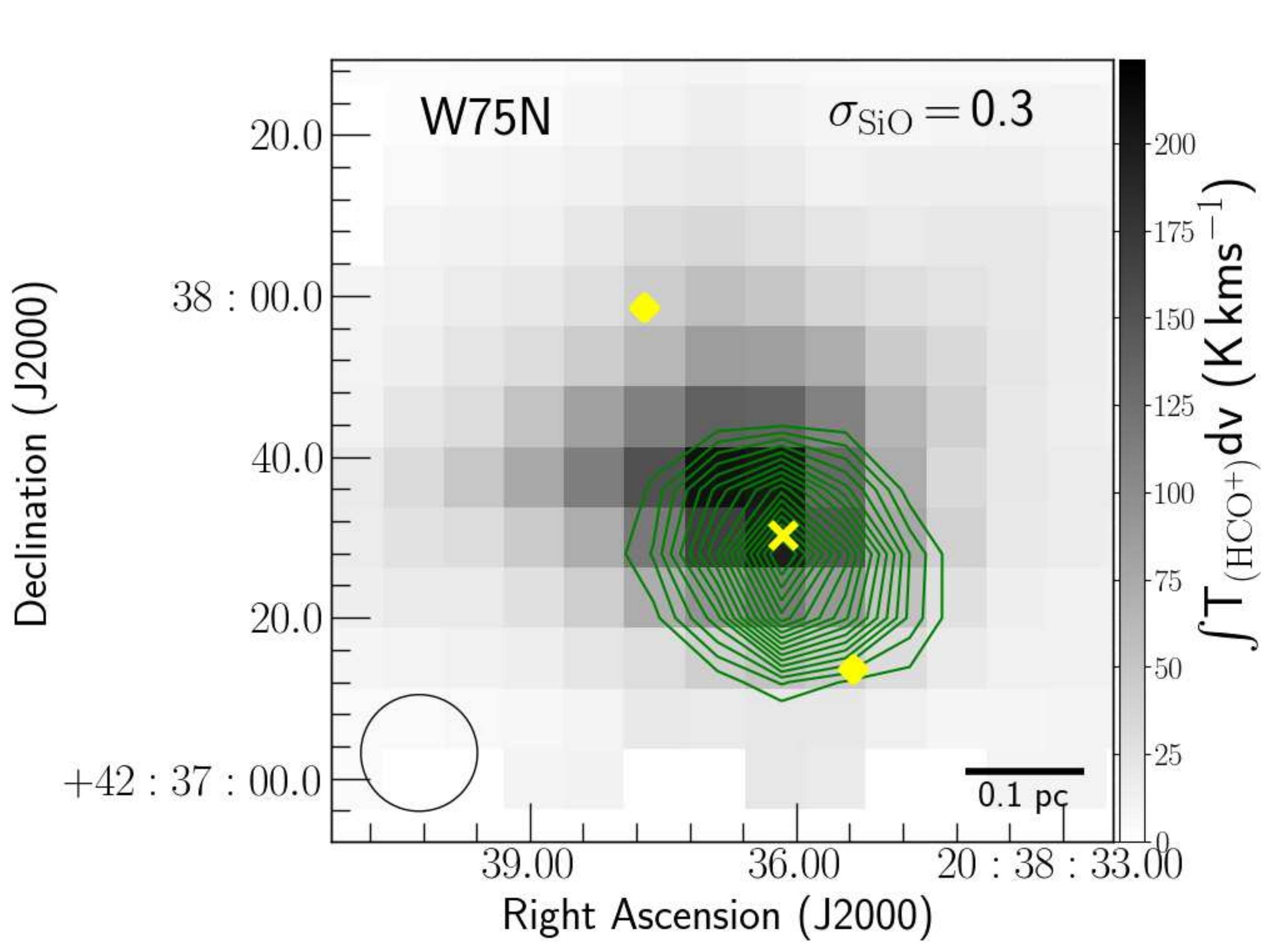}
\includegraphics[width=0.49\textwidth]{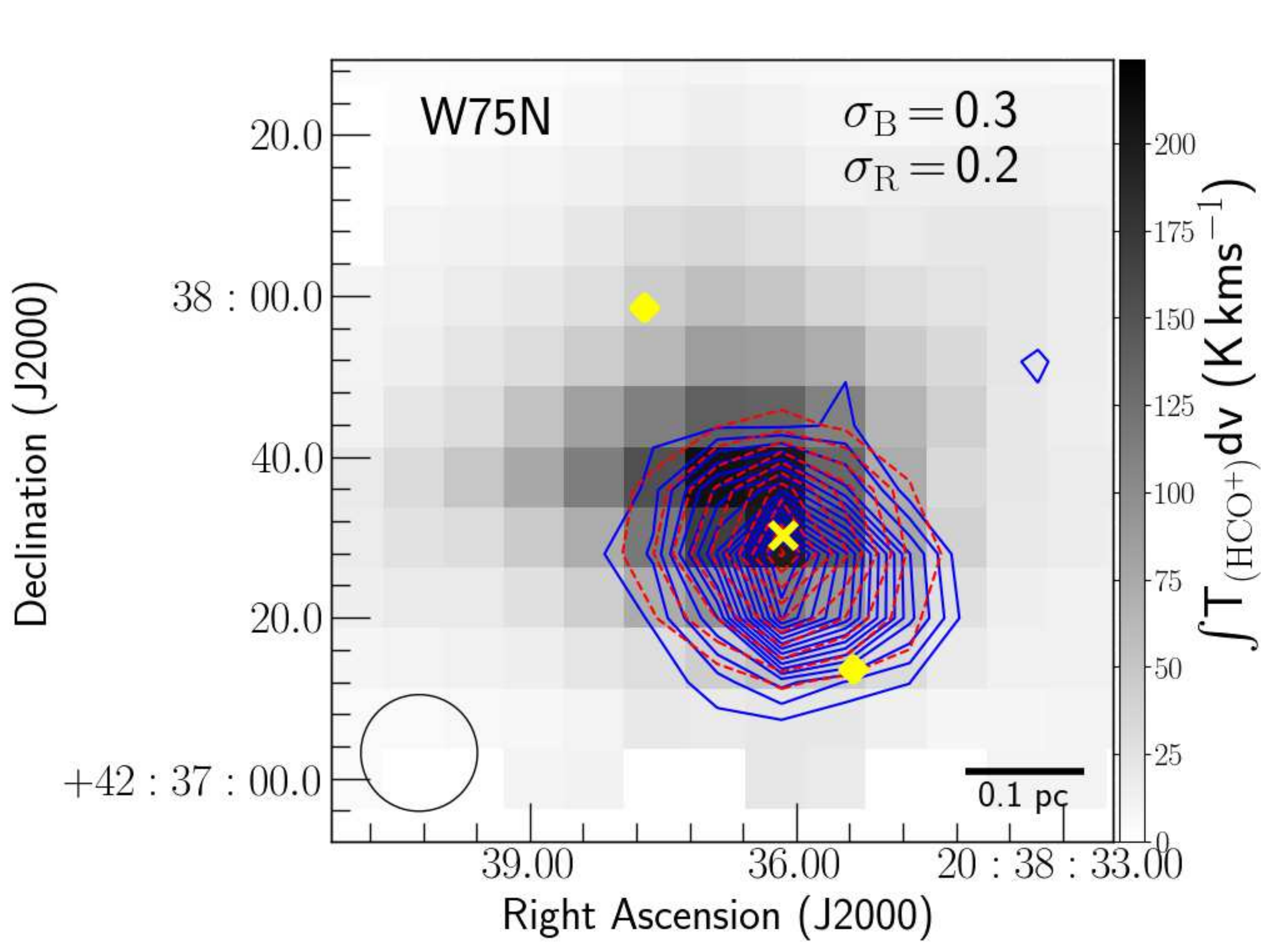}
\includegraphics[width=0.49\textwidth]{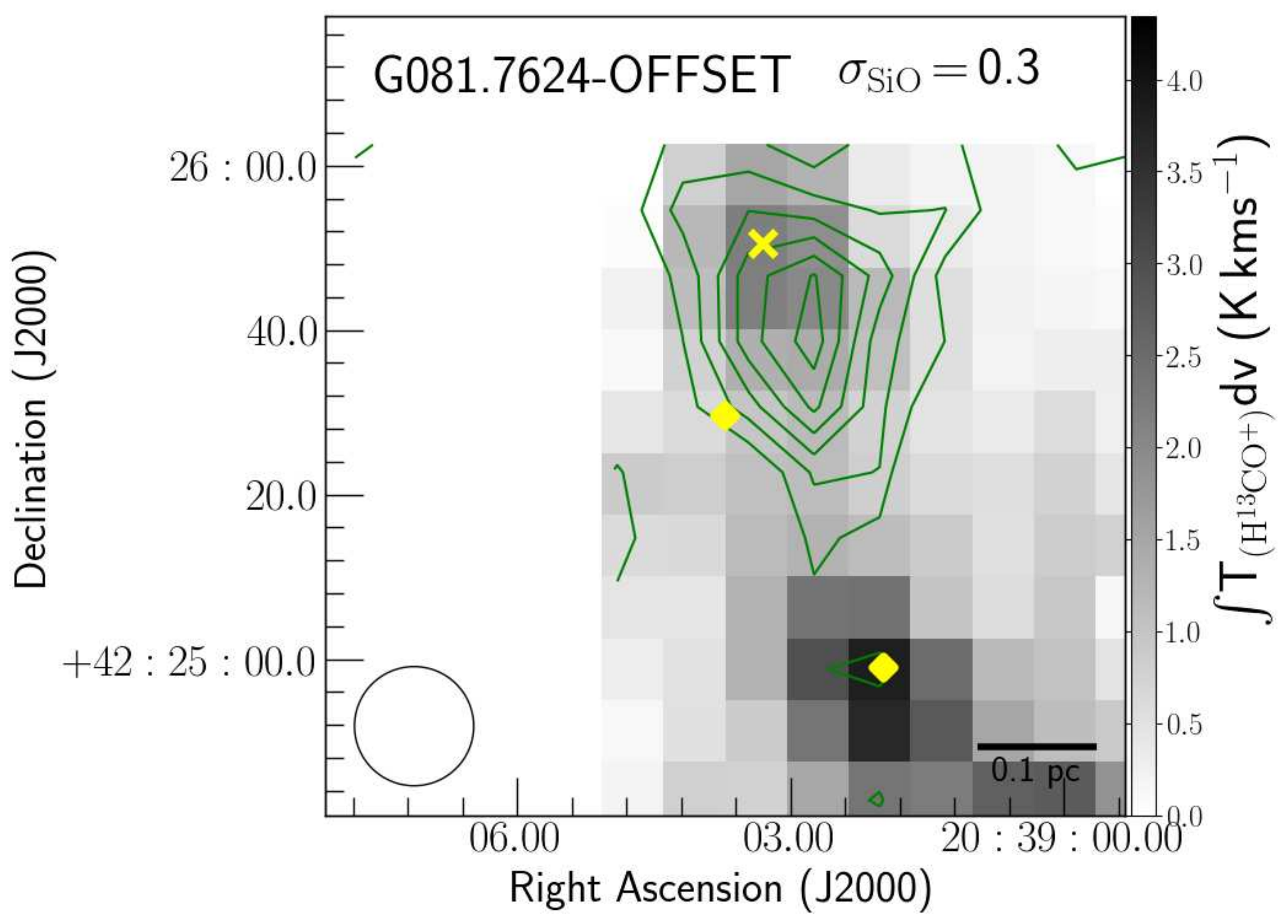}
\includegraphics[width=0.49\textwidth]{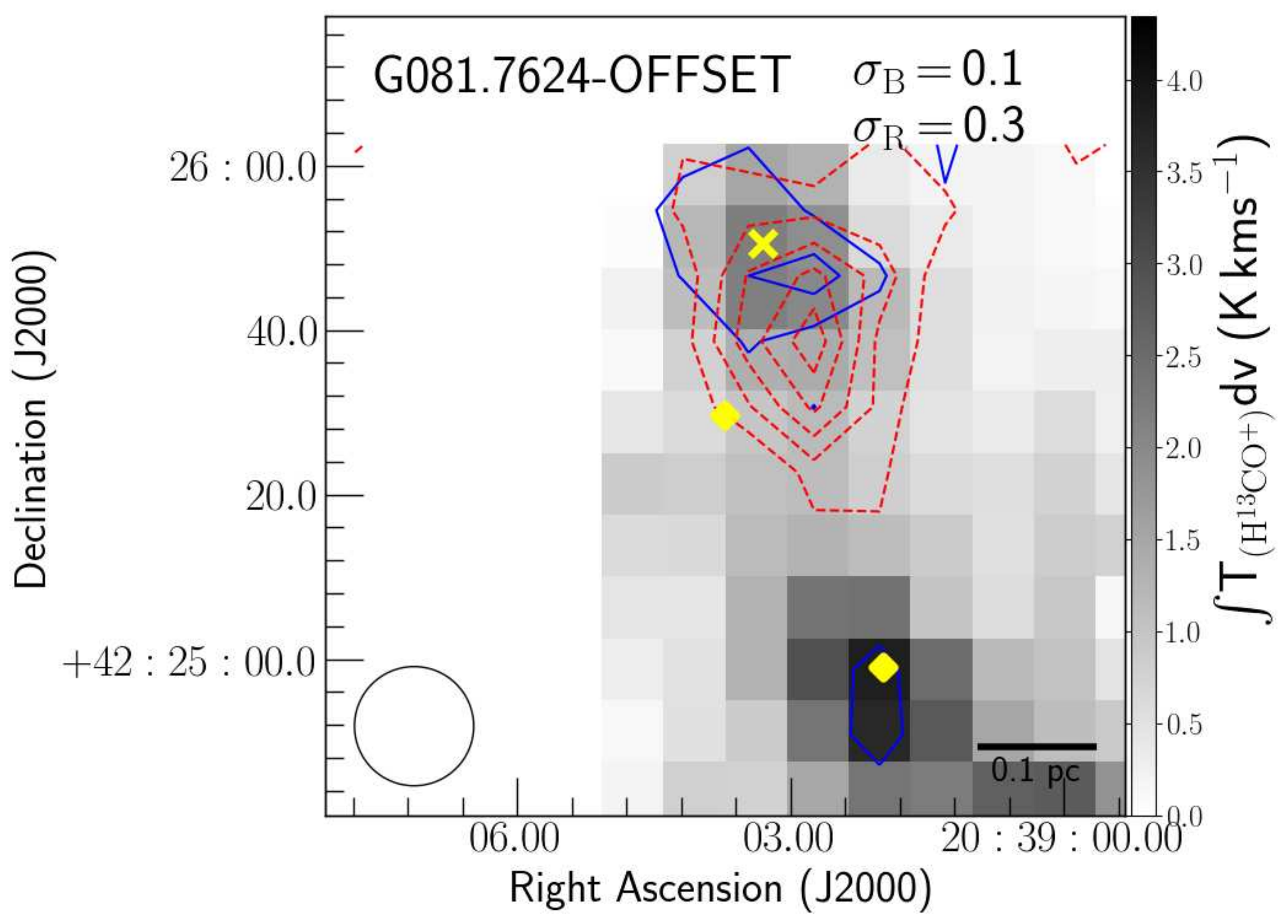}

\contcaption{--\,The velocity ranges used for the total integrated SiO emission are -16.6\,--\,8.6\,\kms for G081.7220, -10\,--\,21.9\,\kms for W75N and -8.2\,--\,22.0\,\kms for G081.7624-OFFSET.}
\end{figure*}

\begin{figure*}
\includegraphics[width=0.49\textwidth]{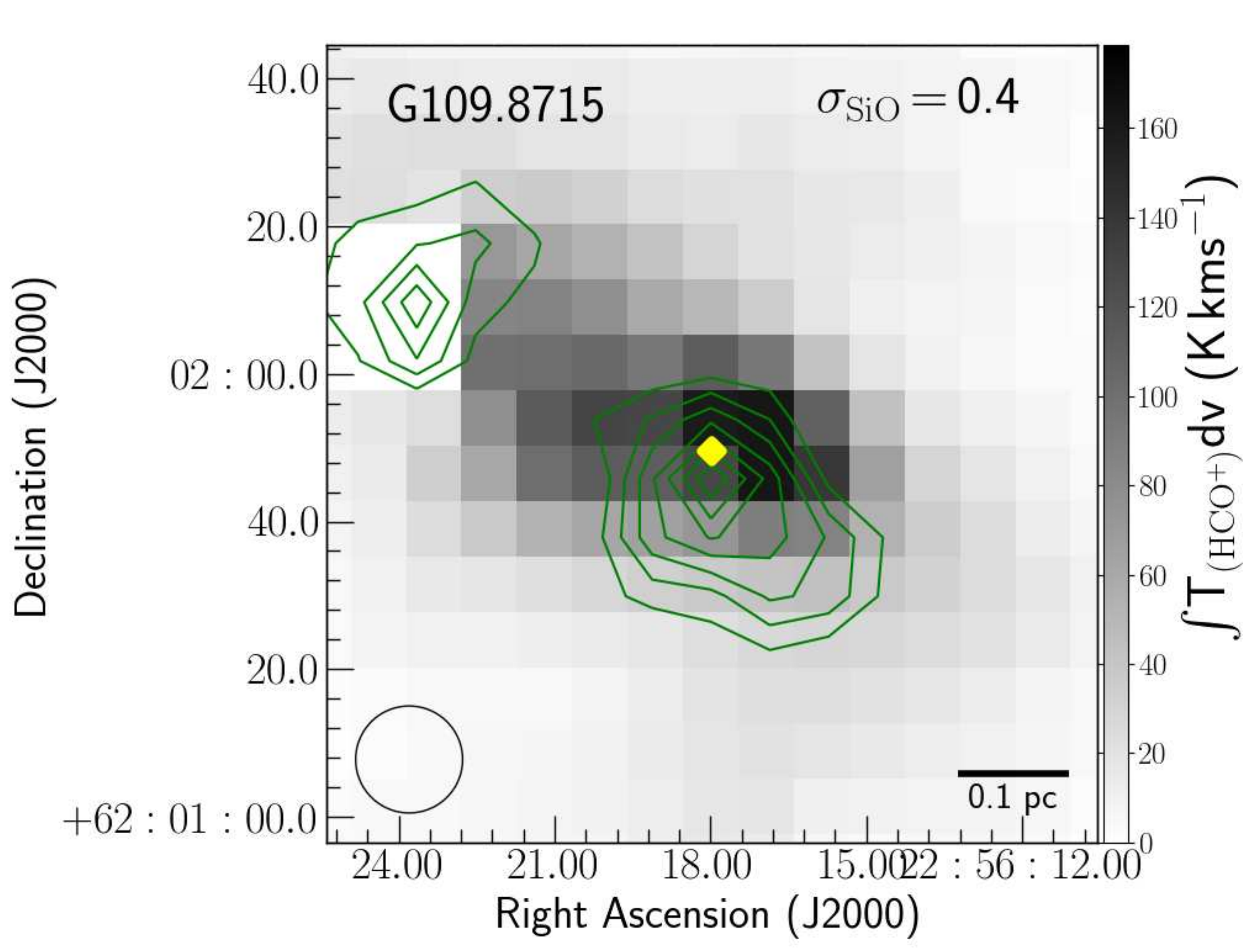}
\includegraphics[width=0.49\textwidth]{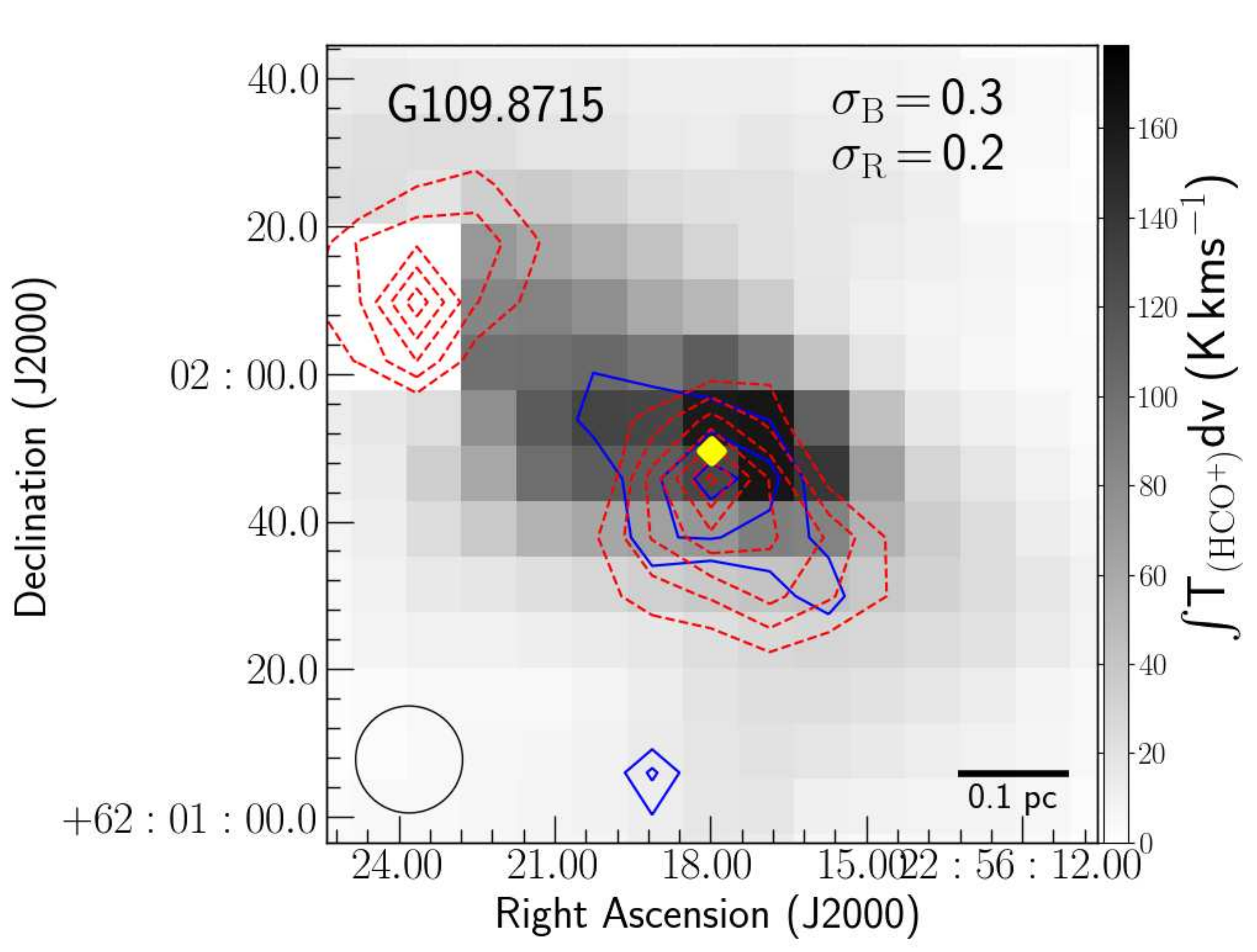}
\includegraphics[width=0.49\textwidth]{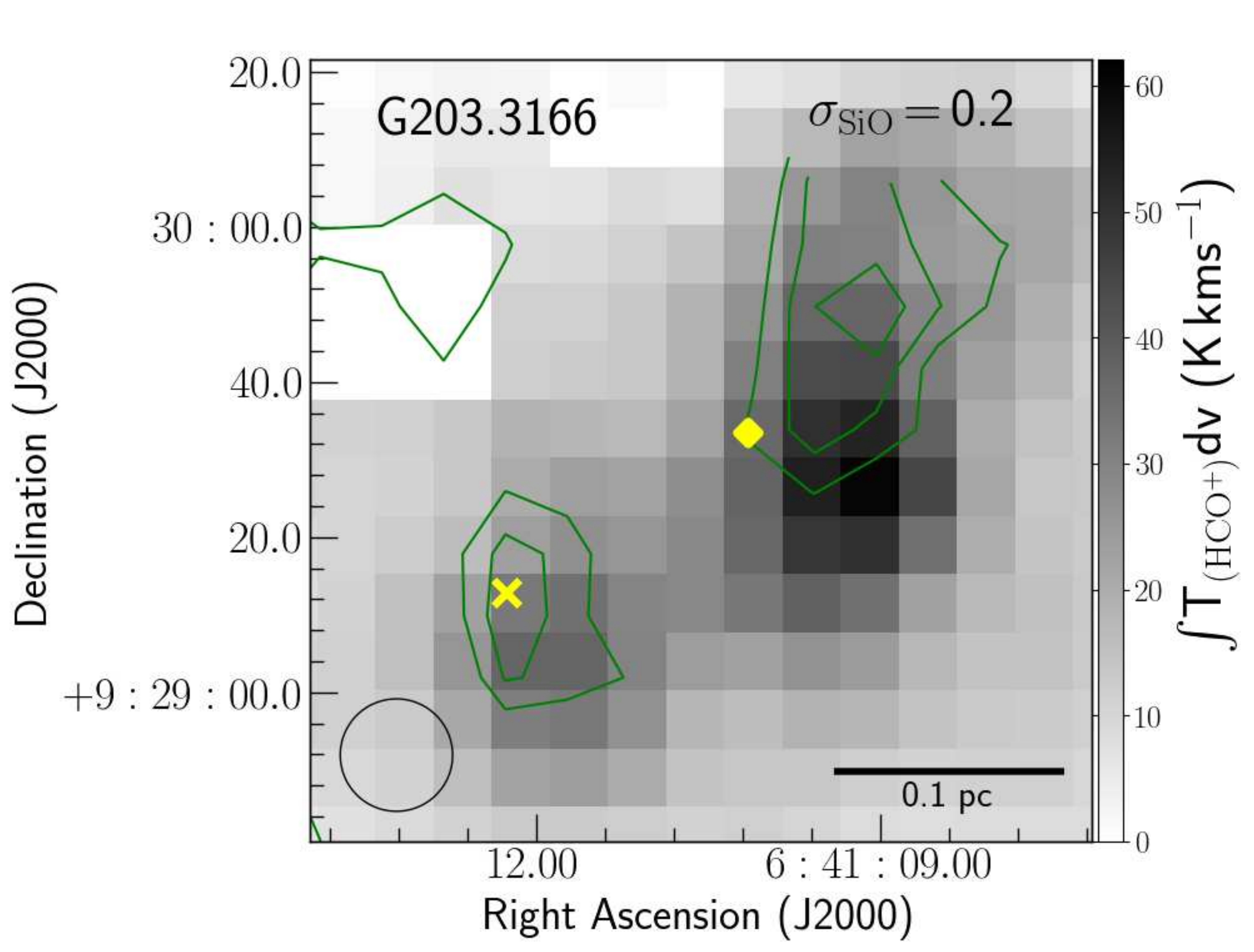}
\includegraphics[width=0.49\textwidth]{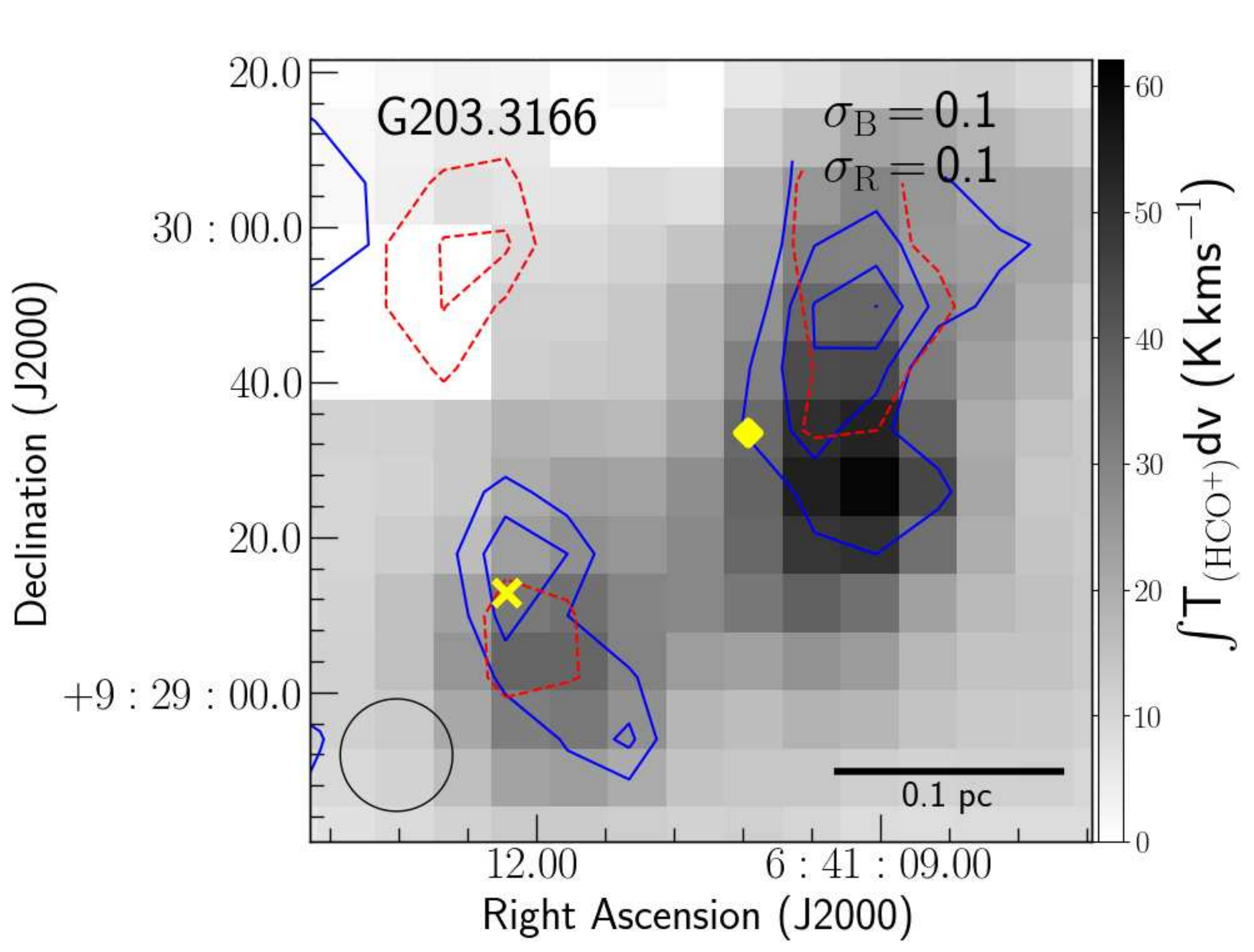}
\includegraphics[width=0.49\textwidth]{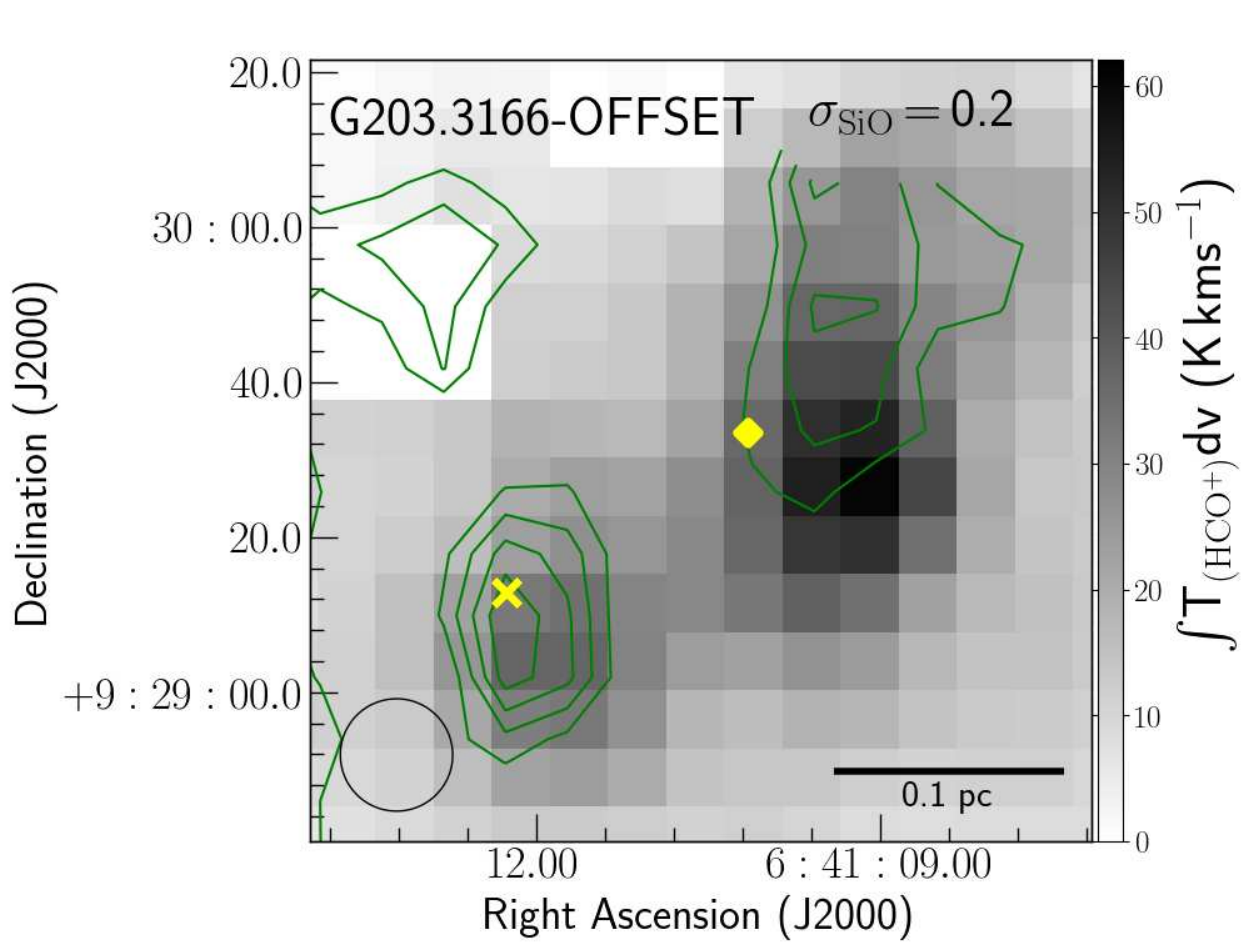}
\includegraphics[width=0.49\textwidth]{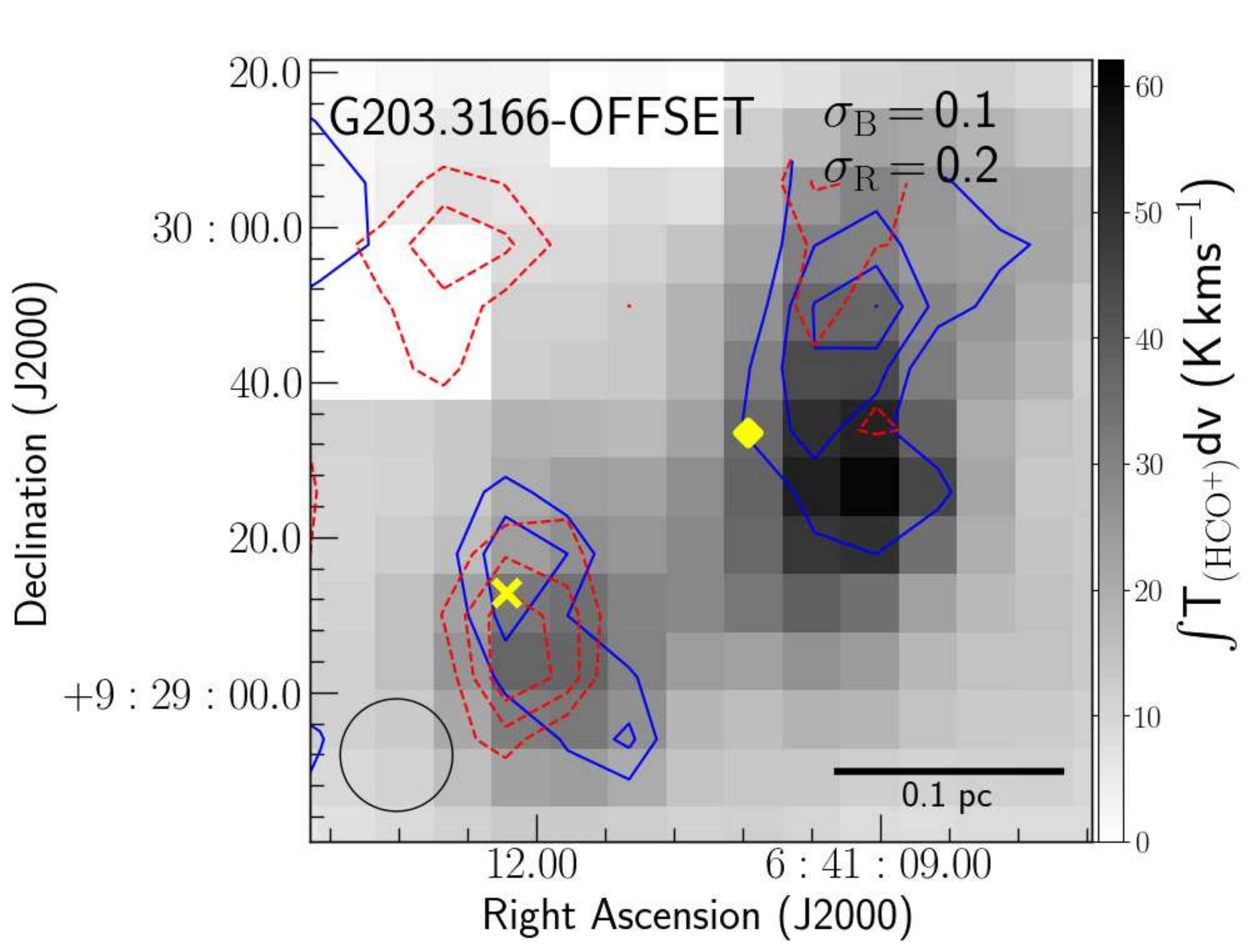}

\contcaption{--\, The velocity ranges used for the total integrated SiO emission are -18.2\,--\,0.3\,\kms for G109.8715, 1.6\,--\,15.1\,\kms for G203.3166, and 3.3\,--\,25.6\,\kms for G203.3166-OFFSET.}
\end{figure*}

\begin{figure*}
\includegraphics[width=0.49\textwidth]{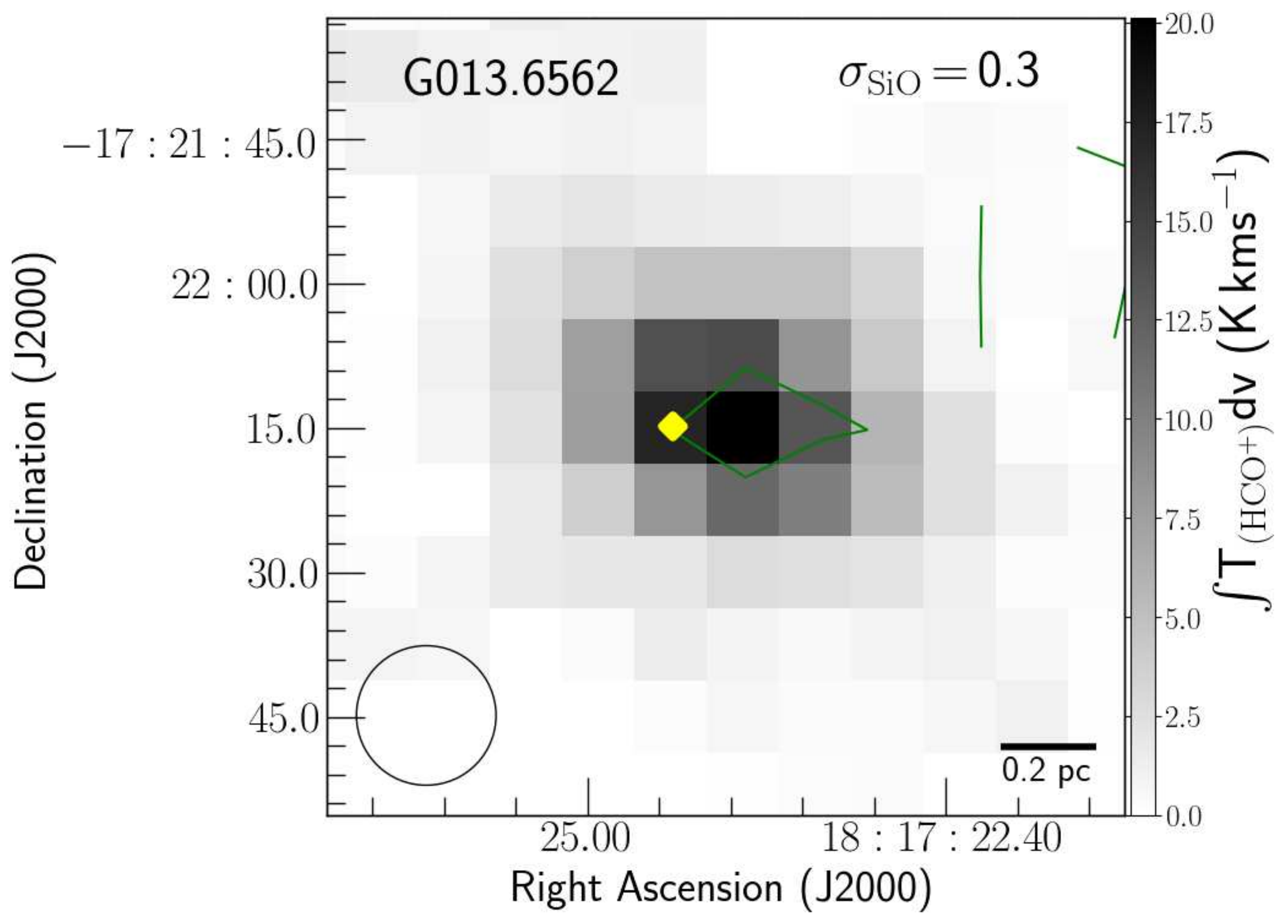}
\includegraphics[width=0.49\textwidth]{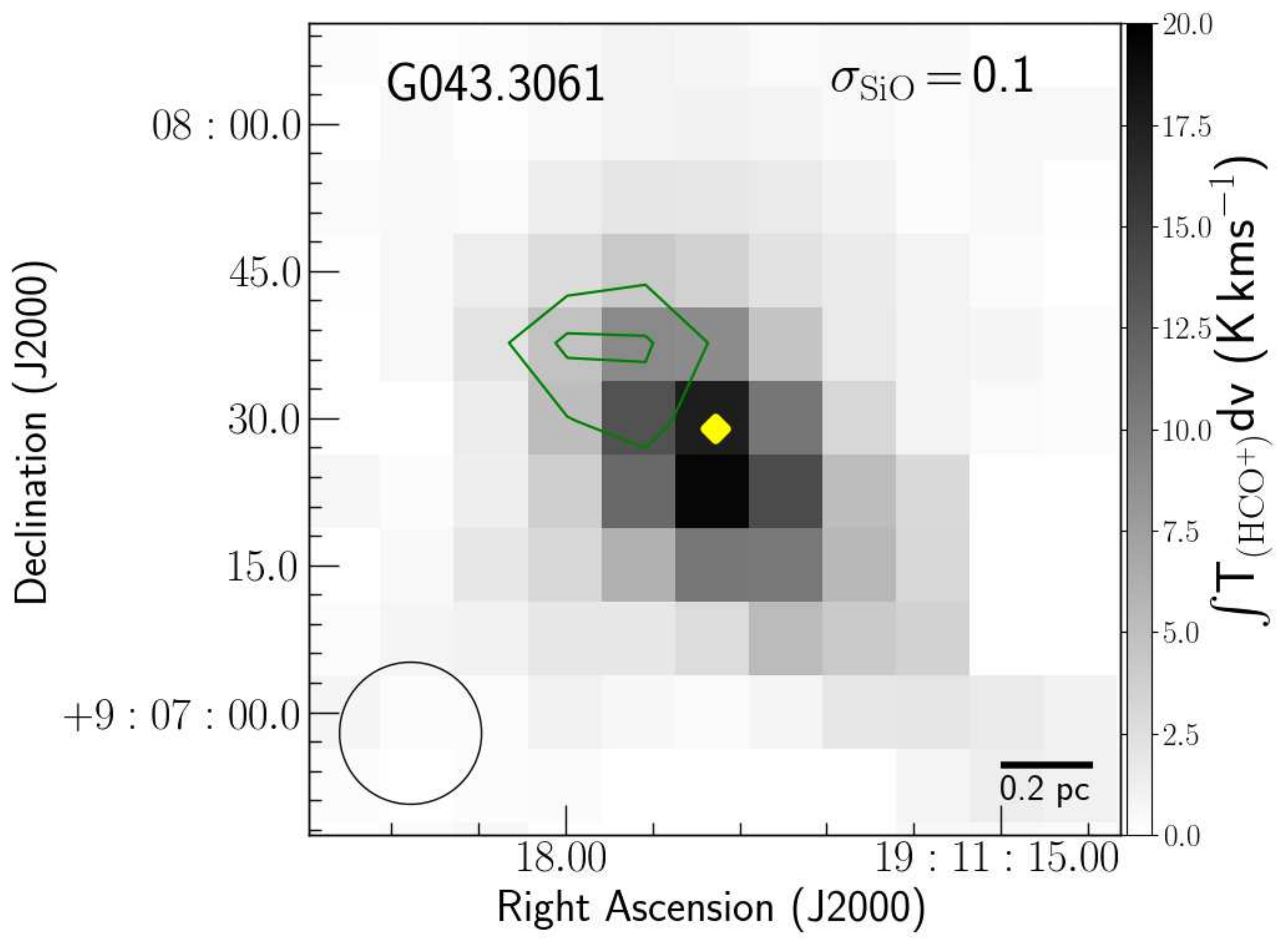}
\includegraphics[width=0.49\textwidth]{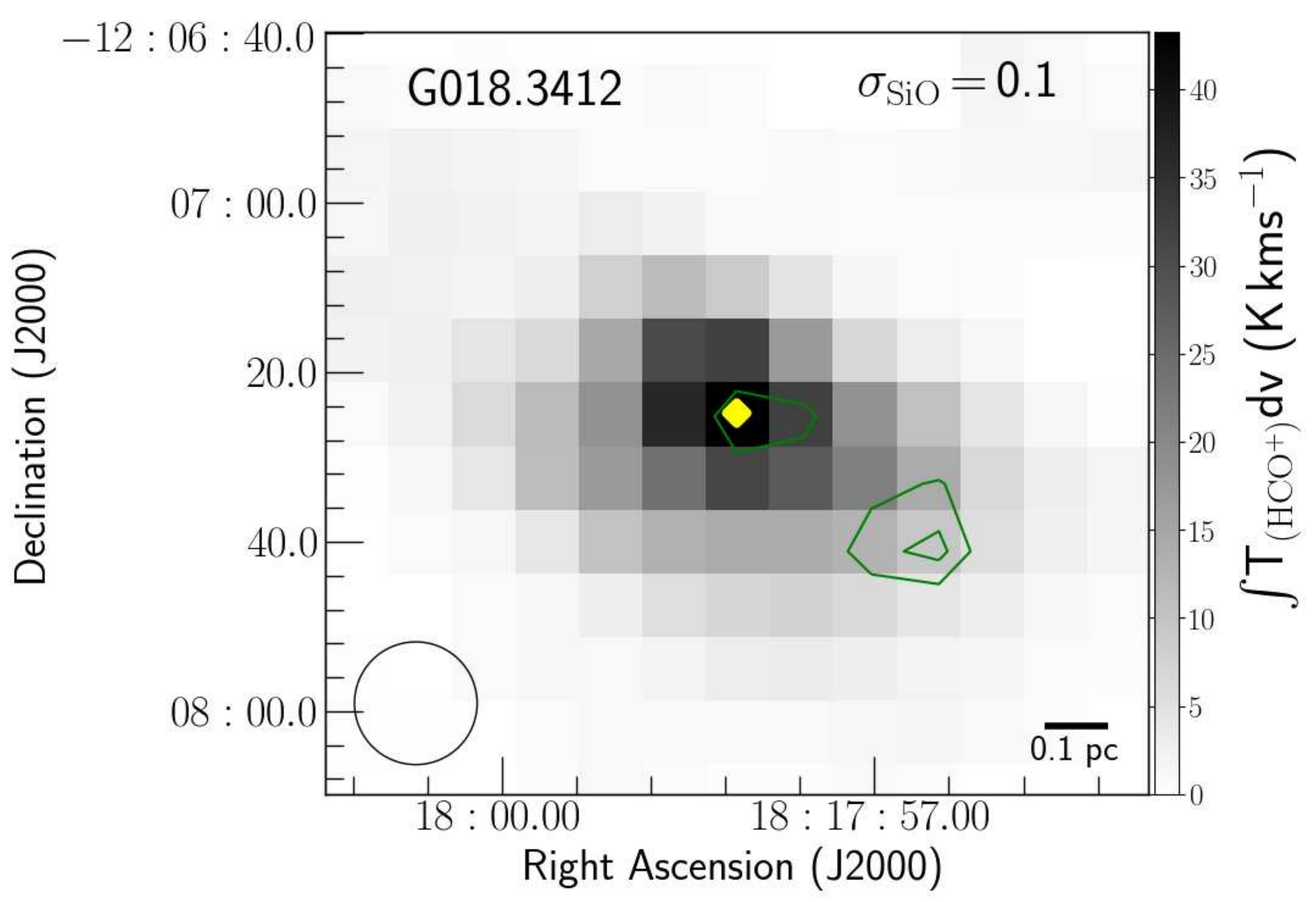}
\includegraphics[width=0.49\textwidth]{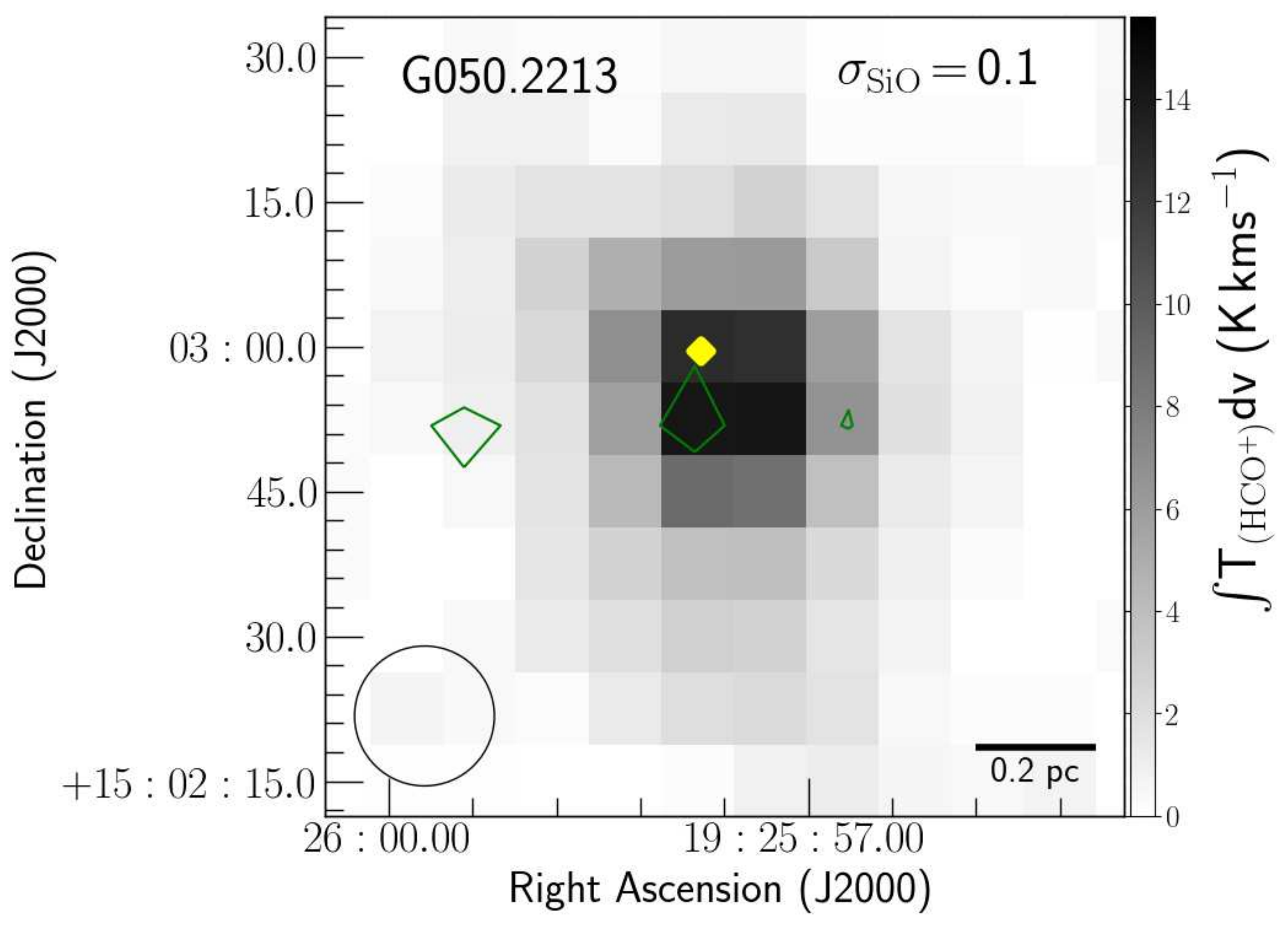}
\includegraphics[width=0.49\textwidth]{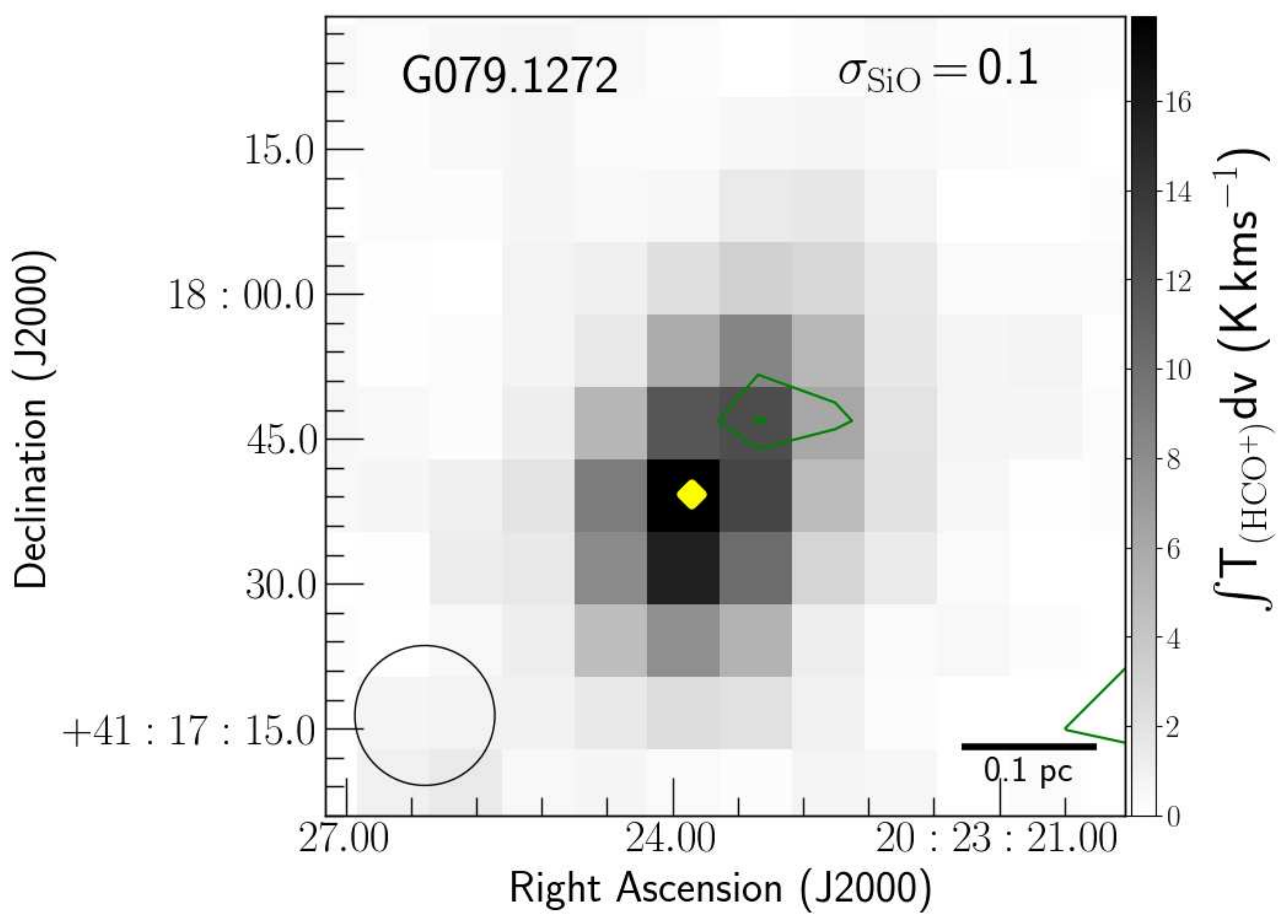}
\includegraphics[width=0.49\textwidth]{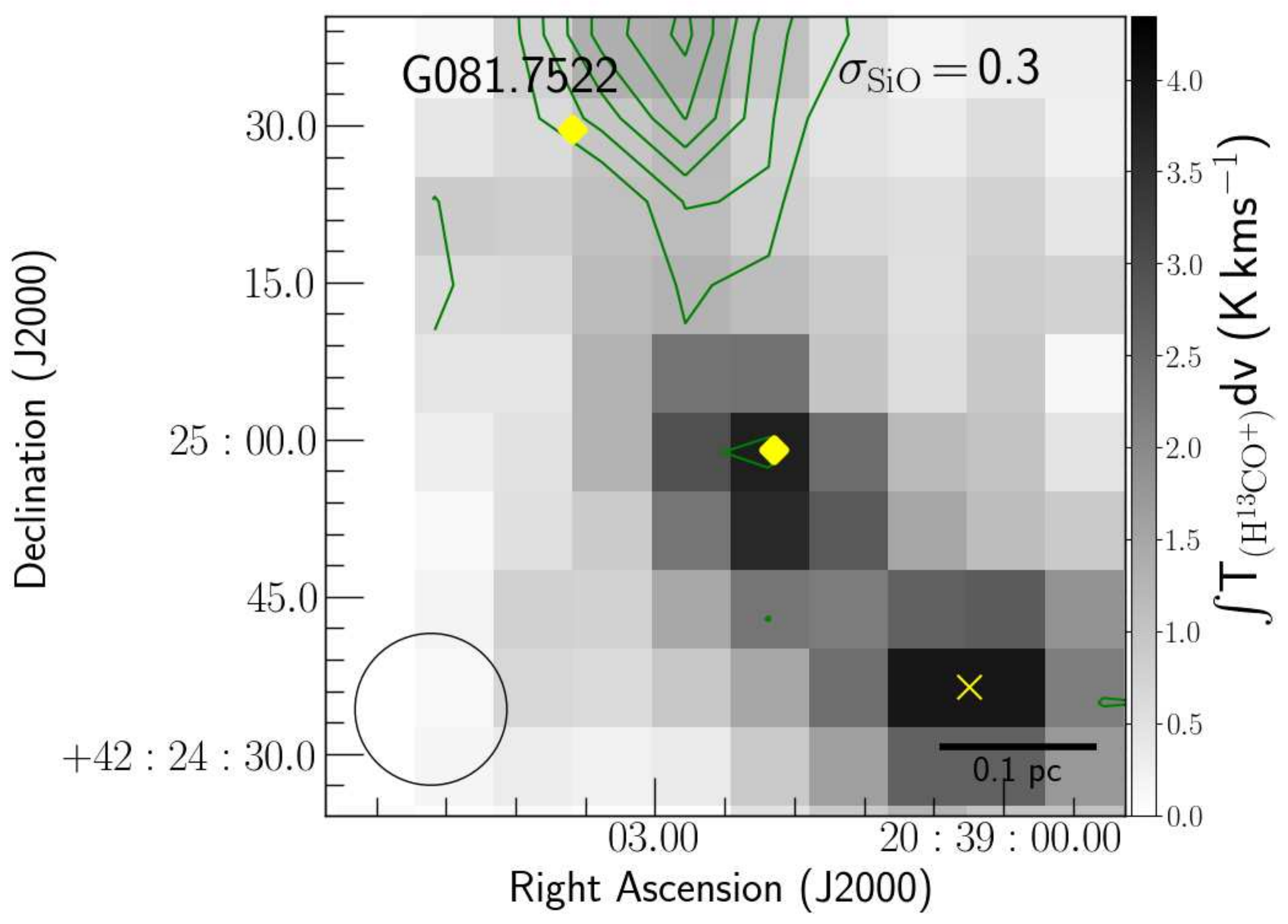}
\contcaption{--\,The velocity ranges used to integrate the emission are 33.0\,--\,43.6\,\kms for G013.6562, 30.5\,--\,37.3\,\kms for G018.3412, 57.5\,--\,64.2\,\kms for G043.3061, 37.0\,--\,42.0\,\kms for G050.2213, -1.5\,--\,5.2\,\kms for G079.1272 and -8.2\,--\,5.2\,\kms for G081.7522. It should be noted that for G081.7522, the velocity range used is the same as for G081.7624-OFFSET as the SiO emission towards G081.7522 is weaker it is difficult to estimate the SiO velocity range for this source.}
\end{figure*}

\begin{figure*}
\includegraphics[width=0.49\textwidth]{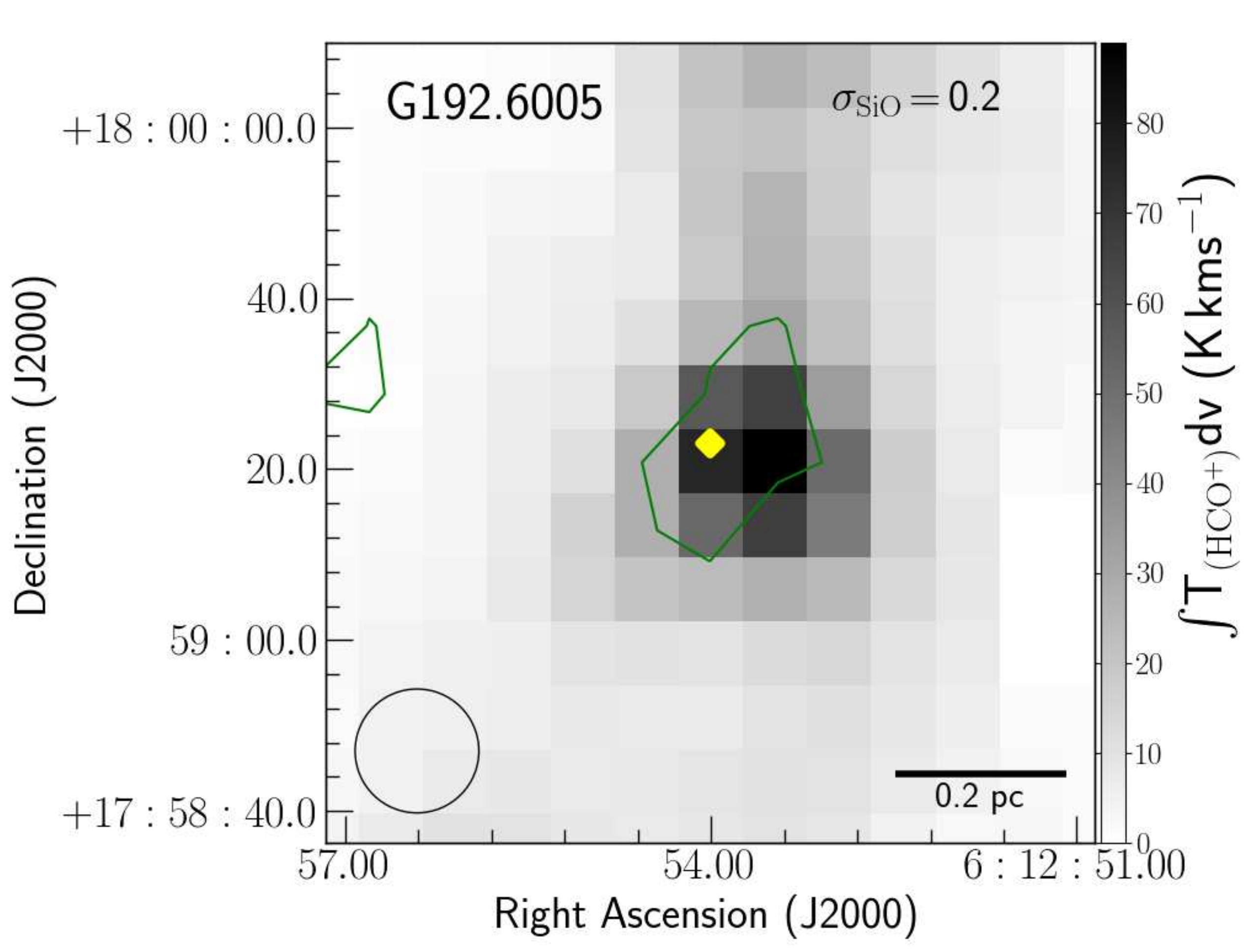}
\includegraphics[width=0.49\textwidth]{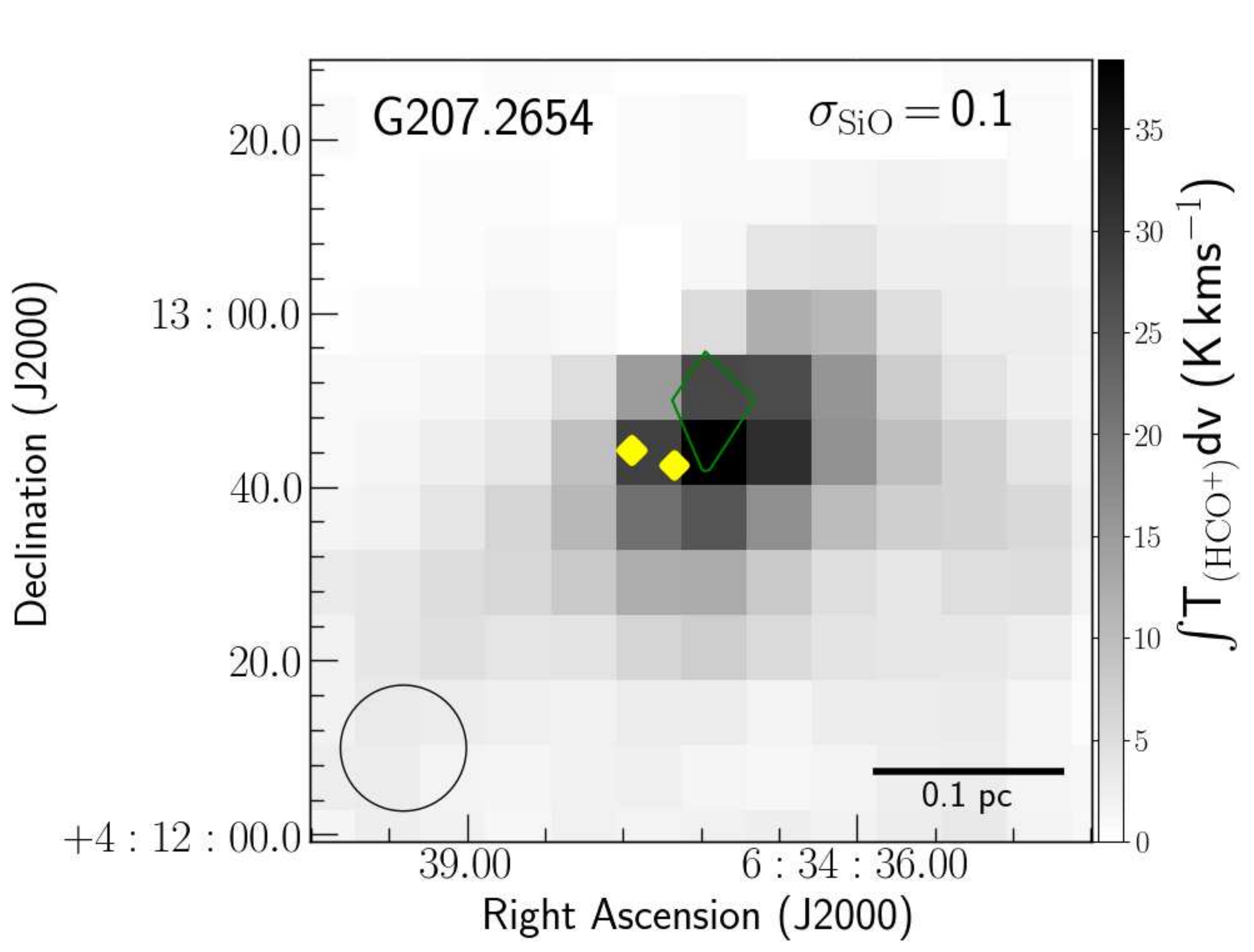}

\contcaption{--\,The velocity ranges used to integrate the emission are 6.8\,--\,11.9\,\kms for G192.6005, and 11.8\,--\,15.2\,\kms for G207.2654.}
\end{figure*}

%%%%%%%%%%%%%%%%%%%%%%%%%%%%%%%%%%%%%%%%%%%%%%%%%%

% Don't change these lines
\bsp	% typesetting comment
\label{lastpage}
\end{document}